\documentclass[11pt]{article}
\usepackage[left=1in,top=1in,right=1in,bottom=1in,head=.1in,nofoot]{geometry}

\usepackage{setspace,url,bm,amsmath} %

\usepackage{titlesec} %
\titlelabel{\thetitle.\quad} %

\titleformat*{\section}{\bf\Large\center}

\usepackage{graphicx} %
\usepackage{bbm}
\usepackage{latexsym}
\usepackage{caption}
\usepackage[margin=20pt]{subcaption}
\usepackage{hyperref}
\usepackage{booktabs}
\usepackage{multirow}

\usepackage{enumitem}

\usepackage[table]{xcolor}

\newcommand{\GG}[1]{}

\usepackage{amsthm}
\usepackage{amssymb}
\usepackage{amsmath}
\usepackage{color}
\usepackage{mathrsfs}

\usepackage{comment}
\theoremstyle{definition}
\newtheorem{assumption}{Assumption}
\newtheorem*{theorem*}{Theorem}
\newtheorem{theorem}{Theorem}
\newtheorem*{rmk*}{Remark}

\newtheorem{proposition}{Proposition}
\newtheorem{lemma}{Lemma}

\newtheorem{remark}{Remark}
\newtheorem{corollary}{Corollary}
\newtheorem*{corollary*}{Corollary}

\def\rank{\text{r}}

\def\c{\complement}

\usepackage{natbib} %
\bibpunct{(}{)}{;}{a}{}{,} %

\usepackage{etoolbox} %
\apptocmd{\sloppy}{\hbadness 10000\relax}{}{} %

\usepackage{color}
\usepackage{listings}

\DeclareMathOperator*{\argmin}{arg\,min}
\DeclareMathOperator*{\argmax}{arg\,max}

\def\ind{\begin{picture}(9,8)
         \put(0,0){\line(1,0){9}}
         \put(3,0){\line(0,1){8}}
         \put(6,0){\line(0,1){8}}
         \end{picture}
        }

\def\rank{\textup{rank}}

\def\RS{\mathcal{S}}
\def\Pr{\mathbb{P}}

\def\Var{\textup{Var}}
\def\Cov{\textup{Cov}}

\def\converge{\stackrel{}{\rightarrow}}
\def\convergeas{\stackrel{\textup{a.s.}}{\longrightarrow}}
\def\converged{\stackrel{d}{\longrightarrow}}
\def\convergep{\stackrel{\Pr}{\longrightarrow}}

\def\I{\mathbbm{1}}
\def\E{\mathbb{E}}
\def\t{\texttt{t}}
\def\c{\texttt{c}}

\def\RKS{\textup{RKS}}
\def\KS{\textup{KS}}

\def\thre{b}
\def\bs{\boldsymbol}

\def\limsup{\overline{\lim}}
\def\liminf{\underline{\lim}}
\def\co{\textup{co}}

\def\limsup{\overline{\lim}}
\def\liminf{\underline{\lim}}

\RequirePackage[normalem]{ulem}

\usepackage{soul}

\usepackage{subcaption}
\usepackage[textsize=tiny, textwidth = 2cm, shadow]{todonotes}

\begin{document}

\onehalfspacing

\title{\bf 
Sharp Design-based Inference for Quantiles of Individual Treatment Effects
}
\author{
Jiangchuan Du and Xinran Li
\footnote{Jiangchuan Du, Department of Statistics, University of Chicago, Chicago, IL (e-mail: \href{mailto:jiangchuand@uchicago.edu}{jiangchuand@uchicago.edu}). 
Xinran Li, Department of Statistics, University of Chicago, Chicago, IL (e-mail: \href{mailto:xinranli@uchicago.edu}{xinranli@uchicago.edu}). 
}    
}
\date{}
\maketitle

\begin{abstract}
Understanding the distribution of individual treatment effects (ITEs) is a central problem in causal inference. In general, however, this distribution is only partially identifiable from observed data because it depends on the unidentifiable joint distribution of the two potential outcomes. In this paper, we develop design-based inference for the distribution of ITEs whose validity relies solely on the randomization of treatment assignment, treating the potential outcomes as arbitrary fixed constants. Building on recent randomization tests for ITE quantiles, we propose tests based on either a rank-based Kolmogorov--Smirnov (KS) statistic or the usual KS statistic, with the latter being more conservative in the presence of ties. We show that, under certain regularity conditions, the confidence bounds obtained by test inversion are asymptotically no worse than the sharp bounds on ITE quantiles given the marginal distributions of the potential outcomes. We further derive a closed-form expression for the confidence bounds from the KS-based randomization test and show that they can be equivalently constructed by combining sharp bounds on ITE quantiles with uncertainty of empirical marginal distributions. This equivalence connects randomization-based inference with partial identification and clarifies the special role of the KS statistic in achieving sharp inference. Under additional distributional assumptions, we characterize the power of the proposed randomization tests under local alternatives. 
Finally, we discuss how to incorporate covariates to further sharpen inference. 
We illustrate the properties of the proposed methods using both simulations and a real-data application. 
\end{abstract}


\section{Introduction}
A central goal of causal inference is to understand how a treatment affects each unit in a population of interest. This, however, is challenging because at most one potential outcome can be observed for each unit.
The average treatment effect is the primary estimand in most causal analyses, but it can mask substantial heterogeneity in individual effects.
In many applications, investigators care about the distribution of treatment effects across units, rather than the average alone.
For example, 
a positive average treatment effect may suggest that the treatment benefits the population as a whole, even when the corresponding individual treatment effects are negative for a majority of the population.
In addition, a small proportion of units experiencing severe negative effects, such as adverse reactions to a vaccine, may still be crucial for decision-making.
\citet{heckman_making_1997} first formally proposed studying the distribution and quantiles of individual treatment effects (ITEs), emphasizing their importance for evaluating social programs, especially when program benefits are nontransferable or constraints on feasible redistribution prevent the achievement of distributional objectives.
The distribution and quantiles of ITEs provide comprehensive summaries of potentially heterogeneous individual effects, supplement the usual average treatment effect, and are less sensitive to extreme individual effects.

Inference for the distribution and quantiles of ITEs is more challenging than inference for average effects. The distribution and quantiles of ITEs depend on the joint distribution of the two potential outcomes, whereas observed data reveal only their marginal distributions.
Consequently, these causal estimands are generally only partially identifiable, even in randomized experiments; see, for example, \citet{fan2010sharp}, \citet{Lu2018}, \citet{zhang2024bounds}, \citet{WP2026} and \citet{he2026causal}.
Much of the existing literature on distributions and quantiles of ITEs adopts the usual superpopulation framework with independent and identically distributed (i.i.d.) units, and some approaches also require large-sample approximations under certain regularity conditions; see, e.g., 
\citet{fan2010sharp, fan2012confidence, Lee2024, fava2024predicting, ruiz2022high}.

For randomized experiments, \citet{fisher1935design} and \citet{Neyman:1923} proposed design-based inference, also called randomization-based or finite population inference, for analyzing treatment effects.
This framework relies solely on randomness in the treatment assignment and views all potential outcomes as fixed attributes of the experimental units.
Once the experimental design is known, the resulting inference is nearly assumption-free, making it especially appealing to researchers seeking robust causal conclusions under minimal assumptions.
Design-based inference has seen increasing adoption and substantial development in recent decades across statistics, economics, political science, clinical research, and related fields \citep{imbens2015causal, Greenbook, rosenberger2015randomization}.

Recent work has expanded the scope of design-based inference to distributions and quantiles of ITEs. \citet{caughey2023randomisation} proposed randomization tests for null hypotheses on ITE quantiles under completely randomized experiments (CREs), and \citet{su2024treatment} extended this approach to stratified randomized experiments and matched observational studies.
\citet{CL24quantile} improved these procedures by reinterpreting them as inference for ITEs among treated or control units, which can then be combined to infer ITEs among all units.
These methods provide finite-sample valid simultaneous confidence bands for the entire quantile function, or equivalently the distribution function, of the ITEs.
An important question remains: Are these confidence bands sharp?
If not, they may be overly conservative and may not fully use the information in the observed data.
Our paper answers this question affirmatively for CREs, with an appropriate choice of test statistic.

Specifically, we propose sharp design-based inference for the entire quantile function of ITEs in completely randomized experiments. By sharpness, we mean that, as the sample size tends to infinity, the finite-sample valid simultaneous lower (or upper) confidence band 
is asymptotically no less (or no greater) than the sharp, or best possible, lower (or upper) bound on the true ITE quantile function given only the marginal distributions of the potential outcomes.
We achieve sharp inference by using the Kolmogorov--Smirnov (KS) statistic after a rank transformation with random tie breaking.
In particular, we first develop randomization tests for ITE quantiles using either a rank-based KS statistic or the usual KS statistic. By test inversion, these tests yield a finite-sample valid simultaneous confidence band for the ITE quantile function. We then establish the sharpness of the confidence band under finite population asymptotics and certain regularity conditions \citep{fcltxlpd2016}.
We further derive explicit forms of the prediction and confidence bounds obtained from the randomization test based on the usual KS statistic. These forms reveal an equivalent direct construction that combines the sharp bounds on ITE quantiles with uncertainty quantification for the marginal distributions of the potential outcomes. This equivalence connects randomization-based inference with partial identification and clarifies the special role of the KS statistic in achieving sharp inference.
We next study the power of the proposed randomization tests under local alternatives and additional distributional assumptions. This analysis characterizes the convergence rate of confidence bounds to the sharp bounds on the true ITE quantiles.
Finally, we discuss how to incorporate covariates into design-based inference and improve precision by attaining covariate-adjusted sharp bounds on ITE quantiles.

The remainder of the paper is organized as follows. Section \ref{sec:dist_quantile_ite} introduces the framework and notation, and Section \ref{sec:review} reviews existing randomization-based procedures for inference on ITE quantiles. Section \ref{sec:rks} develops the proposed rank-based KS tests and the associated prediction and confidence intervals. 
Section \ref{sec:rks_consistency_fin} establishes the sharpness and consistency of the proposed randomization tests. Section \ref{sec:direct_connection} simplifies the bounds obtained by inverting the KS-based randomization test, presents a direct construction, and explains the equivalence between the two approaches. Section \ref{sec:rks_power} studies local power. 
Section \ref{sec:cov_adj} develops covariate adjustment.
Section \ref{sec:illust} presents simulation studies and a real-data analysis. 
Section \ref{sec:con} concludes with a discussion.
The supplementary material contains simulation studies and the proofs of all the theorems, propositions and corollaries. 
An R package implementing the proposed methods is publicly available at 
\url{https://github.com/jiangchuanduz-del/sharpQITE}.

\section{Framework and notation}\label{sec:dist_quantile_ite}

Consider an experiment with $n$ units indexed by $i=1,2,\ldots,n$ and a binary treatment $z\in\{1,0\}$, where $z=1$ and $z=0$ denote the treatment and control assignments, respectively. 
We invoke the potential outcome framework \citep{Neyman:1923, Rubin:1974} and impose the stable unit treatment value assumption \citep{Rubin:1980}. 
Let $Y_i(1)$ and $Y_i(0)$ denote, respectively, the potential outcomes of unit $i$ under treatment and control. 
The ITE for unit $i$ is $\tau_i = Y_i(1) - Y_i(0)$.
Let $Z_i$ denote the treatment assignment for unit $i$, where $Z_i = 1$ if the unit receives treatment and $0$ otherwise. 
The observed outcome for unit $i$ is then $Y_i = Z_iY_i(1) + (1 - Z_i)Y_i(0)$, one of the potential outcomes determined by the treatment assignment.
Let $\bs{Y}(1) = (Y_1(1), \ldots, Y_n(1))^\top$, $\bs{Y}(0) = (Y_1(0), \ldots, Y_n(0))^\top$, $\bs{\tau} = (\tau_1, \ldots, \tau_n)^\top$,  $\bs{Z} = (Z_1, \ldots, Z_n)^\top$ and $\bs{Y} = (Y_1, \ldots, Y_n)^\top$ denote the vectors of potential outcomes, ITEs, treatment assignments and observed outcomes for all $n$ units.

We focus on design-based inference for treatment effects \citep{fisher1935design, Neyman:1923}. All potential outcomes $\bs{Y}(1)$ and $\bs{Y}(0)$, as well as the ITEs $\bs{\tau}$, are viewed as fixed constants, or equivalently are conditioned on.
The randomness in the observed data $(\bs{Z}, \bs{Y})$ comes solely from the random assignment $\bs{Z}$; note that $\bs{Y}$ is random only through its dependence on $\bs{Z}$. 
Throughout the paper, we focus on the CRE with fixed positive integers $n_1$ and $n_0 = n - n_1$, under which $n_1$ units are randomly assigned to treatment and the remaining $n_0$ units are assigned to control. 
Under the CRE, the probability that the treatment assignment vector $\bs{Z}$ takes value $\bs{z} = (z_1, z_2, \ldots, z_n)^\top \in \{0,1\}^n$ is $n_1 ! n_0 ! /n!$ if $\sum_{i=1}^n z_i =n_1$, and zero otherwise.

We are interested in inference for the distribution and quantiles of ITEs under the CRE.
Let $F_{n\tau}(c) = n^{-1} \sum_{i=1}^n \I(\tau_i \le c)$ for $c\in \mathbb{R}$ denote the distribution function of ITEs among the $n$ experimental units, 
and $F_{n\tau}^{-1}(\beta)$ for $\beta \in (0,1)$ be the corresponding quantile function. 
Let $\tau_{(1)} \le \tau_{(2)} \le \ldots \le \tau_{(n)}$ denote the sorted ITEs for the $n$ units. 
For any $\beta\in (0,1)$, we have $F_{n\tau}^{-1}(\beta) = \tau_{(k)}$ with $k = \lceil n\beta \rceil$. 

We further introduce the distribution and quantile functions of ITEs among treated or control units only.
These quantities are useful intermediate targets for inference on the distribution of ITEs among all units and may also be estimands of direct interest, for example, when the effects received by the treated units are of primary concern.
Let $F^\t_{n\tau}(c) = n_1^{-1} \sum_{i=1}^n Z_i \I(\tau_i \le c)$ be the distribution of ITEs among treated units, $F_{n\tau}^{\t,-1}(\beta)$ be the corresponding quantile function, 
and $\tau_{\t(1)} \le  \ldots \le \tau_{\t(n_1)}$ be the sorted ITEs among treated units. 
Define analogously $F^\c_{n\tau}(c)$, $F_{n\tau}^{\c,-1}(\beta)$ and $\tau_{\c(1)} \le  \ldots \le \tau_{\c(n_0)}$ for ITEs among control units.

We consider the following null hypothesis on ITE quantiles: 
\begin{align}\label{eq:H_beta_c_T}
    H^\t_{\beta, c}: 
    F_{n\tau}^{\t,-1}(\beta) \le c 
    \Longleftrightarrow 
    1 - F_{n\tau}^{\t}(c) \le  1 - \beta, 
    \quad (0 < \beta < 1)
\end{align}
which is equivalent to the following one on the sorted ITEs with $k = \lceil n_1 \beta \rceil$:
\begin{align}\label{eq:quantile_null}
    H^\t_{k, c}: \tau_{\t(k)} \leq c 
    \Longleftrightarrow 
    \sum_{i=1}^{n} Z_i\I(\tau_i > c) \leq n_1 - k, 
    \quad (1 \le k \le n_1). 
\end{align}
The null hypotheses in \eqref{eq:H_beta_c_T} and \eqref{eq:quantile_null} differ from traditional null hypotheses because they depend on the observed treatment assignments and are therefore random.
For a random null hypothesis $H$, a p-value $p$ is said to be valid for testing $H$
if and only if $\Pr(p\le \alpha \textup{ and } H  \textup{ holds}) \le \alpha$ for $\alpha \in (0,1)$ \citep{jin2023selection, CL24quantile}.
Importantly, with such a p-value, 
we can still conduct standard test inversion to construct valid prediction sets, as shown later.

We can also test null hypotheses $F_{n\tau}^{\t,-1}(\beta) \ge c$ or $\tau_{\t(k)} \geq c$ by flipping the outcome signs, and can test treatment effects among control units by switching the treatment labels.
Moreover, as discussed later, we can combine the inference for treatment effects among treated and control units to infer treatment effects among all units. 
Therefore, throughout the paper, we focus mainly on the quantiles of ITEs for the treated group, as in \eqref{eq:H_beta_c_T} and \eqref{eq:quantile_null}.

\section{Review of existing approaches}\label{sec:review}

Under the CRE, \citet{caughey2023randomisation} first proposed valid randomization tests for ITE quantiles among all experimental units,
and \citet{CL24quantile} extended the approach to ITE quantiles among treated, control or all units, yielding substantial power gains.
These approaches build on a worst-case extension of the classical Fisher randomization test, which was originally proposed for sharp null hypotheses, and utilize distribution-free rank statistics. Such statistics both facilitate the calculation of worst-case p-values and provide robustness to extreme ITEs.
Below we briefly review the testing procedure for the null hypothesis in \eqref{eq:quantile_null}.

We first introduce the class of distribution-free statistics $t(\bs{z}, \bs{y})$ under the CRE,  with inputs $\bs{z}\in \{0,1\}^n$ and $\bs{y} \in \mathbb{R}^n$, such that $t(\bs{Z}, \bs{y})$ follows the same distribution as $t(\bs{Z}, \bs{y}')$ for any constant vectors $\bs{y}, \bs{y}^\prime \in \mathbb{R}^n$. 
For example, \citet{caughey2023randomisation} and \citet{CL24quantile} considered the following class of rank sum statistics: 
\begin{align}\label{eq:rank_sum}
    t(\bs{z}, \bs{y}) = \sum_{i=1}^n z_i \phi( \rank_i (\bs{y})), 
\end{align}
where $\phi(\cdot)$ is a rank transformation and $\rank_i (\bs{y})$ denotes the rank of the $i$th coordinate of $\bs{y}$. 
To ensure the distribution-free property of \eqref{eq:rank_sum},
we break ties in ranking using the ``random'' method,
or equivalently the ``first'' method after randomly shuffling the ordering of the units before analysis.
That is, $\rank_i(\bs{y}) < \rank_j(\bs{y})$ if and only if $y_i < y_j$, or $y_i = y_j$ and $i < j$.

We next consider the null hypothesis $H^\t_{k, c}$ in \eqref{eq:quantile_null} for any given $1 \le k\le n_1$ and $c\in \mathbb{R}$.
Let $\mathcal{H}^\t_{k,c}$ denote the set of all possible ITE vectors in $\mathbb{R}^n$ that satisfy $H_{k,c}^\t$.  
That is,  $\mathcal{H}^\t_{k,c} = \{\bs{\delta} \in \mathbb{R}^n: \delta_{\t(k)} \le c\} = \{ \bs{\delta} \in \mathbb{R}^n: \sum_{i=1}^n Z_i\I(\delta_i > c) \le n_1 - k\}$, 
where $\bs{\delta} = (\delta_1, \delta_2, \ldots, \delta_n)^\top$ and $\delta_{\t(k)}$ denotes the coordinate of $\bs{\delta}$ at rank $k$ among $\{i:Z_i = 1\}$. 
Extending \citet{caughey2023randomisation}, 
\citet{CL24quantile} proposed the following valid $p$-value for testing $H_{k,c}^\t$ under the CRE, using a distribution-free test statistic $t(\cdot, \cdot)$: 
\begin{align}\label{eq:rand_p_val}
    p_{k,c}^\t = G_n \Big( \inf_{\bs \delta \in \mathcal{H}_{k,c}^\t} t(\bs Z, \bs Y - \bs{Z} \circ \bs{\delta}) \Big) 
    \textup{ with } 
    G_n(c) \equiv \Pr\{ t(\bs{Z}, \bs{y}) \ge c \} \textup{ for } c \in \mathbb{R},  \bs{y}\in \mathbb{R}^n,
\end{align}
where $\circ$ denotes element-wise product. 
The definition of $G_n(\cdot)$ is the same for any constant $\bs{y}\in \mathbb{R}^n$ because $t(\cdot, \cdot)$ is distribution-free under the CRE.
As shown in \citet{CL24quantile},
the p-value in \eqref{eq:rand_p_val} is the supremum of the Fisher randomization p-value over all sharp null hypotheses satisfying $H_{k,c}^\t$ in \eqref{eq:quantile_null}, where $\bs{\delta}$ in \eqref{eq:rand_p_val} represents a hypothesized vector of ITEs.

Furthermore,
\citet{caughey2023randomisation} and \citet{CL24quantile} showed that, when using the rank sum statistic in \eqref{eq:rank_sum} with a monotone nondecreasing rank transformation $\phi(\cdot)$, the infimum of the test statistic in \eqref{eq:rand_p_val} has a closed-form solution, so that the resulting p-value $p_{k,c}^\t$ can be efficiently computed. 
Specifically, for any $1 \le k \le n_1$ and $c\in \mathbb{R}$, the infimum of $t(\bs Z, \bs{Y} - \bs{Z} \circ \bs{\delta})$ over $\bs \delta \in \mathcal{H}_{k,c}^\t$ is achieved when the treated units with the largest $n_1-k$ observed outcomes (where we break ties in the same way as for the rank statistic) have infinite ITEs, while the remaining treated units have ITEs equal to $c$. 
The value of the statistic $t(\bs Z, \bs Y - \bs{Z} \circ \bs{\delta})$ does not depend on the hypothesized ITEs $\delta_i$ for control units, which may therefore be set arbitrarily.
That is,
$\inf_{\bs \delta \in \mathcal{H}_{k,c}^\t} t(\bs Z, \bs Y - \bs{Z} \circ \bs{\delta}) = t(\bs Z, \bs Y - \bs{Z} \circ \bs{\xi}_{k,c}^\t)$, 
where $\bs{\xi}_{k,c}^\t = (\xi_{k,c, 1}^\t, \xi_{k,c, 2}^\t, \ldots, \xi_{k,c, n}^\t)^\top$ and 
\begin{equation}\label{eq:xi_kc}
    \xi_{k,c, i}^\t
    = 
    \begin{cases}
        +\infty, & \textup{if } Z_i = 1 \textup{ and }\rank_i^\t(\bs{Y}) > k, 
        \\
        c, & \textup{otherwise},
    \end{cases}
    \qquad (1\le i \le n)
\end{equation}
where $\rank^\t_i(\bs{Y} )$
denotes the rank of $Y_i$ within the treated group $\{Y_i: Z_i = 1, 1 \leq i \leq n\}$, with ties broken in the same way as in ranking among all units. 

By inverting the randomization p-value in \eqref{eq:rand_p_val}, \citet{CL24quantile} constructed lower prediction bounds for the sorted ITEs in the treated group. Importantly, these prediction bounds are simultaneously valid across all quantiles.
Analogously,
by switching the treatment labels and changing the outcome signs,
we can construct simultaneous lower prediction bounds for all sorted ITEs in the control group.
Finally, we can combine the prediction bounds from the treated and control groups to obtain simultaneous confidence bounds for the ITE quantiles among all units, using a Bonferroni correction with a factor of $2$; see also Section \ref{sec:ci_effect}.

\section{Randomization test with the Kolmogorov-Smirnov statistic for quantiles of individual treatment effects}\label{sec:rks}
Although the confidence bounds of \citet{CL24quantile} are valid, it is unclear whether they can be improved nontrivially. In other words, as the sample size $n$ tends to infinity, do the confidence bounds converge in some sense to the ``best'' possible bounds?
The testing procedure in Section \ref{sec:review} considers a worst-case scenario in which some units are hypothesized to have infinitely large ITEs, which may appear conservative.
As we demonstrate, however, the test in \eqref{eq:rand_p_val} can be sharp in a suitable sense when the test statistic is chosen appropriately---in particular, when it is the KS statistic.

In this section, we develop finite-sample valid design-based inference for ITE quantiles using the KS statistic.
The traditional KS test is one of the most popular nonparametric tests for two-sample comparisons
and is consistent against all alternatives when the samples in the two groups are independent draws from different distributions.
This property motivates our use of the KS statistic for design-based inference.
We emphasize, however, that the classical KS test operates under a superpopulation framework with i.i.d. samples and tests the equality of distributions. By contrast, we conduct design-based inference with fixed potential outcomes and random treatment assignments, and focus on the distribution of ITEs.

\subsection{Kolmogorov-Smirnov statistic and its rank-based version}\label{sec:rks_properties}

For any possible treatment assignment vector $\bs{z} \in \{0,1\}^n$ under the CRE and any outcome vector $\bs{y} \in \mathbb{R}^n$, 
the one-sided KS statistic has the following form: 
\begin{align}\label{eq:KS_statistic}
    t_{\KS}(\bs{z},\bs{y}) 
    = 
    \sup_{\thre \in \mathbb{R}} 
    \Big\{ 
    \frac{1}{n_0} \sum_{i:z_i=0} \I(y_i \le \thre) 
    - 
    \frac{1}{n_1} \sum_{i:z_i=1} \I(y_i \le \thre) 
    \Big\},
\end{align}
which is the supremum of the difference between the empirical distribution functions of the outcomes in the control and treated groups.
The KS statistic in \eqref{eq:KS_statistic} depends only on the relative ordering between treated outcomes $\{y_i:z_i=1\}$ and control outcomes $\{y_i:z_i=0\}$, and is therefore essentially rank-based.
In the presence of ties among the outcomes, however, the KS statistic in \eqref{eq:KS_statistic} is generally not distribution-free under the CRE and thus cannot be used directly in the randomization test of Section \ref{sec:review}.
To ensure the distribution-free property,
we introduce the following rank-based KS (RKS) statistic:
\begin{align}\label{eq:RKS_statistic}
    t_{\RKS}(\bs{z},\bs{y}) = t_{\KS}(\bs{z},\rank(\bs{y})), 
\end{align}
which is the original KS statistic applied to the ranks of the outcomes rather than their values.
In \eqref{eq:RKS_statistic}, $\rank(\bs{y}) = (\rank_1(\bs{y}), \rank_2(\bs{y}), \ldots, \rank_n(\bs{y}))^\top$, recalling that $\rank_i(\bs{y})$ denotes the rank of the $i$th coordinate of $\bs{y}$. 
As in Section \ref{sec:review}, we break ties using the unit indices after randomly shuffling the ordering of the units before analysis.

The following proposition summarizes key properties of the RKS statistic that will be used in our later inference on quantiles of ITEs.

\begin{proposition}\label{prop:rks_properties}
Consider the KS statistic in  \eqref{eq:KS_statistic} and its ranked version in \eqref{eq:RKS_statistic}. 
\begin{enumerate}[label=(\roman*), topsep=1ex,itemsep=-0.3ex,partopsep=1ex,parsep=1ex]

    \item $t_{\RKS}(\bs{Z},\bs{y})$ is distribution-free when $\bs{Z}$ is from a CRE. That is, for any constant vectors $\bs{y}, \bs{y}^\prime\in \mathbb{R}^n$, $t_{\RKS}(\bs{Z},\bs{y})$ and $t_{\RKS}(\bs{Z},\bs{y}^\prime)$ follow the same distribution. 

   \item 
   Both $t_{\KS}(\bs{z}, \bs{y})$ and $t_{\RKS}(\bs{z}, \bs{y})$ are effect-increasing. That is,    
   their values weakly increase when treated outcomes weakly increase and control outcomes weakly decrease.

    \item 
    $t_{\RKS}(\bs{z},\bs{y}) \ge t_{\KS}(\bs{z},\bs{y})$ for all $\bs{y}\in\mathbb{R}^n$, with equality if the coordinates of $\bs{y}$ are distinct.

    \item When $\bs{Z}$ is from a CRE, and the coordinates of $\bs{W}$ are i.i.d.\ from some continuous distribution, 
    then $t_{\RKS}(\bs{Z},\bs{y})$ and $t_{\KS}(\bs{a},\bs{W})$ follow the same distribution for any constant $n$-dimensional vector $\bs{y}$ and $\bs{a} \in \{0,1\}^n$ such that $\sum_{i=1}^n a_i = \sum_{i=1}^n Z_i$.
\end{enumerate}
\end{proposition}

Proposition \ref{prop:rks_properties}(i) and (ii) show that the RKS statistic is both effect-increasing and distribution-free under the CRE, as are the rank sum statistics in \eqref{eq:rank_sum} with monotone nondecreasing rank transformations \citep{caughey2023randomisation}.
Part (iii) shows that the RKS statistic coincides with the usual KS statistic when there are no ties among the outcomes and is generally bounded from below by the KS statistic.
Part (iv) shows that the randomization distribution of the RKS statistic is the same as the distribution of the KS statistic under the superpopulation setting, where the treatment assignments are fixed and the units in the treated and control groups are i.i.d.~samples from a continuous distribution.
As a result, the classical asymptotic result of the null distribution of the KS statistic under i.i.d.~sampling can be used for the randomization distribution of the RKS statistic. This fact is useful in the later large-sample analysis.

\subsection{Randomization tests for ITE quantiles using RKS statistic}
We now study inference for ITE quantiles using the RKS statistic,
following the approaches in \citet{caughey2023randomisation} and \citet{CL24quantile}.
Specifically, we consider again the p-value in \eqref{eq:rand_p_val}, but choose the test statistic $t(\cdot, \cdot)$ to be the RKS statistic $t_{\RKS}(\cdot, \cdot)$ in \eqref{eq:RKS_statistic}. 
By the distribution-free property of $t_{\RKS}(\cdot, \cdot)$ in Proposition \ref{prop:rks_properties}(i),
this p-value is well-defined and valid for testing the null hypothesis $H_{k,c}^\t$ in \eqref{eq:quantile_null}.
Moreover, the corresponding infimum of the test statistic $t_{\RKS}(\bs{Z}, \bs{Y} - \bs{Z} \circ \bs{\delta})$ over $\bs{\delta} \in \mathcal{H}_{k,c}^\t$ is also achieved at $\bs{\delta} = \bs{\xi}_{k,c}^\t$ defined in \eqref{eq:xi_kc}. 
This result is intuitive because the RKS statistic can be viewed as the supremum of a collection of rank sum statistics with rank transformation $-\I\{r \le b\}$
for a given rank $r$ and $b\in \mathbb{R}$ (up to certain shifting and scaling), each of which is minimized at $\bs{\delta} = \bs{\xi}_{k,c}^\t$ \citep{kim2026randomization}. 
We summarize the results in the following theorem.

\begin{theorem}\label{thm:pval_RKS}
    For any $1\le k \le n_1$ and $c\in \mathbb{R}$, the p-value $p_{k,c}^\t$ in \eqref{eq:rand_p_val} with test statistic $t$ being the RKS statistic  $t_{\RKS}$ in \eqref{eq:RKS_statistic} 
    has the following equivalent forms:
    \begin{align}\label{eq:rand_p_val_rks}
    p_{k,c}^\t = G_n \Big( \inf_{\bs \delta \in \mathcal{H}_{k,c}^\t} t_{\RKS}(\bs Z, \bs Y - \bs{Z} \circ \bs{\delta}) \Big) 
    = G_n \Big( t_{\RKS}(\bs Z, \bs Y - \bs{Z} \circ \bs{\xi}_{k,c}^\t) \Big), 
\end{align}
    where $G_n(c) \equiv \Pr\{ t_{\RKS}(\bs{Z}, \bs{y}) \ge c \}$ for any $c \in \mathbb{R}$ and $\bs{y}\in \mathbb{R}^n$, 
    and $\bs{\xi}_{k,c}^\t$ is defined as in \eqref{eq:xi_kc}. 
    Moreover, this p-value in \eqref{eq:rand_p_val_rks} is valid for testing the null hypothesis $H_{k,c}^\t$ in \eqref{eq:quantile_null}. 
\end{theorem}

Proposition \ref{prop:rks_properties}(iii) immediately yields the following corollary for a valid p-value constructed directly from the usual KS statistic.
Define 
\begin{align}\label{eq:rand_p_val_ks}
    \overline{p}_{k,c}^\t & \equiv G_n \Big( \inf_{\bs \delta \in \mathcal{H}_{k,c}^\t} t_{\KS}(\bs Z, \bs Y - \bs{Z} \circ \bs{\delta}) \Big) 
\end{align}
where $G_n(\cdot)$ is
defined in Theorem \ref{thm:pval_RKS}.
Let 
$\hat F_{n}^\t(\cdot)$ and $\hat F_{n}^\c(\cdot)$ denote the empirical distribution functions of observed outcomes in treated and control groups:
\begin{align}
\label{eq:empir_distrib}
    \hat F_{n}^\t(\thre) & = 
    \frac{1}{n_1} \sum_{i=1}^n Z_i \I(Y_i \le \thre), 
    \qquad 
    \hat F_{n}^\c(\thre)
    = \frac{1}{n_0} \sum_{i=1}^n (1-Z_i) \I(Y_i \le \thre), 
\end{align}
and let $\hat{F}_{n, k/n_1}^\t(\cdot)$ denote a ``shifted'' distribution function of the observed outcomes in the treated group, in which the largest $n_1 - k$ treated outcomes are shifted to $-\infty$:
\begin{align}\label{eq:shift_CDF}
    \hat{F}_{n, k/n_1}^\t(\thre) \equiv \min \{ 1, 1 - k/n_1 + \hat F_{n}^\t(\thre) \}.
\end{align}

\begin{corollary}\label{cor:pval_KS}
    For any $1\le k \le n_1$ and $c\in \mathbb{R}$, 
    the p-value $\overline{p}_{k,c}^\t$ in \eqref{eq:rand_p_val_ks} is bounded from below by $p_{k,c}^\t$ in \eqref{eq:rand_p_val_rks}, i.e., 
    $\overline{p}_{k,c}^\t \ge p_{k,c}^\t$,
    and it is thus also valid for testing the null hypothesis $H_{k,c}^\t$ in \eqref{eq:quantile_null}.
    Moreover, $\overline{p}_{k,c}^\t$ in \eqref{eq:rand_p_val_ks} has the following equivalent forms:
    \begin{align}\label{eq:rand_p_val_ks_equi}
    \overline{p}_{k,c}^\t & 
    = G_n \Big( t_{\KS}(\bs Z, \bs Y - \bs{Z} \circ \bs{\xi}_{k,c}^\t) \Big) 
    = 
    G_n \Big( \sup_b \{\hat F_{n}^\c(b) - \hat F_{n, k/n_1}^\t(b + c) \} \Big),
    \end{align}
    where $t_{\KS}(\cdot, \cdot)$ is defined in \eqref{eq:KS_statistic}, $G_n(\cdot)$ is defined the same as in  Theorem \ref{thm:pval_RKS}, and $\hat F_{n}^\c(\cdot)$ and $\hat F_{n, k/n_1}^\t(\cdot)$ are defined in \eqref{eq:empir_distrib} and \eqref{eq:shift_CDF}. 
\end{corollary}

In Corollary \ref{cor:pval_KS}, $G_n(\cdot)$ denotes the randomization distribution of the RKS statistic rather than the KS statistic, where the latter may not be distribution-free due to potential ties.
The p-value $\overline{p}_{k,c}^\t$ in \eqref{eq:rand_p_val_ks_equi} is generally more conservative than $p_{k,c}^\t$ in \eqref{eq:rand_p_val_rks}.
The two p-values coincide, or approximately coincide, when there are no ties, or few ties, among the coordinates of $\bs Y - \bs{Z} \circ \bs{\xi}_{k,c}^\t$, aside from those at $-\infty$.
In particular, $\overline{p}_{k,c}^\t = p_{k,c}^\t$ when there are no ties among the imputed outcomes except for ties at $-\infty$.

For convenience, we call the randomization test in Theorem \ref{thm:pval_RKS} the RKS test for ITE quantiles and the test in Corollary \ref{cor:pval_KS} the KS test for ITE quantiles.

\begin{remark}
\label{rmk:KS_equiv}
    When $k=n_1$ and there are no ties in the shifted observed outcomes, the KS test from Corollary \ref{cor:pval_KS} shares the same procedure as the one-sided KS test under the usual superpopulation inference, where the control units are i.i.d.~samples from a certain distribution and the treated units are i.i.d.~samples from the same distribution but with a location shift of size $c$; see, e.g., \citet[][p. 239]{gibbons2014nonparametric}.
    Despite this similarity, the justifications behind them are quite different: our KS test is driven by randomization whereas the usual KS test is driven by exchangeability in the outcomes. 
    Moreover, our KS test is valid for testing the null hypothesis that the
    treatment potential outcomes are uniformly bounded above by the control
    potential outcomes. Relatedly, \citet{bugni2025testing} use the KS test
    to test for stochastic dominance between two distributions.
\end{remark}

\subsection{Prediction and confidence intervals for ITE quantiles}
\label{sec:ci_effect}

Using the p-values from Theorem \ref{thm:pval_RKS} and Corollary \ref{cor:pval_KS},
we can invert the tests to construct simultaneously valid prediction intervals for ITE quantiles among treated or control units. These intervals can then be combined to obtain confidence intervals for ITE quantiles among all units.
The results follow immediately from \citet[][Section 3.2]{CL24quantile}.
We summarize them below for completeness
and refer readers to \citet{CL24quantile} for details.

\begin{corollary}\label{cor:pred_interval_pool}
    Consider a CRE with $n_1$ treated and $n_0$ control, and any $0 < \alpha < 1$. 
    \begin{enumerate}[label=(\roman*), topsep=1ex,itemsep=-0.3ex,partopsep=1ex,parsep=1ex]
    \item For $1\le k \le n_1$, 
    $\mathcal{I}^\t_{k, \alpha} \equiv \{ c\in \mathbb{R}: p_{k,c}^\t > \alpha \}$ is a one-sided interval of the form $[\underline{c}, \infty)$ or $(\underline{c}, \infty)$. Moreover, they are simultaneous prediction intervals for ITE quantiles $\tau_{\t(k)}$s among treated units, i.e.,
    \begin{align*}
        \Pr\Big( \tau_{\t(k)} \in \mathcal{I}^\t_{k, \alpha} \textup{ for all } 1\le k \le n_1 \Big) \ge 1-\alpha.
    \end{align*}
    
    \item Construct intervals $\mathcal{I}^\c_{k, \alpha}$, $1\le k \le n_0$, in the same way as (i), after switching treatment labels and changing outcome signs. 
    Then $\mathcal{I}^\c_{k, \alpha}$s are simultaneous prediction intervals for treatment effect quantiles $\tau_{\c(k)}$s among control units. 

    \item Pool all the $n_1 + n_0=n$ prediction intervals $\mathcal{I}^\t_{k, \alpha}$s and $\mathcal{I}^\c_{k, \alpha}$s from (i) and (ii), 
    and sort and denote them by $\mathcal{I}_{1,\alpha} \supset \mathcal{I}_{2,\alpha} \supset \ldots \supset \mathcal{I}_{n,\alpha}$. 
    These intervals are then $1-2\alpha$ simultaneous confidence intervals for ITE quantiles $\tau_{(k)}$s among all units, i.e., 
    \begin{align*}
        \Pr\big( \tau_{(k)} \in \mathcal{I}_{k, \alpha} \textup{ for all } 1\le k \le n \big) \ge 1-2\alpha.
    \end{align*}
    \end{enumerate}
\end{corollary}

Corollary \ref{cor:pred_interval_pool} provides simultaneous lower prediction or confidence bounds for ITE quantiles among treated, control and all units.
As discussed at the end of Section \ref{sec:dist_quantile_ite}, we can also
construct upper prediction and confidence bounds by simply flipping the outcome signs.
We can further combine the lower and upper bounds to construct two-sided prediction or confidence intervals for ITE quantiles using the Bonferroni method.

\begin{remark}\label{rmk:comp_p_RKS_KS}
In addition to the result in Corollary \ref{cor:pval_KS}, we have $\overline{p}_{k,c-\eta}^\t \le p_{k,c}^\t \le \overline{p}_{k,c}^\t$ for any $1\le k\le n_1$, $c\in\mathbb{R}$, and $\eta>0$.
This implies that the prediction intervals from RKS test and KS test, $\mathcal{I}^\t_{k, \alpha} \equiv \{ c\in \mathbb{R}: p_{k,c}^\t > \alpha \}$ and $\overline{\mathcal{I}}^\t_{k, \alpha} \equiv \{ c\in \mathbb{R}: \overline{p}_{k,c}^\t > \alpha \}$,  have the same lower limit and can differ only in its inclusion, with $\mathcal{I}^\t_{k, \alpha} \subseteq \overline{\mathcal{I}}^\t_{k, \alpha}$.
\end{remark}

\section{Consistency of the design-based KS test for quantiles of individual treatment effects}
\label{sec:rks_consistency_fin}

We now study the large-sample properties of the randomization tests for ITE quantiles proposed in Section \ref{sec:rks}.
We first discuss the sharp bounds on quantiles of the ITEs given only the marginal distributions of the potential outcomes, and then demonstrate the sharpness of our test and the resulting prediction and confidence bounds.

\subsection{Sharp bounds on ITE quantiles}
Because the two potential outcomes cannot be observed simultaneously for any unit, their joint distribution (particularly their copula) is generally not identifiable from observed data.
We therefore consider sharp bounds on ITE quantiles based on the marginal distributions of the potential outcomes, which can be estimated unbiasedly and consistently from the observed data through their empirical analogues in \eqref{eq:empir_distrib}.
By sharpness, we mean that the bound for a given ITE quantile can be attained by some joint distribution of potential outcomes with the specified marginals.

Formally, let $F_{n1}(\cdot)$ and $F_{n0}(\cdot)$ denote the marginal distributions of the potential outcomes among all the experimental units:
\begin{align}\label{eq:finite_population_dist_PO}
    F_{n1}(\thre) = \frac{1}{n}\sum_{i=1}^n \I\{ Y_i(1) \leq \thre\}, \quad F_{n0}(\thre) = \frac{1}{n}\sum_{i=1}^n \I\{Y_i(0) \leq \thre\}, 
\end{align}
and $F_{n1}^{-1}(\cdot)$ and $F_{n0}^{-1}(\cdot)$ denote the corresponding quantile functions for the two potential outcomes with $F_{n1}^{-1}(0)=-\infty$. 
\citet{fan2010sharp} and \citet{zhang2024bounds} derived the sharp bounds for the ITE quantiles based on the marginal distributions of the two potential outcomes; see references therein about bounding joint distributions based on marginal distributions. 
For example, they showed that, for any $\beta\in (0,1)$, a pointwise sharp lower bound on the $\beta$ quantile of ITEs, $\tau_{(\lceil n \beta \rceil)}=F_{n\tau}^{-1}(\beta)$, with given marginal distributions $F_{n1}(\cdot)$ and $F_{n0}(\cdot)$ in \eqref{eq:finite_population_dist_PO} is
\begin{align}\label{eq:sharp_bound_effect_quantile_orig}
    \underline{\tau}_{n,\beta} \equiv 
    \sup_{u\in[0,\beta]} \big\{ F_{n1}^{-1}(u)-F_{n0}^{-1}(u+1-\beta) \big\}. 
\end{align}

We next give an equivalent form of the sharp bound in \eqref{eq:sharp_bound_effect_quantile_orig}, which is more convenient for the later asymptotic analyses.
In addition, we show that this sharp bound can be attained by a finite population with the specified marginal distributions of the potential outcomes.

\begin{theorem}\label{thm:tau_nbeta}
Consider any marginal distributions in \eqref{eq:finite_population_dist_PO} and any $\beta \in (0,1)$. 
\begin{enumerate}[label=(\roman*), topsep=1ex,itemsep=-0.3ex,partopsep=1ex,parsep=1ex]
    \item The sharp lower bound in \eqref{eq:sharp_bound_effect_quantile_orig} of the $\beta$ quantile of ITEs, $F_{n\tau}^{-1}(\beta)$, with given marginals in \eqref{eq:finite_population_dist_PO} has the following equivalent form: 
    \begin{align}\label{eq:sharp_bound_effect_quantile}
    \underline{\tau}_{n,\beta} 
    =
    \inf \big\{ d \in \mathbb{R} : \ \sup_{\thre \in \mathbb{R}} \big\{ F_{n0}(\thre) - F_{n1, \beta}(\thre + d) \big\} = 0 \big\}, 
\end{align}
where $F_{n1, \beta}(\thre) = \min \{ 1, 1 - \beta + F_{n1}(\thre) \}$ denotes a shifted distribution similar to \eqref{eq:shift_CDF}. 
    \item The sharp lower bound in \eqref{eq:sharp_bound_effect_quantile_orig} and \eqref{eq:sharp_bound_effect_quantile} can be obtained by some finite population of $n$ units, whose marginal distributions of potential outcomes coincide with those in \eqref{eq:finite_population_dist_PO}.
    In particular, it is attained by pairing the smallest $k=\lceil n\beta\rceil$ treatment potential outcomes with the largest $k$ control potential outcomes; specifically, $Y_{(i)}(1)$ is paired with $Y_{(n-k+i)}(0)$ for $1 \le i \le k$\footnote{The pairing of the remaining units can be arbitrary, since it will not affect the $\beta$ quantile of ITEs. This is because these $k$ units have the smallest ITEs.},
    where $Y_{(1)}(z)\leq Y_{(2)}(z)\leq\cdots\leq Y_{(n)}(z)$ denote the sorted potential outcomes under treatment assignment $z\in\{0,1\}$.
    
\end{enumerate}
\end{theorem}

We first clarify a subtle point in Theorem \ref{thm:tau_nbeta}(ii). Existing sharp bounds, such as those in \citet{fan2010sharp} and \citet{zhang2024bounds}, are derived in an infinite superpopulation setting and show that the bound can be attained
by some joint distribution of potential outcomes with given marginals. Theorem \ref{thm:tau_nbeta}(ii) further ensures that such a joint distribution can correspond to a valid empirical joint distribution for a finite population of $n$ units.

In the equivalent form in Theorem \ref{thm:tau_nbeta}(i),
the quantity $\sup_{\thre \in \mathbb{R}} \{ F_{n0}(\thre) - F_{n1, \beta}(\thre + d)\}$ in \eqref{eq:sharp_bound_effect_quantile} is a population analogue of $\sup_{b\in \mathbb{R}} \{\hat F_{n}^\c(b) - \hat F_{n, k/n_1}^\t(b + c) \}$ in the p-value $\overline{p}_{k,c}^\t$ in \eqref{eq:rand_p_val_ks_equi}.
As shown below, this correspondence is key to establishing the sharpness of the randomization tests in Theorem \ref{thm:pval_RKS} and Corollary \ref{cor:pval_KS}.

\begin{figure}[htbp]
    \centering
    \includegraphics[width=\textwidth]{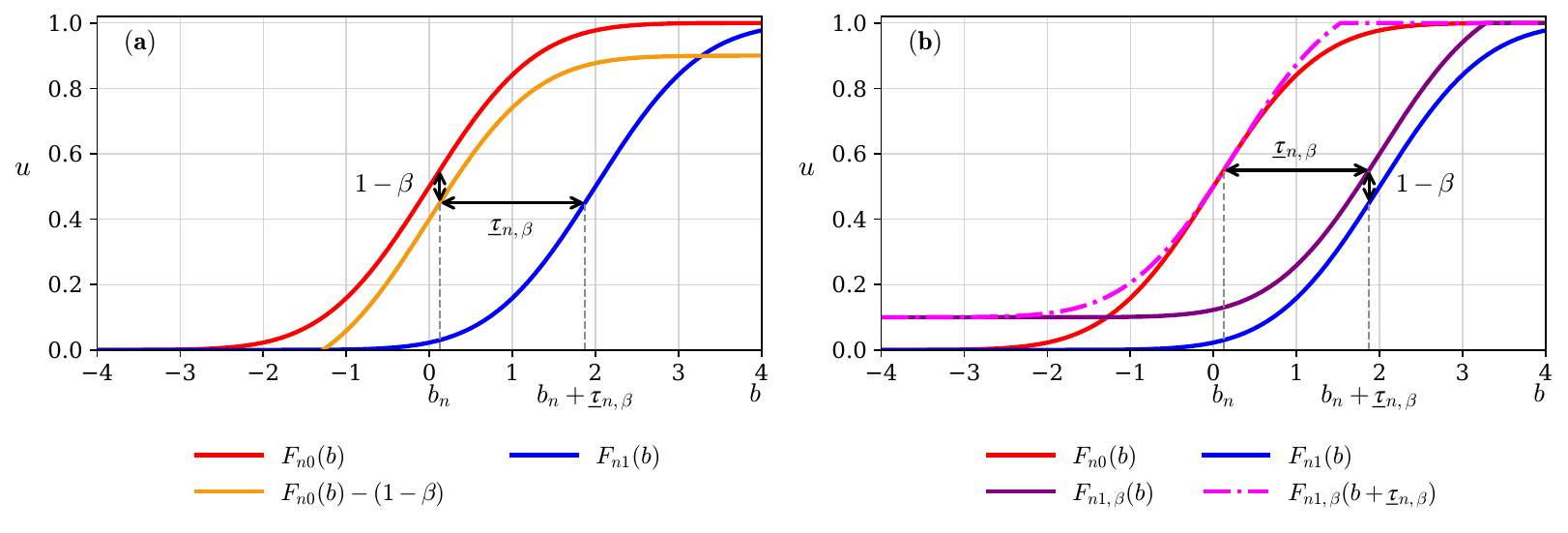}
    \caption{
    	Illustration of the equivalence between the two expressions in \eqref{eq:sharp_bound_effect_quantile_orig} and \eqref{eq:sharp_bound_effect_quantile}, which characterize the sharp lower bound $\underline{\tau}_{n,\beta}$ of the $\beta$-quantile of ITEs.
    }
    \label{fig:sharp_bound_equivalence}
\end{figure}

We provide a graphical illustration of the equivalence between \eqref{eq:sharp_bound_effect_quantile_orig} and \eqref{eq:sharp_bound_effect_quantile}.
In Figure~\ref{fig:sharp_bound_equivalence}, the blue and red curves represent the marginal distribution functions of the treatment and control potential outcomes, respectively. Figure~\ref{fig:sharp_bound_equivalence}(a) visualizes the expression in \eqref{eq:sharp_bound_effect_quantile_orig}.  
Specifically, by definition, $F_{n0}^{-1}(u + 1 - \beta) = \inf \left\{ y : F_{n0}(y) - (1 - \beta) \ge u \right\}$ is the inverse of the function $F_{n0}(\cdot) - (1 - \beta)$. This function is obtained by shifting the red curve downward by $1 - \beta$ and is shown as the orange curve in Figure~\ref{fig:sharp_bound_equivalence}(a).
The expression in \eqref{eq:sharp_bound_effect_quantile_orig} then corresponds to the maximal horizontal difference between the blue curve (inverse of $F_{n1}$) and the orange curve (inverse of $F_{n0} - (1 - \beta)$) over $u \in [0, \beta]$; this is denoted by the horizontal double-headed arrow in Figure~\ref{fig:sharp_bound_equivalence}(a). 

Figure~\ref{fig:sharp_bound_equivalence}(b) visualizes the expression in \eqref{eq:sharp_bound_effect_quantile}.  
The purple curve represents the function $F_{n1,\beta}(b) = \min\{1, 1 - \beta + F_{n1}(b)\}$, obtained by shifting the blue curve upward by $1 - \beta$ and truncating at 1. 
To obtain $F_{n1,\beta}(b + d)$, we further shift the purple curve leftward by $d$.  
The second expression is the smallest $d$ such that the shifted purple curve lies entirely above the red curve. This value of $d$ is equivalent to the maximal horizontal distance between the purple and red curves and is denoted by the horizontal double-headed arrow in Figure~\ref{fig:sharp_bound_equivalence}(b).

From the above, both expressions describe the maximal horizontal gap between the distribution functions of the treatment and control potential outcomes, after a certain shift. 
In Figure~\ref{fig:sharp_bound_equivalence}(a), we shift the control distribution downward by $1 - \beta$, while in Figure~\ref{fig:sharp_bound_equivalence}(b), we shift the treatment distribution upward by the same amount. 
Because these two shifts lead to the same relative positioning of the two distributions, the resulting maximal horizontal gap is identical, which implies the equivalence between \eqref{eq:sharp_bound_effect_quantile_orig} and \eqref{eq:sharp_bound_effect_quantile}.

\subsection{Consistency of the design-based KS test}\label{sec:consistency_fp}

We now study the consistency of the design-based KS test in Section \ref{sec:rks} for inferring ITE quantiles among treated units; as discussed later, this also implies consistency for ITE quantiles among control and all units.
Specifically, we conduct a finite population asymptotic analysis by embedding the finite population of size $n$ into a sequence of finite populations with increasing sizes \citep{fcltxlpd2016}.
Along this sequence of finite populations, we will consider a sequence of null hypotheses of the form \eqref{eq:H_beta_c_T} for the same quantile $\beta\in (0,1)$ but with possibly sample-size-varying threshold $c$. 
That is, for each finite population of size $n$, we consider the null hypothesis 
$
H^\t_{\beta, c_n}: 
F_{n\tau}^{\t,-1}(\beta) \le c_n,
$
for any fixed $\beta \in (0,1)$ and some constant sequence $\{c_n\}$.
By randomization and under some regularity conditions, the true effect quantile $F_{n\tau}^{\t,-1}(\beta)$ among treated units will be close to the effect quantile $F_{n\tau}^{-1}(\beta)$ among all units, which is bounded below by $\underline{\tau}_{n, \beta}$ as shown in Theorem \ref{thm:tau_nbeta}.
Therefore, when $c_n$ is strictly less than $\underline{\tau}_{n, \beta}$ with a gap that is non-vanishing or vanishing at a slow rate, 
we expect the null hypothesis $H^\t_{\beta, c_n}$ to fail with high probability. 
In this case, we want the test to reject the null hypothesis $H^\t_{\beta, c_n}$ with high probability. As shown in the remainder of this subsection, the design-based KS test proposed in Section \ref{sec:rks} has this property under appropriate regularity conditions.

More specifically, we consider the valid p-value $\overline{p}_{\lceil n_1 \beta \rceil,c_n}^\t$ from Corollary \ref{cor:pval_KS} for testing $H^\t_{\lceil n_1 \beta \rceil, c_n}$ in \eqref{eq:quantile_null}, or equivalently $H^\t_{\beta, c_n}$ in \eqref{eq:H_beta_c_T}. We show that it converges almost surely to zero as $n\rightarrow \infty$ under certain regularity conditions, provided that there is a sufficient gap between $c_n$ and $\underline{\tau}_{n, \beta}$.
In other words, our design-based KS test is consistent for detecting any violation of the partially identified bounds on quantiles of ITEs. 
Moreover, because the p-value in Corollary \ref{cor:pval_KS} is more conservative than that in Theorem \ref{thm:pval_RKS}, the consistency of the design-based KS test implies the consistency of the design-based RKS test as well.

To facilitate the discussion, we introduce some notation.
Recall the empirical distribution functions $\hat{F}^\t_{n}$ and  $\hat{F}^\c_{n}$ of the observed outcomes in \eqref{eq:empir_distrib} and the finite population distribution functions $F_{n1}$ and $F_{n0}$ of potential outcomes in \eqref{eq:finite_population_dist_PO}. 
Let 
\begin{align}\label{eq:notation_psin}
    \psi_{n,\beta}(b, d) & = F_{n0}(b) - F_{n1, \beta}(b + d) 
    =
    F_{n0}(b) -
    \min \{ 1, 1 - \beta + F_{n1}(b+d) \}, 
    \\
    \hat{\psi}_{n,\beta}(b, d) & = \hat{F}^\c_{n}(b) - \hat{F}^\t_{n, \frac{\lceil n_1\beta \rceil}{n_1}}(b + d)
    = 
    \hat{F}^\c_{n}(b) -
    \min \Big\{ 1, 1 - \frac{\lceil n_1\beta \rceil}{n_1} + \hat{F}^\t_n (b+d) \Big\}
    \nonumber
\end{align}
denote, respectively, the difference between the distribution function of the control potential outcomes and that of the shifted treatment potential outcomes,
and its empirical analogue;
see the previous Figure \ref{fig:sharp_bound_equivalence}(b) for a graphical illustration. 

We further define the maxima of these differences over all $b\in \mathbb{R}$:
\begin{equation}
\label{eq:notation_xin}
\xi_{n,\beta}(d) = \max_{b\in \mathbb{R}} \psi_{n,\beta}(b, d) \ge 0, \qquad \hat{\xi}_{n,\beta}(d) = \max_{b\in \mathbb{R}} \hat{\psi}_{n,\beta}(b, d) \ge 0,     
\end{equation}
where the nonnegativity follows from the fact that 
both $\psi_{n,\beta}(b, d)$ and $\hat{\psi}_{n,\beta}(b, d)$ approach $0$ as $b\rightarrow \infty$, 
and each maximum is well-defined because it can be attained at some finite $b \in \mathbb{R}$; see the supplementary material for the proof.

From Theorem \ref{thm:tau_nbeta} and by definition, we then have $\underline{\tau}_{n,\beta} = \inf \{ d \in \mathbb{R} : \xi_{n,\beta}(d) = 0 \}$. 
Moreover, by \eqref{eq:rand_p_val_ks_equi}, the valid p-value $\overline{p}_{\lceil n_1 \beta \rceil,c_n}^\t$ for testing $H^\t_{\beta, c_n}$ in \eqref{eq:H_beta_c_T} can be written equivalently as
$
\overline{p}_{\lceil n_1 \beta \rceil,c_n}^\t = G_n( \hat \xi_{n,\beta}(c_n)). 
$
Thus, to understand the limiting behavior of this p-value, 
it suffices to study the limiting behavior of (i) the null tail probability $G_n(\cdot)$ defined as in \eqref{eq:rand_p_val} for the RKS statistic $t_{\RKS}$
and (ii) the maximal difference $\hat{\xi}_{n,\beta}(c_n)$ between the empirical distribution functions of the control outcomes and the shifted treated outcomes.

Suppose that  the proportions of treated and control units have positive limits as $n\rightarrow\infty$.
From Proposition \ref{prop:rks_properties}(iv) and the theory for the classical KS test under the superpopulation setting \citep[see, e.g.,][p. 239]{gibbons2014nonparametric}, 
\begin{align}
\label{eq:KS_asymptotic}
\lim_{{n \to \infty}} G_n\Bigg(
\sqrt{ \frac{n}{n_1n_0} }
 \ x \Bigg) 
 = e^{-2x^2}
 \quad \textup{for any } x\ge 0. 
\end{align}
In addition,
the finite population Glivenko--Cantelli theorem \citep{motoyama2024extended} implies that
\begin{align}\label{eq:uniform_converge_xi}
    n^{\kappa} \sup_{d \in \mathbb{R}} |\hat{\xi}_{n,\beta}(d)-\xi_{n,\beta}(d)| \convergeas 0 \quad \textup{as } n\rightarrow \infty, 
    \qquad (\textup{for any $0\le \kappa < 1/2$}); 
\end{align}
see the supplementary material for the proof. 
These results imply the following 
theorem.

\begin{theorem}
\label{thm:sharp_bound_consist_fin}
Assume that $n_1/n$ has a limit in $(0,1)$ as $n\rightarrow \infty$, and consider any $\beta \in (0,1)$ and any constant sequence $\{c_n\}$. 
As $n\rightarrow \infty$, if $n^{\kappa} \xi_{n, \beta}(c_n) \rightarrow \infty$
for some $0 < \kappa < 1/2$, then the p-values $\overline{p}_{\lceil n_1 \beta \rceil,c_n}^\t$ in \eqref{eq:rand_p_val_ks} and $p_{\lceil n_1 \beta \rceil,c_n}^\t$ in \eqref{eq:rand_p_val_rks} for testing the null hypothesis $H^\t_{\beta, c_n}$ in \eqref{eq:H_beta_c_T} both converge almost surely to zero. 
\end{theorem}

\begin{remark}\label{rmk:pre_int_cn}
    Let $\mathcal{I}^\t_{k, \alpha}=\{c \in \mathbb{R}:p_{k,c}^\t>\alpha\}$ be the one-sided prediction interval for $\tau_{\t(k)}$ in Section \ref{sec:ci_effect}. 
    For any $\alpha\in (0,1)$, $p_{\lceil n_1 \beta \rceil,c_n}^\t \le \alpha$ if and only if $\mathcal{I}^\t_{\lceil n_1 \beta \rceil, \alpha} \subseteq (c_n, \infty)$.
    Therefore, 
    once the convergence in Theorem \ref{thm:sharp_bound_consist_fin} holds, the lower prediction bound will be no less than $c_n$ with probability tending to 1; that is, $\Pr(\inf \mathcal{I}^\t_{\lceil n_1 \beta \rceil, \alpha} \ge c_n) \rightarrow 1$ as $n \rightarrow \infty$. 
\end{remark}

In Theorem \ref{thm:sharp_bound_consist_fin}, $\xi_{n,\beta}(c_n)=\max_{b\in\mathbb{R}}\{F_{n0}(b)-F_{n1,\beta}(b+c_n)\}$ is a population-level one-sided KS difference between the control outcomes and the shifted treated outcomes, which measures the signal against the null hypothesis. By \eqref{eq:uniform_converge_xi}, the uniform approximation difference between $\hat{\xi}_{n,\beta}$ and $\xi_{n,\beta}$ vanishes almost surely at a rate arbitrarily close to $n^{-1/2}$. Hence, the condition $n^\kappa \xi_{n,\beta}(c_n)\rightarrow\infty$ for $0<\kappa<1/2$ implies $n^\kappa \hat \xi_{n,\beta}(c_n)\convergeas\infty$ and therefore $\overline{p}_{\lceil n_1 \beta \rceil,c_n}^\t = G_n( \hat \xi_{n,\beta}(c_n))\convergeas0$ based on limiting null tail probability in \eqref{eq:KS_asymptotic}.

The following proposition provides a simpler sufficient condition for 
the consistency of the design-based KS test.  Recall that the sharp bound 
can be written equivalently as $\underline{\tau}_{n,\beta} = \inf\big\{d\in\mathbb{R}:\xi_{n,\beta}(d)=0\big\}$. Below we also characterize the behavior of $\xi_{n,\beta}(\cdot)$ near $\underline{\tau}_{n,\beta}$.

\begin{proposition} \label{prop:xi_nbeta}
Consider $\xi_{n,\beta}(\cdot)$ defined in \eqref{eq:notation_xin} for any $n\ge 1$ and $\beta\in (0, 1)$.  
\begin{enumerate}[label=(\roman*), topsep=1ex,itemsep=-0.3ex,partopsep=1ex,parsep=1ex]
\item 
   $\xi_{n,\beta}(\cdot)$  is a nonnegative and monotone nonincreasing function. 
Moreover,  
\begin{align*}
\xi_{n,\beta}(c)> 0 \textup{ for any } c < \underline{\tau}_{n,\beta}, \textup{ and } 
\xi_{n,\beta}(c)=0 \textup{ for any } c \ge \underline{\tau}_{n,\beta}.
\end{align*}

\item 
For any $\kappa\in \mathbb{R}$,
     $n^{\kappa} \xi_{n, \beta}(c_n) \to \infty$ as $n\rightarrow \infty$ if   
	there exists  a sequence of real numbers $\{b_{n}\}$ such that $\psi_{n,\beta}(b_n,\underline{\tau}_{n,\beta})\equiv F_{n0}(b_n)-F_{n1,\beta}(b_n+\underline{\tau}_{n,\beta})=0$ for all $n$ and 
    \begin{align*}
        \lim_{n\rightarrow \infty}n^{\kappa-1}\sum_{i=1}^n \I\{b_n + c_n < Y_i(1) < d_n \} = \infty,
    \end{align*}
    where
 $d_n=b_n+\underline{\tau}_{n,\beta}$ if $F_{n0}(b_n)<1$ and $d_n=F_{n1}^{-1}(\beta)$ if $F_{n0}(b_n)=1$. 
\end{enumerate}
\end{proposition}

\begin{figure}[htbp]
    \centering
    \includegraphics[width=\textwidth]{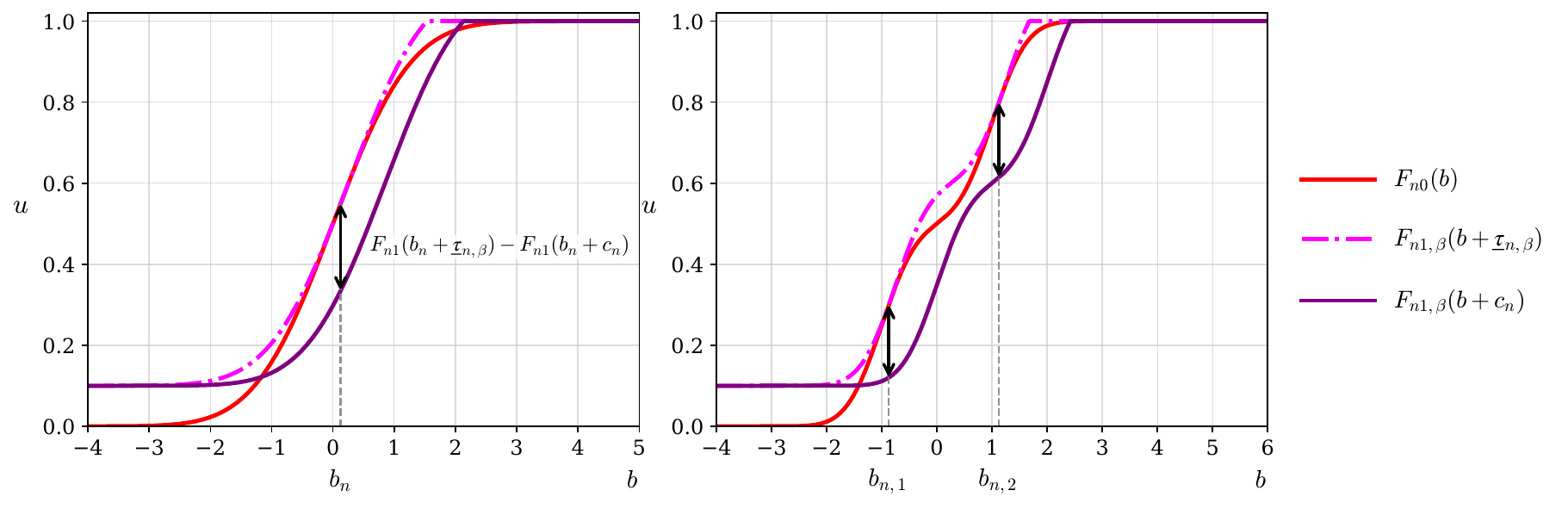}
    \caption{Illustration of $b_n$ such that $\psi_{n,\beta}(b_n,\underline{\tau}_{n,\beta})=F_{n0}(b_n) - F_{n1,\beta}(b_n+\underline{\tau}_{n,\beta})$ reaches its maximum value $0$ and $F_{n0}(b_n)  < 1$.
    The gap between $F_{n1}(b_n+\underline{\tau}_{n,\beta})$ and $F_{n1}(b_n+c_n)$ is a lower bound of $\xi_{n, \beta}(c_n)$.
    The left panel shows one value of $b_n$ satisfying the desired condition. The right panel shows two values, denoted by $b_{n,1}$ and $b_{n,2}$, satisfying this condition.
    }
    \label{fig:bn_plots}
\end{figure}

Below we provide intuition for the condition in Proposition \ref{prop:xi_nbeta}(ii). 
As demonstrated in the supplementary material, the set of points $b_n$ such that $\psi_{n,\beta}(b_n,\underline{\tau}_{n,\beta})=0$
is nonempty. 
For such $b_n$, 
we can have the following lower bound for $\xi_{n,\beta}(c_n)=\max_{b \in \mathbb{R}} \psi_{n,\beta}(b,c_n)$:
\begin{align*}
\psi_{n,\beta}(b_n,c_n)
& =F_{n0}(b_n)-F_{n1,\beta}(b_n+c_n)
=F_{n1,\beta}(b_n+\underline{\tau}_{n,\beta})-F_{n1,\beta}(b_n+c_n)\\
& \ge 
\min\{F_{n1}(b_n+\underline{\tau}_{n,\beta}),\beta\}-F_{n1}(b_n+c_n)\\
& = 
\begin{cases}
	F_{n1}(b_n+\underline{\tau}_{n,\beta})-F_{n1}(b_n+c_n), & \text{if } F_{n0}(b_n)<1, \\
	\beta-F_{n1}(b_n+c_n), 
    & \text{if } F_{n0}(b_n)=1,
\end{cases}
\end{align*}
which then implies Proposition \ref{prop:xi_nbeta}(ii). 
Figure \ref{fig:bn_plots} illustrates one and two representative values of $b_n$ satisfying $\psi_{n,\beta}(b_n,\underline{\tau}_{n,\beta})=0$ and $F_{n0}(b_n)<1$, respectively, with the empirical distribution functions represented by continuous curves for visual clarity.
The vertical double-headed arrows indicate the lower bound of $\xi_{n,\beta}(c_n)$.

\subsection{Sufficient condition for the consistency of the KS test based on limiting distributions}\label{sec:fp_simp}

We next derive another sufficient condition for the consistency of the design-based KS test under the assumption that the finite-population distributions of potential outcomes converge to some limiting distributions. 
We formally introduce the assumption below.

\begin{assumption}\label{asmp:fp_limit}
The following hold for some $\kappa \in (0, 1/2)$ and any $z=0,1$. 
\begin{enumerate}[label=(\roman*), topsep=1ex,itemsep=-0.3ex,partopsep=1ex,parsep=1ex]
\item 
As $n\rightarrow \infty$, the distribution functions $F_{n1}$ and $F_{n0}$ of potential outcomes have limiting distribution functions $F_{\infty,1}$ and $F_{\infty,0}$, in the sense that $\sup_{b\in\mathbb{R}} n^\kappa | F_{nz}(b) - F_{\infty, z}(b)| \to 0$.  

\item $F_{\infty,z}$ is strictly increasing on $\big(F^{-1}_{\infty,z}(0+),F^{-1}_{\infty,z}(1)\big)$.

\end{enumerate}
\end{assumption}

In Assumption \ref{asmp:fp_limit}, condition (i) requires the potential outcome distributions to converge to their limits at a sufficiently fast rate,
and condition (ii) requires the limiting distributions to be strictly increasing on their supports,
where $F^{-1}_{\infty,z}(0+)=\lim_{u\downarrow 0}F^{-1}_{\infty,z}(u)$ is the lower endpoint of the distribution's support and may equal $-\infty$, and $F^{-1}_{\infty,z}(1)$ is the upper endpoint and may equal $+\infty$, for $z\in\{0,1\}$.

\begin{remark}\label{rmk:iid_sampling}
	As verified in the supplementary material,
	when the potential outcomes $\{Y_i(z)\}_{i=1}^n$, for $z=0,1$, are i.i.d.~samples from a distribution whose support is an interval, either bounded or unbounded, and whose distribution function is strictly increasing on its support,
	Assumption \ref{asmp:fp_limit} holds almost surely.
	Thus, under i.i.d.~sampling of units, the conditions in Assumption \ref{asmp:fp_limit} have a relatively simple justification.
\end{remark}

The following theorem establishes the limit of the sharp lower bound and demonstrates the consistency of our design-based KS test under Assumption \ref{asmp:fp_limit}. 

\begin{theorem}
\label{thm:consistency_finite_high_Level}
Under Assumption \ref{asmp:fp_limit}, the following results hold for any $\beta\in (0,1)$. 
\begin{enumerate}[label=(\roman*), topsep=1ex,itemsep=-0.3ex,partopsep=1ex,parsep=1ex]
\item 
As $n\rightarrow \infty$, the sharp lower bound on the $\beta$ quantile of the ITEs, $\underline{\tau}_{n,\beta}$, converges to 
\begin{align*}
    \underline{\tau}_{\infty,\beta} = \sup_{u\in[0,\beta]} \big\{ F_{\infty,1}^{-1}(u)-F_{\infty,0}^{-1}(u+1-\beta) \}. 
\end{align*}
\item For any constant sequence $\{c_n\}$ such that 
$c_n \le \underline{\tau}_{\infty,\beta} - \delta$ for all $n$ and some fixed $\delta>0$,
we have 
$n^\kappa \xi_{n,\beta}(c_n) \to \infty$ as $n\rightarrow \infty$
for the given $\kappa$ in Assumption \ref{asmp:fp_limit}.
If further  $n_1/n$ has a limit in $(0,1)$ as $n\rightarrow \infty$, 
then the p-values $\overline{p}_{\lceil n_1 \beta \rceil,c_n}^\t$ in \eqref{eq:rand_p_val_ks} and $p_{\lceil n_1 \beta \rceil,c_n}^\t$ in \eqref{eq:rand_p_val_rks} for testing the null hypothesis $H^\t_{\beta, c_n}$ in \eqref{eq:H_beta_c_T} both converge almost surely to zero. 
\end{enumerate}
\end{theorem}

For any $\beta\in (0,1)$, 
by the same logic as in Remark \ref{rmk:pre_int_cn}, Theorem \ref{thm:consistency_finite_high_Level} implies that, for any $\delta>0$,
$\Pr(\inf \mathcal{I}^\t_{\lceil n_1 \beta \rceil, \alpha} \ge \underline{\tau}_{\infty,\beta} - \delta) \rightarrow 1$ as $n\rightarrow\infty$.
Therefore, the proposed prediction interval for the $\beta$ quantile of ITEs among treated units is asymptotically sharp. Similarly, under appropriate conditions and by the same argument as in Theorem \ref{thm:consistency_finite_high_Level}, the prediction interval for the $\beta$ quantile of ITEs among control units is also asymptotically sharp, satisfying that, for any $\delta>0$,
$\Pr(\inf \mathcal{I}^\c_{\lceil n_0 \beta \rceil, \alpha} \ge \underline{\tau}_{\infty,\beta} - \delta) \rightarrow 1$ as $n\rightarrow\infty$.

Finally, we can verify that 
$F_{n\tau}^{-1}(\beta) \ge \min\{\tau_{\t(\lceil n_1\beta\rceil)},\tau_{\c(\lceil n_0\beta\rceil)}\}$
and its confidence interval in Corollary \ref{cor:pred_interval_pool}(iii) satisfies
$\inf \mathcal{I}_{\lceil n\beta\rceil,\alpha}\ge \min\{\inf \mathcal{I}^\t_{\lceil n_1\beta\rceil,\alpha},\inf \mathcal{I}^\c_{\lceil n_0\beta\rceil,\alpha}\}$. Therefore, the confidence interval for 
the $\beta$ quantile of ITEs among all units is also asymptotically sharp.

\section{Simplified randomization-based bounds and their connection to a direct construction}\label{sec:direct_connection}

Using properties derived in Section \ref{sec:rks_consistency_fin}, we further simplify the prediction and confidence bounds in Section \ref{sec:ci_effect}. The resulting expressions reveal a close connection to the sharp bounds on ITE quantiles in \eqref{eq:sharp_bound_effect_quantile_orig}. They also admit a direct construction that combines these sharp bounds with uncertainty about the marginal distributions of the potential outcomes.

\subsection{Simplification of the prediction and confidence bounds}

Let $g_{n, \alpha}$ denote the $1-\alpha$ quantile of the distribution of $t_{\RKS}(\bs{Z}, \bs{y})$ under the CRE:
\begin{align}\label{eq:RKS_quantile}
    g_{n, \alpha} & \equiv \inf\big\{c: \Pr\{ t_{\RKS}(\bs{Z}, \bs{y})\le c \} \ge 1-\alpha \big\} = \inf\big\{c: G_n(c+) \le \alpha \big\}, 
\end{align}
where $G_n(\cdot)$ is defined as in Theorem \ref{thm:pval_RKS}
and $G_n(c+) = \lim_{b\downarrow c} G_n(b)$.
Define the following sample analogue of the sharp bound in \eqref{eq:sharp_bound_effect_quantile_orig}\footnote{In the expression below and in the later \eqref{eq:sharp_adj_PO}, we will define $-\infty - (-\infty)$ as $-\infty$.}: 
\begin{align}\label{eq:sharp_bound_effect_quantile_emp}
    \hat{\underline{\tau}}_{n,\beta} 
    \equiv
    \sup_{u\in[0,\beta]} \big\{ \hat F_{n}^{\t, -1}(u)-\hat F_{n}^{\c, -1}(u+1-\beta) \big\}, 
\end{align}
where $\hat F_{n}^{\t, -1}(\cdot)$ and $\hat F_{n}^{\c, -1}(\cdot)$ denote the empirical quantile functions of the observed outcomes in the treatment and control groups, respectively.
For convenience, set $\hat{\underline{\tau}}_{n,\beta} = -\infty$ for $\beta<0$.
For any $d\in \mathbb{R}$ and $\beta \in \mathbb{R}$, further define
\begin{align*}
    \hat{\xi}_{n,\beta}'(d) = \sup_{b\in \mathbb{R}} 
    \big\{ \hat{F}^\c_{n}(b) -
    \min \{ 1, 1 - \beta + \hat{F}^\t_n (b+d) \} \big\} \ge 0 
\end{align*}
which is 
analogous to $\hat{\xi}_{n,\beta}(d)$ in \eqref{eq:notation_psin} but without the ceiling operation.
Let $\overline{\mathcal{I}}^\t_{k, \alpha}$, $\overline{\mathcal{I}}^\c_{k, \alpha}$, and $\overline{\mathcal{I}}_{k, \alpha}$ denote the prediction and confidence intervals constructed as in Corollary \ref{cor:pred_interval_pool}, but using the more conservative p-value from the design-based KS test in Corollary \ref{cor:pval_KS}.

\begin{theorem}\label{thm:PI_simp}
    \begin{enumerate}[label=(\roman*), topsep=1ex,itemsep=-0.3ex,partopsep=1ex,parsep=1ex]
        \item For any $1\le k \le n_1$ and $\beta = k/n_1$, the $1-\alpha$ prediction interval $\overline{\mathcal{I}}^\t_{k, \alpha}$ for the $\beta$ quantile of ITEs among treated units has the following equivalent expressions:
        \begin{equation*}%
            \overline{\mathcal{I}}^\t_{k, \alpha} 
            \equiv \{ c\in \mathbb{R}: \overline{p}_{k,c}^\t > \alpha \} 
            = \{ c\in \mathbb{R}: \hat{\xi}'_{n,\beta}(c) \le g_{n, \alpha} \}
            = \{ c\in \mathbb{R}: \hat{\xi}'_{n,\beta-g_{n, \alpha}}(c) = 0 \}.
        \end{equation*}
        The corresponding lower prediction bound can then be written equivalently as
        \begin{align}\label{eq:pred_limit}
            \inf \overline{\mathcal{I}}^\t_{k, \alpha} = \inf \{ c\in \mathbb{R}: \hat{\xi}'_{n,\beta - g_{n, \alpha}}(c) = 0 \} = \hat{\underline{\tau}}_{n,\beta - g_{n, \alpha}}. 
        \end{align}
        \item For any $1\le k \le n_0$ and $\beta = k/n_0$, 
        the $1-\alpha$ prediction interval $\overline{\mathcal{I}}^\c_{k, \alpha}$ for the $\beta$ quantile of ITEs among control units
        has the following equivalent expressions: 
        \begin{align*}
            \overline{\mathcal{I}}^\c_{k, \alpha} 
            \equiv \{ c\in \mathbb{R}: \overline{p}_{k,c}^\c > \alpha \} 
            = \{ c\in \mathbb{R}: \hat{\xi}'_{n,\beta - g_{n, \alpha}}(c) = 0 \}
            \quad \text{with} \quad 
            \inf \overline{\mathcal{I}}^\c_{k, \alpha} = \hat{\underline{\tau}}_{n,\beta - g_{n, \alpha}}.
        \end{align*}

        \item For any $1\le k \le n$ and $\beta = k/n$, the $1-2\alpha$ confidence interval $\overline{\mathcal{I}}_{k, \alpha}$ for the $\beta$ quantile of ITEs among all units satisfies
        $\inf \overline{\mathcal{I}}_{k, \alpha} = \hat{\underline{\tau}}_{n,\beta' - g_{n, \alpha}}$, where $\beta' = \min_{z\in \{0,1\}}\lceil n_z \beta \rceil / n_z$. 
        Equivalently, $\inf \overline{\mathcal{I}}_{k, \alpha} = \min\{ \inf \overline{\mathcal{I}}^\t_{\lceil n_1 \beta \rceil, \alpha} , \inf \overline{\mathcal{I}}^\c_{\lceil n_0 \beta \rceil, \alpha} \}$. 
    \end{enumerate}
\end{theorem}

Theorem \ref{thm:PI_simp} shows that the prediction and confidence bounds for ITE quantiles obtained from the design-based KS test can be written in the form of the  empirical sharp lower bound in \eqref{eq:sharp_bound_effect_quantile_emp}.
Below we provide some intuition for Theorem \ref{thm:PI_simp}.
In (i), the equivalent representations of $\overline{\mathcal{I}}^\t_{k, \alpha}$ follow from the definition and some algebra. The representation of the lower prediction limit in \eqref{eq:pred_limit} follows from the same reasoning that establishes the equivalence between the two expressions for the sharp bound on ITEs in \eqref{eq:sharp_bound_effect_quantile_orig} and \eqref{eq:sharp_bound_effect_quantile}. Thus, \eqref{eq:pred_limit} is essentially the sample analogue of that equivalence. (ii) shows that the prediction intervals for ITE quantiles among control units have exactly the same form as those among treated units. This equivalence follows from the fact that both the worst-case test statistic and its randomization null distribution are invariant to the switch of the treatment and control labels. (iii) then follows immediately from (i) and (ii).

From Remark \ref{rmk:comp_p_RKS_KS}, the results in Theorem \ref{thm:PI_simp} about lower prediction and confidence limits also apply to intervals obtained from the design-based RKS test.

\subsection{A direct construction based on the sharp bound}

The equivalent representations of the prediction and confidence bounds suggest a direct construction based on the sharp bounds on ITE quantiles in \eqref{eq:sharp_bound_effect_quantile_orig}, together with uncertainty quantification for the finite-population distributions of the potential outcomes.
In particular, as shown in the supplementary material,
\begin{equation}\label{eq:ks_dist}
    \sup_{c\in \mathbb{R}} \{ F_{n1}(c) - \hat{F}^\t_n(c) \}
    \le r_0 t_{\RKS}(\bs{Z}, \bs{Y}(1)),
    \sup_{c\in \mathbb{R}} \{ \hat{F}^\c_n(c) - F_{n0}(c) \}
    \le r_1 t_{\RKS}(\bs{Z}, \bs{Y}(0)),
\end{equation}
where $r_z = n_z/n$ for $z=0,1$, and both inequalities become equalities when there are no ties among the corresponding potential outcomes.
Importantly, because the RKS statistic is distribution-free under the CRE, the two discrepancies in \eqref{eq:ks_dist} can be controlled using a known randomization distribution.
Recall that $g_{n,\alpha}$ in \eqref{eq:RKS_quantile} denotes the quantile of the RKS statistic.
By \eqref{eq:ks_dist} and a union bound, with probability at least $1-2\alpha$,
$
    \sup_{c\in\mathbb{R}}
    \{F_{n1}(c)-\hat F_n^\t(c)\}
    \le r_0 g_{n,\alpha}
$
and 
$
    \sup_{c\in\mathbb{R}}
    \{\hat F_n^\c(c)-F_{n0}(c)\}
    \le r_1 g_{n,\alpha}.
$
On this event, the corresponding quantile functions satisfy, for every $u\in[0,1]$,
$
    F_{n1}^{-1}(u)
    \ge
    \hat F_n^{\t,-1}
    (
        u-r_0 g_{n,\alpha}
    )
$
and 
$
    F_{n0}^{-1}(u)
    \le
    \hat F_n^{\c,-1}
    (
        u+r_1 g_{n,\alpha}
    ).
$
Substituting these inequalities into the sharp lower bound
$\underline{\tau}_{n,\beta}$ in
\eqref{eq:sharp_bound_effect_quantile_orig}, we then have 
\begin{align*}
    \underline{\tau}_{n,\beta}
    &\ge
    \sup_{u\in[0,\beta]}
    \big\{
        \hat F_n^{\t,-1}
        \left(
            u-r_0 g_{n,\alpha}
        \right)
        -
        \hat F_n^{\c,-1}
        \left(
            u+ r_1 g_{n,\alpha}+1-\beta
        \right)
    \big\} = \hat{\underline{\tau}}_{n,\beta-g_{n,\alpha}},
\end{align*}
where the equality follows from a reparameterization detailed in the supplementary material.

In addition, up to a constant scaling factor, the differences between the empirical and finite-population distributions in \eqref{eq:ks_dist} are equivalent to differences between the observed and counterfactual outcome distributions in the two treatment groups. We can therefore construct, for example, a simultaneous upper prediction band for the quantile function of the control potential outcomes among treated units. Applying the sharp bounds in \eqref{eq:sharp_bound_effect_quantile_orig} within the treated group then yields prediction bounds on ITE quantiles among only the treated group.

Thus, $\hat{\underline{\tau}}_{n,\beta-g_{n,\alpha}}$ can be viewed as lower confidence and prediction bounds directly obtained by combining the sharp partial-identification bound with uncertainty bounds for the distributions of the potential outcomes.
We summarize the results below.

\begin{theorem}\label{thm:direct_construct}
Combining the sharp bound in \eqref{eq:sharp_bound_effect_quantile_orig} with the uncertainty bounds from \eqref{eq:ks_dist} yields the following direct construction of simultaneous confidence and prediction bounds.
\begin{enumerate}[label=(\roman*), topsep=1ex,itemsep=-0.3ex,partopsep=1ex,parsep=1ex]
\item For $\beta \in (0,1)$, a $1-2\alpha$ simultaneous lower confidence bound on the $\beta$ quantile of ITEs among all units is $\hat{\underline{\tau}}_{n,\beta - g_{n, \alpha}}$. 

\item For $\beta \in (0,1)$, a $1-\alpha$ simultaneous lower prediction bound on the $\beta$ quantile of ITEs among treated (or control) units is $\hat{\underline{\tau}}_{n,\beta - g_{n, \alpha}}$. 
\end{enumerate}
\end{theorem}

A direct construction of confidence bounds 
similar to that in Theorem \ref{thm:direct_construct}(i) has also been considered in the superpopulation setting \citep[see, e.g.,][]{ruiz2022high, fava2024predicting}. That approach uses the Dvoretzky--Kiefer--Wolfowitz inequality to bound the difference between empirical and population distributions under i.i.d.~sampling. In contrast, our construction uses the inequalities in \eqref{eq:ks_dist} and the distribution-free property of the RKS statistic.

We can also combine the prediction intervals in (ii), following Corollary \ref{cor:pred_interval_pool}(iii), to construct confidence intervals for ITE quantiles among all units.
The resulting confidence bounds can be slightly more informative than those in Theorem \ref{thm:direct_construct}(i).
Specifically, Theorem \ref{thm:PI_simp}(iii) shows that, for any $1\le k \le n$, the combined lower confidence bound for the $\beta = k/n$ quantile of ITEs among all units is $\inf \overline{\mathcal{I}}_{k, \alpha} = \hat{\underline{\tau}}_{n,\beta' - g_{n, \alpha}}$, with $\beta' = \min_{z\in \{0,1\}}\lceil n_z \beta \rceil / n_z \in [\beta, \beta + n^{-1} ] $.

\subsection{Connection}
Although the prediction and confidence bounds in Theorems \ref{thm:PI_simp} and \ref{thm:direct_construct} have equivalent forms, they arise from quite different arguments. The intervals in Theorem \ref{thm:PI_simp} are obtained by inverting worst-case Fisher randomization tests, whereas those in Theorem \ref{thm:direct_construct} combine the sharp bounds on ITE quantiles with confidence bands for the marginal distributions of the potential outcomes.
This equivalence generally fails if other test statistics are used in the randomization test or if uncertainty in the marginal distributions of the potential outcomes is quantified by other methods \citep[see, e.g.,][]{FREY20094126}.
The key to the equivalence is the sample analogue of the two equivalent expressions for the sharp bounds on ITE quantiles in \eqref{eq:sharp_bound_effect_quantile_orig} and \eqref{eq:sharp_bound_effect_quantile}, which is also central to the asymptotic sharpness of the randomization test in Section \ref{sec:rks_consistency_fin}.
The connection established by Theorems \ref{thm:PI_simp} and \ref{thm:direct_construct} further implies that the asymptotic analyses in Sections \ref{sec:rks_consistency_fin} and \ref{sec:rks_power} for the design-based KS test and its inverted intervals apply more generally to prediction and confidence intervals obtained from the direct construction.

\section{Power of the design-based KS test for quantiles of individual treatment effects under local alternatives}
\label{sec:rks_power}

In Sections \ref{sec:rks} and \ref{sec:rks_consistency_fin}, we study the finite-sample validity and large-sample consistency of the KS test under the design-based framework, with fixed potential outcomes and randomness arising solely from the treatment assignment.
In this section, we study the power of the proposed KS test under local alternatives, where the threshold $c$ in the null hypothesis \eqref{eq:H_beta_c_T} can approach the sharp lower bound on the ITE quantile
as the sample size tends to infinity.
When $c$ approaches the sharp bound at an appropriate rate, the power of the KS test converges to a value strictly between $0$ and $1$.
This result provides a more accurate approximation to finite-sample power and helps clarify how the power for inferring ITE quantiles depends on the distributions of the potential outcomes and the quantile levels.

\subsection{Superpopulation formulation}

To facilitate the analysis, we assume that the units are i.i.d.~samples from a superpopulation.
This assumption simplifies the power analysis under local alternatives
while retaining useful insights into the properties of the proposed KS test.

\begin{assumption}\label{asmp:iid}
    The potential outcomes $\{(Y_i(1), Y_i(0))\}_{i=1}^n$ are i.i.d.~samples from some superpopulation with marginal distributions $F_1$ and $F_0$ and marginal densities $f_1$ and $f_0$. 
    The marginal densities are continuous, positive and bounded on the whole real line.
    
\end{assumption}

Under Assumption \ref{asmp:iid}, let
$F_{\tau}(\cdot)$ denote the distribution function of the ITE $\tau_i = Y_i(1)-Y_i(0)$ under the superpopulation,
and let $F_{\tau}^{-1}(\cdot)$ denote the corresponding quantile function.
Analogous to Theorem \ref{thm:tau_nbeta}(i), the following proposition gives sharp lower bounds on the superpopulation ITE quantiles using only the marginal distributions 
of the potential outcomes.
\begin{proposition}
\label{prop:tau_beta}
For $\beta \in (0,1)$, 
a pointwise sharp lower bound on $F_{\tau}^{-1}(\beta)$ is 
\begin{equation}
\label{eq:quantile_lower_bound}
\underline{\tau}_{\beta} \equiv  \sup_{u\in[0,\beta]} \{ F_1^{-1}(u)-F_0^{-1}(1+u-\beta) \} = \inf \{ d \in \mathbb{R} : \sup_{b \in \mathbb{R}} \{ F_0(b) - F_{1, \beta}(b + d) \} = 0 \}
\end{equation}
with $F_{1, \beta}(b) = \min \{ 1, 1 - \beta + F_1(b) \}$.
Moreover, this sharp lower bound can be obtained by some joint distribution of the two potential outcomes with marginal distributions $F_1$ and $F_0$.
In particular, it is achieved by pairing the lowest $\beta$ fraction of treatment potential outcomes with the highest $\beta$ fraction of control potential outcomes\footnote{For example, let $U$ and $V$ be two uniform random variables on $(0,1)$ such that $V = U + 1 - \beta$ when $0 < U \le \beta$. The remaining values of $U$ and $V$ can be paired arbitrarily. Then the sharp lower bound for the $\beta$ quantile of the ITE is achieve when $(Y(1), Y(0))$ follows the joint distribution of $(F_1^{-1}(U), F_0^{-1}(V))$.}.
 	
\end{proposition}

The first expression for $\underline{\tau}_{\beta}$ in \eqref{eq:quantile_lower_bound} is from \citet{fan2010sharp} and \citet{zhang2024bounds}. The second, equivalent expression is more convenient for our asymptotic analysis because it corresponds directly to the sample analogue in the randomization p-value in \eqref{eq:rand_p_val_ks_equi}.

\subsection{Consistency of the randomization-based KS test}\label{sec:consis_super}

As in Sections \ref{sec:consistency_fp} and \ref{sec:fp_simp}, we now study the consistency of the design-based KS test under the superpopulation formulation. The results are consistent with those in Section 
\ref{sec:fp_simp}.
For any $\beta\in (0,1)$, when $c$ is strictly less than and bounded away from $\underline{\tau}_{\beta}$, the null hypothesis $H^\t_{\beta, c}$ fails with high probability, since the treated units are also i.i.d.~samples from the superpopulation in Assumption \ref{asmp:iid}. 
We show that our KS test is able to reject $H^\t_{\beta, c}$ with probability approaching $1$.

We first introduce some notation. 
For any $b, d\in \mathbb{R}$, define 
\begin{align}\label{eq:notation_xi}
    \psi_\beta(b, d) \equiv F_0(b) - F_{1, \beta}(b + d), \quad \xi_\beta(d) \equiv \sup_{b \in \mathbb{R}} \psi_\beta(b, d). 
\end{align}
The quantities in \eqref{eq:notation_xi} are essentially the superpopulation analogue of those in \eqref{eq:notation_psin} and \eqref{eq:notation_xin}. 
From Proposition \ref{prop:tau_beta}, the sharp bound in \eqref{eq:quantile_lower_bound} can be equivalently written as $\underline{\tau}_{\beta} = \inf \{ d \in \mathbb{R} : \xi_\beta(d) = 0 \}$.
By the same logic as in Proposition \ref{prop:xi_nbeta}, and as demonstrated in the supplementary material,
$\xi_\beta(\cdot)$ is a nonnegative and monotone nonincreasing function. Moreover, $\xi_\beta(c) > 0$ for $c<\underline{\tau}_\beta$ and $\xi_\beta(c) =0$ for $c\ge \underline{\tau}_\beta$. 
In addition, similar to \eqref{eq:uniform_converge_xi} and using the Glivenko--Cantelli theorem, 
when the proportions of treated and control units have positive limits as $n\rightarrow \infty$, 
we can show that 
$ n^\kappa\sup_{d \in \mathbb{R}} |\hat{\xi}_{n,\beta}(d) - \xi_\beta(d)| \convergeas 0$ as $n\rightarrow \infty$
for any $0\le \kappa < 1/2$. 
As a result, $\sqrt{n}\hat{\xi}_{n,\beta}(c)$ converges almost surely to infinity for $c<\underline{\tau}_{\beta}$, and the p-value
$
\overline{p}_{\lceil n_1 \beta \rceil,c}^\t = G_n( \hat \xi_{n,\beta}(c)) 
$
converges to zero almost surely by \eqref{eq:KS_asymptotic}.
This establishes the consistency of our design-based KS test as in Theorem \ref{thm:sharp_bound_consist_fin}.

\begin{theorem}
\label{thm:sharp_bound_consist}
Consider any $\beta \in (0,1)$ and any $c$ strictly less than the sharp lower bound $\underline{\tau}_{\beta}$ of the $\beta$ quantile of the ITE in \eqref{eq:quantile_lower_bound}. 
As $n\rightarrow \infty$, 
if $n_1/n \to \lambda \in (0,1)$ and Assumption \ref{asmp:iid} holds, 
then the p-values $\overline{p}_{\lceil n_1 \beta \rceil,c}^\t$ in \eqref{eq:rand_p_val_ks} and $p_{\lceil n_1 \beta \rceil,c}^\t$ in \eqref{eq:rand_p_val_rks}, for testing the null hypothesis $H^\t_{\beta, c}$ in \eqref{eq:H_beta_c_T}, are equal almost surely and both converge to zero almost surely. 
\end{theorem}

The discussion in Section \ref{sec:fp_simp} shows that the assumptions on the superpopulation distributions can be weakened further.
In particular, it suffices to require the marginal distributions $F_1$ and $F_0$ to be strictly increasing on their supports.
For conciseness, we retain Assumption \ref{asmp:iid}, which is needed for the subsequent power analysis under local alternatives.
Moreover, similar to the discussion in Remark \ref{rmk:pre_int_cn}, Theorem \ref{thm:sharp_bound_consist} implies the sharpness of the corresponding prediction and confidence bounds from Corollary \ref{cor:pred_interval_pool}.

\subsection{Power of the design-based KS test under local alternatives}
For power analysis, we consider the following null hypothesis and local alternative for any fixed $h>0$:
\begin{align}\label{eq:local_alter}
    H^\t_{\beta,c_n}: 
    F_{n\tau}^{\t,-1}(\beta)
    \le c_n 
    \ \ \ \textup{vs} \ \ \  
    \overline{H}^\t_{\beta,c_n}: 
    F_{n\tau}^{\t,-1}(\beta) 
    > c_n, 
    \ \ \ \ 
    \textup{where} \ \ c_n = \underline{\tau}_{\beta} - n^{-1/2} h. 
\end{align}
In \eqref{eq:local_alter}, the threshold $c_n$ varies with the sample size and converges to the sharp lower bound $ \underline{\tau}_{\beta}$ at the $n^{-1/2}$ rate.
The null hypothesis in \eqref{eq:local_alter} concerns ITE quantiles among treated units in the CRE, which are also i.i.d.~samples from the superpopulation.
By classical asymptotic theory for sample quantiles and under some regularity conditions,
$F_{n\tau}^{\t,-1}(\beta)$
is approximately Gaussian with mean greater than or equal to $\underline{\tau}_{\beta}$ and standard deviation of order $n^{-1/2}$.
Therefore, we expect the null hypothesis to fail with non-negligible probability, and this probability increases with $h$.
We investigate the asymptotic rejection probability, or equivalently the asymptotic distribution of the p-value, under the design-based KS test as $n\rightarrow \infty$. As shown below, this probability is determined by the value of $h$ in \eqref{eq:local_alter}, the marginal distributions $F_1$ and $F_0$ of the potential outcomes, and the quantile $\beta$ of interest.

\begin{remark}\label{rmk:power_random_null}
    Similar to the discussion in Section \ref{sec:dist_quantile_ite}, the null hypothesis in \eqref{eq:local_alter} is random, due to randomness in both treatment assignment and superpopulation sampling. 
    From the definition of type I error, 
    we could define the power of a test for a random null hypothesis $H$ as $\Pr(\textup{Reject } H \textup{ \& } H \textup{ fails})$. For a valid test, this power is bounded from below by 
    \begin{align*}
        \Pr(\textup{Reject } H \textup{ \& } H \textup{ fails})
        & = 
        \Pr(\textup{Reject } H ) - 
        \Pr(\textup{Reject } H \textup{ \& } H \textup{ holds}) \ge \Pr(\textup{Reject } H ) - \alpha, 
    \end{align*}
    where the inequality follows from the validity of the test. 
    Therefore, the probability of rejection $\Pr(\textup{Reject } H )$ that we will study throughout this section can provide a lower bound for the power of a test similar to the traditional sense.     
    For convenience, in the following we will simply call $\Pr(\textup{Reject } H )$ the power of the test. 

    In addition, the asymptotic analysis under local alternative can also help us characterize the asymptotic properties of the prediction and confidence bounds in Corollary \ref{cor:pred_interval_pool} from test inversion; see Remark \ref{rmk:bound_rate} for more details.
\end{remark}

Recall the randomization p-value in \eqref{eq:rand_p_val_ks_equi} of the form $\overline p^\t_{\lceil n_1\beta \rceil,c_n} = G_n ( \hat{\xi}_{n,\beta}(c_n))$ under our design-based KS test. 
Because the randomization distribution $G_n$ has the limit in \eqref{eq:KS_asymptotic}, the main challenge
is to understand the asymptotic behavior of $\hat{\xi}_{n,\beta}(c_n)$ defined in \eqref{eq:notation_xin}.
To facilitate the analysis,
we introduce the following contact set for any $\beta \in (0,1)$: 
\begin{align}\label{eq:S_beta}
    \RS_{\beta}^\star = \{ b \in \mathbb{R} : \psi_\beta(b, \underline{\tau}_{\beta}) = 0 \}, 
\end{align}
which contains the values of $b\in \mathbb{R}$ at which $F_0(b) - F_{1, \beta}(b + \underline{\tau}_{\beta})$ attains its maximum value of zero over the real line.
This contact set is non-empty and may contain multiple, or even infinitely many, contact points, but is always compact; 
see the supplementary material for an example where $\RS_{\beta}^\star$ contains infinitely many points.

The following theorem gives the asymptotic distribution of $\hat{\xi}_{n,\beta}(c_n)$, in which the contact set $\RS_{\beta}^\star$ plays a crucial role.
Recall that 
$F_1$ and $F_0$ are the distribution functions of the treatment and control potential outcomes, 
and $f_1$ is the density function of the treatment potential outcome. 
Let $\mathscr{B}_1(\cdot)$ and $\mathscr{B}_0(\cdot)$ be two independent Brownian bridges.

\begin{theorem}
\label{thm:xin_beta_convergence}
Consider any $\beta\in (0, 1)$, and 
the sequence $\{c_n\}$ defined in \eqref{eq:local_alter} with any $h>0$. 
If $n_1/n \to \lambda \in (0,1)$ and Assumption \ref{asmp:iid} holds, then we have, as $n\rightarrow \infty$,
\begin{align*}
\sqrt{n} \, \hat \xi_{n,\beta}(c_n)  \converged \max \big\{ 0, \sup_{b \in \RS_\beta^\star} \mathscr{A}(b, \underline{\tau}_{\beta}) \big\},
\end{align*}
where $\hat \xi_{n,\beta}(\cdot)$ is defined in \eqref{eq:notation_xin}, 
$\RS_{\beta}^\star$ is the contact set defined in \eqref{eq:S_beta}, and  
\begin{align}\label{eq:GP_power}
  \mathscr{A}(b, \underline{\tau}_{\beta}) \equiv \frac{1}{\sqrt{1-\lambda}} \mathscr{B}_0(F_0(b)) - \frac{1}{\sqrt{\lambda}} \mathscr{B}_1(F_1(b + \underline{\tau}_{\beta})) + h \cdot f_1(b + \underline{\tau}_{\beta}).    
\end{align}
\end{theorem}

In \eqref{eq:GP_power}, we can equivalently replace $f_1(b + \underline{\tau}_{\beta})$ by $f_0(b)$, as they are identical for $b\in \RS_{\beta}^\star$; see the supplementary material for proof.
Theorem \ref{thm:xin_beta_convergence} immediately implies the following corollary for the power of the design-based KS test under the local alternative in \eqref{eq:local_alter}.
The proof of Corollary \ref{cor:asymptotic_power} uses the compactness of the contact set $\RS_{\beta}^\star$ to ensure that $\sup_{b \in \RS_\beta^\star} \mathscr{A}(b, \underline{\tau}_{\beta})$ has a continuous distribution.

\begin{corollary}
\label{cor:asymptotic_power}
Consider any $\beta\in (0, 1)$ and 
the sequence $\{c_n\}$ in \eqref{eq:local_alter} with any $h>0$. 
If $n_1/n \to \lambda \in (0,1)$ and Assumption \ref{asmp:iid} holds, 
then for any level $\alpha \in (0,1)$,
as $n\rightarrow \infty$
\begin{align}
\label{eq:asy_power_closed_form}
\Pr(\textup{reject } H^\t_{\beta,c_n}) = \Pr(\overline p^\t_{\lceil n_1\beta \rceil,c_n}\leq \alpha) \to 
\Pr\Big(\sqrt{2\lambda(1-\lambda)}  \sup_{b \in \RS_\beta^\star} \mathscr{A}(b, \underline{\tau}_{\beta}) > \sqrt{-\log\alpha}\Big).
\end{align}
\end{corollary}

By Corollary \ref{cor:asymptotic_power}, the local power of the design-based KS test depends crucially on the Gaussian process $\mathscr{A}(b, \underline{\tau}_{\beta})$ in \eqref{eq:GP_power}.
When the contact set $\RS_\beta^\star$ is a singleton, the local power of the test can be characterized by a Gaussian tail probability. More generally, $\RS_\beta^\star$ may contain multiple, or even infinitely many, elements. In these cases, the local power is characterized by the tail probability of either the maximum of finitely many jointly Gaussian random variables or the supremum of a Gaussian process over an infinite, possibly uncountable, contact set.
For any fixed $b\in \RS_\beta^\star$,
$\mathscr{A}(b, \underline{\tau}_{\beta})$ is Gaussian with mean and variance:
\begin{align}
\label{eq:Gaussian_process_mean_variance}
\E\{ \mathscr{A}(b, \underline{\tau}_{\beta}) \} & = h \cdot f_1(b + \underline{\tau}_{\beta})
= h \cdot f_0(b),
\\
\Var\{\mathscr{A}(b, \underline{\tau}_{\beta})\}
& = \frac{1}{1-\lambda} F_0(b) \big( 1-F_0(b)\big) + \frac{1}{\lambda} F_1(b + \underline{\tau}_{\beta})\big( 1-F_1(b + \underline{\tau}_{\beta}) \big) 
\nonumber
\\
&= \frac{1}{1-\lambda} F_0(b) \big( 1-F_0(b)\big) + \frac{1}{\lambda} \big(F_0(b)+\beta-1\big)\big(2-\beta-F_0(b)  \big),
\nonumber
\end{align}
and for any fixed $b \le b'$ in the contact set $\RS_\beta^\star$, 
\begin{align*}
\Cov\big\{ \mathscr{A}(b,\underline{\tau}_{\beta}),
\mathscr{A}(b',\underline{\tau}_{\beta})\big\}
&=\frac{1}{1-\lambda}F_0(b)\big(1-F_0(b')\big) +\frac{1}{\lambda}
F_1(b+\underline{\tau}_{\beta})
\big(1 - F_1(b'+\underline{\tau}_{\beta})\big). 
\end{align*}
A larger control density near the contact point gives a stronger local signal and higher local power. 
The variance arises from two independent sources: fluctuations of the empirical distribution of the control outcomes $\hat F_{n}^\c$ in \eqref{eq:empir_distrib} at $b$,  
and fluctuations of the empirical distribution of the treated outcomes $\hat F_{n}^\t$ in \eqref{eq:empir_distrib} at $b+\underline{\tau}_{\beta}$. 
Moreover, the mean in \eqref{eq:Gaussian_process_mean_variance}, particularly its dependence on $f_1(b+\underline{\tau}_{\beta})$, agrees with the assumption in Proposition \ref{prop:xi_nbeta}(ii) for the consistency of the design-based KS 
test,
which requires  a substantial proportion of units with treatment potential outcomes near $b+\underline{\tau}_{\beta}$ for some $b$ in the contact set. 

\begin{remark}\label{rmk:bound_rate}
From Remark \ref{rmk:comp_p_RKS_KS} and by the same argument as in Remark \ref{rmk:pre_int_cn}, under the regularity conditions of Corollary \ref{cor:asymptotic_power}, the prediction interval constructed as in Corollary \ref{cor:pred_interval_pool} by p-value $\overline{p}_{\lceil n_1 \beta \rceil,c}^\t$ in \eqref{eq:rand_p_val_ks} for the $\beta$ quantile of ITEs among treated units satisfies, as $n\rightarrow\infty$, 	\begin{align*}
	\Pr \{ \inf \mathcal{I}^\t_{\lceil n_1\beta \rceil, \alpha} \ge \underline{\tau}_{\beta} - n^{-1/2} h \}
	& 
	\to 
	\Pr\Big(\sqrt{2\lambda(1-\lambda)}  \sup_{b \in \RS_\beta^\star} \mathscr{A}(b, \underline{\tau}_{\beta}) > \sqrt{-\log\alpha}\Big). 
\end{align*}
This nondegenerate local-power limit shows that the lower prediction bound approaches (or at least exceeds) the sharp lower bound on the $\beta$ quantile of the ITE at the usual $n^{-1/2}$ rate. 
An analogous argument yields the corresponding result for the prediction interval for the $\beta$ quantile of ITEs among control units. Moreover, because $\inf\mathcal{I}_{\lceil n\beta\rceil,\alpha}\ge\min\{\inf\mathcal{I}^\t_{\lceil n_1\beta\rceil,\alpha},\inf\mathcal{I}^\c_{\lceil n_0\beta\rceil,\alpha}\}$, these results further yield an asymptotic lower bound for $\Pr\{\inf\mathcal{I}_{\lceil n\beta\rceil,\alpha}\ge\underline{\tau}_{\beta}-n^{-1/2}h\}$.
\end{remark}

In the supplementary material, we 
present a numerical example in which $\RS_\beta^\star$ contains two elements and the supremum in \eqref{eq:asy_power_closed_form} is necessary for the asymptotic power approximation.

\begin{remark}
The local power in Corollary \ref{cor:asymptotic_power} is closely related to that of the traditional KS test. 
Specifically, 
consider a special case where $F_0 = F_1$ and suppose we are interested in the $\beta = 100\%$ quantile (maximum) of ITEs.
We can verify that the sharp lower bound $\underline{\tau}_{\beta} = 0$, the contact set $\RS_\beta^\star = \mathbb{R}$, and 
the expression for the local power in \eqref{eq:asy_power_closed_form} reduces to\footnote{Here we are not claiming that Corollary \ref{cor:asymptotic_power} remains valid when $\beta=1$. In general, when $\beta$ equals $0$ or $1$, the sharp bound $\underline{\tau}_{\beta}$ may take an infinite value.}
$$
    \Pr \big(\sqrt{2\lambda(1-\lambda)}\,\sup_{b \in \mathbb{R}} \mathscr{A}(b, 0) 
    > \sqrt{-\log\alpha} \big), 
$$
which coincides with the local power of the one-sided two-sample KS test under a location-shift alternative \citetext{see, e.g., 
\citealp[][p.~290]{van2000asymptotic} and related results in \citealp[][p.~779]{lehmann2022testing}}. 
This is intuitive, since, for the $100\%$ quantile of ITEs among treated units, the procedure of our design-based KS test coincides with that of usual permutation test. 
\end{remark}

\begin{remark}
In the supplementary material, we study Gaussian potential outcomes, for which $\RS_\beta^\star$ is a singleton, and derive a closed-form expression for asymptotic local power when the two potential outcomes have equal variances. We also examine how power varies with the quantile level and potential outcome variances.
\end{remark}

\section{Covariate adjustment}\label{sec:cov_adj}

We now study how pretreatment covariate information can be leveraged to improve inference on ITE quantiles.
Let $X_i$ denote the covariate vector for unit $i$, $1\le i \le n$. We consider the following covariate-adjusted potential and observed outcomes:
\begin{align}\label{eq:adj_PO}
	\tilde{Y}_i(1)=Y_i(1)-s(X_i), \quad \tilde{Y}_i(0)=Y_i(0)-s(X_i), \quad \tilde{Y}_i=Y_i-s(X_i).
\end{align}
Importantly, this adjustment does not change the ITE because
$\tilde{\tau}_i \equiv \tilde{Y}_i(1) - \tilde{Y}_i(0) = Y_i(1)-Y_i(0) = \tau_i$. 
When $s(\cdot)$ is prespecified or estimated independently of the data,
we can apply the methods in Section \ref{sec:rks} directly, replacing the original potential outcomes with the adjusted potential outcomes in \eqref{eq:adj_PO}.
When $s(\cdot)$ is estimated using the experimental data, sample splitting can be employed. For example, one half of the sample can be used to estimate the adjustment function, which is then applied to the other half, and the resulting inference from the two splits can be combined using the Bonferroni method.

Specifically, consider testing the null hypothesis in \eqref{eq:H_beta_c_T} that the $\beta$ quantile of ITEs among treated units is bounded by a threshold $c$.
We conduct the design-based RKS test as in Section \ref{sec:rks}, using the adjusted outcomes $\tilde{Y}_i$ in \eqref{eq:adj_PO} in place of the original outcomes.
With $k = \lceil n_1 \beta \rceil$, the resulting p-value can be written as
\begin{align}\label{eq:rand_p_val_Rks_equi_covariates}
	p_{k,c}^{\t, \co} & 
	= G_n \big( t_{\RKS}(\bs Z, \tilde{\bs Y} - \bs{Z} \circ \tilde{\bs{\xi}}_{k,c}^\t) \big), 
\end{align}
where $\tilde{\bs{Y}} = (\tilde{Y}_{1}, \tilde{Y}_{2}, \ldots, \tilde{Y}_{n})$ and $\tilde{\bs{\xi}}_{k,c}^\t$ is defined in the same way as \eqref{eq:xi_kc} but based on the adjusted outcomes $\tilde{\bs Y}$.
Define also a more conservative p-value as 
\begin{align}\label{eq:rand_p_val_ks_equi_covariates}
	\overline{p}_{k,c}^{\t, \co} & 
	= G_n \big( t_{\KS}(\bs Z, \tilde{\bs Y} - \bs{Z} \circ \tilde{\bs{\xi}}_{k,c}^\t) \big)
    \ge p_{k,c}^{\t, \co},
\end{align}
where the inequality follows from Proposition \ref{prop:rks_properties}(iii). 
By the same logic as Theorem \ref{thm:pval_RKS}, both p-values are valid for the null hypothesis $H^\t_{k,c}$ in \eqref{eq:quantile_null}. We emphasize that this validity holds for arbitrary prespecified covariate adjustment, including cases where the adjustment may take positive infinite values; see Remark \ref{rmk:adjustment_infty}.

A natural next question is how to choose an appropriate covariate adjustment to sharpen inference on ITE quantiles and, ideally, attain the sharp lower bounds given covariate information. Under a superpopulation framework with observed covariates, \cite{fan2010sharp} derived the following sharp lower bound for the $\beta$ quantile of the ITE distribution with covariate information for any $\beta\in(0,1)$:
\begin{align}\label{eq:quantile_lower_bound_covariate}
    \underline{\tau}_{\beta}^{\textup{co}}=
    \bar{F}^{-1}(\beta) 
    \quad \textup{ with } \ 
        \bar{F}(y)=1+\E\big[\inf_b\big\{F_1(b \mid X)-F_0(b-y \mid X)\big\}\big],    
\end{align}
where $\bar{F}^{-1}(\beta) = \inf\{y\in \mathbb{R}: \bar{F}(y) \ge \beta\}$, and 
$F_z(\cdot \mid X)$ denotes the conditional distribution function of the potential outcome $Y(z)$ given the covariate $X$ for $z=0,1$.
In \eqref{eq:quantile_lower_bound_covariate}, $\bar{F}(\cdot)$ is the Fr\'echet--Hoeffding sharp upper bound on the ITE distribution function with covariate information. 
The sharp lower bound on ITE quantiles in \eqref{eq:quantile_lower_bound_covariate} with covariate information is at least as large as the bound in \eqref{eq:quantile_lower_bound} without covariate information, demonstrating the potential gain from covariate information.
\citet{fava2024predicting} showed that the sharp bound $\bar{F}(y)$ in \eqref{eq:quantile_lower_bound_covariate} on the ITE distribution can be attained by applying a similar bound directly to the covariate-adjusted potential outcomes in \eqref{eq:adj_PO}.
Below we further extend the results of \cite{fava2024predicting} to sharp bounds on ITE quantiles and relax the boundedness assumption on the potential outcomes.

To facilitate analyses, we consider a superpopulation setting with the following conditions.

\begin{assumption}\label{asmp:iid_covariate}
	Suppose $\{(Y_i(1), Y_i(0),X_i)\}_{i=1}^n$ are i.i.d.~samples from some distribution. 
	Moreover, 
	the conditional distribution functions of potential outcomes given covariates, $F_1(\cdot|X)$ and $F_0(\cdot|X)$, are almost surely continuous and strictly increasing. 
\end{assumption}

For any $y\in \mathbb{R}$, 
define the covariate adjustment function
\begin{align}\label{eq:cov_adj}
	\bar{s}(x;y) \in \argmin_{b \in \mathbb{R}\cup\{+\infty\}} \big\{F_1(b \mid x)-F_0(b-y \mid x)\big\}
\end{align}
as any element of $\mathbb{R}\cup\{+\infty\}$ that minimizes the difference between the (shifted) conditional distribution functions of treated and control potential outcomes given covariate $x$.
We emphasize that $+\infty$ must be included in the feasible region because the minimum need not be attained at a finite $b$; 
see Remark \ref{rmk:adjustment_normal}.
Moreover, such infinite adjustment would not affect the validity of our KS test with adjusted outcomes; see the following remark.

\begin{remark}\label{rmk:adjustment_infty}
    When the adjustment $s(X_i)=+\infty$ for some units, both adjusted potential outcomes 
    in \eqref{eq:adj_PO} are $-\infty$, under which the adjusted ITE $\tilde{\tau}_i
    =\tilde{Y}_i(1)-\tilde{Y}_i(0)$ is not well-defined. Nevertheless, the RKS test and the KS test using adjusted outcomes, with p-values in \eqref{eq:rand_p_val_Rks_equi_covariates} and \eqref{eq:rand_p_val_ks_equi_covariates}, still remain valid; see the supplementary material for the proof.
\end{remark}

The following theorem demonstrates the optimality of the covariate adjustment in \eqref{eq:cov_adj} for inferring the ITE distribution and quantile functions. 
Define further 
\begin{align}\label{eq:CDF_adj_PO}
	F_{y, 1} (c)= \Pr\{ Y_i(1)-\bar{s}(X_i;y) \le c\}
	\quad 
	\textup{and}
	\quad 
	F_{y, 0} (c)= \Pr\{ Y_i(0)-\bar{s}(X_i;y) \le c\}
\end{align}
as the distribution functions of the adjusted treatment and control potential outcomes, 
and 
$F_{y, z}^{-1}(\cdot)$ as the corresponding quantile function
for $z=0,1$.

\begin{theorem}\label{thm:bar_F_covariate_adjustment}
	Suppose Assumption \ref{asmp:iid_covariate} holds.
    \begin{enumerate}[label=(\roman*), topsep=1ex,itemsep=-0.3ex,partopsep=1ex,parsep=1ex]
    \item The sharp upper bound $\bar{F}(\cdot)$ in \eqref{eq:quantile_lower_bound_covariate} on the ITE distribution function is a continuous distribution function and can be equivalently written as 
	\begin{align}\label{eq:sharp_adj_PO_dist}
		\bar{F}(y)=1+\inf_b\big\{F_{y,1}(b)-F_{y,0}(b-y )\big\},
	\end{align}
	where $F_{y,1}$ and $F_{y,0}$ are defined in \eqref{eq:CDF_adj_PO}.
    \item The sharp lower bound $\underline{\tau}_{\beta}^{\textup{co}}$ in \eqref{eq:quantile_lower_bound_covariate} on ITE quantiles can be equivalently written as
    \begin{align}\label{eq:sharp_adj_PO}
		\underline{\tau}_{\beta}^{\textup{co}} =  \sup_{u\in[0,\beta]} \{ F_{\underline{\tau}_{\beta}^{\textup{co}}, 1}^{-1}(u)-F_{\underline{\tau}_{\beta}^{\textup{co}}, 0}^{-1}(1+u-\beta) \},   
	\end{align}
    where $F_{\underline{\tau}_{\beta}^{\textup{co}}, z}^{-1} (\cdot)$ is  defined as $F_{y, z}^{-1}(\cdot)$ with $y=\underline{\tau}_{\beta}^{\textup{co}}$  for $z=0,1$.
    \end{enumerate}
\end{theorem}

The equivalent forms in \eqref{eq:sharp_adj_PO_dist} and \eqref{eq:sharp_adj_PO} coincide with the sharp bounds on ITE distribution and quantile functions applied directly to the covariate-adjusted potential outcomes,
which implies optimality of the adjustment in \eqref{eq:cov_adj}. 
The equivalent form in \eqref{eq:sharp_adj_PO_dist} has been derived in \citet{fava2024predicting}, but with bounded potential outcomes.  
The equivalent form in \eqref{eq:sharp_adj_PO} is newly derived here
and is preferable for inferring ITE quantiles, as discussed below.

\begin{remark}\label{rmk:adjustment_normal}
    When $Y(z)\mid X\sim \mathcal{N}(\mu_z(X),\sigma^2(X))$ for $z=0,1$, the adjustment $\bar{s}(x;y)$ in \eqref{eq:cov_adj} can be chosen as, up to an additive constant\footnote{A constant shift of the adjustment would not affect the bounds in \eqref{eq:sharp_adj_PO_dist}--\eqref{eq:sharp_adj_PO} and the p-values in \eqref{eq:rand_p_val_Rks_equi_covariates}--\eqref{eq:rand_p_val_ks_equi_covariates}.},
    $\{\mu_1(x)+\mu_0(x)\}/{2}$ if $\mu_1(x)-\mu_0(x)\ge y$ and $+\infty$ otherwise.
    In other words, the adjustment is a truncated average of the conditional means of the two potential outcomes given covariates. The truncation depends on whether the conditional average treatment effect exceeds $y$, which is set to be the sharp bound on the ITE quantiles in Theorem \ref{thm:bar_F_covariate_adjustment}(ii).
\end{remark}

Below we apply the above optimal adjustment to our design-based inference. 
Specifically, to test
$H^\t_{\beta,c}:F_{n\tau}^{\t,-1}(\beta)\leq c \Longleftrightarrow F_{n\tau}^{\t}(c) \ge \beta$ in \eqref{eq:H_beta_c_T}, we use the adjustment $\bar s(X_i;c)$ or $\bar s(X_i;\underline{\tau}_{\beta}^{\textup{co}})$. Applying this adjustment yields a sharp design-based KS test for ITE quantiles.

\begin{theorem}\label{thm:sharp_bound_consist_covariate}
    For any $\beta \in (0,1)$, 
    suppose that $c<\underline{\tau}_{\beta}^{\textup{co}}=\bar{F}^{-1}(\beta)$, where $\underline{\tau}_{\beta}^{\textup{co}}$ is the covariate-informed sharp lower bound on the $\beta$ quantile of the ITE distribution in \eqref{eq:quantile_lower_bound_covariate}.
	As $n\rightarrow \infty$, 
	if $n_1/n \to \lambda \in (0,1)$ and Assumption \ref{asmp:iid_covariate} holds, then the p-value $\overline{p}_{\lceil n_1 \beta \rceil,c}^{\t, \co}$ in \eqref{eq:rand_p_val_ks_equi_covariates} 
    based on covariate adjustment of either $s(X_i)=\bar{s}(X_i;c)$ or $s(X_i)=\bar{s}(X_i;\underline{\tau}_{\beta}^{\textup{co}})$, with the form of $\bar{s}$ in \eqref{eq:cov_adj},
    for testing the null hypothesis $H^\t_{\beta, c}$ in \eqref{eq:H_beta_c_T}
    converges almost surely to zero.
\end{theorem}

Note that the optimal covariate adjustment depends on the conditional distributions of potential outcomes given covariates, including the sharp bound itself. These quantities may be estimated using external data or sample splitting. We emphasize, however, that consistent estimation is not required for the finite-sample validity of the proposed test, since the test remains valid under arbitrary covariate adjustment. The primary role of Theorem \ref{thm:sharp_bound_consist_covariate} is to characterize the ideal optimal adjustment. When this optimal adjustment can be estimated consistently, the theorem further suggests the sharpness of the design-based KS test.

From Theorem \ref{thm:sharp_bound_consist_covariate}, we can apply two forms of covariate adjustment to achieve sharp inference. One uses $s(X_i)=\bar{s}(X_i;c)$ and depends on the threshold $c$ in the null hypothesis, whereas the other uses $s(X_i)=\bar{s}(X_i;\underline{\tau}_{\beta}^{\textup{co}})$ and depends on the quantile level $\beta$. Although both adjustments yield sharp inference, each is preferable for a different inferential objective when test inversion is used to construct prediction or confidence intervals.

Specifically, to construct a prediction set for the ITE quantile $F_{n\tau}^{\t,-1}(\beta)$, we recommend using the adjustment $s(X_i)=\bar{s}(X_i;\underline{\tau}_{\beta}^{\textup{co}})$. By test inversion, the resulting acceptance region is guaranteed to be a one-sided interval and is also a valid prediction interval for $F_{n\tau}^{\t,-1}(\beta)$. In contrast, these properties may not hold when using the adjustment $s(X_i)=\bar{s}(X_i;c)$. Analogously, to construct a prediction set for the ITE distribution function $F_{n\tau}^{\t}(c)$, we recommend using the adjustment $s(X_i)=\bar{s}(X_i;c)$, which, by test inversion, yields a valid one-sided prediction interval for $F_{n\tau}^{\t}(c)$; see the supplementary material for the details and proofs.

These prediction intervals for the ITE distribution and quantile functions are pointwise, rather than simultaneously, valid because different thresholds $c$ or quantile levels $\beta$ may yield different covariate adjustments.
In practice, we can construct simultaneous intervals using a common covariate adjustment based on, for example, a single choice of $\beta$; see Section \ref{sec:illust}.

\section{Illustration: simulation and application}\label{sec:illust}

\subsection{Simulation studies comparing test statistics and evaluating covariate adjustment}\label{ssec:simu_stat_compare}
We conduct simulation studies to compare inference based on various rank statistics
and to examine the gain from covariate adjustment.
We consider two data-generating processes (DGPs) for the potential outcomes and pretreatment covariates: 
\begin{alignat}{3}\label{eq:simu_stat_compare_dgp}
    &\text{DGP I}: \quad
    &Y_i(1) &= 4X_i + 3e^{X_i/4}\varepsilon_{i1},
    \qquad
    &Y_i(0) &= 3X_i + e^{-X_i/4}\varepsilon_{i0};
    \\
    &\text{DGP II}: \quad
    &Y_i(1) &= 6X_i + e^{X_i}\varepsilon_{i1},
    \qquad
    &Y_i(0) &= \varepsilon_{i0}.
    \nonumber
\end{alignat}
In both DGPs, $X_i\overset{\text{i.i.d.}}{\sim} \text{Unif}(-1,1)$, $\varepsilon_{i1}\overset{\text{i.i.d.}}{\sim} \mathcal{N}(0,1)$, $\varepsilon_{i0}\overset{\text{i.i.d.}}{\sim} \mathcal{N}(0,1)$, and they are mutually independent.
We consider sample size 
$n\in\{10^2,10^3,10^5\}$ and completely randomize half of the units into treatment. 
For each simulation, we conduct 300 replications, where we regenerate the assignments, potential outcomes and covariates in each replication. 
We focus on the simultaneous lower prediction bounds for $\beta\in\{0.02,0.04,\ldots,0.98\}$ quantile of ITEs among the treated units.

We compare four rank-based test statistics: 
the Wilcoxon rank-sum statistic,
the Stephenson rank-sum statistic with its parameter set to $6$ \citep{stephenson1985two}, 
the RKS statistic, 
and the covariate-adjusted RKS statistic. 
For covariate adjustment, we generate an independent training dataset in the same way as the inference dataset, 
estimate the conditional distribution functions of potential outcomes given covariates using kernel estimation with a Gaussian kernel \citep{nadaraya1964estimating,watson1964smooth}, 
and 
substitute these estimates into the covariate-informed sharp lower bound $\underline{\tau}^{\mathrm{co}}_{\beta_0}$ in \eqref{eq:quantile_lower_bound_covariate} and the associated optimal adjustment $\bar s(x;\underline{\tau}^{\mathrm{co}}_{\beta_0})$ in \eqref{eq:cov_adj},
with $\beta_0$ fixed at $1/2$.
That is, we apply the estimated optimal covariate adjustment for the median of ITEs as the common adjustment for all quantiles of ITEs. This ensure the simultaneous validity of the resulting prediction bounds.

\begin{figure}[htbp]
    \centering
    \includegraphics[width=\textwidth]{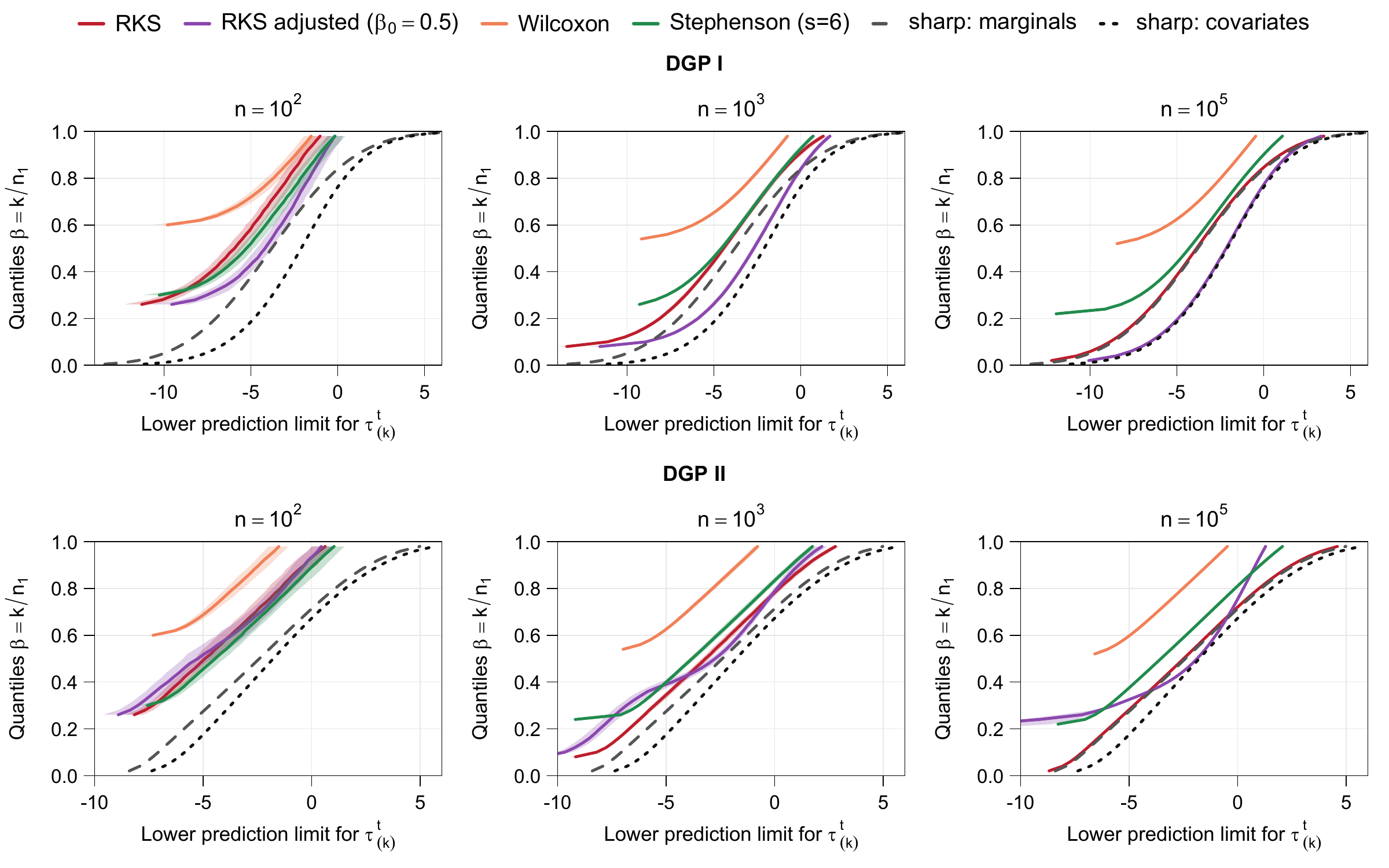}
    \caption{Comparison of simultaneous $95\%$ lower prediction bounds for the treated-group ITE quantiles under DGP I and DGP II in \eqref{eq:simu_stat_compare_dgp}
    for sample sizes $n=10^2,10^3,10^5$.
    The solid colored curves show the medians of the lower prediction bounds across $300$ replications, and the shaded regions show their empirical interquartile ranges. The dashed and dotted black curves are the marginal and covariate-informed sharp lower bounds, respectively.}
    \label{fig:simu_stat_compare_dgps}
\end{figure}

The top row in Figure \ref{fig:simu_stat_compare_dgps} summarizes the simultaneous lower prediction bounds from the simulation under DGP I. The horizontal axis shows the lower bound and the vertical axis shows the quantile level $\beta$.
At each fixed quantile level, a curve farther to the right represents a more informative lower bound. 
The solid colored curves show the medians of the lower prediction bounds, and the shaded regions show the corresponding interquartile ranges. The dotted and dashed  black curves are the sharp lower bounds with and without the covariate information; we call them marginal sharp bounds and covariate-informed sharp bounds, respectively. 
From Figure \ref{fig:simu_stat_compare_dgps}, the bounds from the unadjusted RKS statistic approach the marginal sharp bounds as $n$ increases, and their variability across replications decreases. The bounds from Wilcoxon and Stephenson rank-sum statistics remain visibly smaller than the marginal sharp bounds. Accordingly, in large samples, the RKS statistic provides more informative prediction bounds than the Wilcoxon and Stephenson rank-sum statistics. This finding confirms the consistency of the design-based inference using the RKS statistic established in Sections \ref{sec:consistency_fp} and \ref{sec:consis_super}.
Moreover, the bounds from the adjusted RKS statistic approach the covariate-informed sharp bound for the median ITE as $n$ increases, which are consistent with the results in Section \ref{sec:cov_adj}.
Interestingly, the adjustment also improves the bounds for most quantiles, with substantially smaller gaps from the covariate-informed sharp bounds at many other quantiles. 

The bottom row of Figure \ref{fig:simu_stat_compare_dgps} shows analogous results under DGP II. One notable difference from DGP I is that covariate adjustment worsens inference for most quantiles, although it helps attain the covariate-informed sharp bound for the median ITE.
This occurs because, under DGP II, the optimal adjustment $\bar s(x;\underline{\tau}^{\mathrm{co}}_{\beta_0})$ varies substantially with the quantile level $\beta_0$. Under DGP I, by contrast, the optimal adjustments are similar across quantile levels up to additive constants. See the supplementary material for a graphical illustration. In practice, we can use quantile-specific optimal adjustments for pointwise inference or choose a common adjustment that balances gains and losses across quantiles.

\subsection{A randomized experiment on vocabulary and mathematics training}
\label{ssec:educational_training}

We analyze the experiment in \citet{keller2025new}, where $2200$ participants were randomized to vocabulary training (labeled as treatment) or mathematics training (labeled as control). 
To illustrate our methods, 
we randomly divide the $2200$ participants into equally sized training and inference samples, and focus on the effect of vocabulary rather than mathematics training on posttest vocabulary scores.

We first
conduct inference for 
the $1100$ inference samples using Wilcoxon rank-sum, Stephenson rank-sum (with its parameter set to $6$), RKS, and covariate-adjusted RKS statistics. 
For the covariate adjustment, we consider three choices of $s(X_i)$ in \eqref{eq:adj_PO}: the pretest score, the estimated optimal adjustment for the median ITE based on the pretest score alone, and the corresponding adjustment based on both the pretest score and sex. The estimation of the optimal adjustment uses the training sample and follows the procedure in Section \ref{ssec:simu_stat_compare}. 
Panel A in Table \ref{tab:real_data} summarizes $90\%$ simultaneous lower confidence limits for quantiles of ITEs under different  test statistics and covariate adjustments. 
Among the unadjusted statistics, RKS provides the most informative bounds. 
The pretest-score adjustment improves the bounds for all displayed quantiles except the $90\%$ and $100\%$ quantiles. 
The estimated optimal adjustment based on the pretest score alone further improves the bounds for several quantiles, although the bound at the targeted median remains the same. 
Additionally including sex in the adjustment yields a  small further improvement at the $70\%$ quantile.

We then analyze the full sample, using sample splitting for the two covariate-adjusted statistics that require estimating the optimal adjustment as discussed in Section \ref{sec:cov_adj}.
The results are reported in Panel B of Table \ref{tab:real_data} and are similar to those in Panel A. Here, compared to the simple pretest-score adjustment, the gains from the estimated optimal adjustments are partially offset by the Bonferroni correction. In addition, as shown in the supplementary material, the simple pretest-score adjustment is already close to the estimated optimal pretest-score adjustment across several quantiles, including the median, in this application.

\begin{table*}
\caption{$90\%$ Simultaneous lower confidence bounds for ITE quantiles}
\label{tab:real_data}
\centering
\scriptsize
\setlength{\tabcolsep}{2.5pt}
\renewcommand{\arraystretch}{1.15}
\begin{tabular}{@{}l*{10}{r}@{}}
\hline
Method
 & 10\% & 20\% & 30\% & 40\% & 50\%
 & 60\% & 70\% & 80\% & 90\% & 100\% \\
\hline
\multicolumn{11}{@{}l}{\textit{Panel A: Inference sample ($\text{sample size}=1100$)}} \\
\hline
Wilcoxon
 & $-\infty$ & $-\infty$ & $-\infty$ & $-\infty$ & $-\infty$
 & $-6.00$ & $-3.00$ & $-1.00$ & 1.00 & 2.00 \\
Stephenson (s=6)
 & $-\infty$ & $-\infty$ & $-9.00$ & $-6.00$ & $-4.00$
 & $-3.00$ & $-2.00$ & $-1.00$ & 1.00 & 2.00 \\
RKS
 & $-15.00$ & $-10.00$ & $-7.00$ & $-5.00$ & $-3.00$
 & $-2.00$ & $-1.00$ & 1.00 & 2.00 & 3.00 \\
RKS + naive (pretest)
 & $-11.00$ & $-6.00$ & $-4.00$ & $-3.00$ & $-1.00$
 & $-1.00$ & 0.00 & 1.00 & 1.00 & 2.00 \\
RKS + optimal (pretest)
 & $-9.00$ & $-5.00$ & $-4.00$ & $-2.00$ & $-1.00$
 & 0.00 & 0.00 & 1.00 & 2.00 & 2.00 \\
RKS + optimal (pretest + sex)
 & $-9.00$ & $-5.00$ & $-4.00$ & $-2.00$ & $-1.00$
 & 0.00 & 1.00 & 1.00 & 2.00 & 2.00 \\
\hline
\multicolumn{11}{@{}l}{\textit{Panel B: Full sample ($\text{sample size}=2200$)}} \\
\hline
Wilcoxon
 & $-\infty$ & $-\infty$ & $-\infty$ & $-\infty$ & $-\infty$
 & $-6.00$ & $-3.00$ & $-1.00$ & 1.00 & 2.00 \\
Stephenson (s=6)
 & $-\infty$ & $-\infty$ & $-9.00$ & $-6.00$ & $-4.00$
 & $-3.00$ & $-2.00$ & $-1.00$ & 1.00 & 2.00 \\
RKS
 & $-13.00$ & $-9.00$ & $-7.00$ & $-5.00$ & $-3.00$
 & $-2.00$ & 0.00 & 1.00 & 2.00 & 3.00 \\
RKS + naive (pretest)
 & $-11.00$ & $-6.00$ & $-4.00$ & $-3.00$ & $-2.00$
 & $-1.00$ & 0.00 & 1.00 & 1.00 & 2.00 \\
RKS + optimal (pretest)
 & $-12.00$ & $-6.00$ & $-4.00$ & $-3.00$ & $-2.00$
 & $-1.00$ & 0.00 & 1.00 & 2.00 & 2.00 \\
RKS + optimal (pretest + sex)
 & $-12.00$ & $-6.00$ & $-4.00$ & $-3.00$ & $-2.00$
 & $-1.00$ & 0.00 & 1.00 & 2.00 & 2.00 \\
\hline
\end{tabular}
\end{table*}

\section{Conclusion}\label{sec:con}

We introduce a design-based approach to infer ITE quantiles using the (rank-based) KS statistic. The approach yields simultaneous confidence and prediction bounds that are valid in finite samples and can be asymptotically sharp. 
With appropriate covariate adjustment, these bounds can also be sharp given the covariate information.

It would be interesting to extend the proposed approach to other designs beyond CREs, such as stratified and matched experiments, to experiments with sample attrition \citep{li2025randomization}, and to observational studies \citep{su2024treatment, CL24quantile}.

\section*{Acknowledgments}
An earlier version of this paper was completed in 2025 as the first author's master's paper. The authors used ChatGPT and Claude to assist with proofreading, editing, and coding during the preparation of this manuscript. All errors are our own.

\bibliographystyle{abbrvnat}
\bibliography{paper-ref.bib}

\newpage
\setcounter{equation}{0}
\setcounter{section}{0}
\setcounter{figure}{0}
\setcounter{example}{0}
\setcounter{proposition}{0}
\setcounter{corollary}{0}
\setcounter{theorem}{0}
\setcounter{lemma}{0}
\setcounter{table}{0}
\setcounter{condition}{0}

\renewcommand {\theproposition} {A\arabic{proposition}}
\renewcommand {\theexample} {A\arabic{example}}
\renewcommand {\thefigure} {A\arabic{figure}}
\renewcommand {\thetable} {A\arabic{table}}
\renewcommand {\theequation} {A\arabic{equation}}
\renewcommand {\thelemma} {A\arabic{lemma}}
\renewcommand {\thesection} {A\arabic{section}}
\renewcommand {\thetheorem} {A\arabic{theorem}}
\renewcommand {\thecorollary} {A\arabic{corollary}}
\renewcommand {\thecondition} {A\arabic{condition}}
\renewcommand {\thermk} {A\arabic{rmk}}

\renewcommand {\thepage} {A\arabic{page}}
\setcounter{page}{1}

\begin{center}
	\bf \LARGE 
	Supplementary Material 
\end{center}

Appendix \ref{ssec:interval_covariate} studies the prediction and confidence intervals for ITE quantiles and distribution functions given covariate information.

Appendix \ref{ssec:Gaussian_local_power} studies the asymptotic local power of the RKS statistic for Gaussian potential outcomes.

Appendix \ref{sec:numerical_illustration} provides numerical illustration including various examples, simulation for the asymptotic power and details about covariate adjustments in simulation and application.

Appendix \ref{ssec:rks} studies the properties of the RKS statistic, including the proofs of Proposition \ref{prop:rks_properties}.

Appendix \ref{ssec:pval_rks} studies the p-values of the RKS statistic, including the proofs of Theorem \ref{thm:pval_RKS}, Corollary \ref{cor:pval_KS} and Corollary \ref{cor:pred_interval_pool}.

Appendix \ref{ssec:consist_fin} studies the consistency of the RKS statistic in the finite population, including the proofs of Theorem \ref{thm:tau_nbeta}, Theorem \ref{thm:sharp_bound_consist_fin}, Proposition \ref{prop:xi_nbeta} and Theorem \ref{thm:consistency_finite_high_Level}.

Appendix \ref{ssec:bounds_direct_proof} studies the simplified randomization-based bounds and their connection to a direct construction, including the proofs of Theorem \ref{thm:PI_simp} and Theorem \ref{thm:direct_construct}.

Appendix \ref{ssec:consist_super} studies the consistency of the RKS statistic in the superpopulation, including the proofs of Proposition \ref{prop:tau_beta} and Theorem \ref{thm:sharp_bound_consist}.

Appendix \ref{ssec:example_infinite_contact_points} provides an example of a contact set \eqref{eq:S_beta} with infinitely many elements.

Appendix \ref{ssec:asymptotic_power} studies the asymptotic power of the RKS statistic, including the proofs of Theorem \ref{thm:xin_beta_convergence}, Corollary \ref{cor:asymptotic_power}, Theorem \ref{thm:asymptotic_power_normal} and Proposition \ref{prop:local_power_maximum}.

Appendix \ref{ssec:covariate_adjust} studies the consistency of the RKS statistic after covariate adjustment, including the proofs of 
Remark \ref{rmk:adjustment_infty}, Theorem \ref{thm:bar_F_covariate_adjustment}, Remark \ref{rmk:adjustment_normal} and Theorems \ref{thm:sharp_bound_consist_covariate}.

Appendix \ref{ssec:proof_covariate_interval} provides the proofs of Theorem \ref{thm:ci_adj_ITE} and Theorem \ref{thm:ci_adj_quanitle}.

\section{Confidence and prediction intervals for ITE quantile and distribution functions under covariate adjustment} 
\label{ssec:interval_covariate}

In this section, we construct prediction and confidence sets by inverting the randomization p-value in \eqref{eq:rand_p_val_Rks_equi_covariates}. The choice of covariate adjustment depends on the estimand for test inversion. To infer a fixed quantile level $\beta$ of treatment effects, we require that the adjustment does not depend on $c$. Conversely, to infer the distribution function of treatment effects at a fixed threshold $c$, we  require the adjustment does not depend on $\beta$. This constraint preserves the monotonicity of the p-value, so that the inverted set is a one-sided interval and can be more efficiently calculated.

We first consider prediction sets for ITE quantiles using covariate adjustment $s(x)=\bar{s}(x;\underline{\tau}_{\beta}^{\textup{co}})$.

\begin{theorem}\label{thm:ci_adj_ITE}
	For any fixed $\beta \in (0,1)$, construct the set $\mathcal{I}_{\lceil n_1 \beta \rceil,\alpha}^{\t,\co}=\{c\in \mathbb{R}:p_{\lceil n_1 \beta \rceil,c}^{\t, \co}>\alpha\}$ with randomization p-value $p_{\lceil n_1 \beta \rceil,c}^{\t, \co}$ in \eqref{eq:rand_p_val_Rks_equi_covariates}, where the adjustment $s(x)$ may depend on the quantile level $\beta$ but not on $c$. Then, the p-value $p_{\lceil n_1 \beta \rceil,c}^{\t, \co}$ is nondecreasing in $c$, and     
    $\mathcal{I}_{\lceil n_1 \beta \rceil,\alpha}^{\t,\co}$ is  an individually valid $1-\alpha$ one-sided prediction interval for $\tau_{\t(\lceil n_1 \beta \rceil)}$; that is, $\Pr\{\tau_{\t(\lceil n_1 \beta \rceil)}\in\mathcal{I}_{\lceil n_1 \beta \rceil,\alpha}^{\t,\co}\}\geq 1-\alpha$. 
    Moreover, under Assumption \ref{asmp:iid_covariate} and the adjustment $s(x)=\bar{s}(x;\underline{\tau}_{\beta}^{\textup{co}})$, this interval is asymptotically sharp; that is, for any fixed $c<\underline{\tau}_{\beta}^{\textup{co}}$, $\Pr\{c \le  \inf  \mathcal{I}_{\lceil n_1 \beta \rceil,\alpha}^{\t, \co} \}\to 1$ as $n \to \infty$ if $n_1/n\to\lambda\in(0,1)$.
\end{theorem}

It is crucial here that the adjustment depends on the fixed quantile level $\beta$, rather than on the candidate threshold $c$ being inverted. 
Although each test remains valid with  adjustment depending on $c$, the p-value need not be monotone in $c$, and the inverted set may not be a one-sided interval and can be hard for computation.

By Theorem \ref{thm:ci_adj_ITE}, an analogous prediction interval for the ITE quantile among the control units can be constructed by switching the treatment and control labels. 
Note that $F_{n\tau}^{-1}(\beta) \ge \min\{\tau_{\t(\lceil n_1\beta\rceil)},\tau_{\c(\lceil n_0\beta\rceil)}\}$. 
Combining the treated and control prediction intervals 
can then yield confidence intervals for ITE quantiles among all units.

We then consider prediction sets for the distribution function of ITEs. %
Recall that $F_{n\tau}^\t(c)=n_1^{-1} \sum_{i=1}^n Z_i\mathbbm{1}(\tau_i \le c)$
denotes the proportion of treated units whose ITEs do not exceed $c$.

\begin{theorem}\label{thm:ci_adj_quanitle}
    For any fixed $c \in \mathbb{R}$, construct the set $\mathcal{J}^{\t,\co}_{c,\alpha}=\{\beta \in [0,1]:p_{\lceil n_1 \beta \rceil,c}^{\t, \co}>\alpha\}$ with randomization p-value $p_{\lceil n_1 \beta \rceil,c}^{\t, \co}$ in \eqref{eq:rand_p_val_Rks_equi_covariates} and $p_{0,c}^{\t, \co}\equiv1$, where the adjustment $s(x)$ may depend on the threshold $c$ but not on $\beta$. Then, 
    the p-value $p_{\lceil n_1 \beta \rceil,c}^{\t, \co}$ is nonincreasing in $\beta$, and     
    $\mathcal{J}^{\t,\co}_{c,\alpha}$ is a valid $1-\alpha$ one-sided prediction interval for $F_{n\tau}^\t(c)$; that is $\Pr\{F_{n\tau}^\t(c)\in\mathcal{J}_{c,\alpha}^{\t,\co}\}\geq 1-\alpha$. 
    Moreover, under Assumption \ref{asmp:iid_covariate} and the adjustment $s(x)=\bar{s}(x;c)$,
    this interval is asymptotically sharp; that is, for any fixed $\beta> \bar{F}(c)$, $\Pr\{\beta \ge \sup  \mathcal{J}_{c,\alpha}^{\t,\co}\}\to 1$ as $n \to \infty$ if $n_1/n\to\lambda\in(0,1)$.
\end{theorem}

Analogously, we can construct a $1-\alpha$ prediction interval for the proportion of control units whose ITEs do not exceed $c$,
$F^\c_{n\tau}(c)=n_0^{-1}\sum_{i=1}^n(1-Z_i)\mathbbm{1}(\tau_i\le c)$.
Note that $F_{n\tau}(c) = n_1/n\cdot F^\t_{n\tau}(c) + n_0/n\cdot F^\c_{n\tau}(c)$. 
Combining the treated and control prediction intervals can then yield confidence intervals for the ITE distribution among all units.

We emphasize that the validity of prediction or confidence intervals in Theorems \ref{thm:ci_adj_ITE} and \ref{thm:ci_adj_quanitle} holds for any prespecified covariate adjustment, which may depend on the quantile level $\beta$ or the threshold $c$, respectively. The particular adjustments $s(x)=\bar{s}(x;\underline{\tau}_{\beta}^{\textup{co}})$ and $\bar{s}(x;c)$ additionally give asymptotic sharpness guarantees.

\section{Asymptotic local power under Gaussian potential outcomes} \label{ssec:Gaussian_local_power}

We consider the asymptotic local power in Corollary \ref{cor:asymptotic_power} in the special case where both potential outcomes follow Gaussian distributions.
We show that in this case $\RS_\beta^\star$ is a singleton for any $\beta \in (0,1)$, so the power of the design-based KS test reduces to a Gaussian tail probability.
Numerical methods can be used to find $\underline{\tau}_{\beta}$ and $\RS_\beta^\star$ and then calculate the power of the test.
When the two potential outcomes have equal variances, the sharp lower bound $\underline{\tau}_{\beta}$, the element of $\RS_\beta^\star$ and the power of the KS test all have closed-form expressions.
We summarize these results in the following theorem.
Let $\phi(\cdot)$, $\Phi(\cdot)$ and  $\Phi^{-1}(\cdot)$ denote the density, distribution and quantile functions of the standard Gaussian distribution.

\begin{theorem}
\label{thm:asymptotic_power_normal}
Suppose that $Y(z) \sim \mathcal{N}(\mu_z, \sigma^2_z)$ for $z=0,1$, where $\mu_1, \mu_0 \in \mathbb{R}$ and $\sigma_1, \sigma_0> 0$,
and consider the $\beta$ quantile of the ITE for any $\beta \in (0,1)$. 
\begin{enumerate}
    \item[(i)] The contact set $\RS_\beta^\star$
defined in \eqref{eq:S_beta}
is always a singleton, and its element $b_0$, together with  $\underline{\tau}_{\beta}$,  is the unique solution to the following equations: 
\begin{align*}
F_0(b_0) = 1-\beta+F_1(b_0+\underline{\tau}_{\beta}), \quad f_0(b_0) = f_1(b_0+\underline{\tau}_{\beta}).
\end{align*}

    \item[(ii)] If further $\sigma_1 = \sigma_0 = \sigma$, then the sharp lower bound $\underline{\tau}_\beta=\mu_1-\mu_0+2\sigma \Phi^{-1}(\beta/2)$, the contact set $\RS_\beta^\star=\{ \mu_0-\sigma\Phi^{-1}(\beta/2) \}$, and
the asymptotic power in \eqref{eq:asy_power_closed_form} of the KS test has the following equivalent form: 
\begin{align}
\label{eq:pbeta_eqvar_main}
\Phi \left( \frac{\sqrt{2\lambda(1-\lambda)} \cdot h/\sigma \cdot \phi\big( \Phi^{-1}(\beta/2) \big) - \sqrt{ - \log \alpha}}{\sqrt{\beta (1-\beta/2)}} \right).
\end{align}

    \item[(iii)] The asymptotic power in \eqref{eq:pbeta_eqvar_main} is increasing in $\beta$ 
    and decreasing in $\sigma$.
\end{enumerate}
\end{theorem}

The local power in \eqref{eq:pbeta_eqvar_main} does not depend on $\mu_1$ and $\mu_0$. More generally, it is invariant to location shifts of the potential outcome distributions. This is because the gap between the sharp lower bound and the null threshold in \eqref{eq:local_alter} is unaffected by such shifts. Thus, without loss of generality, we may set $\mu_1 = \mu_0 = 0$.
As shown in Theorem \ref{thm:asymptotic_power_normal}(iii), the local power in \eqref{eq:pbeta_eqvar_main} is increasing in $\beta$ and decreasing in $\sigma$. The latter property is intuitive because $\sigma$ is a common scaling factor for both potential outcomes and the sharp lower bound. Thus, local power depends on the standardized gap between $c_n$ and $\underline{\tau}_{\beta}$, explaining why $(h, \sigma)$ enters only through the ratio $h/\sigma$.
To understand the monotonicity in $\beta$, note that when $-\log \alpha$ is close to zero, the term inside the Gaussian distribution function in \eqref{eq:pbeta_eqvar_main} is approximately proportional to the reciprocal of the asymptotic standard deviation of the $\beta/2$ sample quantile of standard Gaussian observations, which is minimized at $\beta = 1$.

\begin{figure}[h!]
	\centering
    \includegraphics[width=\textwidth]{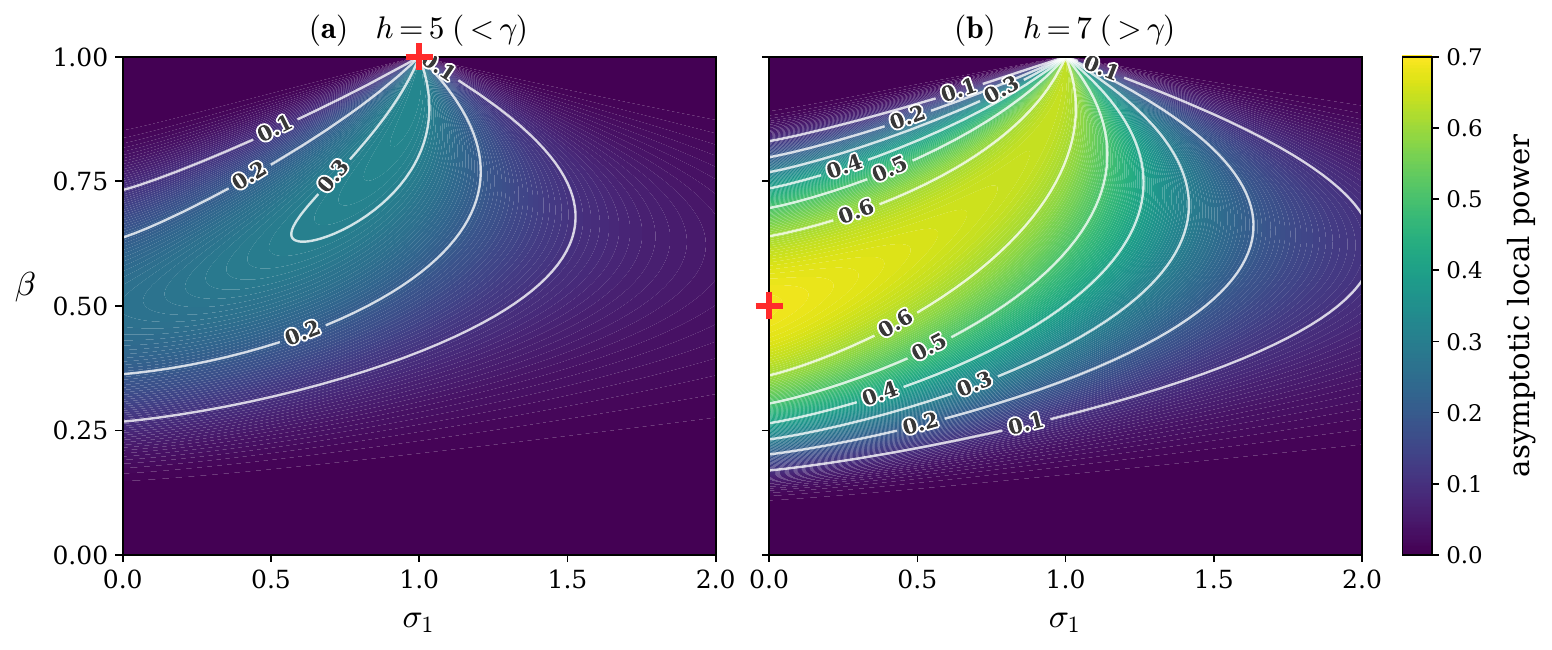}
	\caption{The relationship between the asymptotic local power of the design-based KS test and the quantile level $\beta$ and standard deviation $\sigma_1$ for two cases: (a) $h = 5 <\gamma=6.136$ and (b) $h = 7 >\gamma=6.136$, where $\gamma$ is approximately $\sigma_0\sqrt{-\pi \log \alpha }/\sqrt{\lambda(1-\lambda)}$. The red cross in each heatmap indicates the parameter combination that maximizes the power.}
	\label{fig:asym_power_analysis} 
\end{figure}

In more general settings with unequal potential outcome variances $Y(z) \sim \mathcal{N}(\mu_z, \sigma^2_z)$ for $z=0,1$, local power has a more complex dependence on $\beta$ and $(\sigma_1, \sigma_0)$.
Figure \ref{fig:asym_power_analysis} shows the local power of the KS test under varying $\beta$ and $\sigma_1$, with $\lambda$ fixed at $0.5$,
$\sigma_0$ fixed at $1$ and $h$ fixed at $5$ and $7$, respectively.
Both plots in Figure \ref{fig:asym_power_analysis} show similar patterns. 
When $\sigma_1 =1$ (i.e., the two potential outcomes have equal variances), the power increases with $\beta$ and achieves its maximum at $\beta=1$, which is consistent with Theorem \ref{thm:asymptotic_power_normal}(iii).
When $\sigma_1 \ne 1$, so that the two potential outcomes are heteroscedastic, the power becomes nonmonotone.
As $\beta$ increases,
the power first increases and then decreases, attaining its maximum at some $\beta$ between $0.5$ and $1$. Moreover, the maximizing value decreases as the heteroscedasticity of the two potential outcomes increases, or equivalently as $\sigma_1$ moves further from $1$.
In the extreme case where $\sigma_1$ is close to $0$, the maximum is achieved at $\beta = 0.5$. 
This behavior is unsurprising because, when $\sigma_1 = 0$, the ITE is equivalent to the negative control potential outcome up to some constant shift and its distribution is point identified. As discussed earlier, for Gaussian distributions, the sample median has the smallest asymptotic variance among all sample quantiles.

In addition, there is a noticeable difference between the two plots in Figure  \ref{fig:asym_power_analysis}. As $\sigma_1$ decreases from $1$ to $0$, the maximum power over all $\beta\in (0,1)$ tends to decrease in (a), while it tends to increase in (b). 
As verified in the supplementary material, across all values of $(\sigma_1, \beta)$, the power is maximized at $(\sigma_1, \beta) \to (1, 1)$ or $(0, 0.5)$, depending on whether $h$ exceeds
$\sigma_0\sqrt{-\pi \log \alpha }/\sqrt{\lambda(1-\lambda)}$, which is approximately $6.136$ in the example in Figure  \ref{fig:asym_power_analysis} with $\alpha=0.05$. 
In Figure  \ref{fig:asym_power_analysis}(a), $h<6.136$, and the maximum power is achieved as $(\sigma_1, \beta)\to(1, 1)$, while in Figure  \ref{fig:asym_power_analysis}(b),  $h>6.136$, and the maximum power is achieved as $(\sigma_1, \beta)\to(0, 0.5)$.
A more formal result is established in the following proposition.

\begin{proposition}\label{prop:local_power_maximum}
    Under the Gaussian distributions setting in Theorem \ref{thm:asymptotic_power_normal}, for any fixed $\sigma_0>0$ and $h>0$, let $\gamma = \sigma_0\sqrt{-\pi \log \alpha }/\sqrt{\lambda(1-\lambda)}$ and $\Gamma_\beta$ be the local power defined in \eqref{eq:asymptotic_power}.
    \begin{enumerate}[label=(\roman*), topsep=1ex,itemsep=-0.3ex,partopsep=1ex,parsep=1ex]
        \item If $h  < \gamma$, then $\Gamma_{\beta}$ is maximized when $(\sigma_1, \beta) \to (\sigma_0,1^-)$ along a certain path.
        \item If $h > \gamma$, then $\Gamma_{\beta}$ is maximized when $(\sigma_1,\beta) \to (0^+,0.5)$ along a certain path.
    \end{enumerate}
\end{proposition}

\section{Additional numerical illustration} \label{sec:numerical_illustration}

\subsection{An example of a contact set with infinitely many elements}\label{sec:infinite_contact_points}

\begin{figure}[htbp]
    \centering
    \includegraphics[width=0.8\textwidth]{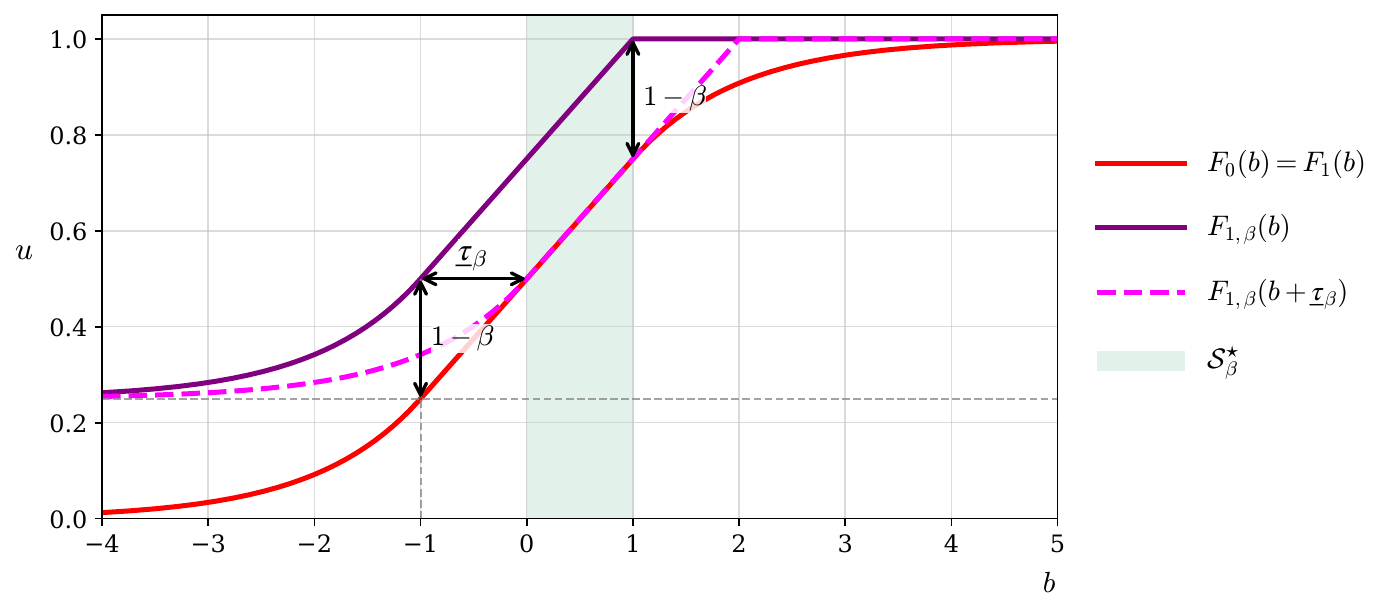}
    \caption{Illustration of the case in which the contact set $\RS_{\beta}^\star=[0,1]$ in Appendix \ref{sec:infinite_contact_points}.
    }
    \label{fig:contact_set_plots}
\end{figure}

Below we give an example of $\RS_{\beta}^\star$ with infinitely many contact points. 
Suppose that the potential outcome distributions $F_1=F_0$ have density $f(x)=1/4$ for $|x|\le 1$ and $f(x)=\exp{-(|x|-1)}/4$ for $|x|\ge 1$. As verified in the supplementary material, 
for $\beta=3/4$, we have $\underline{\tau}_{\beta}=-1$ and $F_0(b)=F_{1,\beta}(b+\underline{\tau}_{\beta})=(b+2)/4$ for all $b \in [0,1]$. These imply $\RS_{\beta}^\star=[0,1]$. 
See also Figure \ref{fig:contact_set_plots} for a graphical illustration.

\subsection{An example illustrating the necessity of the supremum in the asymptotic local power}

Let the control potential outcome follow a mixture Gaussian distribution $0.5\cdot\mathcal{N}(-5,1)+0.5\cdot\mathcal{N}(5,1)$, and the treatment potential outcome follow a Gaussian distribution $\mathcal{N}(2.216,7.398^2)$. 
For the $\beta=75\%$ quantile of the ITE, the sharp lower bound $\underline{\tau}_{\beta}=0$, and the corresponding $\RS_\beta^\star=\{b_1=-3.223,\ b_2=6.729\}$. 
With half of the units randomly assigned to treatment and control and $h$ in \eqref{eq:local_alter} being $50$, the asymptotic power in \eqref{eq:asy_power_closed_form} at significance level $\alpha = 0.05$ is 
$\Pr(\max\{\mathscr{A}(b_1, \underline{\tau}_{\beta}),\mathscr{A}(b_2, \underline{\tau}_{\beta})\} \geq l)= 0.554$, 
where $l = \sqrt{-\log\alpha /\{ 2\lambda(1-\lambda) \}}$. 
This is substantially greater than both $\Pr(\mathscr{A}(b_1, \underline{\tau}_{\beta}) \geq l)= 0.338$ and $\Pr(\mathscr{A}(b_2, \underline{\tau}_{\beta}) \geq l)= 0.377$, evaluated only at one of the maximizers, and provides more accurate approximations for the power of the KS test. 
For example, when sample size $n = 10^3, 10^4, 10^5, 10^6$, and $10^7$, the empirical power from $10^4$ Monte Carlo simulations is, respectively, $0.692, 0.643, 0.608, 0.582,$ and $0.542$.

\subsection{Simulation for the asymptotic local power of the design-based KS test}

We conduct simulations to investigate the asymptotic local power approximation of the design-based KS test. 
We generate the potential outcome vectors $(Y_i(1), Y_i(0))$ as i.i.d.~draws from the following model
and assign equal proportions of units to treatment and control:
\begin{align}
	Y_i(1) \sim \mathcal{N}(\mu_1, \sigma_1^2), 
	\quad 
	Y_i(0) \sim \mathcal{N}(\mu_0, \sigma_0^2), 
	\quad 
	Y_i(1) \ind Y_i(0).
\end{align}
We choose $\mu_1 = \mu_0 = 0, \sigma_1 = 3$ and $\sigma_0 = 1$, and set the significance level at $\alpha = 0.05$. 
We consider inference for the $\beta = 0.5$ quantile of ITEs among treated units.
From Theorem \ref{thm:asymptotic_power_normal}(i), $\RS_\beta^\star$ contains only a single point $b_0$ that can be numerically calculated. 
Accordingly, $\sup_{b \in \RS_\beta^\star} \mathscr{A}(b, \underline{\tau}_\beta) = \mathscr{A}(b_0, \underline{\tau}_\beta)$ follows a Gaussian distribution with mean and variance as in \eqref{eq:Gaussian_process_mean_variance}, 
and the asymptotic local power in \eqref{eq:asy_power_closed_form} can be equivalently written as a Gaussian tail probability.  
We vary the sample size from $n=10^3$ to $10^7$. 
For each sample size, we conduct $10^4$ simulation replications. We use the asymptotic null distribution from \eqref{eq:KS_asymptotic} rather than the actual permutation distribution to reduce computational cost.

\begin{figure}[h!]
    \centering
    \includegraphics[width=0.7\textwidth]{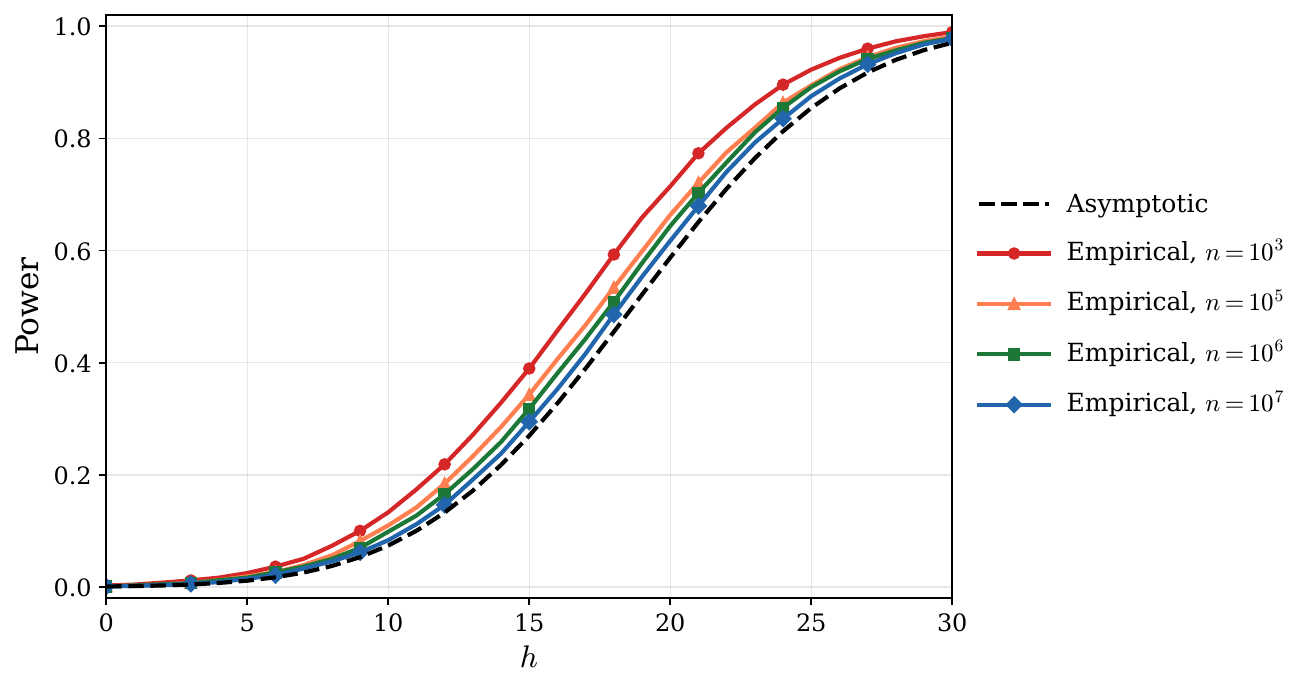}
    \caption{Empirical and asymptotic power of the design-based KS test under local alternatives. Solid curves show empirical rejection probabilities averaged over $10^4$ replications. The dashed curve is the asymptotic local power from Corollary \ref{cor:asymptotic_power}. 
    }
    \label{fig:power_convergence}
\end{figure}

Figure \ref{fig:power_convergence} shows the empirical power of the design-based KS test under the local alternative in \eqref{eq:local_alter} with varying values of $h$ at each sample size, along with the asymptotic approximation from \eqref{eq:asy_power_closed_form}. 
As the sample size increases, the empirical local power steadily approaches the asymptotic local power, in agreement with Corollary \ref{cor:asymptotic_power}.
The convergence nevertheless requires a very large sample size, which
is likely due to the slow uniform convergence to the Brownian bridge in the proof of this corollary.

\subsection{More about optimal covariate adjustments in the simulation studies in Section \ref{ssec:simu_stat_compare}}

\begin{figure}[htbp]
    \centering
    \includegraphics[width=\textwidth]{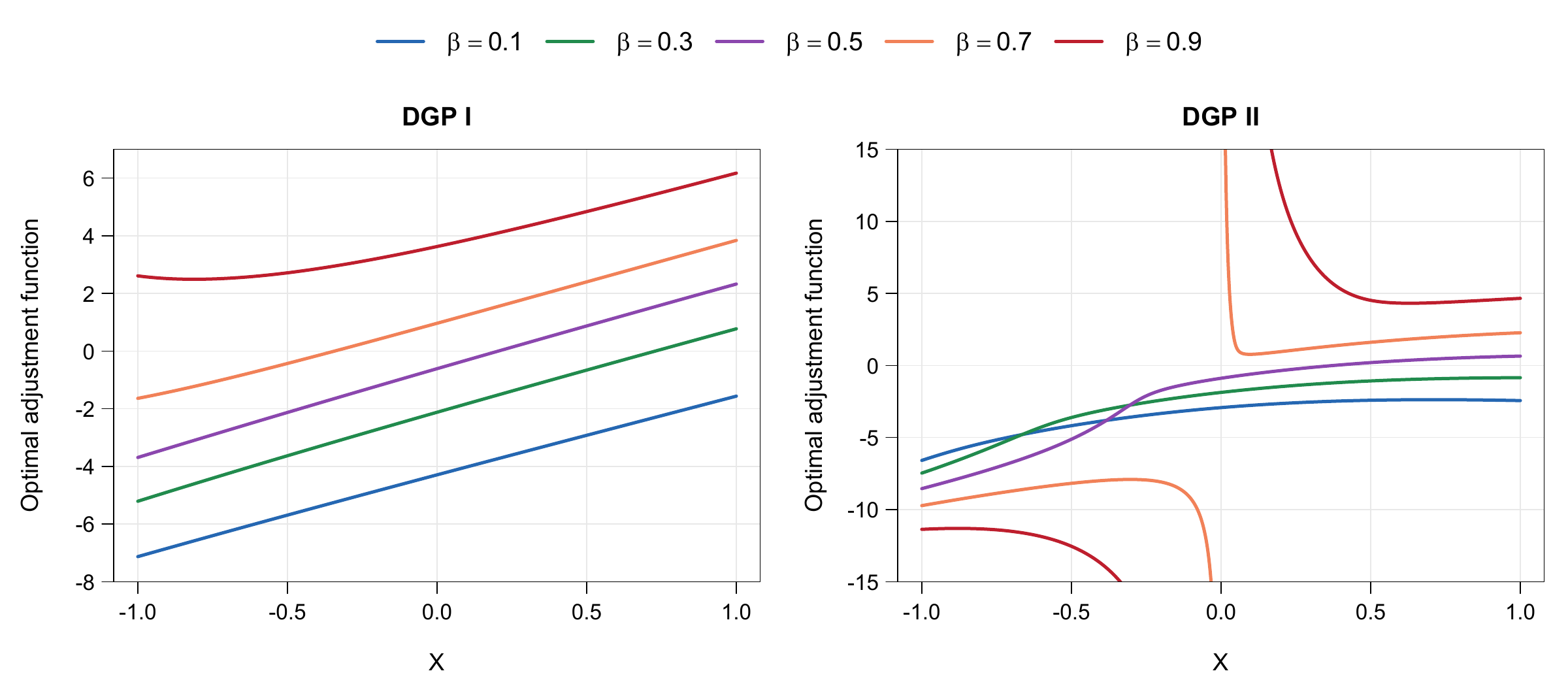}
    \caption{Optimal covariate adjustment functions \eqref{eq:cov_adj} for different quantile levels under DGP I and DGP II in \eqref{eq:simu_stat_compare_dgp}.
    }
    \label{fig:simu_adjust_compare}
\end{figure}

Figure \ref{fig:simu_adjust_compare} plots the optimal covariate adjustment
$\bar s(x;\underline{\tau}^{\mathrm{co}}_{\beta_0})$ as a function of $x$
for quantile levels $\beta_0 \in \{0.1,0.3,0.5,0.7,0.9\}$ under the two
DGPs in \eqref{eq:simu_stat_compare_dgp}.
Under DGP I, the optimal adjustment functions are approximately parallel
across quantile levels, which helps explain why the optimal adjustment
for the median also improves inference at other quantiles.
In contrast, under DGP II, the optimal adjustment functions vary
substantially across quantile levels, which helps explain why the optimal adjustment for the median may worsen inference at other quantiles.
Note that under DGP II, both potential outcomes have the conditional distribution
$\mathcal{N}(0,1)$ given $X=0$.
By \eqref{eq:cov_adj}, we therefore have
$\bar s(0;\underline{\tau}^{\mathrm{co}}_{\beta})=+\infty$
whenever $\underline{\tau}^{\mathrm{co}}_{\beta}>0$.
This explains why the optimal adjustment diverges near $x=0$ for $\beta \in \{0.7,0.9\}$ but not for $\beta\le 0.5$, since $\underline{\tau}^{\mathrm{co}}_{0.9} \ge \underline{\tau}^{\mathrm{co}}_{0.7}=0.3>0$ and $\underline{\tau}^{\mathrm{co}}_{0.1} \le \underline{\tau}^{\mathrm{co}}_{0.3} \le \underline{\tau}^{\mathrm{co}}_{0.5}=-1.77<0$.

\subsection{Estimated optimal adjustments in the application in Section \ref{ssec:educational_training}}

\begin{figure}[htbp]
    \centering
    \includegraphics[width=\textwidth]{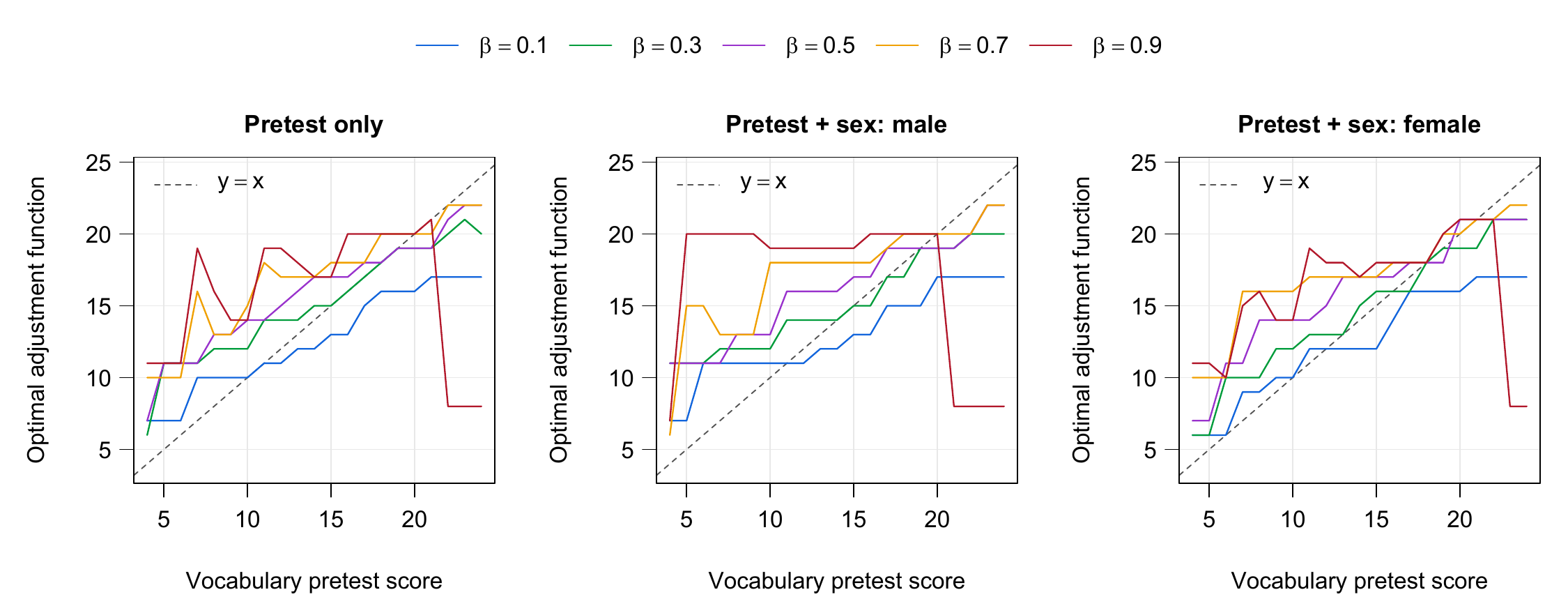}
    \caption{Estimated optimal adjustment functions at different quantile levels using pretest score alone or both pretest score and sex, for the application in Section \ref{ssec:educational_training}.
    }
    \label{fig:simu_adjust_real}
\end{figure}

Figure \ref{fig:simu_adjust_real} plots the estimated adjustment functions
using pretest score alone and using both pretest score and sex, for
$\beta \in \{0.1,0.3,0.5,0.7,0.9\}$. The curves connect estimated
adjustment values at integer pretest scores.
The dashed line $y=x$ represents the simple pretest-score adjustment.
Except for $\beta=0.9$, the curves exhibit an increasing trend similar
to that of the simple adjustment. This similarity helps explain why the adjustment
optimized for the median also improves inference at other quantile
levels in this application, and why the estimated adjustment performs
similarly to the simple adjustment.
Moreover, including sex has little effect on the estimated adjustment,
consistent with the limited additional gains reported in
Table \ref{tab:real_data}.
Overall, among the covariates considered, pretest score captures most of the useful information for adjustment, and the simple adjustment appears close to the optimal one.

\section{Proof of Proposition \ref{prop:rks_properties}} 
\label{ssec:rks}

To prove Proposition \ref{prop:rks_properties}, 
we need the following six lemmas. 
For any possible treatment assignment vector $\bs{z}$ and outcome vector $\bs{y}$, 
we use $\rank^\t_i(\bs{y})$ to denote the rank of $y_i$ for a treated unit $i$ within the treated group $\{y_l: z_l = 1, 1 \leq l \leq n\}$, 
and $\rank^\c_j(\bs{y})$ to denote the rank of $y_j$ for a control unit $j$ within the control group $\{y_l: z_l = 0, 1 \leq l \leq n\}$.
Again, we break ties in the same way as we rank all units. 
Specifically, assuming that the order of units has been randomly shuffled before analysis, we break ties by the  ``first'' method using the units' indices. 
Note that, throughout the lemmas in this section, we allow elements of $\bs{y}$ and $\bs{y}'$ to take positive or negative infinite values, treating equal infinite values as ties in the ranking, unless otherwise stated. 
In addition, in the following lemmas, we assume there are positive numbers of treated and control units in the CRE, and the vector $\bs{z}\in \{0,1\}^n$ always satisfies $0< \sum_{i=1}^n z_i < n$.

\begin{lemma}
\label{lem:dist-free}
$t_{\RKS}(\bs{Z},\bs{y})$ in (\ref{eq:RKS_statistic}) is distribution free when $\bs{Z}$ is from a CRE. That is, for any $n$-dimensional constant vectors $\bs{y}, \bs{y}^\prime$, $t_{\RKS}(\bs{Z}, \bs{y})$ and $t_{\RKS}(\bs{Z}, \bs{y}^\prime)$
follow the same distribution.
\end{lemma}

\begin{proof}[\bf Proof of Lemma \ref{lem:dist-free}]
Because we break ties in ranking, there must exist a permutation $\pi(\cdot)$ of $\{1,2,\ldots,n\}$ such that $\rank_{\pi(i)}(\bs{y}) = i$ for all $1 \leq i \leq n$. This implies that
\begin{align}
t_{\RKS}(\bs{Z},\bs{y}) &= t_{\KS}(\bs{Z},\rank(\bs{y})) \notag\\
&= \sup_{r} \Big\{ \frac{1}{n_0} \sum_{i=1}^n (1-Z_i) \I\{\rank_{i}(\bs{y}) \leq r\} - \frac{1}{n_1} \sum_{i=1}^n Z_i \I\{\rank_{i}(\bs{y}) \leq r\} \Big\} \notag \\
&= \sup_{r} \Big\{ \frac{1}{n_0} \sum_{i=1}^n (1-Z_{\pi(i)}) \I\{\rank_{\pi(i)}(\bs{y})\leq r\} - \frac{1}{n_1} \sum_{i=1}^n Z_{\pi(i)} \I\{\rank_{\pi(i)}(\bs{y})\leq r\} \Big\} \notag \\
&= \sup_{r} \Big\{ \frac{1}{n_0} \sum_{i=1}^n (1-Z_{\pi(i)}) \I\{i \leq r\} - \frac{1}{n_1} \sum_{i=1}^{n} Z_{\pi(i)} \I\{i \leq r\} \Big\} \label{eq:ks_distribution-free}.
\end{align}
Under the CRE, the treatment assignments are exchangeable across all units, which implies that 
$\bs{Z} \sim \bs{Z}_{\pi} \equiv (Z_{\pi(1)}, Z_{\pi(2)}, \ldots, Z_{\pi(n)})$. 
Consequently, 
from \eqref{eq:ks_distribution-free}, 
we can know that 
$t_{\RKS}(\bs{Z},\bs{y})$
follows the same distribution as 
\begin{align*} 
\sup_{r} \Big\{ \frac{1}{n_0} \sum_{i=1}^n (1-Z_i) \I\{i \leq r\} - \frac{1}{n_1} \sum_{i=1}^n Z_i \I\{i \leq r\} \Big\},
\end{align*} 
which does not depend on the outcome vector $\bs{y}$. Therefore, the statistic $t_{\RKS}(\bs{Z},\bs{y})$ is distribution free under the CRE, i.e., Lemma \ref{lem:dist-free} holds.  
\end{proof} 

\begin{lemma}
\label{lem:effect-increase_KS}
The KS statistic $t_{\KS}(\bs{z},\bs{y})$ defined in \eqref{eq:KS_statistic} is effect increasing. That is, for any $ \bs{z}\in \{0,1\}^n$ and any $n$-dimensional constant vectors $\bs{y}, \bs{y}'$ satisfying that 
$y_i^\prime \geq y_i$ if $z_i = 1$ and $y_i^\prime \leq y_i$ if $z_i = 0$, 
we have $t_{\KS}(\bs{z},\bs{y}') \ge t_{\KS}(\bs{z},\bs{y})$. 
\end{lemma}

\begin{proof}[\bf Proof of Lemma \ref{lem:effect-increase_KS}]
From the relation between $\bs{y}$ and $\bs{y}'$, 
for any $b \in \mathbb{R}$, we have 
\begin{align*}
\I\{y_i^\prime \leq b\} \leq \I\{y_i \leq b\} \text{ if } z_i = 1, 
\ \ \ \text{and} \ \ \ 
\I\{y_i^\prime \leq b\} \geq \I\{y_i \leq b\} \text{ if } z_i = 0.
\end{align*}
These then imply that 
\begin{align*}
t_{\KS}(\bs{z},\bs{y}^\prime) 
&= \sup_b \Big\{  \frac1{n_0} \sum_{i=1}^n (1-z_i) \I(y_i^\prime \leq b) -  \frac1{n_1} \sum_{i=1}^n z_i \I(y_i^\prime \leq b) \Big\} \\
&\geq \sup_b \Big\{  \frac1{n_0} \sum_{i=1}^n (1-z_i) \I(y_i \leq b) -  \frac1{n_1} \sum_{i=1}^n z_i \I(y_i \leq b) \Big\} \\
& = t_{\KS}(\bs{z},\bs{y}).
\end{align*}
Therefore,  Lemma \ref{lem:effect-increase_KS} holds.
\end{proof}

\begin{lemma}
\label{lem:rank_comparison}
Consider any $\bs{z}\in \{0,1\}^n$ and 
any $n$-dimensional constant vectors $\bs{y}, \bs{y}'$ satisfying that 
$y_i^\prime \geq y_i$ if $z_i = 1$ and $y_i^\prime \leq y_i$ if $z_i = 0$. 
Define $\pi(\cdot)$ to be the permutation of $\{1,2,\ldots, n\}$ such that, for any $1\le l \le n$, 
\begin{align*}
\rank^\t_l(\bs{y}) = \rank^\t_{\pi(l)}(\bs{y}^\prime ) \text{ if } z_l = 1
\ \ \text{ and } \ \ 
\rank^\c_l(\bs{y}) = \rank_{\pi(l)}^\c(\bs{y}^\prime ) \text{ if } z_l = 0, 
\end{align*}
recalling that $\rank^\t_l(\bs{y})$ (or $\rank^\c_l(\bs{y})$) denotes the rank of a treated (or control) unit within the treated (or control) group. 
That is, the within-group rank of unit $\pi(l)$ under the outcome $\bs{y}'$ is the same as that of unit $l$ under the outcome $\bs{y}$. 
Then, for any $1\le i,j\le n$ with $z_i = 1, z_j = 0$ and $\rank_j(\bs{y}) < \rank_i(\bs{y})$, 
we must have $\rank_{\pi(j)}(\bs{y}^\prime) < \rank_{\pi(i)}(\bs{y}^\prime)$.
\end{lemma}

\begin{proof}[\bf Proof of Lemma \ref{lem:rank_comparison}]
By the definition of $\bs{y}^\prime$, 
we know that $y_l^\prime \geq y_l$ if $z_l = 1$ and $y_l^\prime \leq y_l$ if $z_l = 0$. 
Thus, for any $1\le i,j\le n$ with $z_i = 1, z_j = 0$ and $\rank_j(\bs{y}) \leq \rank_i(\bs{y})$, 
we must have $y^\prime_{\pi(j)} \leq y_j \leq y_i \leq y^\prime_{\pi(i)}$. 
When $y^\prime_{\pi(j)} < y^\prime_{\pi(i)}$, we immediately have $\rank_{\pi(j)}(\bs{y}^\prime) \leq \rank_{\pi(i)}(\bs{y}^\prime)$. 
To prove Lemma \ref{lem:rank_comparison}, it suffices to consider the case where $y^\prime_{\pi(j)} = y^\prime_{\pi(i)}$. 

Suppose that $y^\prime_{\pi(j)} = y^\prime_{\pi(i)}$. Then we must have $y^\prime_{\pi(j)} = y_j = y_i = y^\prime_{\pi(i)}$. 
Below we first show that $\pi(i)\ge i$. 
We prove this by contradiction. 
Assume that $\pi(i)< i$. Then for any treated unit $l$ with a larger rank than unit $i$ under outcome vector $\bs{y}$, i.e., $z_l=1$ and $\rank_l^\t(\bs{y}) > \rank_i^\t(\bs{y})$, 
we have either $y_l > y_i$, or $y_l=y_i$ and $l>i$. 
Because $y_l^\prime\ge y_l$ and $y_i = y^\prime_{\pi(i)}$, 
we then have either $y_l^\prime > y^\prime_{\pi(i)}$, or $y_l^\prime = y^\prime_{\pi(i)}$ and $l>i>\pi(i)$, which implies that 
$\rank_l^\t(\bs{y}^\prime) > \rank_{\pi(i)}^\t(\bs{y}^\prime)$. 
Thus, any treated unit with a larger rank than unit $i$ under outcome $\bs{y}$ must also have a larger rank than unit $\pi(i)$ under outcome $\bs{y}^\prime$. 
In addition, 
because $y_i^\prime\ge y_i = y^\prime_{\pi(i)}$ and $i>\pi(i)$, 
we also have $\rank_i^\t(\bs{y}^\prime) > \rank_{\pi(i)}^\t(\bs{y}^\prime)$. 
Consequently, we must have 
\begin{align*}
    n_1 - \rank^\t_i(\bs{y})
    & = 
    \sum_{l = 1}^n z_l \I\{\rank^\t_l(\bs{y}) > \rank^\t_i(\bs{y})\}
    =
    \sum_{l \ne i} z_l \I\{\rank^\t_l(\bs{y}) > \rank^\t_i(\bs{y})\}\\
    & 
    \le \sum_{l \ne i} z_l \I\{\rank^\t_l(\bs{y}^\prime) > \rank^\t_{\pi(i)}(\bs{y}^\prime)\} 
    \\
    & < \sum_{l \ne i} z_l \I\{\rank^\t_l(\bs{y}^\prime) > \rank^\t_{\pi(i)}(\bs{y}^\prime)\} + 1
    \\
    & = \sum_{l=1}^n z_l \I\{\rank^\t_l(\bs{y}^\prime) > \rank^\t_{\pi(i)}(\bs{y}^\prime)\} 
    = n_1 - \rank^\t_{\pi(i)}(\bs{y}^\prime ),
\end{align*}
which leads to a contradiction with the definition of $\pi(\cdot)$. 
Therefore, we must have $\pi(i)\ge i$. By the same logic, we can show that $\pi(j) \le j$.
Because $y_j = y_i$ and $\rank_j(\bs{y}) \leq \rank_i(\bs{y})$, we then have 
$\pi(j) \le j < i \le \pi(i)$. 
Because  $y^\prime_{\pi(j)}= y^\prime_{\pi(i)}$, this further implies that $\rank_{\pi(j)}(\bs{y}^\prime) < \rank_{\pi(i)}(\bs{y}^\prime)$.

From the above, 
Lemma \ref{lem:rank_comparison} holds.
\end{proof}

\begin{lemma}
\label{lem:effect-increase_RKS}
The test statistic $t_{\RKS}(\bs{z},\bs{y})$ in \eqref{eq:RKS_statistic} is effect-increasing, i.e., 
$t_{\RKS}(\bs{z},\bs{y}') \ge t_{\RKS}(\bs{z},\bs{y})$ for
any $n$-dimensional constant vectors $\bs{y}, \bs{y}'$ satisfying that 
$y_i^\prime \geq y_i$ if $z_i = 1$ and $y_i^\prime \leq y_i$ if $z_i = 0$. 
\end{lemma}

\begin{proof}[\bf Proof of Lemma \ref{lem:effect-increase_RKS}]
From Lemma \ref{lem:rank_comparison}, there exists a permutation $\pi(\cdot)$ such that, for $i = 1,2,\dots n$, 
\begin{enumerate}[label=(\roman*), topsep=1ex,itemsep=-0.3ex,partopsep=1ex,parsep=1ex]
\item $z_{\pi(i)} = 1 \text{ and } \rank^\t_i(\bs{y}) = \rank^\t_{\pi(i)}(\bs{y}^\prime ) \quad \text{if } z_i = 1$,  
\item $z_{\pi(i)} = 0 \text{ and } \rank^\c_i(\bs{y}) = \rank_{\pi(i)}^\c(\bs{y}^\prime ) \quad \text{if }  z_i = 0$.
\end{enumerate}
Because $y_i^\prime \geq y_i$ for all $i$ with $z_i = 1$, we must have $ y_{\pi(i)}^\prime \geq y_i$ for all $i$ with $z_i = 1$. Similarly, because $ y_j^\prime \leq y_j$ for all $j$ with $z_j = 0$, we have $y_{\pi(j)}^\prime \leq y_j $ for all $j$ with $z_j = 0$.

Consider first any treated unit $i$ with $z_i=1$. We have
\begin{align*}
    \rank_i({\bs{y}}) &= \sum_{k=1}^{n} z_k \I\{\rank_k(\bs{y}) \leq \rank_i(\bs{y})\} + \sum_{k=1}^{n} (1 - z_k) \I\{\rank_k(\bs{y}) \leq \rank_i(\bs{y})\} \\
    &= \rank^\t_i(\bs{y}) + \sum_{k=1}^{n} (1 - z_k) \I\{\rank_k(\bs{y}) < \rank_i(\bs{y})\}  \\
    &\leq \rank^\t_{\pi(i)}(\bs{y}^\prime ) + \sum_{k=1}^{n} (1 - z_k) \I\{\rank_{\pi(k)}(\bs{y}^\prime) < \rank_{\pi(i)}(\bs{y}^\prime)\} 
    \\
    & = 
    \rank^\t_{\pi(i)}(\bs{y}^\prime ) + \sum_{k=1}^{n} (1 - z_k) \I\{\rank_{k}(\bs{y}^\prime) < \rank_{\pi(i)}(\bs{y}^\prime)\} 
    \\
    &= \rank_{\pi(i)}(\bs{y}^\prime),
\end{align*}
where the inequality follows from the construction of $\pi(\cdot)$ and Lemma  \ref{lem:rank_comparison}, and the second last equality follows from the fact that $\pi(\cdot)$ permutes units within the same treatment group.

Consider then any control unit $i$ with $z_i=0$.  
By the same logic, 
we have 
\begin{align*}
    \rank_i({\bs{y}}) &= \sum_{k=1}^{n} z_k \I\{\rank_k(\bs{y}) \leq \rank_i(\bs{y})\} + \sum_{k=1}^{n} (1 - z_k) \I\{\rank_k(\bs{y}) \leq \rank_i(\bs{y})\} \\
    &= \sum_{k=1}^{n} z_k \I\{\rank_k(\bs{y}) \leq \rank_i(\bs{y})\} + \rank^\c_i(\bs{y})   \\
    &\geq \sum_{k=1}^{n} z_k \I\{\rank_{\pi(k)}(\bs{y}^\prime) \leq \rank_{\pi(i)}(\bs{y}^\prime)\} + \rank^\c_{\pi(i)}(\bs{y}^\prime ) \\
    &\geq \sum_{k=1}^{n} z_k \I\{\rank_{k}(\bs{y}^\prime) \leq \rank_{\pi(i)}(\bs{y}^\prime)\} + \rank^\c_{\pi(i)}(\bs{y}^\prime ) \\
    &= \rank_{\pi(i)}(\bs{y}^\prime).
\end{align*}

From the above, 
\begin{align*}
t_\RKS(\bs{z}, \bs{y}^\prime) 
&= \sup_r \Big\{ \frac{1}{n_0} \sum_{i=1}^n (1-z_i) \I\{\rank_{\pi(i)}(\bs{y}^\prime) \leq r\} - \frac{1}{n_1} \sum_{i=1}^n z_i \I\{\rank_{\pi(i)}(\bs{y}^\prime) \leq r\}\Big\}\\
&\geq \sup_r \Big\{ \frac{1}{n_0} \sum_{i=1}^n (1-z_i) \I\{\rank_i(\bs{y}) \leq r\} - \frac{1}{n_1} \sum_{i=1}^n z_i \I\{\rank_i(\bs{y}) \leq r\}\Big\}\\
&= t_\RKS(\bs{z}, \bs{y}).
\end{align*}
Therefore, we derive Lemma \ref{lem:effect-increase_RKS}. 
\end{proof}

\begin{lemma}
\label{lem:RKS_KS_inequality}
\begin{enumerate}
    \item[(a)] The RKS statistic $t_{\RKS}(\bs{z},\bs{y})$ is bounded below by the original KS statistic $t_{\KS}(\bs{z},\bs{y})$, in the sense that $t_{\RKS}(\bs{z},\bs{y}) \geq t_{\KS}(\bs{z},\bs{y})$ for all $\bs{z} \in \{0,1\}^n$ and $\bs{y}$. 
    \item[(b)] Suppose that the coordinates of $\bs{y}$ are all strictly less than infinity (i.e., their values are either finite or minus infinity). If the coordinates of $\bs{y}$ are all different across all units, or the ties occur among only treated units with outcomes strictly less than those of all control units,  
 then $t_{\KS}(\bs{z},\bs{y}) = t_{\RKS}(\bs{z},\bs{y})$.
\end{enumerate}

\end{lemma}

\begin{proof}[\bf Proof of Lemma \ref{lem:RKS_KS_inequality}]

We first prove (a). 
Let $R_i=\rank_i(\bs{y})$ for $1\le i \le n$, and define
\[
D_r=
\frac{1}{n_0}\sum_{i=1}^n(1-z_i)\I\{R_i\leq r\}
-
\frac{1}{n_1}\sum_{i=1}^n z_i\I\{R_i\leq r\}, 
\qquad (r=0,\ldots,n).
\]
We can verify that $D_0=D_n=0$ and
$t_{\RKS}(\bs{z},\bs{y})=\max_{0\leq r\leq n}D_r$.
For any finite $b$, let $m(b)=\#\{i:y_i\leq b\}$.
We can verify that, for any $1\le i \le n$ and $b\in \mathbb{R}\cup\{\pm \infty\}$, $\I\{y_i\leq b\}=\I\{R_i\leq m(b)\}$. 
Consequently, by definition, 
\begin{align*}
    t_{\KS}(\bs{z},\bs{y})
    & =
    \sup_b \Big\{  \frac1{n_0} \sum_{i=1}^n (1-z_i) \I(y_i \leq b) -  \frac1{n_1} \sum_{i=1}^n z_i \I(y_i \leq b) \Big\}
    \\
    & = \sup_b \Big\{  \frac1{n_0} \sum_{i=1}^n (1-z_i) \I\{R_i\leq m(b)\} -  \frac1{n_1} \sum_{i=1}^n z_i \I\{R_i\leq m(b)\} \Big\}\\
    & = \sup_{b\in\mathbb{R}}D_{m(b)}
\leq \max_{0\leq r\leq n}D_r
=t_{\RKS}(\bs{z},\bs{y}).
\end{align*}

We then prove (b). 
Because $D_0=D_n=0$, there must exist a maximizer
$r_*\in\{1,\ldots,n\}$ such that $D_{r_*} = \max_{0\leq r\leq n}D_r \ge 0$.
Moreover, the outcome or the coordinate of $\bs{y}$ at rank $r_*$ cannot be tied with other coordinates. 
Otherwise, for any coordinate $i$ whose rank is less than or equal to $r_*$, we must have $z_i=1$ and $y_i$ strictly less than all control units, which then implies that $D_{r_*} = - r_*/n_1 < 0$, a contradiction. 
Below we find a finite threshold $b_*$ such that $m(b_*) = r_*$. 
If the outcome at rank $r_*$ is finite, then we can let $b_*$ take this outcome value; 
otherwise, the outcome at rank $r_*$ is minus infinity, $r_*=1$, and we can take $b_*$ be to any finite value strictly less than all the finite outcomes. 
Therefore, we have
\[
t_{\RKS}(\bs{z},\bs{y})
=D_{r_*}
=D_{m(b_*)}
\leq \sup_{b\in\mathbb{R}}D_{m(b)}
=t_{\KS}(\bs{z},\bs{y}).
\]
From (a), this then implies that $t_{\RKS}(\bs{z},\bs{y}) = t_{\KS}(\bs{z},\bs{y}).$

From the above, Lemma \ref{lem:RKS_KS_inequality} holds.
\end{proof}

\begin{lemma}
\label{lem:null_distribution_equivalence}
When $\bs{Z}$ is from a CRE with $n_1$ units receiving treatment, and the coordinates of $\bs{W}$ are i.i.d.\ from some continuous distribution, we then have 
$t_{\KS}(\bs{a},\bs{W}) \sim t_{\RKS}(\bs{Z},\bs{y})$, for any $n$-dimensional constant vector $\bs{y}$ and $\bs{a} = (a_1, a_2, \ldots, a_n)^\top \in \{0,1\}^n$ satisfying that $\sum_{i=1}^n a_i = n_1$. 
\end{lemma}

\begin{proof}[\bf Proof of Lemma \ref{lem:null_distribution_equivalence}]
Because the coordinates of $\bs{W}$ are i.i.d. from a continuous distribution, they are distinct almost surely. Let $\sigma_W$ denote the random permutation satisfying
\begin{align*}
W_{\sigma_W(1)}
<
W_{\sigma_W(2)}
<
\cdots
<
W_{\sigma_W(n)}.
\end{align*}
By exchangeability, $\sigma_W$ is uniformly distributed over all permutations of $\{1,\ldots,n\}$. Consequently, the treatment-label sequence ordered by $\bs{W}$,
\begin{align*}
L_W
=
\bigl(
a_{\sigma_W(1)},
\ldots,
a_{\sigma_W(n)}
\bigr),
\end{align*}
is uniformly distributed over
\begin{align*}
\mathcal{A}_{n_1}
=
\left\{
\ell\in\{0,1\}^n:
\sum_{r=1}^n \ell_r=n_1
\right\}.
\end{align*}
From Lemma \ref{lem:RKS_KS_inequality}, we have, almost surely, 
\begin{align*}
    t_{\KS}(\bs{a},\bs{W})
    & = t_{\KS}(\bs{a},\rank( \bs{W}) )
    = t_{\KS}(L_W, \bs{v}), 
\end{align*}
where $\bs{v} = (1, 2, \ldots, n)$.

Next, let $\sigma_y$ be the unique permutation satisfying
\begin{align*}
\rank_{\sigma_y(r)}(\bs{y})=r,
\qquad
1\le r\le n,
\end{align*}
where ties are broken according to the prescribed rule. Because $\bs{Z}$ follows a completely randomized experiment and $\sigma_y$ is fixed, the ordered treatment-label sequence
\begin{align*}
L_y
=
\bigl(
Z_{\sigma_y(1)},
\ldots,
Z_{\sigma_y(n)}
\bigr)
\end{align*}
is also uniformly distributed over $\mathcal{A}_{n_1}$, 
and 
\begin{align*}
    t_{\RKS}(\bs{Z},\bs{y})
    & = 
    t_{\KS}(\bs{Z},\rank(\bs{y}) )
    = 
    t_{\KS}(L_y, \bs{v} ),
\end{align*}
where $\bs{v} = (1, 2, \ldots, n)$ is defined the same as before.

From the above, we then have 
\begin{align*}
    t_{\KS}(\bs{a},\bs{W})
    = t_{\KS}(L_W, \bs{v})
    \sim t_{\KS}(L_y, \bs{v} )
    = t_{\RKS}(\bs{Z},\bs{y}).
\end{align*}
Therefore, Lemma \ref{lem:null_distribution_equivalence} holds.
\end{proof}

\begin{proof}[\bf Proof of Proposition \ref{prop:rks_properties}]
    (i) follows from Lemma \ref{lem:dist-free}, 
    (ii) follows from Lemmas \ref{lem:effect-increase_KS} and \ref{lem:effect-increase_RKS}, 
    (iii) follows from Lemma \ref{lem:RKS_KS_inequality}, 
    and (iv) follows from Lemma \ref{lem:null_distribution_equivalence}. 
\end{proof} 

\section{Proofs of Theorem \ref{thm:pval_RKS}, Corollary \ref{cor:pval_KS}, Remark \ref{rmk:comp_p_RKS_KS} and Corollary \ref{cor:pred_interval_pool}} 

\label{ssec:pval_rks}

\begin{proof}[\bf Proof of Theorem \ref{thm:pval_RKS}]
First, for any fixed $r \in \mathbb{R}$, define the rank-sum statistic $t_r(\bs{z}, \bs{y}) = \sum_{i=1}^n z_i \phi_r(\rank_{i}(\bs{y}))$,  where $\phi_r(x) = \I\{x>r\}$ is monotone nondecreasing in $x$. 
From \citet[Theorem 1]{CL24quantile},  which applies \citet[Theorem 3]{caughey2023randomisation}, we have 
\begin{align*}
    \inf_{\bs \delta \in \mathcal{H}_{k,c}^\t} t_r (\bs Z, \bs{Y}_{\bs{\delta}}) = t_r(\bs Z, \bs Y - \bs{Z} \circ \bs{\xi}_{k,c}^\t). 
\end{align*}
Here $\bs{Y}_{\bs{\delta}}=\bs Y - \bs{Z} \circ \bs{\delta}$ and $\bs{\xi}_{k,c}^\t$ is defined in \eqref{eq:xi_kc}.
As a result, for any $\bs \delta \in \mathcal{H}_{k,c}^\t$, we have
\begin{align*}
    t_\RKS(\bs{Z}, \bs{Y}_{\bs \delta}) 
    &= \sup_r \Big\{ \frac{1}{n_0} \sum_{i=1}^n (1-Z_i) \I\{\rank_{i}(\bs{Y}_{\bs \delta}) \leq r\} - \frac{1}{n_1} \sum_{i=1}^n Z_i \I\{\rank_{i}(\bs{Y}_{\bs \delta}) \leq r\}\Big\}\\
    &= \sup_r \Big\{ \frac{1}{n_0} \sum_{i=1}^n  \I\{\rank_{i}(\bs{Y}_{\bs \delta}) \leq r\} - \left( \frac{1}{n_0} + \frac{1}{n_1} \right) \sum_{i=1}^n Z_i \I\{\rank_{i}(\bs{Y}_{\bs \delta}) \leq r\} 
    \Big\}\\
    &= \sup_r \Big\{ \frac{1}{n_0} \sum_{i=1}^n  \I\{i \leq r\} - \frac{n}{n_1 n_0} \big(n_1- t_r(\bs{Z}, \bs{Y}_{\bs \delta})\big) 
    \Big\} \\
    &\ge \sup_r \Big\{ \frac{1}{n_0} \sum_{i=1}^n  \I\{i \leq r\} - \frac{n}{n_1 n_0} \big(n_1- t_r(\bs{Z}, \bs Y - \bs{Z} \circ \bs{\xi}_{k,c}^\t)\big) 
    \Big\} \\
    &= t_\RKS(\bs{Z}, \bs Y - \bs{Z} \circ \bs{\xi}_{k,c}^\t).
\end{align*}
Note that, although $\bs\xi_{k,c}^\t$ contains $+\infty$ coordinates, replacing
them by a sufficiently large finite constant preserves the
ranks of the treated units up to a permutation and consequently the value of the test statistic. Thus, the above
lower bound is attainable as an infimum over the null space.
Consequently, we have
\begin{align*}
    p_{k,c}^\t = G_n \Big( \inf_{\bs \delta \in \mathcal{H}_{k,c}^\t} t_{\RKS}(\bs Z, \bs Y - \bs{Z} \circ \bs{\delta}) \Big) 
    = G_n \Big( t_{\RKS}(\bs Z, \bs Y - \bs{Z} \circ \bs{\xi}_{k,c}^\t) \Big),
\end{align*}
and $p_{k,c}^\t \ge  G_n \Big( t_{\RKS}(\bs Z, \bs Y - \bs{Z} \circ \bs{\tau}) \Big)$ if $\bs{\tau} \in \mathcal{H}_{k,c}^\t$ or equivalently the null hypothesis $H^\t_{k, c}$ holds.
From the validity of the standard Fisher randomization test, $G_n ( t_{\RKS}(\bs Z, \bs Y - \bs{Z} \circ \bs{\tau}))= G_n ( t_{\RKS}(\bs Z, \bs Y(0)))$ is stochastically no less than the uniform distribution on $(0,1)$. Therefore, for any $\alpha \in (0, 1),$
\begin{align*}
\Pr\Big(p_{k,c}^\t \le \alpha, H^\t_{k, c} \text{ holds} \Big) 
& \le
\Pr\Big(G_n \big( t_{\RKS}(\bs Z, \bs Y - \bs{Z} \circ \bs{\tau}) \big)  \le 
\alpha, H^\t_{k, c} \text{ holds} \Big) \\
&
\le 
\Pr\Big(G_n \big( t_{\RKS}(\bs Z, \bs Y - \bs{Z} \circ \bs{\tau}) \big) \le 
\alpha  \Big) \le
\alpha, 
\end{align*}
which implies that the p-value $p_{k,c}^\t$ is valid for testing $H_{k,c}^\t$.
From the above, Theorem \ref{thm:pval_RKS} holds.
\end{proof}

\begin{proof}[\bf Proof of Corollary \ref{cor:pval_KS}]

Because $t_{\RKS}(\bs{z}, \bs{y}) \geq t_{\KS}(\bs{z}, \bs{y})$ by Lemma \ref{lem:RKS_KS_inequality} and $G_n$ is monotone nonincreasing,
 we have 
\begin{align*}
\overline{p}_{k,c}^\t =  G_n\Big(t_{\KS}(\bs{Z}, \bs Y - \bs{Z} \circ \bs{\xi}_{k,c}^\t) \Big) \geq G_n \Big (t_{\RKS}(\bs{Z}, \bs Y - \bs{Z} \circ \bs{\xi}_{k,c}^\t ) \Big) = p_{k,c}^\t.
\end{align*}
From Theorem \ref{thm:pval_RKS}, this implies that $\overline{p}_{k,c}^\t$ is also a valid p-value for $H^\t_{k, c}$. 
In the remainder of the proof, we will demonstrate the equivalent forms of $\overline{p}_{k,c}^\t$ in \eqref{eq:rand_p_val_ks_equi}.

First, due to the effect-increasing property in Lemma \ref{lem:effect-increase_KS}, 
the infimum of
$t_\KS(\bs Z, \bs Y - \bs Z \circ \bs \delta)$ over $\bs \delta \in \mathcal{H}^\t_{k,c}$ must be achieved 
at some $\bs{\delta}$ in the set $\mathcal{S}_{k,c}^\t$, 
where $\mathcal{S}_{k,c}^\t = \{ \bs{\delta} \in \{c,+\infty\}^n : \sum_{i=1}^{n} \I\{\delta_i = +\infty\} = \sum_{i=1}^{n} Z_i\I\{\delta_i = +\infty\} = n_1-k\}$\footnote{Note that the value of the test statistic at any $\bs{\delta}\in\mathcal{S}_{k,c}^{\t}$ can also be
attained at some element of $\mathcal{H}_{k,c}^{\t}$, obtained by replacing the $+\infty$
coordinates of $\bs{\delta}$ with a sufficiently large constant.}.
Define 
$\hat{F}_{n}^\t(b; \bs{\delta}) = n_1^{-1} \sum_{i=1}^n Z_i \mathbbm{1} \{ Y_i - \delta_i \leq b \}$, and 
$Y^\t_{(i)}$ as the $i$th smallest observed outcome among treated units for $1\le i \le n_1$.
By definition, 
for any $\bs{\delta} \in \mathcal{S}_{k,c}^\t$ and $b \in \mathbb{R}$, 
we have 
\begin{align}\label{eq:bound_F_n_t_KS}
    \hat{F}_{n}^\t(b; \bs{\delta}) &\equiv 
    \frac{1}{n_1} \sum_{i=1}^n Z_i \mathbbm{1} \{ Y_i - \delta_i \leq b \}
    \nonumber
    \\
    &= \frac{1}{n_1} \sum_{i=1}^n Z_i \mathbbm{1} \{ Y_i - \delta_i \leq b,  \delta_i=c \} + \frac{1}{n_1} \sum_{i=1}^n Z_i \mathbbm{1} \{ Y_i - \delta_i \leq b, \delta_i = + \infty \}
    \nonumber
    \\
    &= \frac{1}{n_1} \sum_{i=1}^n Z_i \mathbbm{1}\{\delta_i=c\} \mathbbm{1} \{ Y_i \leq b+c\} + \frac{n_1-k}{n_1}
    \nonumber
    \\
    & \le \frac{1}{n_1} \sum_{j=1}^k \mathbbm{1} \{ Y^\t_{(j)} \leq b+c\} + \frac{n_1-k}{n_1}, \nonumber
    \\
    &= \hat{F}_{n}^\t(b; \bs{\xi}_{k,c}^\t). 
\end{align}
Consequently, we have, for any  $\bs{\delta} \in \mathcal{H}_{k,c}^\t$, 
\begin{align*} 
t_{\KS}(\bs{Z}, \bs Y - \bs{Z} \circ \bs{\delta}) 
& = 
\sup_b \{ \hat{F}_{n}^\c (b ) - \hat{F}_{n}^\t(b; \bs{\delta}) \}
\\
& \ge 
\sup_b \{ \hat{F}_{n}^\c (b ) - \hat{F}_{n}^\t(b; \bs{\xi}_{k,c}^\t) \}
= t_{\KS}(\bs{Z}, \bs Y - \bs{Z} \circ \bs{\xi}_{k,c}^\t) . 
\end{align*} 
By the definition in \eqref{eq:rand_p_val_ks}, 
$\overline{p}_{k,c}^\t$ is equivalently 
\begin{align*}
\overline{p}_{k,c}^\t & \equiv G_n \Big( \inf_{\bs \delta \in \mathcal{H}_{k,c}^\t} t_{\KS}(\bs Z, \bs Y - \bs{Z} \circ \bs{\delta}) \Big)  =  G_n\{t_{\KS}(\bs{Z}, \bs Y - \bs{Z} \circ \bs{\xi}_{k,c}^\t) \}\\
& 
=
G_n \Big( 
\sup_b \{ \hat{F}_{n}^\c (b) - \hat{F}_{n}^\t(b; \bs{\xi}_{k,c}^\t) \} \Big).
\end{align*}

Second, we prove that 
\begin{align}\label{eq:F_delta^* simplification}
\hat{F}_{n}^\t(b; \bs{\xi}_{k,c}^\t) = \min \Big \{ 1, 1 - \frac{k}{n_1} + \hat{F}_{n}^\t(b + c) \Big \} \equiv \hat F_{n, k/n_1}^\t(b + c). 
\end{align}
We consider the following two cases for $c$. 
\begin{itemize}
    \item Consider the case where $ c \geq Y_{(k)}^\t - b$. 
In this case, 
using \eqref{eq:bound_F_n_t_KS}, we have 
\begin{align*}
    \frac{1}{n_1} \sum_{j=1}^k \I \{ Y_{(j)}^\t \leq b + c\} = \frac{k}{n_1}, \ \text{ and }\  \hat{F}_{n}^\t(b; \bs{\xi}_{k,c}^\t) =  \frac{k}{n_1} + \frac{n_1 - k}{n_1}  = 1.     
\end{align*}
In addition, because 
\begin{align*}
\hat{F}^\t_{n}(b+c) = \frac{1}{n_1} \sum_{i=1}^n Z_i \I \{ Y_i  \leq b + c\} \geq \frac{1}{n_1} \sum_{j=1}^k \I \{ Y_{(j)}^\t \leq b + c\} = \frac{k}{n_1}, 
\end{align*}
we have
\begin{align*}
\hat F_{n, k/n_1}^\t(b + c)\equiv
\min \Big \{ 1, 1 - \frac{k}{n_1} + \hat{F}_{n}^\t(b + c) \Big \}
= 1.
\end{align*}
Thus, \eqref{eq:F_delta^* simplification} holds in this case.

\item 
Consider the case where $c < Y_{(k)}^\t - b$. 
From \eqref{eq:bound_F_n_t_KS}, 
\begin{align*}
   \hat{F}_{n}^\t(b; \bs{\xi}_{k,c}^\t) &= \frac{1}{n_1} \sum_{j=1}^k \I \{ Y_{(j)}^\t \leq b + c\} + \frac{n_1 - k}{n_1} 
   = \frac{1}{n_1} \sum_{j=1}^{n_1} \I \{ Y_{(j)}^\t \leq b + c\} + \frac{n_1 - k}{n_1} \\
    &= \frac{1}{n_1} \sum_{i=1}^{n} Z_i\I \{ Y_i \leq b +c\} + \frac{n_1 - k}{n_1} 
    = \hat{F}^\t_{n}(b+c) + 1 - \frac{k}{n_1},
\end{align*}
where the second equality holds because $Y_{(j)}^\t  \geq Y_{(k)}^\t  > b+c$ for all $j>k$. In addition,
because 
\begin{align*}
\hat{F}^\t_{n}(b+c) & = \frac{1}{n_1} \sum_{i=1}^{n} Z_i\I \{ Y_i  \leq b + c\} = \frac{1}{n_1} \sum_{j=1}^{n_1} \I \{ Y_{(j)}^\t  \leq b +c\} = \frac{1}{n_1} \sum_{j=1}^{k} \I \{ Y_{(j)}^\t  \leq b + c\} \\
& \leq \frac{k}{n_1}, 
\end{align*}
we have
\begin{align*}
\hat{F}_{n, k/n_1}^\t(b + c)
\equiv
\min \Big \{ 1, 1 - \frac{k}{n_1} + \hat{F}_{n}^\t(b + c) \Big \}
= 1 - \frac{k}{n_1} + \hat{F}_{n}^\t(b + c ) . 
\end{align*}
Thus, \eqref{eq:F_delta^* simplification} holds in this case as well. 
\end{itemize}

From the above, 
\begin{align*}
\overline{p}_{k,c}^\t &=  G_n\Big( t_{\KS}(\bs{Z}, \bs{Y}-\bs{Z} \circ \bs{\xi}_{k,c}^\t) \Big) 
= G_n \Big(\sup_b \{ \hat{F}_{n}^\c (b) - \hat{F}_{n}^\t(b; \bs{\xi}_{k,c}^\t) \}\Big) \\
&= G_n \Big( \sup_b \{ \hat{F}_{n}^\c(b) - \hat{F}_{n, k/n_1}^\t(b + c) \} \Big).
\end{align*}
This completes the proof of Corollary \ref{cor:pval_KS}.
\end{proof}

\begin{proof}[\bf Comment on when $\overline{p}_{k,c}^\t = p_{k,c}^\t$]
   If $\bs Y - \bs{Z} \circ \bs{\xi}_{k,c}^\t$ has ties only among treated units with outcomes of negative infinity, then $t_{\KS}(\bs{Z}, \bs Y - \bs{Z} \circ \bs{\xi}_{k,c}^\t) = t_{\RKS}(\bs{Z}, \bs Y - \bs{Z} \circ \bs{\xi}_{k,c}^\t)$ by Lemma \ref{lem:RKS_KS_inequality} and therefore $\overline{p}_{k,c}^\t = p_{k,c}^\t$.
\end{proof}

\begin{proof}[\bf Proof of Corollary \ref{cor:pred_interval_pool}]
Corollary \ref{cor:pred_interval_pool} follows by the same logic as Theorems 1 and 2 in \citet{CL24quantile}. We omit its proof for conciseness. 
\end{proof}

\begin{proof}[\bf Comment on the pooled intervals]\label{rmk:pred_interval_pool}
    In Corollary \ref{cor:pred_interval_pool}(iii), we can verify that the $k$th pooled interval admits the equivalent representation $\mathcal{I}_{k, \alpha}=\bigcap_{j \in J_k} \{\mathcal{I}^\t_{j, \alpha} \cup \mathcal{I}^\c_{k+1-j, \alpha}\}$, where $J_k=\{
    j: \max\{1, k-n_0\} \le j \le \min\{k, n_1+1\}\}$ with the conventions $\mathcal{I}^\t_{n_1+1, \alpha}\equiv\emptyset$ and $\mathcal{I}^\c_{n_0+1, \alpha}\equiv\emptyset$.
    Intuitively, this is the interval-valued analogue of finding the $k$th order statistic by merging two sorted lists.
\end{proof}

\begin{proof}[\bf Technical details for Remark \ref{rmk:comp_p_RKS_KS}]
Fix any $1\le k\le n_1$ and any finite $c'<c$.
Below we prove that $
    \overline{p}_{k,c'}^\t
    \le
    p_{k,c}^\t.
$
Let $\bs{v} = \bs Y-\bs Z\circ\bs{\xi}_{k,c}^{\t}$ and
$\bs{v}' = \bs Y-\bs Z\circ\bs{\xi}_{k,c'}^{\t}$.
We then have $v_i=v_i'-Z_i(c-c')$ for $1\le i\le n$.
From the proof of Lemma \ref{lem:RKS_KS_inequality},
there exists $r_*\in\{1,\ldots,n\}$ such that
\begin{align*}
    t_{\RKS}(\bs Z,\bs{v})
    =
    \frac{1}{n_0}\sum_{i=1}^n(1-Z_i)\I\{\rank_i(\bs{v})\le r_*\}
    -
    \frac{1}{n_1}\sum_{i=1}^n Z_i\I\{\rank_i(\bs{v})\le r_*\}.
\end{align*}
Moreover, the coordinate of $\bs{v}$ at rank $r_*$ must correspond
to a control unit; otherwise, reducing $r_*$ by one would strictly
increase the above expression, leading to a contradiction.
Let $b$ denote the value of the coordinate of $\bs{v}$ at rank $r_*$,
which equals the observed outcome of a control unit and is therefore finite.
Let $\mathcal{J}=\{i:\rank_i(\bs{v})\le r_*,\,1\le i\le n\}$.
We then have
\begin{align*}
t_{\KS}(\bs Z,\bs{v}')
&\ge
\frac{1}{n_0}\sum_{i=1}^n(1-Z_i)\I\{v_i'\le b\}
-\frac{1}{n_1}\sum_{i=1}^n Z_i\I\{v_i'\le b\}\\
&=
\frac{1}{n_0}\sum_{i=1}^n(1-Z_i)\I\{v_i\le b\}
-\frac{1}{n_1}\sum_{i=1}^n Z_i\I\{v_i\le b-(c-c')\}\\
&\ge
\frac{1}{n_0}\sum_{i\in\mathcal{J}}(1-Z_i)
-\frac{1}{n_1}\sum_{i\in\mathcal{J}}Z_i
=t_{\RKS}(\bs Z,\bs{v}),
\end{align*}
where the first inequality holds by definition,
the first equality holds by construction,
the second inequality holds because $c-c'>0$,
$v_i\le b$ for every $i\in\mathcal{J}$, and
$v_i\ge b$ for every $i\notin\mathcal{J}$,
and the last equality follows from the construction of $\mathcal{J}$.
Since $G_n$ is nonincreasing, this implies that
$
    \overline{p}_{k,c'}^\t
    =
    G_n\bigl(t_{\KS}(\bs Z,\bs{v}')\bigr)
    \le
    G_n\bigl(t_{\RKS}(\bs Z,\bs{v})\bigr)
    =
    p_{k,c}^\t.
$
\end{proof}

\section{Proofs of Theorem \ref{thm:tau_nbeta}, Theorem \ref{thm:sharp_bound_consist_fin}, Proposition \ref{prop:xi_nbeta}, Remark \ref{rmk:iid_sampling} and Theorem \ref{thm:consistency_finite_high_Level}}
\label{ssec:consist_fin}

To prove Theorem \ref{thm:tau_nbeta}, we need the following two lemmas. 

\begin{lemma}
\label{lem:quntile_property}
Suppose $F:\overline{\mathbb{R}} \to [0,1]$ is a distribution function with $F(+\infty) = 1$ and $F(-\infty) = 0$, where $\overline{\mathbb{R}} = \mathbb{R} \cup \{\pm \infty\}$. Define its quantile function $F^{-1}(q)$ for $q \in [0,1]$ by 
\begin{align*}
F^{-1}(q) = \inf\{x \in \overline{\mathbb{R}}:q\leq F(x)\}. 
\end{align*}
We then have the following results:
\begin{enumerate}[label=(\roman*), topsep=1ex,itemsep=-0.3ex,partopsep=1ex,parsep=1ex]

    \item $F(F^{-1}(q)) \geq q$, for $q \in [0,1]$,

    \item $F^{-1}(q) \leq x \iff q \leq F(x)$, for $q \in [0,1]$ and $x \in \overline{\mathbb{R}}$, 

   \item $F^{-1}(F(x)) \leq x$, for $x \in \overline{\mathbb{R}}$, 

   \item $F(F^{-1}(q)-) \leq q$, for $q \in (0,1]$,

   \item If $F^{-1}$ is continuous at $q \in (0,1)$, then $F^{-1}(q) < x$ implies $F(x) > q$,

   \item $F^{-1}(q) \ge x \iff F(z)<q$ for any $z<x$.

\end{enumerate}
\end{lemma}

\begin{proof}[\bf Proof of Lemma \ref{lem:quntile_property}]

(i)--(iv) follow from \citep[][p. 304, Lemma 21.1]{van2000asymptotic}, where we can verify that these statements also hold when $q = 0,1$ or $x = \pm \infty$. 

For (v), by continuity, if $F^{-1}(q) < x$, then there exists $q^\prime>q$ such that $F^{-1}(q^\prime) < x$. By (ii), this further implies that $q < q^\prime \le F(x)$.

For (vi), suppose first that $F^{-1}(q) \ge x$. Then for any $z<x\le F^{-1}(q)$, by the definition of $F^{-1}(q)$, we must have $F(z) < q$. 
Suppose then that $F(z) < q$ for any $z<x$. 
By (ii), for any $z<x$, we then have $z < F^{-1}(q)$. This then implies that $x \le F^{-1}(q)$.

From the above, Lemma \ref{lem:quntile_property} holds.
\end{proof}

\begin{lemma} \label{lem:tau_nbeta_sharp}
    Suppose $F_{n1}$ and $F_{n0}$ are distribution functions of $\{Y_i(1)\}_{i=1}^n$ and $\{Y_i(0)\}_{i=1}^n$. 
    For any $\beta\in (0,1)$, 
    there exists a pairing of the $n$ treatment potential outcomes and the $n$ control potential outcomes such that the corresponding $\beta$-quantile of the paired individual treatment effects is exactly 
    $\underline{\tau}_{n,\beta}\equiv \sup_{u\in[0,\beta]}\left\{F_{n1}^{-1}(u)-F_{n0}^{-1}(u+1-\beta)\right\}$.
\end{lemma}

\begin{proof}[\bf Proof of Lemma \ref{lem:tau_nbeta_sharp}]
    Define order statistics $Y_{(1)}(1)\le \cdots \le Y_{(n)}(1)$ and $Y_{(1)}(0)\le \cdots \le Y_{(n)}(0)$.
    Define further $Y_{(0)}(1)$ and $Y_{(0)}(0)$ as $-\infty$. 
    We then have $F_{n1}^{-1}(u)=Y_{(\lceil nu\rceil)}(1)$ and $F_{n0}^{-1}(u)=Y_{(\lceil nu\rceil)}(0)$ for $u\in[0,1]$. 
    
    Let $k=\lceil n\beta\rceil$, which implies $n\beta \in (k-1,k]$. 
    By definition, 
    \begin{align*}
        \underline{\tau}_{n,\beta} 
        &= \sup_{u\in[0,\beta]}\left\{F_{n1}^{-1}(u)-F_{n0}^{-1}(u+1-\beta)\right\} 
        = \sup_{u\in(0,k/n]}\left\{F_{n1}^{-1}(u)-F_{n0}^{-1}(u+1-\beta)\right\}, 
    \end{align*}
    where the last equality holds because 
    $
        F_{n1}^{-1}(0)-F_{n0}^{-1}(0+1-\beta) = -\infty, 
    $
    and for $u \in [\beta, k/n]$,
    \begin{align*}
        F_{n1}^{-1}(u)-F_{n0}^{-1}(u+1-\beta) = F_{n1}^{-1}(\beta)-F_{n0}^{-1}(u+1-\beta) 
        \le  F_{n1}^{-1}(\beta)-F_{n0}^{-1}(\beta+1-\beta),
    \end{align*}
    with the convention that $F_{n0}^{-1}(q)=+\infty$ for $q>1$.
    We then have 
    \begin{align*}
        \underline{\tau}_{n,\beta} 
        &= \max_{1\le j\le k} \sup_{u\in ((j-1)/n,\ j/n]}\left\{F_{n1}^{-1}(u)-F_{n0}^{-1}(u+1-\beta)\right\}.
    \end{align*}
    For any $1\le j \le k$ and $u\in ((j-1)/n,\ j/n]$, we have $\lceil nu \rceil=j$ and $ n(u+1-\beta) > j-1+n-k.$ The latter implies that $\lceil n(u+1-\beta) \rceil \ge n-k+j$, where the equality can be achieved when $u \in ((j-1)/n,\ j/n+\beta-k/n ]$. 
    Consequently, we have 
    \begin{align*}
        \sup_{u\in ((j-1)/n,\ j/n]}\left\{F_{n1}^{-1}(u)-F_{n0}^{-1}(u+1-\beta)\right\}
        & = Y_{(j)}(1)-Y_{(n-k+j)}(0), \quad 
        (1\le j \le k)
    \end{align*}
    which further implies that 
    \begin{align*}
        \underline{\tau}_{n,\beta} 
        &= \max_{1\le j\le k} \left\{ Y_{(j)}(1)-Y_{(n-k+j)}(0) \right\}.
    \end{align*}

    Now define a permutation $\pi$ of $\{1,\dots,n\}$ 
    such that $\pi(i)= n-k+i$ for $1\le i \le k$. 
    For $1\le i\le k$, pair $Y_{(i)}(1)$ with $Y_{(\pi(i))}(0)$, and let
    $\tau_{\pi,i} = Y_{(i)}(1)-Y_{(\pi(i))}(0)$. 
    \begin{align*}
        \tau_{\pi,i} = Y_{(i)}(1)-Y_{(n-k+i)}(0)
        \le \max_{1\le j\le k} \left\{Y_{(j)}(1)-Y_{(j+n-k)}(0)\right\}
        =\underline{\tau}_{n,\beta}  .
    \end{align*}
    Therefore at least $k$ of the $n$ values $\tau_{\pi,1},\dots,\tau_{\pi,n}$ are less than or equal to $\underline{\tau}_{n,\beta}$. For the distribution $F_{\tau,\pi}$ of $\tau_{\pi,1},\dots,\tau_{\pi,n}$, it follows that $F_{\tau,\pi}(\underline{\tau}_{n,\beta})\ge k/n \ge \beta$ and thus
    $F_{\tau,\pi}^{-1}(\beta)\le \underline{\tau}_{n,\beta}$.
    From \citet{fan2010sharp} and \citet{zhang2024bounds}, $F_{\tau,\pi}^{-1}(\beta)\ge \underline{\tau}_{n, \beta}$.
    Combining the two inequalities yields $F_{\tau,\pi}^{-1}(\beta)= \underline{\tau}_{n, \beta}$. 
    Thus the sharp lower bound is attained by the above pairing.
\end{proof}

\begin{proof}[\bf Proof of Theorem \ref{thm:tau_nbeta}]

To prove (i), we will first show that 
\begin{align}
\label{eq:inverse_CDF_sub}
q \leq F_{n1, \beta}(F_{n0}^{-1}(q) + d), \forall q \in [0, 1]  \iff F_{n0}(b) \leq F_{n1, \beta}(b + d), \forall b \in \overline{\mathbb{R}}.    
\end{align}
Suppose the left-hand side of (\ref{eq:inverse_CDF_sub}) holds. For any $b \in \overline{\mathbb{R}}$, letting $q = F_{n0}(b)$, we have $F_{n0}(b) \leq F_{n1, \beta}(F_{n0}^{-1}(F_{n0}(b)) + d) \leq F_{n1, \beta}(b + d)$ by Lemma \ref{lem:quntile_property}(iii). This then implies the right-hand side of (\ref{eq:inverse_CDF_sub}). 
Now suppose the right-hand side of (\ref{eq:inverse_CDF_sub}) holds. For any $q \in [0,1]$, using Lemma \ref{lem:quntile_property}(i) and the right-hand side with $b = F_{n0}^{-1}(q)$, we have $ q \leq F_{n0}(F_{n0}^{-1}(q))  \leq F_{n1, \beta}(F_{n0}^{-1}(q) + d)$. This then implies the left-hand side of (\ref{eq:inverse_CDF_sub}).

By definition, the lower bound $\underline{\tau}_{n,\beta}$ has the following equivalent forms:
\begin{align*}
\underline{\tau}_{n,\beta} & \equiv \sup_{u \in [0, \beta]} \big\{ F_{n1}^{-1}(u) - F_{n0}^{-1}(1 + u - \beta) \big\} 
= \sup_{q \in [1 - \beta, 1]} \big\{  F_{n1}^{-1}(q - (1 - \beta)) - F_{n0}^{-1}(q) \big\} \quad
\\
&= \inf \Big \{ d \in \mathbb{R}: F_{n1}^{-1}(q - (1 - \beta)) - F_{n0}^{-1}(q) \leq  d  \text{ for all } q \in [1 - \beta, 1] \Big \}
\\
&= \inf \Big \{ d \in \mathbb{R}: q - (1 - \beta) \leq  F_{n1}(F_{n0}^{-1}(q) + d)  \text{ for all } q \in [1 - \beta, 1] \Big \} . 
\end{align*}
where the first equality follows from the definition of $\underline{\tau}_{n,\beta}$, 
the second equality follows by the substitution $q = u + (1 - \beta)$, 
the third equality follows by the property of the supremum, 
and 
the fourth equality follows from Lemma \ref{lem:quntile_property}(ii). 
Note that, for any $d\in \mathbb{R}$, 
\begin{align*}
    & q - (1 - \beta) \leq  F_{n1}(F_{n0}^{-1}(q) + d)
    \text{ for all } q \in [1 - \beta, 1]
    \\
    \Longleftrightarrow \ &
    q - (1 - \beta) \leq  F_{n1}(F_{n0}^{-1}(q) + d)
    \text{ for all } q \in [0, 1]
    \\
    \Longleftrightarrow \ &
    q \leq  F_{n1}(F_{n0}^{-1}(q) + d) + (1 - \beta)
    \text{ for all } q \in [0, 1]
    \\
    \Longleftrightarrow \ &
    q \leq  F_{n1, \beta}(F_{n0}^{-1}(q) + d) 
    \text{ for all } q \in [0, 1]
    \\
    \Longleftrightarrow \ &
    F_{n0}(b) \leq F_{n1, \beta}(b + d), \forall b \in \overline{\mathbb{R}}\\
    \Longleftrightarrow \ &
    F_{n0}(b) \leq F_{n1, \beta}(b + d), \forall b \in \mathbb{R},
\end{align*}
where the third last equivalence follows from the definition of $F_{n1, \beta}(\cdot)$, 
the second last equivalence follows from \eqref{eq:inverse_CDF_sub}, 
and the last equivalence follows from fact that 
$F_{n0}(b) \leq F_{n1, \beta}(b + d)$ holds obviously when $b = \pm \infty$. 
From the above, $\underline{\tau}_{n,\beta}$ then has the following equivalent forms:
\begin{align*}
  \underline{\tau}_{n,\beta} 
&= \inf \Big \{ d \in \mathbb{R}: \sup_{b \in \mathbb{R}} \{ F_{n0}(b) - F_{n1, \beta}(b + d) \} \leq 0 \Big \} \\
&= \inf \Big \{ d \in \mathbb{R}: \sup_{b \in \mathbb{R}} \{ F_{n0}(b) - F_{n1, \beta}(b + d) \} = 0 \Big \}, 
\end{align*}
where the last equality holds because 
$\lim_{b \to +\infty} \{ F_{n0}(b) - F_{n1, \beta}(b + d) \} = 0.$

In addition, Theorem \ref{thm:tau_nbeta}(ii) follows from Lemma \ref{lem:tau_nbeta_sharp}. Therefore, Theorem \ref{thm:tau_nbeta} holds.
\end{proof}

The following proposition shows that the quantities in \eqref{eq:notation_xin} are well-defined. 

\begin{proposition}
\label{prop:xi_psi_t_existence}
Consider any $d \in \mathbb{R}$, and $\psi_{n,\beta}(b,d)$ and $\hat{\psi}_{n,\beta}(b,d)$ defined in \eqref{eq:notation_psin}.
\begin{enumerate}[label=(\roman*), topsep=1ex,itemsep=-0.3ex,partopsep=1ex,parsep=1ex]
    \item There exists $b' \in \mathbb{R}$ such that $ \psi_{n,\beta}(b',d) = \sup_{b\in \mathbb{R}} \psi_{n,\beta}(b,d)$. 

    \item There exists $b'' \in \mathbb{R}$ such that $ \hat \psi_{n,\beta}(b'',d) = \sup_{b\in \mathbb{R}} \hat \psi_{n,\beta}(b,d)$. 
\end{enumerate}
\end{proposition}

\begin{proof}[\bf Proof of Proposition \ref{prop:xi_psi_t_existence}]
We prove only (i), since the proof for (ii) follows by the same logic. 
Because $F_{n0}(b)$ and $F_{n1, \beta}(b + d) = \min\{1, 1- \beta + F_{n1}(b+d)\}$, as functions of $b$, are both right-continuous step functions with finitely many steps, their difference $\psi_{n,\beta}(b, d) = F_{n0}(b) - F_{n1, \beta}(b + d)$ must also be a right-continuous step function with finitely many steps. 
Consequently, the supremum of $\psi_{n,\beta}(b, d)$ must be attained at some finite point. 
\end{proof}

To prove Theorem \ref{thm:sharp_bound_consist_fin}, we need the following four lemmas. 

\begin{lemma}
\label{lem:absolute_min} 
\begin{enumerate}[label=(\roman*)]
    \item For all $ u, v \in \mathbb{R}$, $| \min \{1, u\} - \min \{1, v\} | \leq |u - v|$,
    \item $\sup_b | F_{n1, \beta}(b) - \hat{F}_{n,\lceil n_1 \beta \rceil/n_1}^\t(b) | \leq \sup_b | F_{n1}(b) - \hat{F}_n^\t(b)| + \frac{1}{n_1}$,
\end{enumerate}

\end{lemma}

\begin{proof}[\bf Proof of Lemma \ref{lem:absolute_min}]
To prove (i), it suffices to consider the following three cases. 
\begin{enumerate}
    \item[1.] When $u \leq 1, v \leq 1$, we have $| \min \{1, u\} - \min \{1, v\} | = |u - v|$,
    \item[2.] When $u > 1, v \leq 1$, we have $| \min \{1, u\} - \min \{1, v\} | = |1 - v| \leq |u - v|$,
    \item[3.] When $u > 1, v > 1$, we have $| \min \{1, u\} - \min \{1, v\} | = |1 - 1| = 0 \leq |u - v|$.
\end{enumerate}
In conclusion, $| \min \{1, u\} - \min \{1, v\} | \leq |u - v|$ for all $u, v \in \mathbb{R}$.

We now prove (ii). By definition, 
\begin{align*}
F_{n1, \beta}(b) = \min \{ 1, 1 - \beta + F_{n1}(b) \},\quad \hat{F}_{n,\lceil n_1 \beta \rceil/n_1}^\t(b) = \min \{ 1, 1 - \frac{\lceil n_1 \beta \rceil}{n_1} + \hat{F}_{n}^\t(b) \}.
\end{align*}
Using the inequality in (i) with $u = 1 - \beta + F_{n1}(b)$ and $v = 1 - {\lceil n_1 \beta \rceil}/{n_1} + \hat{F}_{n}^\t(b)$, we have
\begin{align*}
\sup_b | F_{n1, \beta}(b) - \hat{F}_{n,\lceil n_1 \beta \rceil/n_1}^\t(b) | & \leq \sup_b \Big|\{ F_{n1}(b) + 1 - \beta \} - \Big\{ \hat{F}_{n}^\t(b) + 1 - \frac{\lceil n_1 \beta \rceil}{n_1}\Big\} \Big|\\
&\leq  \sup_b | F_{n1}(b) - \hat{F}_{n}^\t(b) | + \Big| \beta - \frac{\lceil n_1 \beta \rceil}{n_1} \Big| \\ 
& \leq \sup_b | F_{n1}(b) - \hat{F}_{n}^\t(b) | + \frac{1}{n_1}.
\end{align*}

From the above, Lemma \ref{lem:absolute_min} holds. 
\end{proof}

\begin{lemma}
\label{lem:absolute_sup}
Given two bounded functions $f$ and $g$ from $\mathbb{R}$ to $\mathbb{R}$, the following inequality holds:
\begin{align*}
\Big|\sup_bf(b)-\sup_bg(b)\Big|\leq\sup_b|f(b)-g(b)|.
\end{align*}
\end{lemma}

\begin{proof}[\bf Proof of Lemma \ref{lem:absolute_sup}]
Without loss of generality, assume $\sup_b f(b)\geq\sup_b g(b)$. By definition, for any $\varepsilon > 0$, there exists $b_\varepsilon$ in $\mathbb{R}$ such that
$
\sup_{b} f(b) - \varepsilon \leq f(b_\varepsilon).
$
Since $\sup_bg(b) \geq g(b_\varepsilon)$, we obtain that
\begin{align*}
\Big|\sup_bf(b)-\sup_bg(b)\Big| &= \sup_bf(b)-\sup_bg(b) \leq f(b_\varepsilon) + \varepsilon - g(b_\varepsilon) \leq \sup_b \{f(b) - g(b)\} + \varepsilon \\
& \leq \sup_b |f(b) - g(b)| + \varepsilon.
\end{align*}
Because the above inequality holds for every $\varepsilon > 0$, we must have
$
|\sup_bf(b)-\sup_bg(b)|\leq\sup_b|f(b)-g(b)|,
$
i.e., Lemma \ref{lem:absolute_sup} holds. 
\end{proof}

\begin{lemma} \label{lem:xin_converge}
Consider $\xi_{n,\beta}$ and $\hat{\xi}_{n,\beta}$ defined in \eqref{eq:notation_xin}, and any $\kappa < 1/2$. 
If $\lim_{n \to \infty} n_1/n = \lambda \in (0,1)$, 
then 
\begin{align}\label{eq:sup_diff_xi}
    n^{\kappa}\sup_{d \in \mathbb{R}} |\hat{\xi}_{n,\beta}(d)-\xi_{n,\beta}(d)| \convergeas 0 \quad \text{as } n\rightarrow \infty.
\end{align}
\end{lemma}

\begin{proof}[\bf Proof of Lemma \ref{lem:xin_converge}]

From Lemmas \ref{lem:absolute_min} and \ref{lem:absolute_sup}, the absolute difference between $\hat{\xi}_{n,\beta}(d) $ and $ \xi_{n,\beta}(d)$ can be bounded by
\begin{align*}
    & \quad \ n^\kappa | \hat{\xi}_{n,\beta}(d) - \xi_{n,\beta}(d)|
    \\
    &= n^\kappa \Big|\sup_b \hat{\psi}_{n,\beta}(b, d) - \sup_b \psi_{n,\beta}(b, d)\Big|\\
    &\leq n^\kappa \sup_b |\hat{\psi}_{n,\beta}(b, d) - \psi_{n,\beta}(b, d)| \quad \text{(by Lemma \ref{lem:absolute_sup})}\\
    &= n^\kappa \sup_b |\{\hat{F}_{n}^\c(b) - \hat{F}_{n,\lceil n_1 \beta \rceil/n_1}^\t(b + d)\} - \{F_{n0}(b) - F_{n1, \beta}(b + d)\}|\\
    &\leq n^\kappa \sup_b |\hat{F}_{n}^\c(b)  - F_{n0}(b)| + n^\kappa \sup_b |\hat{F}_{n,\lceil n_1 \beta \rceil/n_1}^\t(b + d))- F_{n1, \beta}(b + d)|\\
    &\leq n^\kappa \sup_b |\hat{F}_{n}^\c(b)  - F_{n0}(b)| + n^\kappa \sup_b |  \hat{F}_{n}^\t(b) - F_{n1}(b)| + \frac{n^\kappa}{n_1} \quad \text{(by Lemma \ref{lem:absolute_min})}.
\end{align*}
Note that  $\lim_{n \to \infty} n_1/n = \lambda \in (0,1)$. 
From the Glivenko-Cantelli theorem for a finite population \citep[see, e.g.,][]{motoyama2024extended}, 
it follows that, as $n \rightarrow \infty$, 
\begin{align*}
n^\kappa \sup_b | \hat{F}_{n}^\t(b) - F_{n1}(b) |  \convergeas 0,\, \ \text{and } \  n^\kappa \sup_b | \hat{F}_{n}^\c(b) - F_{n0}(b) | \convergeas 0.
\end{align*}
From the above, we can then derive \eqref{eq:sup_diff_xi}, 
i.e., Lemma \ref{lem:xin_converge} holds.
\end{proof}

\begin{lemma}
\label{lem:null_limit}
Suppose $\lim_{n \to \infty} n_1/n = \lambda \in (0,1)$. 
For the test statistic $t_{\RKS}(\bs{z},\bs{y})$, the limiting tail probability of the randomization null distribution satisfies 
\begin{align}
\label{eq:KS_asymptotic_supp}
\lim_{{n \to \infty}} G_n\Big(
\sqrt{ \frac{n}{n_1n_0} }
 \ x \Big) 
 = e^{-2x^2}
 \quad \text{for any } x\ge 0. 
\end{align}
\end{lemma}

\begin{proof}[\bf Proof of Lemma \ref{lem:null_limit}]
From Lemma \ref{lem:null_distribution_equivalence}, 
$t_{\RKS}(\bs{Z},\bs{y}) \sim t_{\KS}(\bs{a},\bs{W})$, where $\bs{a} \in \{0,1\}^n$ is a constant vector satisfying that $\sum_{i=1}^n a_i = n_1$, and $\bs{W}$ is a random vector whose coordinates are i.i.d.\ from some continuous distribution. 
From \citet{gibbons2014nonparametric} (p. 239), we know that 
\begin{equation*}
 \lim_{{n \to \infty}} \Pr\Big(\sqrt{\frac{n_1 n_0}{n}}t_{\KS}(\bs{a},\bs{W}) \geq x\Big) = e^{-2x^2}.
\end{equation*}
These immediately imply \eqref{eq:KS_asymptotic_supp}, i.e., Lemma \ref{lem:null_limit} holds. 
\end{proof}

\begin{proof}[\bf Proof of Theorem \ref{thm:sharp_bound_consist_fin}]
Suppose that $n^\kappa  \xi_{n,\beta}(c_n) \to \infty$ as $n\rightarrow \infty$ for some $0< \kappa < 1/2$. 
From Lemma \ref{lem:xin_converge}, $n^\kappa | \hat{\xi}_{n,\beta}(c_n)-\xi_{n,\beta}(c_n)|$ converges almost surely to zero as $n\rightarrow \infty$.
These then imply that $n^{\kappa} \hat{\xi}_{n,\beta}(c_n) \convergeas \infty$ as $n\rightarrow \infty$. It follows that $\sqrt{n_1 n_0/n}\ \hat{\xi}_{n,\beta}(c_n) \convergeas \infty$ since the limit of $n_1/n$ lies in $(0,1)$.

Consider a realization such that $\sqrt{n_1 n_0/n}\ \hat{\xi}_{n,\beta}(c_n) \rightarrow \infty$ as $n\rightarrow \infty$. 
Then, for any constant $M$, there exists $N$ such that $\sqrt{n_1 n_0/n}\ \hat{\xi}_{n,\beta}(c_n) \geq M$ for all $n \geq N$. 
By the definition in \eqref{eq:rand_p_val_ks_equi}, for $n\ge N$, 
\begin{align*}
\overline p^\t_{\lceil n_1 \beta \rceil, c_n} & = G_n \Big( \sup_b \{ \hat{F}_{n}^\c(b) - \hat{F}_{n, \lceil n_1 \beta \rceil/n_1}^\t(b + c_n) \}\Big) = G_n\Big( \hat{\xi}_{n,\beta}(c_n)\Big) \le G_n \Big(\sqrt{\frac{n}{n_1n_0}}M \Big).
\end{align*}
From Lemma \ref{lem:null_limit}, we then have 
$$
\varlimsup_{n \to \infty} \overline p^\t_{\lceil n_1 \beta \rceil, c_n} \le 
\varlimsup_{n \to \infty}G_n \Big(\sqrt{\frac{n}{n_1n_0}}M \Big)
= \exp(-2M^2).
$$
Since the above inequality holds for any $M$, we must have $\overline p^\t_{\lceil n_1 \beta \rceil, c_n} \rightarrow 0$ as $n\rightarrow \infty$.

From the above, for almost every realization, 
we have $\overline p^\t_{\lceil n_1 \beta \rceil, c_n}  \rightarrow 0$ as $n\rightarrow \infty$.
Thus, $\overline p^\t_{\lceil n_1 \beta \rceil, c_n}$ converges almost surely to zero as $n\rightarrow \infty$. 
Because 
$p^\t_{\lceil n_1 \beta \rceil, c_n} \le \overline p^\t_{\lceil n_1 \beta \rceil, c_n}$, $p^\t_{\lceil n_1 \beta \rceil, c_n}$ must also converge almost surely to zero.  
Therefore, Theorem \ref{thm:sharp_bound_consist_fin} holds. 
\end{proof}

To prove Proposition \ref{prop:xi_nbeta}, we need the following two lemmas. 

\begin{lemma}\label{lem:xin_property}
    \begin{enumerate}[label=(\roman*), topsep=1ex,itemsep=-0.3ex,partopsep=1ex,parsep=1ex]
        \item Both $\xi_{n,\beta}(d)$ and $\hat{\xi}_{n,\beta}(d)$ in \eqref{eq:notation_xin} are nonincreasing in $d$ and non-negative for all $d\in \mathbb{R}$. 
        \item $\xi_{n,\beta}(d)$ is a right-continuous function for $d\in \mathbb{R}$.

        \item $\xi_{n,\beta}(c)=0$ for any $c \ge \underline{\tau}_{n,\beta}$, and $\xi_{n,\beta}(c)>0$ for any $c < \underline{\tau}_{n,\beta}$.
    \end{enumerate}
\end{lemma}

\begin{proof}[\bf Proof of Lemma \ref{lem:xin_property}]
For (i), by the property of distribution functions, 
$\psi_{n,\beta}(b,d)=F_{n0}(b) - F_{n1, \beta}(b + d)$ in \eqref{eq:notation_psin} is decreasing in $d$ and $\lim_{b \to +\infty} \psi_{n,\beta}(b,d) = 0$.
These imply that $\xi_{n,\beta}(d)=\sup_{b \in \mathbb{R}} \psi_{n,\beta}(b,d)$ in \eqref{eq:notation_xin} is nonincreasing in $d$ and non-negative for all $d\in \mathbb{R}$. 
By the same logic, $ \hat{\xi}_{n,\beta}(d)$ in \eqref{eq:notation_xin} is also nonincreasing in $d$ and non-negative for all $d\in \mathbb{R}$.

To prove (ii), it suffices to show $\lim_{\Delta \to 0^+} \xi_{n,\beta}(d_0+\Delta) = \xi_{n,\beta}(d_0)$
for any $d_0$ in $\mathbb{R}$.
By Proposition \ref{prop:xi_psi_t_existence}, there exists $b_0 \in \mathbb{R}$ such that $\psi_{n,\beta}(b_0,d_0)$ $ = \xi_{n,\beta}(d_0)$.
For any $\Delta>0$, we have $\xi_{n,\beta}(d_0+\Delta) \geq \psi_{n,\beta}(b_0,d_0+\Delta)$ by definition. Letting $\Delta \to 0^+$, it follows that
\begin{align*}
\varliminf_{\Delta \to 0^+} \xi_{n,\beta}(d_0+\Delta) \geq \varliminf_{\Delta \to 0^+}\psi_{n,\beta}(b_0,d_0+\Delta) = \psi_{n,\beta}(b_0,d_0) = \xi_{n,\beta}(d_0),
\end{align*}
where the last equality follows from the right-continuity of $\psi_{n,\beta}(b,d)$ with respect to $d$. 
Meanwhile, because $\xi_{n,\beta}(d)$ is nonincreasing in $d$, we have $\xi_{n,\beta}(d_0+\Delta) \leq \xi_{n,\beta}(d_0)$ for any $\Delta > 0$. From the above, we must have $\lim_{\Delta \to 0^+} \xi_{n,\beta}(d_0+\Delta) = \xi_{n,\beta}(d_0)$.

For (iii), from Theorem \ref{thm:tau_nbeta}(i), $\underline{\tau}_{n,\beta} = \inf \{ d \in \mathbb{R} : \xi_{n,\beta}(d) = 0 \}$. By the right-continuity in (ii), we can derive that $\xi_{n,\beta}(\underline{\tau}_{n,\beta})=0$. 
From (i), this further implies that $\xi_{n,\beta}(c)=0$ for any $c \ge \underline{\tau}_{n,\beta}$. 
In addition, since $\underline{\tau}_{n,\beta} = \inf_{d \in \mathbb{R}} \{ d : \xi_{n,\beta}(d) = 0 \}$, we must also have $\xi_{n,\beta}(c)>0$ for any $c < \underline{\tau}_{n,\beta}$.

Therefore, Lemma \ref{lem:xin_property} holds.
\end{proof}

\begin{lemma}
\label{lem:xin_ineq}
If $\psi_{n,\beta}(b_n,\underline{\tau}_{n,\beta})=0$ for $\underline{\tau}_{n,\beta}$ in \eqref{eq:sharp_bound_effect_quantile}, then we have  
\begin{align*}
    \xi_{n,\beta}(c_n) \geq \frac{1}{n} \sum_{i=1}^n \I\{b_n + c_n < Y_i(1) < d_n \},
\end{align*}
where $d_n=b_n+\underline{\tau}_{n,\beta}$ if $F_{n0}(b_n)<1$ and $d_n=F_{n1}^{-1}(\beta)$ if $F_{n0}(b_n)=1$.
Moreover, $F_{n1}(\underline{\tau}_{n,\beta}+b_{n}) < \beta$ if and only if $F_{n0}(b_n)<1$.

\end{lemma}

\begin{proof}[\bf Proof of Lemma \ref{lem:xin_ineq}]
From Lemma \ref{lem:xin_property}(iii), $\xi_{n,\beta}(\underline{\tau}_{n,\beta}) = 0$. 
From the condition in the lemma, we also have $\psi_{n,\beta}(b_n,\underline{\tau}_{n,\beta}) = 0$. 
We then have 
\begin{align*}
\xi_{n,\beta}(c_n) &\geq \psi_{n,\beta}(b_n, c_n)  
= \psi_{n,\beta}(b_n, c_n) - \psi_{n,\beta}(b_n, \underline{\tau}_{n,\beta})\\
&= \{F_{n0}(b_n) - F_{n1, \beta}(b_n + c_n)\} - \{F_{n0}(b_n) - F_{n1, \beta}(b_n + \underline{\tau}_{n,\beta})\} \\
&= F_{n1, \beta}(b_n + \underline{\tau}_{n,\beta}) - F_{n1, \beta}(b_n + c_n) \\
&= \min\{1-\beta+F_{n1}(b_n+\underline{\tau}_{n,\beta}),1\} - \min\{1-\beta+F_{n1}(b_n+c_n),1\} \\
&\ge \min\{1-\beta+F_{n1}(b_n+\underline{\tau}_{n,\beta}),1\} - \{1-\beta+F_{n1}(b_n+c_n)\} \\
&= \min\{F_{n1}(b_n+\underline{\tau}_{n,\beta})-F_{n1}(b_n+c_n), \beta-F_{n1}(b_n+c_n)\} \\
&= \min\{F_{n1}(\underline{\tau}_{n,\beta}+b_{n}),\beta\}-F_{n1}(c_n+b_{n}) \\
&\ge \min\{F_{n1}(\underline{\tau}_{n,\beta}+b_{n}),F_{n1}(F_{n1}^{-1}(\beta)-)\}-F_{n1}(c_n+b_{n}), 
\end{align*}
where the first inequality follows from the definition in \eqref{eq:notation_xin}. 
Note that $\xi_{n,\beta}(c_n)$ is nonnegative by Lemma \ref{lem:xin_property}(i). We then have 
\begin{align*}
     \xi_{n,\beta}(c_n) &\geq \frac{1}{n} \sum_{i=1}^n \I\big\{b_n + c_n < Y_i(1) < \min\{F_{n1}^{-1}(\beta), b_{n}+\underline{\tau}_{n,\beta}\} \big\}, 
\end{align*}
where the right-hand side is zero when $b_n+c_n \ge \min\{F_{n1}^{-1}(\beta), b_{n}+\underline{\tau}_{n,\beta}\}$.

By Lemma \ref{lem:quntile_property} (ii), $\underline{\tau}_{n,\beta}+b_{n} < F_{n1}^{-1}(\beta)$ is equivalent to $F_{n1}(\underline{\tau}_{n,\beta}+b_{n}) < \beta$, which holds if and only if $F_{n0}(b_n)<1$ since $F_{n0}(b_n)=\min\{1,1-\beta+F_{n1}(\underline{\tau}_{n,\beta}+b_{n})\}$.
As a result,
$\min\{F_{n1}^{-1}(\beta), b_{n}+\underline{\tau}_{n,\beta}\}$ equals $\underline{\tau}_{n,\beta}+b_{n}$ if $F_{n0}(b_n)<1$, and equals $F_{n1}^{-1}(\beta)$ otherwise. 
Thus, by definition, we have $d_n=\min\{F_{n1}^{-1}(\beta), b_{n}+\underline{\tau}_{n,\beta}\}$.

From the above, Lemma \ref{lem:xin_ineq} holds.
\end{proof}

\begin{proof}[\bf Proof of Proposition \ref{prop:xi_nbeta}]
Proposition \ref{prop:xi_nbeta} follows directly from Lemmas \ref{lem:xin_property} and \ref{lem:xin_ineq}. 
\end{proof}

\begin{proof}[\bf Technical details for Remark \ref{rmk:iid_sampling}]
Using \citet{Chung1949}, we can know that for $z\in\{0,1\}$, 
\[
\limsup_{n\to\infty}
\sqrt{\frac{n}{\log\log n}}\,
\sup_{b\in\mathbb R}|F_{nz}(b)-F_{\infty,z}(b)|
< \infty
\quad\text{almost surely},
\]
where equality holds when $F_{\infty,z}$ is continuous.  
This then implies that, for any $\kappa\in (0,1/2)$, 
\[
n^\kappa\sup_{b\in\mathbb R}|F_{nz}(b)-F_{\infty,z}(b)|
\longrightarrow0
\quad\text{almost surely},
\]
Thus, 
Assumption \ref{asmp:fp_limit}(i) holds.
Assumption \ref{asmp:fp_limit}(ii) follows directly from the assumed
monotonicity of the distribution functions on their supports.
\end{proof}

To prove Theorem \ref{thm:consistency_finite_high_Level}, we need the following five lemmas. 

\begin{lemma}\label{lem:quantile_continuous}
If a distribution function $F$ is strictly increasing on $\big(F^{-1}(0+),F^{-1}(1)\big)$, then $F^{-1}$ is continuous at each point $u_0 \in (0,1)$.
\end{lemma}

\begin{proof}[\bf Proof of Lemma \ref{lem:quantile_continuous}]
Consider any given $u_0 \in (0,1)$ and any $\varepsilon>0$.
Below we find $\delta>0$ such that $|F^{-1}(u)-F^{-1}(u_0)| \le \varepsilon$ for any $|u-u_0|<\delta$.

Let $y_0=F^{-1}(u_0)$. 
From Lemma \ref{lem:quntile_property}(ii) and by the fact that $F$ is strictly increasing on $\big(F^{-1}(0+),F^{-1}(1)\big)$, we can know that $F(y_0 - \varepsilon) < u_0$ and $F(y_0+\varepsilon)> u_0$, where the last inequality holds since $F(y_0+\varepsilon)= 1 > u_0$ when $F(y_0)=1$ and $F(y_0+\varepsilon)> F(y_0) \ge u_0$ when $F(y_0)<1$.    
Let $\delta = \min\{ u_0, u_0 - F(y_0 - \varepsilon), F(y_0+\varepsilon) - u_0, 1-u_0 \}$, which must be positive. 

We now consider any $u$ such that $|u- u_0| < \delta.$
By definition, we can verify that $F(y_0-\varepsilon)<u < F(y_0+\varepsilon)$ and $u\in (0,1)$. 
From Lemma \ref{lem:quntile_property}(ii), this implies that 
$y_0 - \varepsilon < F^{-1}(u) \le y_0 +\varepsilon$. 
Consequently, $|F^{-1}(u)-F^{-1}(u_0)| \le \varepsilon$.

From the above, Lemma \ref{lem:quantile_continuous} holds. 
\end{proof}

\begin{lemma}\label{lem:quantile_uniform}
Let $F$ be a distribution function on $\mathbb{R}$, and let $\{F_n\}$ be a sequence of distribution functions such that $\sup_{x\in\mathbb{R}}|F_n(x)-F(x)|\rightarrow 0$ as $n \rightarrow \infty$.
\begin{enumerate}[label=(\roman*), topsep=1ex,itemsep=-0.3ex,partopsep=1ex,parsep=1ex]
\item For any $q\in (0,1)$ and $\delta \in (0, \min\{q, 1-q\})$, there exists $N>0$ such that for every $n>N$, 
\begin{align*}
F^{-1}(q+\delta) \ge F_n^{-1}(q)\ge F^{-1}(q-\delta),    
\end{align*}
\item If $F^{-1}$ is continuous at $q \in (0,1)$, then $|F_n^{-1}(q)-F^{-1}(q)|\converge 0$ as $n \rightarrow \infty$,
\item If $F^{-1}$ is continuous at every point of $[a,b] \subseteq(0,1)$, then $\sup_{q\in[a,b]}|F_n^{-1}(q)-F^{-1}(q)|\converge 0$ as $n \rightarrow \infty$.
\end{enumerate}
\end{lemma}

\begin{proof}[\bf Proof of Lemma \ref{lem:quantile_uniform}]
First, we prove (i). The uniform convergence implies that there exists $N>0$ such that for every $n>N$, $\sup_x |F_n(x)-F(x)|<\delta$.
Consequently,
\begin{align*}
    F_n(F^{-1}(q+\delta)) \ge F(F^{-1}(q+\delta)) - \sup_x|F_n(x)-F(x)|\ge q+\delta-\delta=q,
\end{align*}
and
\begin{align*}
    F(F_n^{-1}(q)) \ge F_n(F_n^{-1}(q)) - \sup_x|F_n(x)-F(x)|\ge q-\delta,
\end{align*}
which implies 
$
    F^{-1}(q+\delta) \ge F_n^{-1}(q)\ge F^{-1}(q-\delta).
$

Second, we prove (ii). 
From (i), for every sufficiently small $\delta>0$, there exists $N>0$ such that when $n>N$, 
\begin{align*}
   | F_n^{-1}(q)-F^{-1}(q)|\le \max\{F^{-1}(q)-F^{-1}(q-\delta),F^{-1}(q+\delta)-F^{-1}(q)\}.
\end{align*}
This implies that 
\begin{align*}
   \overline{\lim}_{n\rightarrow \infty}| F_n^{-1}(q)-F^{-1}(q)|\le \max\{F^{-1}(q)-F^{-1}(q-\delta),F^{-1}(q+\delta)-F^{-1}(q)\}.
\end{align*}
Because $F^{-1}$ is continuous at $q$, the right-hand side tends to zero as $\delta$ converges to zero. 
Hence, we must have $\lim_{n\rightarrow \infty}| F_n^{-1}(q)-F^{-1}(q)| = 0$.

Third, we prove (iii). Suppose that $F^{-1}$ is continuous at every point of $[a,b] \subseteq(0,1)$. Then $F^{-1}$ is uniformly continuous on  $[a,b]$.
Consider any given $\varepsilon>0$. 
We can choose $\delta \in (0, (b-a)/2)$ such that, 
for any $q_1,q_2 \in[a,b]$ with $|q_1-q_2|\le\delta$, we have $|F^{-1}(q_1)-F^{-1}(q_2)|<\varepsilon$.
By the uniform convergence of $F_n$ and (ii), there exists $N>0$ such that for all $n>N$, 
\begin{align*}
    \sup_x|F_n(x)-F(x)|<\delta, 
    \quad 
    |F^{-1}_n(a)-F^{-1}(a)|<\varepsilon, 
    \quad 
    |F^{-1}_n(b)-F^{-1}(b)|<\varepsilon
\end{align*}
Below we consider three cases.

First, for any $q\in[a+\delta,b-\delta]$ and $n>N$, we have $ F^{-1}(q+\delta) \ge F_n^{-1}(q)\ge F^{-1}(q-\delta)$ by (i) and thus
\begin{align*}
   | F_n^{-1}(q)-F^{-1}(q)|\le \max\{F^{-1}(q)-F^{-1}(q-\delta),F^{-1}(q+\delta)-F^{-1}(q)\} < \varepsilon.
\end{align*}

Second, for any $q\in[a,a+\delta)$ and $n>N$, we have  
$ F^{-1}(q+\delta) \ge F_n^{-1}(q)$ by the same logic as (i) 
and 
$F_n^{-1}(q)\ge F_n^{-1}(a)\ge F^{-1}(a)-\varepsilon$. These imply that  
\begin{align*}
   | F_n^{-1}(q)-F^{-1}(q)|\le \max\{F^{-1}(q)-F^{-1}(a)+\varepsilon,F^{-1}(q+\delta)-F^{-1}(q)\} < 2\varepsilon.
\end{align*}

Third, for any $q\in(b-\delta,b]$ and $n>N$, we have 
$F^{-1}(q-\delta) \le F_n^{-1}(q)$ by the same logic as (i)
and 
$F_n^{-1}(q)\le F_n^{-1}(b)\le F^{-1}(b)+\varepsilon$. These then imply that  
\begin{align*}
   | F_n^{-1}(q)-F^{-1}(q)|\le \max\{F^{-1}(q) - F^{-1}(q-\delta),F^{-1}(b)+\varepsilon-F^{-1}(q)\} < 2\varepsilon.
\end{align*}
Summarizing all these results,  we can obtain that
$\sup_{q\in[a,b]}|F_n^{-1}(q)-F^{-1}(q)| < 2\varepsilon$ for $n>N$. 
Because $\varepsilon$ can be arbitrarily small, we can then derive that 
$
    \sup_{q\in[a,b]}|F_n^{-1}(q)-F^{-1}(q)|\converge 0 
$
as $n \rightarrow \infty.$

From the above, Lemma \ref{lem:quantile_uniform} holds.
\end{proof}

\begin{lemma}\label{lem:tau_infty_convergence}
    If Assumption \ref{asmp:fp_limit} holds, then $\underline{\tau}_{n,\beta}$ in \eqref{eq:sharp_bound_effect_quantile_orig} must have a limit $\underline{\tau}_{\infty,\beta}$, where
\begin{align*}
    \underline{\tau}_{\infty,\beta} = \sup_{u\in[0,\beta]} \big\{ F_{\infty,1}^{-1}(u)-F_{\infty,0}^{-1}(u+1-\beta) \big\},
\end{align*}
\end{lemma}

\begin{proof}[\bf Proof of Lemma \ref{lem:tau_infty_convergence}]
Define, for $u\in[0,\beta]$,
\begin{align*}
G_n(u)=
F_{n1}^{-1}(u)-F_{n0}^{-1}(u+1-\beta),\quad
G_\infty(u)=
F_{\infty,1}^{-1}(u)-
F_{\infty,0}^{-1}(u+1-\beta).
\end{align*}
Then
\begin{align*}
\underline{\tau}_{n,\beta}=
\sup_{u\in[0,\beta]}G_n(u),\quad
\underline{\tau}_{\infty,\beta}=
\sup_{u\in[0,\beta]}G_\infty(u).
\end{align*}

Note that every quantile function $F^{-1}$, defined by
$
F^{-1}(q)
=
\inf\{x\in\overline{\mathbb R}:q\le F(x)\},
$
is left-continuous on $(0,1]$, where continuity at $1$ is understood
in the extended-real sense. Specifically, suppose that $q_m\uparrow q$ for
some $q\in(0,1]$. Because $F^{-1}$ is nondecreasing,
$F^{-1}(q_m)$ has a limit $\ell\le F^{-1}(q)$. If
$\ell<F^{-1}(q)$, we can choose $x\in\mathbb R$ such that $\ell<x<F^{-1}(q)$.
For all sufficiently large $m$, we have $F^{-1}(q_m)\le x$.
Lemma \ref{lem:quntile_property}(ii) then implies $q_m\le F(x)$.
Letting $m\to\infty$ gives $q\le F(x)$, which, again by Lemma
\ref{lem:quntile_property}(ii), implies $F^{-1}(q)\le x$, a
contradiction. Therefore,
$F^{-1}(q_m)\rightarrow F^{-1}(q)$.

Because $F_{\infty,1}^{-1}(0)=-\infty$ and
$F_{\infty,0}^{-1}(1-\beta)\in\mathbb R$, we have
$G_\infty(0)=-\infty$. Moreover, by the left-continuity of the
quantile functions,
\begin{align*}
\lim_{u\uparrow\beta}G_\infty(u)
=
F_{\infty,1}^{-1}(\beta)-
F_{\infty,0}^{-1}(1)
=
G_\infty(\beta),
\end{align*}
where the convergence is understood in the extended-real sense if
$F_{\infty,0}^{-1}(1)=+\infty$. Consequently,
$\sup_{u\in[0,\beta]}G_\infty(u)
=
\sup_{u\in(0,\beta)}G_\infty(u)$.
Furthermore, $\underline{\tau}_{\infty,\beta}$ is finite. 
$G_\infty(\beta/2)\in\mathbb{R}$ ensures that
$\underline{\tau}_{\infty,\beta}>-\infty$. Moreover, for every
$u\in[0,\beta]$,
$
G_\infty(u)
\le
F_{\infty,1}^{-1}(\beta)
-
F_{\infty,0}^{-1}(1-\beta)
<
+\infty,
$
which ensures that $\underline{\tau}_{\infty,\beta}<+\infty$.

We first establish the lower bound
$\liminf_{n\to\infty}\underline{\tau}_{n,\beta}
\ge
\underline{\tau}_{\infty,\beta}$.
For any $\varepsilon>0$, 
because, as shown before, $\sup_{u\in[0,\beta]}G_\infty(u)
=
\sup_{u\in(0,\beta)}G_\infty(u)$, 
there must exist $u_\varepsilon\in(0,\beta)$ such that
$G_\infty(u_\varepsilon)
\ge
\underline{\tau}_{\infty,\beta}-\varepsilon$.
Because both $u_\varepsilon$ and $u_\varepsilon+1-\beta$ belong to $(0,1)$,
Lemmas \ref{lem:quantile_continuous} and
\ref{lem:quantile_uniform}(ii) imply
\begin{align*}
F_{n1}^{-1}(u_\varepsilon)
\rightarrow
F_{\infty,1}^{-1}(u_\varepsilon),\quad
F_{n0}^{-1}(u_\varepsilon+1-\beta)
\rightarrow
F_{\infty,0}^{-1}(u_\varepsilon+1-\beta).
\end{align*}
Therefore,
$G_n(u_\varepsilon)
\rightarrow
G_\infty(u_\varepsilon)$.
It follows that
\begin{align*}
\liminf_{n\to\infty}\underline{\tau}_{n,\beta}
\ge
\liminf_{n\to\infty}G_n(u_\varepsilon)
=
G_\infty(u_\varepsilon)
\ge
\underline{\tau}_{\infty,\beta}-\varepsilon.
\end{align*}
Because $\varepsilon>0$ is arbitrary, we have 
$\liminf_{n\to\infty}\underline{\tau}_{n,\beta}
\ge
\underline{\tau}_{\infty,\beta}$.

We next establish the upper bound
$\limsup_{n\to\infty}\underline{\tau}_{n,\beta}
\le
\underline{\tau}_{\infty,\beta}$.
Fix any $\varepsilon>0$. 
From Lemma \ref{lem:quantile_continuous} and Assumption \ref{asmp:fp_limit}, 
both 
$F_{\infty,0}^{-1}$ 
and 
$F_{\infty,1}^{-1}$ are continuous on $(0,1).$ 
We can choose a sufficiently small 
$\delta\in(0,\beta/2)$ such that
\begin{align*}
F_{\infty,0}^{-1}(1-\beta+\delta)
-
F_{\infty,0}^{-1}(1-\beta)
<
\varepsilon,\quad
F_{\infty,1}^{-1}(\beta)
-
F_{\infty,1}^{-1}(\beta-\delta)
<
\varepsilon.
\end{align*}
Decompose
\begin{align*}
\underline{\tau}_{n,\beta}
=
\max\bigg\{
\sup_{u\in[0,\delta]}G_n(u),\
\sup_{u\in[\delta,\beta-\delta]}G_n(u),\sup_{u\in[\beta-\delta,\beta]}G_n(u)
\bigg\}.
\end{align*}
For the first part, monotonicity of the quantile functions gives
\begin{align*}
\sup_{u\in[0,\delta]}G_n(u)
\le
F_{n1}^{-1}(\delta)
-
F_{n0}^{-1}(1-\beta).
\end{align*}
Because $\delta,1-\beta\in(0,1)$, Lemma
\ref{lem:quantile_uniform}(ii) implies
\begin{align*}
\limsup_{n\to\infty}
\sup_{u\in[0,\delta]}G_n(u)
&\le
F_{\infty,1}^{-1}(\delta)
-
F_{\infty,0}^{-1}(1-\beta)
<
F_{\infty,1}^{-1}(\delta)
-
F_{\infty,0}^{-1}(1-\beta+\delta)
+\varepsilon\\
&=
G_\infty(\delta)+\varepsilon
\le
\underline{\tau}_{\infty,\beta}+\varepsilon.
\end{align*}
For the second part, 
Lemmas
\ref{lem:quantile_continuous} and \ref{lem:quantile_uniform}(iii)
imply
\begin{align*}
& \quad \ \sup_{u\in[\delta,\beta-\delta]}
\left|G_n(u)-G_\infty(u)\right|
\\
&\le
\sup_{u\in[\delta,\beta-\delta]}
\left|
F_{n1}^{-1}(u)-F_{\infty,1}^{-1}(u)
\right|
+
\sup_{u\in[\delta,\beta-\delta]}
\left|
F_{n0}^{-1}(u+1-\beta)
-
F_{\infty,0}^{-1}(u+1-\beta)
\right| \\
&\longrightarrow 0.
\end{align*}
This implies that 
\begin{align*}
\limsup_{n\to\infty}
\sup_{u\in[\delta,\beta-\delta]}G_n(u)
\le
\sup_{u\in[\delta,\beta-\delta]}G_\infty(u)
\le
\underline{\tau}_{\infty,\beta}.
\end{align*}
For the third part, monotonicity gives
\begin{align*}
\sup_{u\in[\beta-\delta,\beta]}G_n(u)
\le
F_{n1}^{-1}(\beta)
-
F_{n0}^{-1}(1-\delta).
\end{align*}
Because $\beta,1-\delta\in(0,1)$, Lemma
\ref{lem:quantile_uniform}(ii) implies
\begin{align*}
\limsup_{n\to\infty}
\sup_{u\in[\beta-\delta,\beta]}G_n(u)
&\le
F_{\infty,1}^{-1}(\beta)
-
F_{\infty,0}^{-1}(1-\delta)
<
F_{\infty,1}^{-1}(\beta-\delta)
-
F_{\infty,0}^{-1}(1-\delta)
+\varepsilon
=
G_\infty(\beta-\delta)+\varepsilon\\
&\le
\underline{\tau}_{\infty,\beta}+\varepsilon.
\end{align*}
Combining the preceding three bounds yields
$\limsup_{n\to\infty}\underline{\tau}_{n,\beta}
\le
\underline{\tau}_{\infty,\beta}+\varepsilon$.
Because $\varepsilon>0$ is arbitrary,
$\limsup_{n\to\infty}\underline{\tau}_{n,\beta}
\le
\underline{\tau}_{\infty,\beta}$.

From the above, we then have 
$\underline{\tau}_{n,\beta}
\rightarrow
\underline{\tau}_{\infty,\beta}$.
Therefore, Lemma \ref{lem:tau_infty_convergence} holds.
\end{proof}

\begin{lemma}\label{lem:xi_psi_infty}
    Under Assumption \ref{asmp:fp_limit}, define $\xi_{\infty,\beta}(d) \equiv \sup_{b \in \mathbb{R}} \psi_{\infty,\beta}(b, d)$ where
    \begin{align*}
        \psi_{\infty,\beta}(b, d)  \equiv F_{\infty,0}(b) - F_{\infty, 1, \beta}(b + d)=F_{\infty,0}(b) - \min\{1,1-\beta+F_{\infty, 1}(b + d)\}.
    \end{align*}
    Then 
    \begin{align}\label{eq:tau_infty_beta}
        \underline{\tau}_{\infty,\beta} \equiv  \sup_{u\in[0,\beta]} \{ F_{\infty,1}^{-1}(u)-F_{\infty,0}^{-1}(1+u-\beta) \} 
        = \inf \{ d \in \mathbb{R} : \xi_{\infty,\beta}(d) = 0 \}
    \end{align}
    and 
    \begin{align}\label{eq:xi_infty_converge}
         \sup_{d \in \mathbb{R}} n^\kappa| \xi_{n,\beta}(d)-\xi_{\infty,\beta}(d)| \converge 0.
    \end{align}
\end{lemma}

\begin{proof}[\bf Proof of Lemma \ref{lem:xi_psi_infty}]
    The proof of \eqref{eq:tau_infty_beta} follows the same argument as the proof of Theorem \ref{thm:tau_nbeta}. To prove \eqref{eq:xi_infty_converge}, note that
    \begin{align*}
        n^\kappa|\xi_{n,\beta}(d) - \xi_{\infty,\beta}(d)| 
        &= n^\kappa|\sup_b \psi_{n,\beta}(b, d) - \sup_b \psi_{\infty,\beta}(b, d)|\\
        &\leq \sup_b n^\kappa |\psi_{n,\beta}(b, d) - \psi_{\infty,\beta}(b, d)| \\
        &= \sup_b n^\kappa|\{ F_{n0}(b) - F_{n1,\beta}(b + d) \} - \{ F_{\infty,0}(b) - F_{\infty,1, \beta}(b + d) \}|\\
        &\leq \sup_b n^\kappa|F_{n0}(b)  - F_{\infty,0}(b)| 
        + \sup_b n^\kappa|{F}_{n1,\beta}(b + d))- F_{\infty,1, \beta}(b + d)|\\
        &\leq \sup_b n^\kappa|{F}_{n0}(b)  - F_{\infty,0}(b)| 
        + \sup_b n^\kappa|  {F}_{n1}(b) - F_{\infty,1}(b)| \\
        & \converge 0 \text{ as } n \rightarrow \infty, 
    \end{align*}
    where the first inequality follows from Lemma \ref{lem:absolute_sup}, the last inequality follows from Lemma \ref{lem:absolute_min}, and the last convergence follows from Assumption \ref{asmp:fp_limit}(i). Hence, Lemma \ref{lem:xi_psi_infty} holds.
\end{proof}

\begin{lemma}\label{lem:xi_infty_beta}
For any $\beta\in(0,1)$, define 
$\psi_{\infty,\beta}(b,d)$ and $\xi_{\infty,\beta}(d)$ as in Lemma \ref{lem:xi_psi_infty}. 
Then the following results hold. 
\begin{enumerate}[label=(\roman*), topsep=1ex, itemsep=-0.3ex, partopsep=1ex, parsep=1ex]
\item
$\xi_{\infty,\beta}(\cdot)$ is nonnegative, monotone nonincreasing, and right-continuous.

\item
For
$\underline{\tau}_{\infty,\beta}$ in \eqref{eq:tau_infty_beta}, 
we have
\begin{align*}
\xi_{\infty,\beta}(d)>0
\quad\text{for all }d<\underline{\tau}_{\infty,\beta},
\qquad
\xi_{\infty,\beta}(d)=0
\quad\text{for all }d\ge\underline{\tau}_{\infty,\beta}.
\end{align*}
\end{enumerate}
\end{lemma}

\begin{proof}[\bf Proof of Lemma \ref{lem:xi_infty_beta}]
For every $d\in\mathbb R$,
\begin{align*}
\lim_{b\to+\infty}\psi_{\infty,\beta}(b,d)=
\lim_{b\to+\infty}
\left\{
F_{\infty,0}(b) - F_{\infty, 1, \beta}(b + d)
\right\}=0.
\end{align*}
Thus, $\xi_{\infty,\beta}(d)\ge 0$. Moreover, because
$F_{\infty, 1, \beta}$ is nondecreasing,
$\psi_{\infty,\beta}(b,d)$ is nonincreasing in $d$ for every fixed $b$.
Consequently, $\xi_{\infty,\beta}(d)$ is nonincreasing in $d$.

To prove right-continuity, 
consider any nonincreasing sequence $\{d_m\}$ that converges to $d$, i.e., $d_m \downarrow d$. By monotonicity,
$
\xi_{\infty,\beta}(d_m)
\le
\xi_{\infty,\beta}(d)
$
for all $m$. 
For any $\varepsilon>0$, choose $b_{\varepsilon}\in\mathbb R$ such that
$
\psi_{\infty,\beta}(b_{\varepsilon},d)
\ge
\xi_{\infty,\beta}(d)-\varepsilon.
$
By the right-continuity of $F_{\infty, 1}$,
$
\psi_{\infty,\beta}(b_{\varepsilon},d_m)
\longrightarrow
\psi_{\infty,\beta}(b_{\varepsilon},d)
$
as $m\rightarrow \infty$. 
Hence,
\begin{align*}
\liminf_{m\to\infty}\xi_{\infty,\beta}(d_m)
&\ge
\lim_{m\to\infty}
\psi_{\infty,\beta}(b_{\varepsilon},d_m)
=
\psi_{\infty,\beta}(b_{\varepsilon},d)
\ge
\xi_{\infty,\beta}(d)-\varepsilon.
\end{align*}
These then imply that $\lim_{m\to\infty}\xi_{\infty,\beta}(d_m) = \xi_{\infty,\beta}(d)$. Hence, $\xi_{\infty,\beta}(\cdot)$ is right-continuous.

The proof of (ii) follows by the same logic as the proof of Lemma \ref{lem:xin_property}(iii).

From the above, Lemma \ref{lem:xi_infty_beta} holds. 
\end{proof}

\begin{proof}[\bf Proof of Theorem \ref{thm:consistency_finite_high_Level}]
(i) follows immediately from Lemma \ref{lem:tau_infty_convergence}. 

For (ii), 
by Lemma \ref{lem:xi_infty_beta}, 
for any $\delta>0$, 
we have $\xi_{\infty,\beta}(\underline{\tau}_{\infty,\beta}-\delta)>0$, which then implies
\begin{align*}
    n^\kappa \xi_{n,\beta}(c_n) &\ge n^\kappa \xi_{n,\beta}(\underline{\tau}_{\infty,\beta}-\delta) 
    \ge  n^\kappa \xi_{\infty,\beta}(\underline{\tau}_{\infty,\beta}-\delta)
    -  n^\kappa \sup_{d\in \mathbb{R}}|\xi_{n,\beta}(d) - \xi_{\infty,\beta}(d)| \\
    &\converge \infty \text{ as } n \rightarrow \infty,
\end{align*}
where the first inequality follows from Proposition \ref{prop:xi_nbeta}(i), and the last convergence follows from Lemma \ref{lem:xi_psi_infty}. 
The remaining parts in Theorem \ref{thm:consistency_finite_high_Level}(ii) then follow from Theorem \ref{thm:sharp_bound_consist_fin}.

From the above, Theorem \ref{thm:consistency_finite_high_Level} holds.
\end{proof}

\begin{proof}[\bf Comment on Assumption \ref{asmp:fp_limit} under i.i.d.~sampling] \label{rmk:iid_limiting_distribution}
     If, for each $n$, the potential outcomes $\{(Y_i(1),Y_i(0))\}_{i=1}^n$ are i.i.d. samples from some joint distribution with marginals $F_{\infty,1}$ and $F_{\infty,0}$, then by the Glivenko–Cantelli theorem, $\sup_{b\in\mathbb{R}} n^\kappa | F_{nz}(b) - F_{\infty, z}(b)| \to 0$ almost surely for any $\kappa \in (0,1/2)$,
     where $F_{nz}(b)$ is the empirical distribution of $\{Y_i(z)\}_{i=1}^n$.
     If further $F_{\infty,1}$ and $F_{\infty,0}$ are both strictly increasing on their supports, $\big(F_{\infty,1}^{-1}(0+),F_{\infty,1}^{-1}(1)\big)$ and $\big(F_{\infty,0}^{-1}(0+),F_{\infty,0}^{-1}(1)\big)$, Assumption \ref{asmp:fp_limit} holds almost surely.
\end{proof}

\section{Proofs of Theorem \ref{thm:PI_simp} and Theorem \ref{thm:direct_construct}}
\label{ssec:bounds_direct_proof}

\begin{lemma}
\label{lemma:tail_quantile}
Let $X$ be a discrete random variable whose support is a finite set $\{x_1, x_2, \ldots, x_m\}$ with 
$x_1 < x_2 < \cdots < x_m$. Define $F(x) = \Pr(X \le x)$ and $G(x) = \Pr(X \ge x)$ for $x \in \mathbb{R}$. For $\alpha \in (0,1)$,  let  
$q_{1-\alpha} = \inf\bigl\{x \in \mathbb{R} : F(x) \ge 1-\alpha\bigr\}$ denote the $1-\alpha$ quantile of the distribution of $X$. 
Then, for any $x \in \mathbb{R}$,
$G(x) > \alpha$ if and only if $x \le q_{1-\alpha}$. 
\end{lemma}
 
\begin{proof}[\bf Proof of Lemma \ref{lemma:tail_quantile}]

We first show that $q_{1-\alpha} = x_k$, where $k = \min\big\{j \in \{1,\dots,m\} : F(x_j) \ge 1-\alpha\big\}$. 
If $x < x_k$, then we have $F(x) = 0 < 1-\alpha$ when $k=1$ and $F(x) \le F(x_{k-1}) < 1 - \alpha$ when $k>1$, where the latter uses the minimality of $k$. 
If $x \ge x_k$, then we have $F(x) \ge F(x_k) \ge 1-\alpha$. 
Thus, we must have $q_{1-\alpha} = x_k$.

We now prove that, for any $x \in \mathbb{R}$,
$G(x) > \alpha$ if and only if $x \le q_{1-\alpha}$. We consider the following two cases.  
    
    Consider first the case in which $x \le q_{1-\alpha} = x_k$. By definition, we have $G(x) = 1 - F(x-) \ge 1 - F(x_k-)$. If $k=1$, then $F(x_k-) = 0$ and $G(x) = 1 > \alpha$. 
    Otherwise, $k>1$ and $F(x) = F(x_{k-1})$ for any $x \in [x_{k-1}, x_k)$, which then implies that $F(x_k-) = F(x_{k-1}) < 1-\alpha$ and $G(x) \ge 1 - F(x_k-) > \alpha$. Thus, in the case, we must have $G(x) > \alpha$. 

    Consider then the case in which $x > q_{1-\alpha} = x_k$. 
    We have $ F(x-) \ge F(x_k) \ge 1-\alpha$. This then implies that $G(x) = 1 - F(x-) \le \alpha$. 
From the above, we must have $G(x) > \alpha \Longleftrightarrow x \le q_{1-\alpha}$. 
Therefore, Lemma \ref{lemma:tail_quantile} holds. 
\end{proof}

\begin{proof}[\bf Proof of Theorem \ref{thm:PI_simp}]
First, we prove the equivalent forms of the prediction intervals in (i). Fix $1\leq k\leq n_1$ and let $\beta=k/n_1$.
By Corollary \ref{cor:pval_KS} and the definition of
$\hat\xi'_{n,\beta}$,
\begin{align*}
\overline p_{k,c}^{\t}
&=
G_n\Big(
\sup_{b\in\mathbb R}
\big\{
\hat F_n^{\c}(b)
-\min\{1,1-\beta+\hat F_n^{\t}(b+c)\}
\big\}
\Big)
=G_n\big( \hat\xi'_{n,\beta}(c)\big).
\end{align*}
Because the RKS statistic can only take finitely many nonnegative values, we have $g_{n,\alpha} \ge 0$, and, by 
Lemma \ref{lemma:tail_quantile}, 
\begin{align*}
\overline p_{k,c}^{\t} > \alpha 
\ \Longleftrightarrow \ 
G_n\big( \hat\xi'_{n,\beta}(c)\big) >\alpha
\ \Longleftrightarrow \ 
\hat\xi'_{n,\beta}(c) \leq g_{n,\alpha}.
\end{align*}
By definition, we further have 
\begin{align*}
    \hat\xi'_{n,\beta}(c) \leq g_{n,\alpha}
    & \ \Longleftrightarrow \ 
    \hat F_n^{\c}(b)
-\min\{1,1-\beta+\hat F_n^{\t}(b+c)\} \le g_{n,\alpha} \ \text{ for any } b\in \mathbb{R}\\
    & \ \Longleftrightarrow \ 
    \hat F_n^{\c}(b)
-\{1-\beta+\hat F_n^{\t}(b+c)\} \le g_{n,\alpha} \ \text{ for any } b\in \mathbb{R} \\
& \ \Longleftrightarrow \ 
    \hat F_n^{\c}(b)
-\{1-(\beta - g_{n, \alpha})+\hat F_n^{\t}(b+c)\} \le 0 \ \text{ for any } b\in \mathbb{R} \\
& \ \Longleftrightarrow \ 
    \hat F_n^{\c}(b)
-\min\{1, 1-(\beta - g_{n, \alpha})+\hat F_n^{\t}(b+c)\} \le 0 \ \text{ for any } b\in \mathbb{R} \\
& \ \Longleftrightarrow \ 
    \hat\xi'_{n,\beta - g_{n, \alpha}}(c)  \le 0 \\
  & \ \Longleftrightarrow \ 
    \hat\xi'_{n,\beta - g_{n, \alpha}}(c)  = 0, 
\end{align*}
where the second equivalence uses the fact that $g_{n,\alpha} \ge 0$, the last equivalence uses the fact that $\hat\xi'_{n,\beta - g_{n, \alpha}}(c) \ge \lim_{b\rightarrow \infty}\big\{ \hat{F}_n^{\c}(b)
-\min\{1, 1-(\beta - g_{n, \alpha})+\hat F_n^{\t}(b+c)\} \big\} \ge 0$, 
and the remaining equivalences follow from the definition or some algebra.

Second, we prove the equivalent forms of the lower prediction bound. 
From the first part, $ \inf \overline{\mathcal{I}}^\t_{k, \alpha} = \inf \{ c\in \mathbb{R}: \hat{\xi}'_{n,\beta - g_{n, \alpha}}(c) = 0 \}$. 
When $\beta - g_{n, \alpha} > 0$, following the same logic as the proof of Theorem \ref{thm:tau_nbeta}(i), we can show that 
$\inf \{ c\in \mathbb{R}: \hat{\xi}'_{n,\beta - g_{n, \alpha}}(c) = 0 \} = \hat{\underline\tau}_{n,\beta - g_{n, \alpha}}$. 
When $\beta - g_{n, \alpha} \le 0$, both $\hat{\underline\tau}_{n,\beta - g_{n, \alpha}}$ and $\inf \{ c\in \mathbb{R}: \hat{\xi}'_{n,\beta - g_{n, \alpha}}(c) = 0 \}$ are $-\infty$, where the former follows from definition and the latter holds because for any $c \in \mathbb{R}$,
\begin{align*}
    \hat{\xi}'_{n,\beta - g_{n, \alpha}}(c)
    = 
    \sup_{b\in \mathbb{R}} \Big\{ \hat{F}_n^{\c}(b)
-\min\big\{1, 1-(\beta - g_{n, \alpha})+\hat F_n^{\t}(b+c)\big\} \Big\}
= \sup_{b\in \mathbb{R}} \big\{ \hat{F}_n^{\c}(b)
-1 \big\} = 0.
\end{align*}
Thus, we always have $ \inf \overline{\mathcal{I}}^\t_{k, \alpha} = \inf \big\{ c\in \mathbb{R}: \hat{\xi}'_{n,\beta - g_{n, \alpha}}(c) = 0 \big\} = \hat{\underline\tau}_{n,\beta - g_{n, \alpha}}$.

Third, we prove (ii). 
Let $Z_i^*=1-Z_i$ and $Y_i^*=-Y_i$, $1\le i \le n$, denote the transformed data after switching the treatment labels and changing the outcome
signs. 
The empirical distribution function of the observed outcomes in the transformed treated group is 
\begin{align*}
    \hat F_n^{*,\t}(b)
    & = \frac{1}{n_0} \sum_{i:Z_i^*=1} 1\{Y_i^* \le b\}
    = \frac{1}{n_0} \sum_{i:Z_i=0} 1\{-Y_i \le b\}
    = 1 - \frac{1}{n_0} \sum_{i:Z_i=0} 1\{Y_i <-b\}\\
    & = 1-\hat F_n^{\c}\big((-b)-\big), 
\end{align*}
where $\hat F_n^{\c}(x-)$ denotes the left limit of $\hat F_n^{\c}$ at $x$. Analogously, we have $\hat F_n^{*,\c}(b)
=1-\hat F_n^{\t}\big((-b)-\big)$. 
It follows that, for any $d\in \mathbb{R},$
\begin{align}\label{eq:label_switch_gap}
\sup_{b\in\mathbb R}
\{\hat F_n^{*,\c}(b)-\hat F_n^{*,\t}(b+d)\}
\nonumber
&=
\sup_{b\in\mathbb R}
\{\hat F_n^{\c}((-b-d)-) -\hat F_n^{\t}((-b)-)\}
\\
& = 
\sup_{x\in\mathbb R}
\{\hat F_n^{\c}(x-)-\hat F_n^{\t}((x+d)-)\} \notag\\
&=
\sup_{x\in\mathbb R}
\{\hat F_n^{\c}(x)-\hat F_n^{\t}(x+d)\}, 
\end{align}
where the second equality follows by letting $x = -b-d$, 
and the last equality uses the property that, for any distribution function $F$, $F(x-) = \lim_{c \uparrow x} F(x)$ and $F(x) = \lim_{c \downarrow x} F(c-)$. 
Define $\hat\xi^{*,\prime}_{n,\beta}(d)$ in the same way as $\hat\xi'_{n,\beta}(d)$ in Section \ref{sec:direct_connection}, but using the transformed data. 
We then have 
\begin{align*}
    \hat\xi'_{n,\beta}(d) & = 
    \sup_{b\in \mathbb{R}} 
    \big\{ \hat{F}^\c_{n}(b) -
    \min \{ 1, 1 - \beta + \hat{F}^\t_n (b+d) \} \big\} \\
    & = \sup_{b\in \mathbb{R}} 
    \max\{ \hat{F}^\c_{n}(b) - 1,\ \hat{F}^\c_{n}(b) -\hat{F}^\t_n (b+d) - (1-\beta) \}\\
    & = \max\big\{ 0,\ \sup_{b\in \mathbb{R}}\{ \hat{F}^\c_{n}(b) -\hat{F}^\t_n (b+d)\} - (1-\beta) \big\}\\
    & = \max\big\{ 0,\ \sup_{b\in\mathbb R}
\{\hat F_n^{*,\c}(b)-\hat F_n^{*,\t}(b+d)\} - (1-\beta) \big\}
= 
\hat\xi^{*,\prime}_{n,\beta}(d). 
\end{align*}
Moreover, by the distribution-free property of the RKS statistic, for any $\bs{y}\in \mathbb{R}^n$ with mutually distinct elements, 
$t_{\RKS}(\bs{Z}, \bs{y})$ follows the same distribution as $t_{\RKS}(\bs{Z}, -\bs{y})$, and 
\begin{align*}
    t_{\RKS}(\bs{Z}, -\bs{y}) & = t_{\KS}(\bs{Z}, -\bs{y}) = \sup_{\thre \in \mathbb{R}} 
    \Big\{ 
    \frac{1}{n_0} \sum_{i:Z_i=0} \I(-y_i \le \thre) 
    - 
    \frac{1}{n_1} \sum_{i:Z_i=1} \I(-y_i \le \thre) 
    \Big\}\\
    & = 
    \sup_{\thre \in \mathbb{R}} 
    \Big\{ 
    \frac{1}{n_1} \sum_{i:Z_i=1} \I(y_i <- \thre) 
    - 
    \frac{1}{n_0} \sum_{i:Z_i=0} \I(y_i <- \thre) 
    \Big\}\\
    & = \sup_{\thre \in \mathbb{R}} 
    \Big\{ 
    \frac{1}{n_1} \sum_{i:Z_i=1} \I(y_i < \thre) 
    - 
    \frac{1}{n_0} \sum_{i:Z_i=0} \I(y_i < \thre) 
    \Big\}\\
     & = \sup_{\thre \in \mathbb{R}} 
    \Big\{ 
    \frac{1}{n_1} \sum_{i:Z_i=1} \I(y_i \le \thre) 
    - 
    \frac{1}{n_0} \sum_{i:Z_i=0} \I(y_i \le \thre) 
    \Big\}
    \\
    & = t_{\KS}(\bs{1}-\bs{Z}, \bs{y}) = t_{\RKS}(\bs{1}-\bs{Z}, \bs{y}),
\end{align*}
where the third equality uses $\I(y_i <- \thre) = 1 - \I(-y_i \le \thre)$, the third last equality follows by the same logic as \eqref{eq:label_switch_gap}, and the first and last equalities use Proposition \ref{prop:rks_properties}(iii). 
Therefore, the randomization null distribution remains the same after treatment label switching. 
As a result, for any $1\le k\le n_0$ and $\beta = k/n_0$, the p-value for the transformed data has the following equivalent forms:
\begin{align*}
    \overline{p}^\c_{k,c} = G_n (\hat\xi^{*,\prime}_{n,\beta}(c)) = G_n (\hat\xi'_{n,\beta}(c)).
\end{align*}
The remaining results in (ii) then follow immediately from (i).

Finally, we prove (iii).
It suffices to show that, for any $1\le k \le n$ and $\beta = k/n$,  $\beta' = \min_{z\in\{0,1\}}
{\lceil n_z\beta\rceil}/{n_z}$ is the $k$th smallest element from the pooled set of $\{j/n_1: 1\leq j\leq n_1\}$ and $\{j/n_0: 1\leq j\leq n_0\}$. 
Let $h(x) = \sum_{j=1}^{n_1} 1\{j/n_1 \le x\} + \sum_{i=1}^{n_0} 1\{i/n_0 \le x\}$ denote the number of elements in the pooled set that are less than or equal to $x$.
For any $x < \beta'$, we have $n_1 x < \lceil n_1\beta\rceil$, $n_0 x < \lceil n_0\beta\rceil$, and thus 
\begin{align*}
    h(x)
    & = \sum_{j=1}^{n_1} 1\{j\le n_1 x\} + \sum_{i=1}^{n_0} 1\{i \le n_0 x\}\le \lceil n_1\beta\rceil - 1+ \lceil n_0\beta\rceil - 1 \le 
    k -1, 
\end{align*}
where the last inequality uses the fact that $\lceil n_1\beta\rceil + \lceil n_0\beta\rceil \le n_1\beta+n_0 \beta + 1 = n\beta+1 = k+1$, since $n_1\beta+n_0\beta=n\beta=k$.
Moreover, by definition, $\beta'= \min_{z\in\{0,1\}}
{\lceil n_z\beta\rceil}/{n_z} \ge \beta$, and 
\begin{align*}
    h(\beta)
    & = \lfloor n_1 \beta \rfloor + \lfloor n_0 \beta \rfloor
    = 
    \begin{cases}
        k, & \text{if $n_1\beta$ and $n_0\beta$ are both integers,}\\
        k-1, & \text{if $n_1\beta$ and $n_0\beta$ are both non-integers.}
    \end{cases}
\end{align*}
Thus, when $n_1\beta$ and $n_0\beta$ are both integers, we have $\beta' = \beta$ and $h(\beta') = k$. 
When $n_1\beta$ and $n_0\beta$ are both non-integers, ${\lceil n_z\beta\rceil}/{n_z}$ is strictly greater than $\beta$ for $z=0,1$. In this case, we must have $h(\beta') \ge h(\beta)+1 = k$, because one additional element in the pooled set, i.e., $\min_{z\in\{0,1\}}
{\lceil n_z\beta\rceil}/{n_z}$, can be bounded by $\beta'$ but not $\beta$. 
From the discussion before, we then have $h(\beta') \ge k$ and $h(x)<k$ for $x<\beta'$. Therefore, $\beta'$ must be the $k$th smallest element from the pooled set. 

From the above, Theorem \ref{thm:PI_simp} holds. 
\end{proof}

\begin{proof}[\bf Comment on the range of $\beta' = \min_{z\in \{0,1\}}\lceil n_z \beta \rceil / n_z$ for $\beta = k/n$]
    Define $h(\cdot)$ the same as in the proof of Theorem \ref{thm:PI_simp}. We have 
\begin{align*}
    h(\beta+1/n) = \lfloor n_1 \beta + n_1/n \rfloor + \lfloor n_0 \beta + n_0/n \rfloor
    \ge n_1 \beta + n_1/n + n_0 \beta + n_0/n - 1 = k. 
\end{align*}
From the proof of Theorem \ref{thm:PI_simp}, $\beta'$ is the $k$th smallest element from the pooled set. Thus, we must have $\beta+1/n \ge \beta'$. Moreover, as shown in the proof of Theorem \ref{thm:PI_simp}, $\beta' \ge \beta$. 
\end{proof}

\begin{proof}[\bf Proof of Theorem \ref{thm:direct_construct}]
First, we prove the inequalities in \eqref{eq:ks_dist}.
For any $x\in\mathbb R$, by definition and some algebra, 
\begin{align*}
F_{n1}(x)-\hat F_n^{\t}(x)
&=
\frac{n_0}{n}
\left\{
\frac{1}{n_0}\sum_{i=1}^n(1-Z_i)\I\{Y_i(1)\leq x\}
-\frac{1}{n_1}\sum_{i=1}^nZ_i\I\{Y_i(1)\leq x\}
\right\},\\
\hat F_n^{\c}(x)-F_{n0}(x)
&=
\frac{n_1}{n}
\left\{
\frac{1}{n_0}\sum_{i=1}^n(1-Z_i)\I\{Y_i(0)\leq x\}
-\frac{1}{n_1}\sum_{i=1}^nZ_i\I\{Y_i(0)\leq x\}
\right\}.
\end{align*}
These imply
\begin{align}\label{eq:emp_pop_ks_diff}
\sup_x\{F_{n1}(x)-\hat F_n^{\t}(x)\}
&=r_0 t_{\KS}(\bs Z,\bs Y(1))
\leq r_0 t_{\RKS}(\bs Z,\bs Y(1)),\notag\\
\sup_x\{\hat F_n^{\c}(x)-F_{n0}(x)\}
&=r_1 t_{\KS}(\bs Z,\bs Y(0))
\leq r_1 t_{\RKS}(\bs Z,\bs Y(0)),
\end{align}
where the inequalities follow from Proposition \ref{prop:rks_properties}(iii). 

Second, define
\begin{align*}
\mathcal E_1
&=\big\{
\sup_x\{F_{n1}(x)-\hat F_n^{\t}(x)\}
\leq r_0 g_{n,\alpha}
\big\},
\ \ 
\mathcal E_0
=\big\{
\sup_x\{\hat F_n^{\c}(x)-F_{n0}(x)\}
\leq r_1 g_{n,\alpha}
\big\}.
\end{align*}
By \eqref{eq:emp_pop_ks_diff} and the definition
of $g_{n,\alpha}$, 
\begin{align*}
\Pr(\mathcal E_1)\geq1-\alpha,
\qquad
\Pr(\mathcal E_0)\geq1-\alpha,
\qquad
\Pr(\mathcal E_1\cap\mathcal E_0)\geq1-2\alpha.
\end{align*}

Third, we prove (i). 
On
$\mathcal E_1\cap\mathcal E_0$, 
we have, for any $u\in [0,1]$, 
\begin{align*}
F_{n1}^{-1}(u)
&= \inf \{ x : F_{n1}(x) \ge u \}
\ge \inf \{
x : \hat F_n^{\t}(x) + r_0 g_{n, \alpha} \ge u
\} \\
& = \inf \{
x : \hat F_n^{\t}(x) \ge u - r_0 g_{n, \alpha}
\} = \hat F_n^{\t, -1}
(u - r_0 g_{n, \alpha} ),
\end{align*}
and
\begin{align*}
F_{n0}^{-1}(u)
&= \inf \{ x : F_{n0}(x) \ge u \} 
\le \inf \{
x : \hat F_n^{\c}(x) - r_1 g_{n,\alpha} \ge u
\} \\
&
= \inf \{
x : \hat F_n^{\c}(x) \ge r_1 g_{n,\alpha} + u
\} = \hat F_n^{\c, -1}
(u + r_1 g_{n,\alpha}).
\end{align*}
These imply that, for any $\beta\in (0,1),$
\begin{align*}
\underline{\tau}_{n,\beta}
&\equiv
\sup_{u\in[0,\beta]}
\{
F_{n1}^{-1}(u)
-
F_{n0}^{-1}(u+1-\beta)
\}
\\
&\ge
\sup_{u\in[0,\beta]}
\big\{
\hat F_n^{\t,-1}
(
u-r_0 g_{n, \alpha}
)
-
\hat F_n^{\c,-1}
(
u+r_1 g_{n,\alpha}+1-\beta
)
\big\}
\\
&=
\sup_{v\in [-r_0 g_{n, \alpha},\,
\beta-r_0 g_{n, \alpha} ]}
\big\{
\hat F_n^{\t,-1}(v)
-
\hat F_n^{\c,-1}
(
v+
r_0 g_{n, \alpha}
+
r_1 g_{n,\alpha}
+1-\beta
)
\big\}
\\
& = 
\sup_{v\in [-r_0 g_{n, \alpha},\,
\beta-r_0 g_{n, \alpha} ]}
\big\{
\hat F_n^{\t,-1}(v)
-
\hat F_n^{\c,-1}
(
v+1-(\beta-g_{n,\alpha})
)
\big\}
\\
&=
\begin{cases}
 \sup_{v\in[0,\beta-g_{n,\alpha}]}
\big\{
\hat F_n^{\t,-1}(v)
-
\hat F_n^{\c,-1}
(
v+1-(\beta-g_{n,\alpha})
)
\big\}, & \text{ if }  \beta-g_{n,\alpha} \ge 0, \\
-\infty, & \text{ otherwise}
\end{cases}
\\
& = 
\hat{\underline{\tau}}_{n,\beta-g_{n,\alpha}},
\end{align*}
where the second last equality holds because the expression inside the supremum equals minus infinity when $v \notin [0,\beta-g_{n,\alpha}].$
Thus, $\hat{\underline{\tau}}_{n,\beta-g_{n,\alpha}}$ is a $1-2\alpha$ simultaneous lower confidence bound on the $\beta$ quantile of ITEs among all units.  

We next prove (ii). 
Let 
\begin{align*}
\check{F}_{n}^{\t}(x)
&=\frac{1}{n_1}\sum_{i=1}^n Z_i\I\{Y_i(0)\leq x\},
&
\check{F}_{n}^{\c}(x)
&=\frac{1}{n_0}\sum_{i=1}^n(1-Z_i)\I\{Y_i(1)\leq x\}.
\end{align*}
denote the empirical distribution functions of the
counterfactual outcomes in the two treatment groups. By definition and from the proof in the first part, 
$
\sup_x\{\hat F_n^{\c}(x)-\check{F}_{n}^{\t}(x)\}
=t_{\KS}(\bs Z,\bs Y(0))
\leq t_{\RKS}(\bs Z,\bs Y(0)).
$
Thus, with probability at least $1-\alpha$,
$\sup_x\{\hat F_n^{\c}(x)-\check{F}_{n}^{\t}(x)\} \le g_{n, \alpha}$. 
On the event that $\sup_x\{\hat F_n^{\c}(x)-\check{F}_{n}^{\t}(x)\} \le g_{n, \alpha}$, by the same logic as before, 
\begin{align*}
    \check{F}_{n}^{\t, -1}(u)
&\leq\hat F_n^{\c,-1}(u+g_{n,\alpha}),
\qquad  u\in [0,1]
\end{align*}
where $\check{F}_{n}^{\t, -1}$ denotes the quantile function corresponding to the distribution function $\check{F}_{n}^{\t}$. 
Applying the sharp lower bound
\eqref{eq:sharp_bound_effect_quantile_orig} within the treated group and by the same logic as before, we have, for any
$\beta\in (0,1)$, 
\begin{align*}
F_{n\tau}^{\t,-1}(\beta)
& \ge 
\sup_{u\in[0,\beta]}
\big\{
\hat F_n^{\t,-1}(u)
-\check{F}_{n}^{\t, -1}(u+1-\beta)
\big\}\\
&\geq
\sup_{u\in[0,\,\beta]}
\big\{
\hat F_n^{\t,-1}(u)
-\hat F_n^{\c,-1}(u+1-\beta+g_{n,\alpha})
\big\}\\
& = 
\begin{cases}
   \sup_{u\in[0,\,\beta - g_{n, \alpha}]}
\big\{
\hat F_n^{\t,-1}(u)
-\hat F_n^{\c,-1}(u+1-(\beta-g_{n,\alpha} ) )
\big\}, & \text{ if } \beta-g_{n,\alpha} \ge 0,\\
-\infty, & \text{ otherwise}
\end{cases}
\\
&=\hat{\underline\tau}_{n,\beta-g_{n,\alpha}}.
\end{align*}
Thus, $\hat{\underline\tau}_{n,\beta-g_{n,\alpha}}$ is 
a $1-\alpha$ simultaneous lower prediction bound on the $\beta$ quantile of ITEs among treated units. 
The proof for control units follows by the same logic and is omitted for conciseness.

From the above, Theorem \ref{thm:direct_construct} holds. 
\end{proof}

\section{Proofs of Proposition \ref{prop:tau_beta} and Theorem \ref{thm:sharp_bound_consist}}
\label{ssec:consist_super}

\begin{lemma} \label{lem:tau_beta}
    Consider any random variables $\big(Y(1),Y(0)\big)$ with marginal distributions $F_1$ and $F_0$, and let $F_\tau$ be the distribution of $\tau=Y(1)-Y(0)$.
    Then, for $\beta \in (0,1)$,
    \begin{align*}
        F^{-1}_\tau(\beta) \ge \underline{\tau}_\beta \equiv \sup_{u \in [0,\beta]} \big\{F_1^{-1}(u)-F_0^{-1}(u+1-\beta)\}
    \end{align*}
    and it is attainable at some copula for $\big(Y(1),Y(0)\big)$.
\end{lemma}

\begin{proof}[\bf Proof of Lemma \ref{lem:tau_beta}]
    By Lemma \ref{lem:quntile_property}(vi), 
    to prove $ F^{-1}_\tau(\beta) \ge \underline{\tau}_\beta$, 
    it suffices to show that $F_\tau(c) < \beta$ for any $c<\underline{\tau}_\beta$. 
    Consider now any $c<\underline{\tau}_\beta$.
    By the definition of $\underline{\tau}_\beta$, there exists $u_0 \in [0, \beta]$ such that $F_1^{-1}(u_0)-F_0^{-1}(u_0+1-\beta)>c$.
    Because $F_1^{-1}(0)=-\infty$, necessarily $u_0\in(0,\beta]$.    
    Let $a=F_1^{-1}(u_0)$ and $b=F_0^{-1}(u_0+1-\beta)$ so that $a-b>c$. Choose $a'<a$ such that $a'-b>c$. Then we have $\{ Y(1)>a'\} \bigcap \{ Y(0)\le b\} \subseteq \{\tau>c\}$, which implies $\{\tau\le c\} \subseteq \{ Y(1)\le a'\} \bigcup \{ Y(0)> b\}$. As a result,
    \begin{align*}
        F_\tau(c) &= \Pr(\tau \le c) \le \Pr\big(Y(1) \le a'\big) + \Pr\big(Y(0) > b\big) = F_1(a') +1-\Pr(Y(0) \le b) \\
        &<u_0 + 1 - (u_0 + 1 - \beta) = \beta,
    \end{align*}
    where we use the facts that $F_1(a') < u_0$ by Lemma \ref{lem:quntile_property}(ii) and $F_0(b) \ge u_0+1-\beta$ by Lemma \ref{lem:quntile_property}(i).  
    Hence $F_\tau(c) < \beta$ for any $c<\underline{\tau}_\beta$. Consequently, $F^{-1}_\tau(\beta) \ge \underline{\tau}_\beta$.

    Next, we show that $\underline{\tau}_\beta$ is achievable. Let $U\sim \mathrm{Unif}(0,1)$, and define 
    \begin{align*}
    V=
        \begin{cases}
            U+1-\beta, & 0<U\le \beta,\\
            U-\beta, & \beta<U<1.
        \end{cases}
    \end{align*}
    By definition, we can verify that $V\sim \mathrm{Unif}(0,1)$. 
    Let $( \tilde{Y}(1), \tilde{Y}(0) )=( F_1^{-1}(U), F_0^{-1}(V))$, which has marginal distributions $F_1$ and $F_0$. 
    Let $\tilde \tau=\tilde Y(1) - \tilde Y(0)$, and let $F_{\tilde \tau}$ denote its distribution function. On the event $\{U\le \beta\}$, we have $V=U+1-\beta$, and
    \begin{align*}
        \tilde \tau= F_1^{-1}(U)-F_0^{-1}(U+1-\beta)
        \le \sup_{u\in[0,\beta]} \{F_1^{-1}(u)-F_0^{-1}(u+1-\beta)\}
        =\underline{\tau}_\beta.
    \end{align*}
    Thus, $F_{\tilde \tau}(\underline{\tau}_\beta)=\Pr(\tilde \tau \le \underline{\tau}_\beta)\ge \Pr(U\le \beta)=\beta$, which implies $F_{\tilde \tau}^{-1}(\beta) \le  \underline{\tau}_\beta$.
    Note that the lower bound proved earlier 
    implies 
    $F_{\tilde \tau}^{-1}(\beta) \ge  \underline{\tau}_\beta$. 
    Hence, we have $F_{\tilde \tau}^{-1}(\beta) =  \underline{\tau}_\beta$, indicating the sharpness of the lower bound $\underline{\tau}_\beta$.

    From the above, Lemma \ref{lem:tau_beta} holds. 
\end{proof}

\begin{proof}[\bf Proof of Proposition \ref{prop:tau_beta}]

The equivalent form in \eqref{eq:quantile_lower_bound} follows by the same argument as in the proof of Theorem \ref{thm:tau_nbeta}(i), and the sharp lower bound follows from Lemma \ref{lem:tau_beta}.
\end{proof}

\begin{proof}[\bf Proof of Theorem \ref{thm:sharp_bound_consist}]
Because $\{Y_i(z)\}_{i=1}^n$ are i.i.d. samples from $F_z$ for $z\in \{0,1\}$ and the marginal distribution functions are strictly increasing on their supports under Assumption \ref{asmp:iid}, Assumption \ref{asmp:fp_limit} holds almost surely by previous comment under i.i.d. sampling at the end of Appendix \ref{ssec:consist_fin}. Theorem \ref{thm:consistency_finite_high_Level} then implies that 
$
\overline{p}_{\lceil n_1\beta\rceil,c}^{\t}
\convergeas 0
$
as $n\rightarrow \infty$. 

We next show that the two p-values are equal almost surely. Under Assumption \ref{asmp:iid}, all finite coordinates of
$
\bs Y-\bs Z\circ\bs{\xi}_{\lceil n_1\beta\rceil,c}^{\t}
$
are distinct almost surely, with  ties occurring only among the
$n_1-\lceil n_1\beta\rceil$ treated units whose imputed outcomes are $-\infty$. 
From Lemma \ref{lem:RKS_KS_inequality}, 
\begin{align*}
t_{\KS}
\left(
\bs Z,
\bs Y-\bs Z\circ\bs{\xi}_{\lceil n_1\beta\rceil,c}^{\t}
\right)
=
t_{\RKS}
\left(
\bs Z,
\bs Y-\bs Z\circ\bs{\xi}_{\lceil n_1\beta\rceil,c}^{\t}
\right)
\quad\text{almost surely}.
\end{align*}
Consequently, $\overline{p}_{\lceil n_1\beta\rceil,c}^{\t}$ equals $p_{\lceil n_1\beta\rceil,c}^{\t}$ almost surely. 

From the above, Theorem \ref{thm:sharp_bound_consist} holds.
\end{proof}

\section{Technical details for Appendix \ref{sec:infinite_contact_points}}
\label{ssec:example_infinite_contact_points}

Note that $F_1=F_0=F$ has density
\begin{align*}
f(x)
=
\begin{cases}
\dfrac{1}{4}, & |x|\leq 1,\\[4pt]
\dfrac{1}{4}\exp\{-(|x|-1)\}, & |x|>1.
\end{cases}
\end{align*}
Integrating the density gives
\begin{align*}
F(b)
=
\begin{cases}
\dfrac{1}{4}e^{b+1}, & b\leq -1,\\[6pt]
\dfrac{1}{4}(b+2), & -1\leq b\leq 1,\\[6pt]
1-\dfrac{1}{4}e^{-(b-1)}, & b\geq 1.
\end{cases}
\end{align*}
Consequently, the quantile function is
\begin{align*}
F^{-1}(u)
=
\begin{cases}
\log(4u)-1, & 0<u\leq \dfrac{1}{4},\\[6pt]
4u-2, & \dfrac{1}{4}\leq u\leq \dfrac{3}{4},\\[6pt]
1-\log\{4(1-u)\}, & \dfrac{3}{4}\leq u<1.
\end{cases}
\end{align*}

For $\beta=3/4$, the sharp lower bound is
\begin{align*}
\underline{\tau}_\beta
&=
\sup_{0\leq u\leq 3/4}
\left\{
F^{-1}(u)-F^{-1}\left(u+\frac{1}{4}\right)
\right\}.
\end{align*}
Define
$
h(u)
=
F^{-1}(u)-F^{-1}(u+1/4).
$
Using the previous piecewise expression for $F^{-1}$, we obtain
\begin{align*}
h(u)
=
\begin{cases}
\log(4u)-4u,
& 0<u\le \dfrac{1}{4},\\[6pt]
-1,
& \dfrac{1}{4}\leq u\leq\dfrac{1}{2},\\[6pt]
4u-3+\log(3-4u),
& \dfrac{1}{2}\le u<\dfrac{3}{4}.
\end{cases}
\end{align*}
For $0<u\le 1/4$, applying $\log x\leq x-1$ with $x=4u$ gives
$
h(u)
=
\log(4u)-4u
\leq
-1,
$
with equality achieved at $u=1/4$. For $1/2\le u<3/4$, let
$x=3-4u\in(0,1]$. By the same logic, 
$
h(u)
=-x+
\log x
\leq
-1,
$
with equality achieved at $u=1/2$. 
Therefore,
\begin{align*}
\underline{\tau}_\beta=-1,
\quad 
\text{and}
\quad 
\argmax_{u\in[0,3/4]}h(u)
=
\left[\frac{1}{4},\frac{1}{2}\right].
\end{align*}

We next determine the contact set. Since $\beta=3/4$ and
$\underline{\tau}_\beta=-1$, 
we have 
\begin{align*}
\psi_\beta(b,\underline{\tau}_\beta)=
F(b)-F_{1,\beta}(b-1)=
F(b)-\min\left\{
1, \ \frac{1}{4}+F(b-1)
\right\}.
\end{align*}
Direct calculation yields
\begin{align*}
\psi_\beta(b,-1)
=
\begin{cases}
\dfrac{1}{4}\left\{(e-1)e^b-1\right\},
& b<-1,\\[6pt]
\dfrac{1}{4}\left( b+1-e^b\right),
& -1\leq b<0,\\[6pt]
0,
& 0\leq b\leq 1,\\[6pt]
\dfrac{1}{4}\left\{2-b-e^{-(b-1)}\right\},
& 1<b<2,\\[6pt]
-\dfrac{1}{4}e^{-(b-1)},
& b\geq 2.
\end{cases}
\end{align*}
If $b<-1$, then
$
(e-1)e^b
<
(e-1)e^{-1}
=
1-e^{-1}
<
1,
$
and hence $\psi_\beta(b,-1)<0$. If $-1\leq b<0$,
then the inequality $e^b\geq 1+b$, with equality only at $b=0$,
implies
$
\psi_\beta(b,-1)<0.
$
If $1<b<2$, 
then the inequality 
$e^{-(b-1)}\ge 1-(b-1)=2-b$, with equality only at $b=1$,
implies $
\psi_\beta(b,-1)<0.
$
If $b\geq 2$, 
then  $
\psi_\beta(b,-1)<0
$ obviously.
It follows that
$\psi_\beta(b,-1)=0$ if and only if $b\in[0,1].$
Therefore, the contact set is
$
\mathcal{S}_\beta^\star=[0,1].
$

\section{Proofs of Theorem \ref{thm:xin_beta_convergence}, Corollary \ref{cor:asymptotic_power} and Theorem \ref{thm:asymptotic_power_normal}}
\label{ssec:asymptotic_power}

In this section, we define the Skorohod space $D[p,q]$ as the set of all functions $z:[p,q] \to \mathbb{R}$ that are right continuous and whose limits from the left exist everywhere in $T=[p,q]$, where $T=[p,q]$ is an interval in the extended real line. 
Define the norm in $D(T)\equiv D[p,q]$ as $\| z \|_T = \sup_{x \in T} |z(x)|$ and $z_n \rightarrow z$ if $\| z_n-z \|_T \to 0$. 
It can be shown that $C[p,q] \subseteq D[p,q] \subseteq l^\infty [p,q]$, see \citep[][p. 257]{van2000asymptotic}, where $C[p, q]$ is the set of all continuous functions $z:[p,q] \to \mathbb{R}$ and $l^\infty [p,q]$ is the set of all functions $z:[p,q] \to \mathbb{R}$ such that $\| z \|_T<\infty$. 

In addition, we assume $n_1/n$ has a limit $\lambda \in (0,1)$, i.e. $\lim_{n \to \infty}n_1/n = \lambda$.

To prove Theorem \ref{thm:xin_beta_convergence}, we need the following ten lemmas.

\begin{lemma}
\label{lem:continuous mapping condition}
Given a bounded interval $T = [p,q] \subseteq \mathbb{R}$ and non-random converging sequence $\{c_n\}$ with $\lim_{n \to \infty}c_n = c_0$, define $h: C[-\infty, +\infty] \to D[p,q]$ and $h_n: D[-\infty, +\infty] \to D[p,q]$ such that for $z_1 \in D[-\infty, +\infty]$ and $z_2 \in C[-\infty, +\infty]$
\begin{align*}
h_n(z_1)(b) = z_1(b + c_n),\quad h(z_2)(b) = z_2(b + c_0)  \text{ where } b \in [p,q].
\end{align*}
Suppose $f_n \in D[-\infty, +\infty]$ and $f \in C[-\infty, +\infty]$ satisfy that $f_n \rightarrow f $ in $ D[-\infty, +\infty]$, then $h_n(f_n) \rightarrow h(f) $ in $ D[p,q]$.

\end{lemma}

\begin{proof}[\bf Proof of Lemma \ref{lem:continuous mapping condition}]

Note the difference between $h_n(f_n)$ and $h(f)$ can be bounded by
\begin{align*}
\| h_n(f_n) - h(f) \|_T &= \sup_{b \in [p,q]} |f_n(b + c_n) - f(b + c_0)| \\
&\le \sup_{b \in [p,q]} |f_n(b + c_n) - f(b + c_n)| + \sup_{b \in [p,q]} |f(b + c_n) - f(b + c_0)| \\
&\le \sup_{b \in \mathbb{R}} |f_n(b) - f(b)| + \sup_{b \in [p,q]} |f(b + c_n) - f(b + c_0)|.
\end{align*}
Since $f_n \rightarrow f $ in $ D[-\infty, +\infty]$, it follows that $\sup_{b \in \mathbb{R}} |f_n(b) - f(b)| \rightarrow 0$ as $n\rightarrow \infty$. 
Moreover,
since $f \in C[-\infty, +\infty]$, $f$ is uniformly continuous on any bounded set. Hence, by $c_n \rightarrow c_0$, we can show that 
$
\sup_{b \in [p,q]} |f(b + c_n) - f(b + c_0)| \rightarrow 0 
$
as $n\rightarrow \infty$. 
From the above, we can obtain that
$
h_n(f_n) \rightarrow h(f) \text{ in } D[p,q],
$
i.e., Lemma \ref{lem:continuous mapping condition} holds. 
\end{proof}

\begin{lemma}
\label{lem:emphirical_cn}
Let $X_1, X_2,\dots$ be i.i.d. random  variables from a continuous distribution $F$, 
and $\hat{F}_n$ be the empirical distribution function of $\{X_i\}_{i=1}^n.$
For a bounded interval $T = [p,q]\subseteq\mathbb{R}$ and non-random converging sequence $\{c_n\}$ with $\lim_{n \to \infty}c_n = c_0$,  the empirical processes $\sqrt{n} \{ \hat{F}_n(b + c_n) - F(b + c_n) \}$ converge in distribution in the space $D[p,q]$ to 
$\mathcal{B}(F(b + c_0))$, i.e.
\begin{align*}
\sqrt{n} \{ \hat{F}_n(b + c_n) - F(b + c_n) \} \converged \mathcal{B}(F(b + c_0)) \text{ in } D[p,q],
\end{align*}
where $\mathcal{B}$ is a Brownian bridge. 
\end{lemma}

\begin{proof}[\bf Proof of Lemma \ref{lem:emphirical_cn}]

By Donsker's Theorem \citep[see, e.g.,][p. 266]{van2000asymptotic}, $\sqrt{n} \{ \hat{F}_n(b) - F(b) \} \converged \mathcal{B}(F(b))$ in $D[-\infty, +\infty]$.
Note that $\mathcal{B}(F(b))$ is a random element in $C[-\infty, +\infty] $ almost surely.
From Lemma \ref{lem:continuous mapping condition} and using the Continuous Mapping Theorem \citep[][p. 259]{van2000asymptotic} for $h_n$ and $h$,  we can show that 
\begin{align*}
\sqrt{n} \{ \hat{F}_n(b + c_n) - F(b + c_n) \} 
&=  \sqrt{n}  \{\hat{F}_n - F\}(b + c_n)  = h_n \big(\sqrt{n} \{ \hat{F}_n - F\}\big)(b) \\
& \converged h(\mathcal{B}(F))(b) = \mathcal{B}(F(b + c_0)) \quad \text{in } D[p,q]. 
\end{align*}
Therefore, Lemma \ref{lem:emphirical_cn} holds. 
\end{proof}

\begin{lemma}
\label{lem:sup converge}
Consider an interval $T = [p,q] \subseteq \overline{\mathbb{R}}$, where $\overline{\mathbb{R}}$ is the extended real line, and a nonempty subset $S \subseteq T$. If random process $Y_n(b)$ converges in distribution to $Y(b)$ in $D[p,q]$, then the sequence of random variables $\sup_{b \in S} Y_n(b)$ converges in distribution to $\sup_{b \in S} Y(b)$.
 
\end{lemma}

\begin{proof}[\bf Proof of Lemma \ref{lem:sup converge}]
Define $h : D[p,q] \to \mathbb{R}$ by $h(f) = \sup_{b \in S} f(b)$.
For any $f_1, f_2 \in D[p,q]$, we have 
\begin{align*}
|h(f_1) - h(f_2)| &= | \sup_{b \in S} f_1(b) - \sup_{b \in S} f_2(b) | 
\leq \sup_{b \in S} |f_1(b) - f_2(b)| \leq \sup_{b \in [p,q]} |f_1(b) - f_2(b)| \\
&= \|f_1 - f_2\|_T, 
\end{align*}
where the first inequality follows from Lemma \ref{lem:absolute_sup}. 
By the Continuous Mapping Theorem \citep[][p. 259]{van2000asymptotic}, we have 
$
\sup_{b \in S} Y_n(b) = h(Y_n) \converged h(Y) = \sup_{b \in S} Y(b).
$
Therefore, Lemma \ref{lem:sup converge} holds. 
\end{proof}

\begin{lemma}
\label{lem:fn to infinity}
Suppose that  Assumption \ref{asmp:iid} holds and $\lim_{n \to \infty}n_1/n = \lambda\in (0,1)$, and consider 
$\hat{\psi}_{n,\beta}(b,d)$ in \eqref{eq:notation_psin}, $\psi_\beta(b,d)$ in \eqref{eq:notation_xi} and $\underline{\tau}_{\beta}$ in \eqref{eq:quantile_lower_bound}. 
Let $T \subseteq \mathbb{R}$ be a set satisfying $\sup_{b \in T} \psi_\beta(b,\underline{\tau}_{\beta}) < 0$.
Then $\sup_{b \in T} \sqrt{n} \hat{\psi}_{n,\beta}(b,c_n)$ converges in probability to $-\infty$ for $c_n = \underline{\tau}_{\beta} - h/\sqrt{n}$, i.e.
\begin{align*}
\sup_{b \in T} \sqrt{n} \hat{\psi}_{n,\beta}(b,c_n) \convergep -\infty.
\end{align*}

\end{lemma}

\begin{proof}[\bf Proof of Lemma \ref{lem:fn to infinity}]
Because the density $f_1$ is bounded by Assumption \ref{asmp:iid},  
there exists a finite constant $U$ such that $f_1(b) < U$ for all $b\in \mathbb{R}$. Consequently, 
\begin{align*}
& \quad \ \sup_{b \in T} \sqrt{n} \hat{\psi}_{n,\beta}(b, c_n)
\\
&= \sup_{b \in T} \bigg \{ \sqrt{n} \Big( \hat{F}_n^\c(b) - \hat{F}_{n,\lceil n_1\beta \rceil/n_1}^\t (b + c_n) \Big) \bigg \} \\
&= \sup_{b \in T} \bigg \{ \sqrt{n} \Big( \hat{F}_n^\c(b) - F_0(b) \Big) - \sqrt{n} \Big( \hat{F}_{n,\lceil n_1\beta \rceil/n_1}^\t (b + c_n)  - F_{1, \beta} (b + c_n) \Big)  \\
&\quad + \sqrt{n} \Big( F_{1, \beta} (b + \underline{\tau}_{\beta}) - F_{1, \beta} (b + c_n) \Big) + \sqrt{n} \Big( F_0(b) - F_{1, \beta} (b + \underline{\tau}_{\beta}) \Big) \bigg \}\\
&\le \sup_{b \in T} \bigg \{ \sqrt{n} \Big( \hat{F}_n^\c(b) - F_0(b) \Big) \bigg \} + \sup_{b \in T} \bigg \{-\sqrt{n} \Big( \hat{F}_{n,\lceil n_1\beta \rceil/n_1}^\t (b + c_n) - F_{1, \beta} (b + c_n) \Big) \bigg \} \\
&\quad + \sup_{b \in T} \bigg \{\sqrt{n} \Big( F_{1, \beta} (b + \underline{\tau}_{\beta}) - F_{1, \beta} (b + c_n) \Big) \bigg \}+ \sup_{b \in T} \bigg \{\sqrt{n} \Big( F_0(b) - F_{1, \beta} (b + \underline{\tau}_{\beta}) \Big)\bigg \}\\
&\le \sup_{b \in \mathbb{R}}  \sqrt{n} \Big| \hat{F}_n^\c(b) - F_0(b) \Big| + \sup_{b \in  \mathbb{R}} \sqrt{n} \Big| \hat{F}_{n,\lceil n_1\beta \rceil/n_1}^\t (b ) - F_{1, \beta} (b) \Big| \\
&\quad + \sup_{b \in T} \sqrt{n} \Big| F_{1, \beta} (b + \underline{\tau}_{\beta}) - F_{1, \beta} (b + c_n) \Big| + \sup_{b \in T} \sqrt{n} \psi_\beta(b,\underline{\tau}_{\beta})\\
&\le \sup_{b \in \mathbb{R}} \sqrt{n} \Big| \hat{F}_n^\c(b) - F_0(b) \Big| + \sup_{b \in  \mathbb{R}}  \sqrt{n} \Big| \hat{F}_{n}^\t (b) - F_1 (b) \Big| + \frac{\sqrt{n}}{n_1} \\
&\quad + \sup_{b \in T} \sqrt{n} \Big| F_1 (b + \underline{\tau}_{\beta}) - F_1 (b + c_n) \Big|  + \sup_{b \in T} \sqrt{n} \psi_\beta(b,\underline{\tau}_{\beta}) \quad \text{by Lemma \ref{lem:absolute_min}} \\
&\le \sup_{b \in \mathbb{R}} \sqrt{n} \Big| \hat{F}_n^\c(b) - F_0(b) \Big| + \sup_{b \in \mathbb{R}} \sqrt{n} \Big| \hat{F}_{n}^\t (b) - F_1 (b) \Big| + \frac{\sqrt{n}}{n_1}
+ h \cdot U + \sup_{b \in T} \sqrt{n} \psi_\beta(b, \underline{\tau}_{\beta}) ,
\end{align*}
where the second last inequality uses Lemma \ref{lem:absolute_min}, and the last inequality is due to 
\begin{align*}
\sqrt{n} \Big| F_1 (b + \underline{\tau}_{\beta}) - F_1 (b + c_n) \Big|  \leq \sqrt{n}  \big(\underline{\tau}_{\beta}-c_n \big) \cdot \sup_{b \in \mathbb{R}} f_1(b) \leq h \cdot U \text{ for all } b\in \mathbb{R}. 
\end{align*}
Note that both  $\sup_{b \in \mathbb{R}} \{ \sqrt{n} | \hat{F}_n^\c(b) - F_0(b) | \}$ and $\sup_{b \in \mathbb{R}} \{ \sqrt{n} | \hat{F}_{n}^\t (b) - F_1 (b) | \}$ converge to the scaled Kolmogorov distribution and are thus of order $O_p(1)$. 
Because $\sup_{b \in T} \psi_\beta(b, \underline{\tau}_{\beta}) < 0$, we must have $\sup_{b \in T} \sqrt{n} \psi_\beta(b, \underline{\tau}_{\beta}) \rightarrow -\infty$ and consequently $\sup_{b \in T} \sqrt{n} \hat{\psi}_{n,\beta}(b,c_n) \convergep -\infty$.
Therefore, Lemma \ref{lem:fn to infinity} holds. 
\end{proof}

\begin{lemma}
\label{lem:pr_inequality}
Let $\{A_n\},\{B_n\}$ and $\{C_n\}$ be three sequences of random variables. 
\begin{enumerate}[label=(\roman*)]
    \item If $\lim_{n\to\infty}\Pr(A_n\geq B_n) = 1$, then
    \begin{align*}
    \varliminf_{n\to\infty}\Pr(A_n\leq C_n) \leq \varliminf_{n\to\infty}\Pr(B_n\leq C_n) \text{ and }     \varlimsup_{n\to\infty}\Pr(A_n\leq C_n) \leq \varlimsup_{n\to\infty}\Pr(B_n\leq C_n),
    \end{align*}
    \item If $\lim_{n\to\infty}\Pr(A_n=B_n) = 1$, then   
    \begin{align*}
    \varliminf_{n\to\infty}\Pr(A_n\leq C_n)= \varliminf_{n\to\infty}\Pr(B_n\leq C_n) \text{ and }     \varlimsup_{n\to\infty}\Pr(A_n\leq C_n)= \varlimsup_{n\to\infty}\Pr(B_n\leq C_n),
    \end{align*}
    \item If $\lim_{n\to\infty}\Pr(A_n=B_n) = 1$ and $\lim_{n\to\infty}\Pr(B_n=C_n) = 1$, then 
    \begin{align*}
    \lim_{n\to\infty}\Pr(A_n=C_n) = 1.
    \end{align*}

\end{enumerate}
\end{lemma}

\begin{proof}[\bf Proof of Lemma \ref{lem:pr_inequality}]
We first prove (i). Note that  
\begin{align*}
\Pr(A_n \leq C_n) & = \Pr(A_n \leq C_n, A_n < B_n) + \Pr(A_n \leq C_n, A_n \geq B_n) \\
&\leq \Pr(A_n < B_n) + \Pr(B_n  \leq C_n),
\end{align*}
On the one hand, 
\begin{align} \label{eq:prob_sup_ineq}
\varlimsup_{n\to\infty} \Pr(A_n \leq C_n) \leq \varlimsup_{n\to\infty}\Pr(A_n < B_n) + \varlimsup_{n\to\infty}\Pr(B_n  \leq C_n).
\end{align}
On the other hand,
\begin{align} \label{eq:prob_inf_ineq}
 \varliminf_{n\to\infty}\Pr(B_n  \leq C_n) &\ge \varliminf_{n\to\infty} \big\{ \Pr(A_n \leq C_n)-\Pr(A_n < B_n)\big\} \notag \\
 &\ge 
 \varliminf_{n\to\infty}  \Pr(A_n \leq C_n)-\varlimsup_{n\to\infty}\Pr(A_n < B_n).
\end{align}
Because $\Pr(A_n < B_n) = 1 - \Pr(A_n \ge B_n) \rightarrow 0$ as $n\rightarrow \infty$, by \eqref{eq:prob_sup_ineq} and \eqref{eq:prob_inf_ineq}, we can then derive the two inequalities in (i). 

We can derive (ii) by applying (i) twice, with the roles of $A_n$ and $B_n$ interchanged.

We finally prove (iii), which follows directly from 
\begin{align*} 
\Pr(A_n \ne C_n) \le \Pr(A_n \ne B_n \text{ or } C_n \ne B_n) \le  \Pr(A_n \ne B_n)+\Pr(C_n \ne B_n) \to 0 \text{ as } n\rightarrow \infty.
\end{align*}

From the above, Lemma \ref{lem:pr_inequality} holds. 
\end{proof}

\begin{lemma}
\label{lem:S0*property}
Suppose that  Assumption \ref{asmp:iid} holds, and consider $\psi_\beta(b,d)$ in \eqref{eq:notation_xi} and $\underline{\tau}_{\beta}$ in \eqref{eq:quantile_lower_bound}.  Then $\RS_{\beta}^\star = \{ b \in \mathbb{R} : \psi_\beta(b, \underline{\tau}_{\beta}) = 0 \}$ defined in \eqref{eq:S_beta} is a subset of $(F_0^{-1}(1 - \beta), F_1^{-1}(\beta) - \underline{\tau}_{\beta})$ and is a non-empty compact set.
\end{lemma}

\begin{proof}[\bf Proof of Lemma \ref{lem:S0*property}]
Under Assumption \ref{asmp:iid}, $h_\beta(u) \equiv F_1^{-1}(u)-F_0^{-1}(1+u-\beta)$ is continuous in $u \in (0,\beta)$ and $\lim_{u \to 0+} h_\beta(u)=\lim_{u \to \beta^-} h_\beta(u)=-\infty$. 
Thus, we can show that there exists $u^* \in (0,\beta)$ such that
\begin{align*}
    \underline{\tau}_\beta= \sup_{u \in (0,\beta)}  \{F_1^{-1}(u)-F_0^{-1}(1+u-\beta)\} = \sup_{u \in (0,\beta)} h_\beta(u)=  h_\beta(u^*).
\end{align*}
Let $b^*= F_0^{-1}(1+u^*-\beta)$.
We then have $\underline{\tau}_\beta=F_1^{-1}(u^*) - b^*$, and thus 
\begin{align*}
    \psi_\beta(b^*, \underline{\tau}_{\beta}) &= F_0(b^*)-\min\{1,1-\beta+F_1(b^*+\underline{\tau}_\beta)\} =(1+u^*-\beta)-\min\{1,1-\beta+u^*\}\\
    & =0. 
\end{align*}
Therefore,  $b^* \in \RS_{\beta}^\star$, and $\RS_{\beta}^\star$ is non-empty. 

We next show that $\RS_{\beta}^\star$ is a subset of $(F_0^{-1}(1 - \beta), F_1^{-1}(\beta) - \underline{\tau}_{\beta})$. 
Consider any $b_0 \in \RS_{\beta}^\star$, i.e., 
$
\psi_\beta(b_0,\underline{\tau}_{\beta}) = F_0(b_0) - F_{1, \beta}(b_0 + \underline{\tau}_{\beta}) = 0. 
$
We must have $F_0(b_0) = 1 - \beta + F_1(b_0 + \underline{\tau}_{\beta}) > 1 - \beta$, which implies $b_0 > F_0^{-1}(1-\beta)$. Meanwhile, $F_1(b_0+\underline{\tau}_{\beta})-\beta =F_0(b_0)-1< 0$, which implies $b_0 < F_1^{-1}(\beta)-\underline{\tau}_{\beta}$. 
These imply that $\RS_{\beta}^\star \subseteq (F_0^{-1}(1 - \beta), F_1^{-1}(\beta) - \underline{\tau}_{\beta})$.

We finally show the compactness of $\RS_{\beta}^\star$.
Because $\psi_\beta(b,d)$ is continuous in $b$ under Assumption \ref{asmp:iid}, 
$\RS_{\beta}^\star = \{ b_0 \in \mathbb{R} : \psi_\beta(b_0, \underline{\tau}_{\beta}) = 0 \}$ must be a closed set. 
Because $\RS_{\beta}^\star$ is also bounded as shown before, $\RS_{\beta}^\star$ must be a compact set.

From the above, Lemma \ref{lem:S0*property} holds. 
\end{proof}

\begin{lemma}
\label{lem:xn equivalent}
Suppose that  Assumption \ref{asmp:iid} holds and $\lim_{n \to \infty}n_1/n = \lambda\in (0,1)$. Consider 
$\hat{\psi}_{n,\beta}(b,d)$ in \eqref{eq:notation_psin}, 
$\hat{\xi}_{n,\beta}(d)$ in \eqref{eq:notation_xin}, 
$\psi_\beta(b,d)$ in \eqref{eq:notation_xi}, $\underline{\tau}_{\beta}$ in \eqref{eq:quantile_lower_bound}, and $\RS_{\beta}^\star$ in \eqref{eq:S_beta}. 
Let $c_n = \underline{\tau}_{\beta} - h/\sqrt{n}$ for some $h>0$. 
For any $ \varepsilon > 0$ and any $M > F_1^{-1}(\beta) - \underline{\tau}_{\beta} + \varepsilon$, define sets $B(\RS_{\beta}^\star, \varepsilon) = \bigcup_{b_0 \in \RS_{\beta}^\star} B(b_0, \varepsilon)$ and $B_M(\RS_{\beta}^\star, \varepsilon) = B(\RS_{\beta}^\star, \varepsilon) \cup (M, +\infty)$, where $B(b_0, \varepsilon) = (b_0-\varepsilon, b_0+\varepsilon)$. 
 Then the following results hold. 
\begin{enumerate}[label=(\roman*)]
    \item $\sup_{b \in B^\complement_M(\RS_{\beta}^\star, \varepsilon)} \psi_\beta(b, \underline{\tau}_{\beta}) < 0$, where  $B^\complement_M(\RS_{\beta}^\star, \varepsilon) = \mathbb{R} \setminus B_M(\RS_{\beta}^\star, \varepsilon)$,
    \item $\mathbb{P} \Big( \sqrt{n} \hat{\xi}_{n,\beta}(c_n) = \sup_{b \in B_M(\RS_{\beta}^\star, \varepsilon)} \sqrt{n} \hat{\psi}_{n,\beta}(b, c_n) \Big) \to 1 \text{ as } n \to \infty$,
    \item $\mathbb{P} \Big(  \sqrt{n} \hat{\xi}_{n,\beta}(c_n) = \max \big \{ \sup_{b \in B(\RS_{\beta}^\star, \varepsilon)} \sqrt{n} \hat{\psi}_{n,\beta}(b, c_n), 0 \big \} \Big) \to 1 \text{ as } n \to \infty$.
\end{enumerate}
\end{lemma}

\begin{proof}[\bf Proof of Lemma \ref{lem:xn equivalent}]
We prove (i) by contradiction. 
Suppose that (i) does not hold, i.e. $\sup_{b \in B^\complement_M(\RS_{\beta}^\star, \varepsilon)} \psi_\beta(b, \underline{\tau}_{\beta}) \geq 0$. By the same logic as Lemma \ref{lem:xi_infty_beta}(ii), we have $\sup_b \psi_\beta(b, \underline{\tau}_{\beta}) = 0$. 
These imply that $\sup_{b \in B^\complement_M(\RS_{\beta}^\star, \varepsilon)} \psi_\beta(b, \underline{\tau}_{\beta}) = 0$.
Note that $\psi_\beta(b, \underline{\tau}_{\beta})$ is continuous in $b$, and 
\begin{align*}
\lim_{b \to -\infty} \psi_\beta(b, \underline{\tau}_{\beta}) &= \lim_{b \to -\infty} \{F_0(b) - F_{1, \beta}(b + \underline{\tau}_{\beta})\} 
= 0 - \min\{1,1-\beta\} 
= \beta - 1 < 0. 
\end{align*}
Thus, there must exist $L \le F_1^{-1}(\beta) - \underline{\tau}_{\beta} < M $ such that $\psi_\beta(b, \underline{\tau}_{\beta}) < (\beta-1)/2 < 0$ for any $b < L$. This implies that
we have
\begin{align*}
0 = \sup_{b \in B^\complement_M(\RS_{\beta}^\star, \varepsilon)} \psi_\beta(b, \underline{\tau}_{\beta}) = \sup_{b \in B^\complement_M(\RS_{\beta}^\star, \varepsilon)\backslash(-\infty,L)} \psi_\beta(b, \underline{\tau}_{\beta}) = \sup_{b \in [L, M] \backslash B(\RS_{\beta}^\star, \varepsilon)} \psi_\beta(b, \underline{\tau}_{\beta}).
\end{align*}
Because $[L, M] \backslash B(\RS_{\beta}^\star, \varepsilon)$ is compact and $\psi_\beta(b,\underline{\tau}_{\beta})$ is continuous in $b$, there exists $b_0 \in [L, M] \backslash B(\RS_{\beta}^\star, \varepsilon)$ such that $\psi_\beta(b_0, \underline{\tau}_{\beta}) = 0$. This then implies that $b_0 \in \RS_{\beta}^\star$, leading to a contradiction. 
Therefore, we must have $\sup_{b \in B^\complement_M(\RS_{\beta}^\star, \varepsilon)} \psi_\beta(b, \underline{\tau}_{\beta}) < 0$. 

We then prove (ii). From (i) and Lemma \ref{lem:fn to infinity}, 
we have 
$
\sup_{b \in B^\complement_M(\RS_{\beta}^\star, \varepsilon)} \sqrt{n} \hat{\psi}_{n,\beta}(b, c_n) \convergep -\infty
$
as $n\rightarrow \infty$. 
Let
\begin{align*}
U_n&=\sup_{b \in B_M(\RS_{\beta}^\star, \varepsilon)} \sqrt{n} \hat{\psi}_{n,\beta}(b, c_n),
\quad 
V_n=\sup_{b \in B^\complement_M(\RS_{\beta}^\star, \varepsilon)} \sqrt{n} \hat{\psi}_{n,\beta}(b, c_n).
\end{align*}
Then, by definition, $\sqrt{n} \hat{\xi}_{n,\beta}(c_n)=\max\{U_n,V_n\}$. Moreover,
$U_n\ge0$ because $(M,\infty)\subseteq B_M(\RS_{\beta}^\star,\varepsilon)$ and
$\lim_{b\to\infty}\hat\psi_{n,\beta}(b,c_n)=0$. Therefore,
\begin{align*}
\mathbb{P} \Big(\sqrt{n} \hat{\xi}_{n,\beta}(c_n) = \sup_{b \in B_M(\RS_{\beta}^\star, \varepsilon)} \sqrt{n} \hat{\psi}_{n,\beta}(b, c_n) \Big) 
&= \Pr(U_n\ge V_n) \geq \Pr(V_n\le0) \to 1,
\end{align*}
where the convergence follows from $V_n\xrightarrow{\mathbb P}-\infty$. This completes the proof of (ii).

We finally prove (iii). Define 
\begin{align*}
X_n = \sup_{b \in B(\RS_{\beta}^\star, \varepsilon)} \sqrt{n} \hat{\psi}_{n,\beta}(b, c_n), \quad Y_n = \sup_{b>M} \sqrt{n} \hat{\psi}_{n,\beta}(b, c_n).
\end{align*}
It suffices to show $\Pr(\max\{X_n,Y_n\}=\max\{X_n,0\})\to1$ by (ii) and Lemma \ref{lem:pr_inequality} (iii).
We first show that 
\begin{align}
\label{claim:shifted_emphirical_1}
\lim_{n\to\infty}\Pr \Big(\hat{F}_{n}^\t(M + c_n) > \beta + \frac{1}{n_1} \Big) = 1.
\end{align}
Because  $M>F_1^{-1}(\beta)-\underline{\tau}_{\beta}+\varepsilon$ and $c_m = \underline{\tau}_{\beta} - {h}/{\sqrt{m}}$, 
there exists $m_0$ such that $M > F_1^{-1}(\beta) - c_{m_0}$ for a sufficiently large $m_0$.  
For any $n\ge m_0$, we then have $c_n \ge c_{m_0}$, 
$\hat{F}_{n}^\t(M + c_n) 
    \ge 
    \hat{F}_{n}^\t(M + c_{m_0})$, 
and 
\begin{align*}
\Pr \Big(\hat{F}_{n}^\t(M + c_n) > \beta + \frac{1}{n_1} \Big) 
& \ge 
\Pr \Big(\hat{F}_{n}^\t(M + c_{m_0}) > \beta + \frac{1}{n_1} \Big) 
= \Pr  \Big(\hat{F}_{n}^\t(M + c_{m_0}) - \frac{1}{n_1} > \beta \Big) 
\nonumber
\\
& \converge 1 \text{ as } n \rightarrow \infty, 
\end{align*}
where the last convergence holds because 
$\hat{F}_{n}^\t(M + c_{m_0})
    \convergeas 
    F_1 (M + c_{m_0}) > \beta$ 
by the  Glivenko–Cantelli theorem. 
Next we show that
\begin{align}
\label{claim:max_Xn_Yn}
\hat{F}_{n}^\t(M + c_n) > \beta + \frac{1}{n_1} \text{ implies } \max\{X_n,Y_n\}=\max\{X_n,0\}.
\end{align}
Suppose that $\hat{F}_{n}^\t(M + c_n) > \beta + {1}/{n_1}$, then we must have 
\begin{align*}
\hat{F}_{n,\lceil n_1\beta \rceil /n_1}^\t (M + c_n) = \min \Big\{ 1, 1 - \frac{\lceil n_1\beta \rceil}{n_1} +\hat{F}_{n}^\t(M + c_n) \Big\} = 1. 
\end{align*}
This then implies that 
\begin{align*}
    Y_n &= \sup_{b > M} \sqrt{n} \hat{\psi}_{n,\beta}(b, c_n) =  \sup_{b > M} \sqrt{n} \Big (\hat{F}_{n}^\c(b) - \hat{F}_{n,\lceil n_1\beta \rceil /n_1}^\t (b + c_n) \Big)\\
    &\leq \sup_{b > M} \sqrt{n} \Big(\hat{F}_{n}^\c(b) - \hat{F}_{n,\lceil n_1\beta \rceil /n_1}^\t (M + c_n) \Big) = \sup_{b > M} \sqrt{n} \Big(\hat{F}_{n}^\c(b) - 1\Big) = 0.
\end{align*}
In addition, by definition, 
\begin{align*}
    Y_n &= \sup_{b > M} \sqrt{n} \hat{\psi}_{n,\beta}(b, c_n) \geq \lim_{b \to +\infty}  \sqrt{n} \Big(\hat{F}_{n}^\c(b) - \hat{F}_{n,\lceil n_1\beta \rceil /n_1}^\t (b + c_n) \Big)= 0.
\end{align*}
Thus, $Y_n = 0$, and $\max\{X_n,Y_n\}=\max\{X_n,0\}$, which proves \eqref{claim:max_Xn_Yn}.
Based on \eqref{claim:shifted_emphirical_1} and \eqref{claim:max_Xn_Yn}, we have
\begin{align*}
\varliminf_{n\to\infty}\Pr(\max\{X_n,Y_n\}=\max\{X_n,0\}) \geq \lim_{n\to\infty}\Pr \Big(\hat{F}_{n}^\t(M + c_n) > \beta + \frac{1}{n_1} \Big) = 1,
\end{align*}
which implies $\lim_{n\to\infty}\Pr(\max\{X_n,Y_n\}=\max\{X_n,0\}) = 1$.

From the above, Lemma \ref{lem:xn equivalent} holds. 
\end{proof}

\begin{lemma}\label{lem:sup fn convergence}
Consider the same setup as Lemmas \ref{lem:S0*property} and \ref{lem:xn equivalent}. 
From Lemma \ref{lem:S0*property}, $\RS_{\beta}^\star$ is compact and a subset of $(F_0^{-1}(1 - \beta), F_1^{-1}(\beta) - \underline{\tau}_{\beta})$, implying that, for sufficiently small $\varepsilon$,  
\begin{align}\label{eq:contact_set_2ep}
    B(\RS_{\beta}^\star, 2\varepsilon) \subseteq (F_0^{-1}(1 - \beta), F_1^{-1}(\beta) - \underline{\tau}_{\beta})
\end{align} 
Define \begin{align}\label{eq:An(b,cn)}
    \mathscr{A}_n(b, c_n)&=\sqrt{n} \Big(\hat{F}_n^\c(b) - F_0(b)\Big) - \sqrt{n} \Big(\hat{F}_{n}^\t(b + c_n) - F_1 (b + c_n)\Big) +  \frac{\sqrt{n}}{n_1}\Big(\lceil n_1\beta \rceil - n_1\beta \Big)  \notag \\
    &\quad+\sqrt{n} \Big(F_1(b + \underline{\tau}_{\beta}) - F_1(b + c_n)\Big),
\end{align}
and $\mathscr{A}(b, \underline{\tau}_{\beta})$ the same as in \eqref{eq:GP_power}: 
\begin{align*}
\mathscr{A}(b, \underline{\tau}_{\beta}) = \frac{1}{\sqrt{1-\lambda}} \mathscr{B}_0(F_0(b)) - \frac{1}{\sqrt{\lambda}} \mathscr{B}_1(F_1(b + \underline{\tau}_{\beta})) + h \cdot f_1(b + \underline{\tau}_{\beta}), 
\end{align*}
where $\mathscr{B}_0$ and $\mathscr{B}_1$ are two independent Brownian bridges.

For any $\varepsilon>0$ such that \eqref{eq:contact_set_2ep} holds, we have 
the following results. 
\begin{enumerate}[label=(\roman*)]
    \item $\lim_{n \to +\infty} \mathbb{P}\left( \sup_{b \in B(\RS_{\beta}^\star, \varepsilon)} \sqrt{n} \hat{\psi}_{n,\beta}(b, c_n) \leq \sup_{b \in B(\RS_{\beta}^\star, \varepsilon)} \mathscr{A}_n(b, c_n) \right) = 1$. 
    \item $\lim_{n \to +\infty} \mathbb{P}\left( \sup_{b \in \RS_{\beta}^\star} \sqrt{n} \hat{\psi}_{n,\beta}(b, c_n) = \sup_{b \in \RS_{\beta}^\star} \mathscr{A}_n(b, c_n)\right) = 1$.
    \item As $n\rightarrow \infty$, $\sup_{b \in \RS_{\beta}^\star}\mathscr{A}_n(b, c_n) \converged  \sup_{b \in \RS_{\beta}^\star}\mathscr{A}(b, \underline{\tau}_{\beta})$,  and $\sup_{b \in B(\RS_{\beta}^\star, \varepsilon)}\mathscr{A}_n(b, c_n) \converged \sup_{b \in B(\RS_{\beta}^\star, \varepsilon)}\mathscr{A}(b, \underline{\tau}_{\beta})$. 
\end{enumerate}
\end{lemma}

\begin{proof}[\bf Proof of Lemma \ref{lem:sup fn convergence}]
Consider any $\varepsilon$ such that $B(\RS_{\beta}^\star,2\varepsilon)$ is a subset of  $(F_0^{-1}(1 - \beta), F_1^{-1}(\beta) - \underline{\tau}_{\beta})$. Then, for any $ b \in B(\RS_{\beta}^\star, \varepsilon)$, $b + \varepsilon< F_1^{-1}(\beta) - \underline{\tau}_{\beta}$, which further implies that $\sup_{ b\in B(\RS_{\beta}^\star, \varepsilon)} F_1(b+\underline{\tau}_{\beta}) \le F_1\big(F_1^{-1}(\beta)-\varepsilon\big) < \beta$. Hence, for any $b\in B(\RS_{\beta}^\star, \varepsilon)$, 
\begin{align*}
F_{1, \beta} (b + \underline{\tau}_{\beta}) = \min\{1,F_1(b + \underline{\tau}_{\beta}) + 1 - \beta\} = F_1(b + \underline{\tau}_{\beta}) + 1 - \beta.
\end{align*}
Moreover, for any  $ b \in B(\RS_{\beta}^\star, \varepsilon)$, because $c_n < \underline{\tau}_{\beta}$, we have $F_1(b + c_n) < F_1(b + \underline{\tau}_{\beta}) < \beta$, and thus 
\begin{align*}
F_{1, \beta} (b + c_n) = \min\{1,F_1(b + c_n) + 1 - \beta\} = F_1(b + c_n) + 1 - \beta.
\end{align*}

We first prove that
\begin{align}
\label{claim:Fnt_uniform_converge}
\Pr\left\{
\sup_{b\in B(\mathcal S_{\beta}^\star,\varepsilon)}
\hat F_n^\t(b+\underline{\tau}_{\beta})<\beta
\right\}
\longrightarrow 1.
\end{align}
By construction of $B(\mathcal S_{\beta}^\star,\varepsilon)$,
$
\sup_{b\in B(\mathcal S_{\beta}^\star,\varepsilon)}
F_1(b+\underline{\tau}_{\beta})\le
F_1\bigl(F_1^{-1}(\beta)-\varepsilon\bigr)
<\beta.
$
Define the fixed positive constant
$
c_\varepsilon
=
\beta-F_1\bigl(F_1^{-1}(\beta)-\varepsilon\bigr)>0.
$
It follows that
\begin{align*}
\sup_{b\in B(\mathcal S_{\beta}^\star,\varepsilon)}
\hat F_n^\t(b+\underline{\tau}_{\beta})-\beta
\le
\sup_{b\in B(\mathcal S_{\beta}^\star,\varepsilon)}
\left|
\hat F_n^\t(b+\underline{\tau}_{\beta})
-
F_1(b+\underline{\tau}_{\beta})
\right|
-c_\varepsilon
\le
\sup_{x\in\mathbb R}
\left|\hat F_n^\t(x)-F_1(x)\right|
-c_\varepsilon.
\end{align*}
By the uniform convergence of $\hat F_n^\t$ to $F_1$,
$
\sup_{x\in\mathbb R}
\left|\hat F_n^\t(x)-F_1(x)\right|
\convergeas 0.
$
Consequently,
\begin{align*}
\Pr\left\{
\sup_{b\in B(\mathcal S_{\beta}^\star,\varepsilon)}
\hat F_n^\t(b+\underline{\tau}_{\beta})<\beta
\right\}\ge
\Pr\left\{
\sup_{x\in\mathbb R}
\left|\hat F_n^\t(x)-F_1(x)\right|
<c_\varepsilon
\right\}
\longrightarrow 1,
\end{align*}
which proves \eqref{claim:Fnt_uniform_converge}.

We then prove (i) and (ii). 
Consider first the case where $\sup_{ b\in B(\RS_{\beta}^\star, \varepsilon)} \hat{F}_{n}^\t(b+\underline{\tau}_{\beta})<\beta$ holds. 
Since $c_n < \underline{\tau}_{\beta}$, we then have $\sup_{ b\in B(\RS_{\beta}^\star, \varepsilon)} \hat{F}_{n}^\t(b+c_n)\le \sup_{ b\in B(\RS_{\beta}^\star, \varepsilon)} \hat{F}_{n}^\t(b+\underline{\tau}_{\beta})<\beta$. Thus, for any $b\in B(\RS_{\beta}^\star, \varepsilon)$, 
\begin{align*}
\hat{F}_{n,\lceil n_1 \beta \rceil/n_1}^\t(b+c_n) = \min \Big \{1,\hat{F}^\t_n(b + c_n)+1 - \frac{\lceil n_1\beta \rceil}{n_1} \Big \} = \hat{F}^\t_{n}(b + c_n)+1 - \frac{\lceil n_1\beta \rceil}{n_1}.
\end{align*}
Consequently, for $b \in B(\RS_{\beta}^\star,\varepsilon)$, we have
\begin{align*} %
\sqrt{n} \hat{\psi}_{n,\beta}(b, c_n) 
&= \sqrt{n} \Big(\hat{F}_n^\c(b) - \hat{F}_{n,\lceil n_1 \beta \rceil/n_1}^\t(b+c_n) \Big) \notag\\
&= \sqrt{n} \Big(\hat{F}_n^\c(b) - F_0(b)\Big) - \sqrt{n} \Big(\hat{F}_{n,\lceil n_1 \beta \rceil/n_1}^\t(b+c_n) - F_{1, \beta} (b + c_n)\Big) \notag\\
&\quad + \sqrt{n} \Big(F_{1, \beta} (b + \underline{\tau}_{\beta}) - F_{1, \beta} (b + c_n)\Big) + \sqrt{n} \Big(F_0(b) - F_{1, \beta} (b + \underline{\tau}_{\beta})\Big) \notag\\
&= \sqrt{n} \Big(\hat{F}_n^\c(b) - F_0(b)\Big) - \sqrt{n} \Big[ 1 - \frac{\lceil n_1\beta \rceil}{n_1} + \hat{F}_{n}^\t(b + c_n) - \Big\{1- \beta + F_1(b + c_n)\Big\}\Big] \notag\\
&\quad + \sqrt{n} \Big(F_1(b + \underline{\tau}_{\beta}) - F_1(b + c_n)\Big) + \sqrt{n} \psi_\beta(b,\underline{\tau}_{\beta})\notag\\
&= \sqrt{n} \Big(\hat{F}_n^\c(b) - F_0(b)\Big) - \sqrt{n} \Big(   \hat{F}_{n}^\t(b + c_n)  -  F_1(b + c_n) \Big) + \frac{\sqrt{n}}{n_1}  \Big(\lceil n_1\beta \rceil - n_1\beta \Big)\notag\\
&\quad + \sqrt{n} \Big(F_1(b + \underline{\tau}_{\beta}) - F_1(b + c_n)\Big) + \sqrt{n} \psi_\beta(b,\underline{\tau}_{\beta})\notag\\
& = \mathscr{A}_n(b, c_n) +  \sqrt{n} \psi_\beta(b,\underline{\tau}_{\beta}).
\end{align*}
Note that $\psi_\beta(b,\underline{\tau}_{\beta}) \leq \sup_{b\in \mathbb{R}} \psi_\beta(b,\underline{\tau}_{\beta}) = \xi_\beta(\underline{\tau}_{\beta}) = 0$  by the same logic as Lemma \ref{lem:xi_infty_beta} and $\psi_\beta(b,\underline{\tau}_{\beta}) = 0$ for $b \in \RS_{\beta}^\star$ by definition. 
From \eqref{claim:Fnt_uniform_converge}, we can then derive (i) and (ii).

We finally prove (iii). Let $T_0 = [F_0^{-1}(1 - \beta), F_1^{-1}(\beta) - \underline{\tau}_{\beta}]$. Since $\lim_{n\to\infty}n_1/n=\lambda$, the following results hold:
\begin{enumerate}
    \item $\sqrt{n}\Big(\lceil n_1\beta \rceil - n_1\beta\Big)\Big/n_1 \to 0$,
    \item $\sqrt{n} \Big(\hat{F}_n^\c(b) - F_0(b)\Big) \converged \frac{1}{\sqrt{1-\lambda}} \mathscr{B}_0(F_0(b)) \text{ in }  D(T_0)$,
    \item $\sqrt{n} \Big(\hat{F}_{n}^\t (b + c_n) - F_1 (b + c_n)\Big) \converged \frac{1}{\sqrt{\lambda}} \mathscr{B}_1(F_1(b + \underline{\tau}_{\beta})) \text{ in }  D(T_0)$ by Lemma \ref{lem:emphirical_cn},
\end{enumerate}
where $\mathscr{B}_0$ and $\mathscr{B}_1$ are two independent Brownian bridges.
Moreover, under Assumption \ref{asmp:iid}, $\hat F_n^\c(\cdot)$ and $\hat F_n^\t(\cdot)$ are independent: conditional on $\bs Z$, they are based on disjoint sets of independent units and are therefore conditionally independent, while their conditional distributions do not depend on the particular realization of $\bs Z$.

In addition, since $c_n=\underline{\tau}_{\beta}-h/\sqrt{n}$
and $f_1$ is uniformly continuous on any bounded set, by the mean value theorem, we can obtain that 
$\sqrt{n} \Big(F_1(b + \underline{\tau}_{\beta}) - F_1(b + c_n)\Big) \Rightarrow f_1(b + \underline{\tau}_{\beta}) \cdot h$ in $ D(T_0)$.
As a result, following \citet[][Chapter 1.4]{van_der_vaart_weak_1996}
and by the continuous mapping theorem \citep[][p. 259]{van2000asymptotic}, we have 
\begin{align*}
\mathscr{A}_n(b, c_n) \converged \mathscr{A}(b, \underline{\tau}_{\beta}) 
\text{ in } D(T_0). 
\end{align*}
Since $ \RS_{\beta}^\star \subseteq B(\RS_{\beta}^\star, \varepsilon) \subseteq T_0$, then by Lemma \ref{lem:sup converge} we have
\begin{align*}
\sup_{b \in \RS_{\beta}^\star}\mathscr{A}_n(b, c_n) \converged  \sup_{b \in \RS_{\beta}^\star}\mathscr{A}(b, \underline{\tau}_{\beta}),\sup_{b \in B(\RS_{\beta}^\star, \varepsilon)}\mathscr{A}_n(b, c_n) \converged \sup_{b \in B(\RS_{\beta}^\star, \varepsilon)}\mathscr{A}(b, \underline{\tau}_{\beta}).
\end{align*}

From the above, Lemma \ref{lem:sup fn convergence} holds. 
\end{proof}

\begin{lemma}
\label{lem:sup g convergence}
Define $\mathscr{A}(b, \underline{\tau}_{\beta})$ the same as in \eqref{eq:GP_power}: 
\begin{align*}
\mathscr{A}(b, \underline{\tau}_{\beta}) = \frac{1}{\sqrt{1-\lambda}} \mathscr{B}_0(F_0(b)) - \frac{1}{\sqrt{\lambda}} \mathscr{B}_1(F_1(b + \underline{\tau}_{\beta})) + h \cdot f_1(b + \underline{\tau}_{\beta}). 
\end{align*}
Under Assumption \ref{asmp:iid}, we have 
\begin{align*}
\sup_{b \in B(\RS_{\beta}^\star, \varepsilon)} \mathscr{A}(b, \underline{\tau}_{\beta}) \xrightarrow{a.s.} \sup_{b \in \RS_{\beta}^\star} \mathscr{A}(b, \underline{\tau}_{\beta}) \text{ as } \varepsilon \to 0^+ 
\end{align*}
\end{lemma}

\begin{proof}[\bf Proof of Lemma \ref{lem:sup g convergence}]
First, for every realization $\omega$,
$\sup_{b \in B(\RS_{\beta}^\star, \varepsilon)} \mathscr{A}(\omega)(b, \underline{\tau}_{\beta}) $ is nondecreasing in $\varepsilon$ and is bounded below by $\sup_{b \in \RS_{\beta}^\star} \mathscr{A}(\omega)(b, \underline{\tau}_{\beta})$. Thus, as $ \varepsilon \to 0^+$, the limit of $\sup_{b \in B(\RS_{\beta}^\star, \varepsilon)} \mathscr{A}(\omega)(b, \underline{\tau}_{\beta})$ exists.

Because Brownian bridge is almost surely continuous, we can know that $\mathscr{A}(\cdot,\underline{\tau}_{\beta})$ is almost surely continuous. 
Consider any realization $\omega$ such that $\mathscr{A}(\omega)(b, \underline{\tau}_{\beta})$ is continuous. Then 
$\mathscr{A}(\omega)(b, \underline{\tau}_{\beta})$ must be uniformly continuous on $[ F_0^{-1}(1 - \beta), F_1^{-1}(\beta) - \underline{\tau}_{\beta} ]$. 
From Lemma \ref{lem:S0*property}, 
for any $\varepsilon^\prime > 0$, there exists $\delta > 0$, such that for any $b \in B(\RS_{\beta}^\star, \delta)$, we can find $b_0 \in \RS_{\beta}^\star$ satisfying $|\mathscr{A}(\omega)(b, \underline{\tau}_{\beta}) - \mathscr{A}(\omega)(b_0, \underline{\tau}_{\beta})| < \varepsilon^\prime.$
Thus, for any $\varepsilon^\prime > 0$, we have 
\begin{align*}
\lim_{\varepsilon \to 0} \sup_{b \in B(\RS_{\beta}^\star, \varepsilon)} \mathscr{A}(\omega)(b, \underline{\tau}_{\beta}) \le \sup_{b \in B(\RS_{\beta}^\star, \delta)} \mathscr{A}(\omega)(b, \underline{\tau}_{\beta}) \le \sup_{b \in \RS_{\beta}^\star} \mathscr{A}(\omega)(b, \underline{\tau}_{\beta}) + \varepsilon^\prime.
\end{align*}
Letting $\varepsilon^\prime \to 0$, we then have  
\begin{align*}
\lim_{\varepsilon \to 0} \sup_{b \in B(\RS_{\beta}^\star, \varepsilon)} \mathscr{A}(\omega)(b, \underline{\tau}_{\beta}) \leq \sup_{b \in \RS_{\beta}^\star} \mathscr{A}(\omega)(b, \underline{\tau}_{\beta}).
\end{align*}
In addition, note that $\sup_{b \in B(\RS_{\beta}^\star, \varepsilon)} \mathscr{A}(\omega)(b, \underline{\tau}_{\beta}) \geq \sup_{b \in \RS_{\beta}^\star} \mathscr{A}(\omega)(b, \underline{\tau}_{\beta})$ for any $\varepsilon>0$. Thus, we must have 
\begin{align*}
\sup_{b \in B(\RS_{\beta}^\star, \varepsilon)} \mathscr{A}(\omega)(b, \underline{\tau}_{\beta}) \rightarrow \sup_{b \in \RS_{\beta}^\star} \mathscr{A}(\omega)(b, \underline{\tau}_{\beta}) \text{ as } \varepsilon \to 0+.
\end{align*}

From the above, Lemma \ref{lem:sup g convergence} holds. 
\end{proof}

\begin{lemma}
\label{lem:xin_beta_convergence}
Consider the same setup as Lemmas \ref{lem:S0*property}-- \ref{lem:sup fn convergence}. 
For almost every
$z \in \mathbb{R}$,  i.e., except for a measure-zero set, 
\begin{align*}
\lim_{n\to\infty}\Pr \big\{ \sqrt{n} \, \hat \xi_{n,\beta}(c_n) \leq z \big\}= \Pr\Big(\max \big\{ 0, \sup_{b \in \RS_{\beta}^\star} \mathscr{A}(b, \underline{\tau}_{\beta}) \big\}\leq z\Big).
\end{align*}    
\end{lemma}

\begin{proof}[\bf Proof of Lemma \ref{lem:xin_beta_convergence}]
For any $\varepsilon>0$, let $C(\mathscr{A},\varepsilon)\subseteq\mathbb{R}$ denote the set of all continuity points of the distribution of $\sup_{b \in B(\RS_{\beta}^\star,\varepsilon)} \mathscr{A}(b, \underline{\tau}_{\beta})$, 
and 
$C(\mathscr{A},0)\subseteq\mathbb{R}$ denote the set of all continuity points of the distribution of $\sup_{b \in \RS_{\beta}^\star} \mathscr{A}(b, \underline{\tau}_{\beta})$. 
For any $\varepsilon\ge 0$, $\mathbb{R}\backslash C(\mathscr{A},\varepsilon)$ has measure zero since the set of discontinuity points of a distribution function is at most countable. Define $C(\mathscr{A}) = \Big(\bigcap_{m=1}^{\infty} C(\mathscr{A},1/m)\Big)\bigcap C(\mathscr{A},0) $. Then $\mathbb{R}\backslash C(\mathscr{A})$ also has measure zero.

Consider any $z\in C(\mathscr{A})$ and $z \ge 0$. For sufficiently large integer $m$, we have 
\begin{align*}
\varliminf_{n \to \infty} \Pr \left( \sqrt{n} \, \hat{\xi}_{n,\beta}(c_n) \leq z \right) &= \varliminf_{n \to \infty} \Pr \Big( \max \big\{ 0, \sup_{b \in B(\RS_{\beta}^\star, 1/m)} \sqrt{n} \hat{\psi}_{n,\beta}(b, c_n) \big\} \leq z \Big)\\
&= \varliminf_{n \to \infty} \Pr \Big( \sup_{b \in B(\RS_{\beta}^\star, 1/m)} \sqrt{n} \hat{\psi}_{n,\beta}(b, c_n) \leq z \Big)\\
&\ge \varliminf_{n \to \infty}  \Pr \left( \sup_{b \in B(\RS_{\beta}^\star, 1/m)} \mathscr{A}_n(b, c_n) \leq z \right)\\
&= \Pr \left( \sup_{b \in B(\RS_{\beta}^\star, 1/m)} \mathscr{A}(b, \underline{\tau}_{\beta}) \leq z \right),
\end{align*}
where the first equality follows from Lemma \ref{lem:xn equivalent}(iii) and Lemma \ref{lem:pr_inequality}(ii), the inequality follows from Lemma \ref{lem:sup fn convergence}(i) and Lemma \ref{lem:pr_inequality}(i), and the last equality follows from Lemma \ref{lem:sup fn convergence}(iii) and the fact that $z \in C(\mathscr{A}) \subseteq C(\mathscr{A},1/m)$.
Note that, by Lemma \ref{lem:sup g convergence},
$\sup_{b \in B(\RS_{\beta}^\star, 1/m)} \mathscr{A}(b, \underline{\tau}_{\beta}) \xrightarrow{a.s.} \sup_{b \in \RS_{\beta}^\star} \mathscr{A}(b, \underline{\tau}_{\beta})$ as $m \to \infty$.  
Letting $m\rightarrow \infty$, 
we then have 
\begin{align*}
\varliminf_{n \to \infty} \Pr \big( \sqrt{n} \hat{\xi}_{n,\beta}(c_n) \leq z \big) \geq \Pr \big( \sup_{b \in \RS_{\beta}^\star} \mathscr{A}(b, \underline{\tau}_{\beta}) \leq z \big), 
\end{align*}
recalling that $z \in C(\mathscr{A}) \subseteq C(\mathscr{A},0)$.

In addition, for any $z\in C(\mathscr{A})$ and $z \ge 0$, we have 
\begin{align*}
\varlimsup_{n \to \infty} \Pr \big( \sqrt{n} \hat{\xi}_{n,\beta}(c_n) \leq z \big) &\leq \varlimsup_{n \to \infty} \Pr \big( \sup_{b \in \RS_{\beta}^\star} \sqrt{n} \hat{\psi}_{n,\beta}(b, c_n) \leq z \big) 
= \varlimsup_{n \to \infty} \Pr \big( \sup_{b \in \RS_{\beta}^\star} \mathscr{A}_n(b, c_n) \leq z \big) \\
&= \Pr \big( \sup_{b \in \RS_{\beta}^\star} \mathscr{A}(b, \underline{\tau}_{\beta}) \leq z \big),
\end{align*}
where the inequality follows from the fact that $\sqrt{n}\hat{\xi}_{n,\beta}(c_n)=\sup_{b} \sqrt{n} \hat{\psi}_{n,\beta}(b, c_n)\geq \sup_{b \in \RS_{\beta}^\star} \sqrt{n} \hat{\psi}_{n,\beta}(b, c_n)$, the first equality is from Lemma \ref{lem:sup fn convergence} (ii) and Lemma \ref{lem:pr_inequality} (ii) and the last equality from Lemma \ref{lem:sup fn convergence} (iii) and $z \in C(\mathscr{A}) \subseteq C(\mathscr{A},0)$.

From the above, 
we then have, 
for any $z \in C(\mathscr{A})$ and $z \ge 0$, $\lim_{n \to \infty} \Pr ( \sqrt{n} \hat{\xi}_{n,\beta}(c_n) \leq z ) = \Pr ( \sup_{b \in \RS_{\beta}^\star} \mathscr{A}(b, \underline{\tau}_{\beta}) \leq z)$. Note that, for $z<0$, $\lim_{n \to \infty} \Pr ( \sqrt{n} \hat{\xi}_{n,\beta}(c_n) \leq z ) = 0$ as $ \hat{\xi}_{n,\beta}$ is nonnegative for all $n$. 
Therefore, for any $z \in C(\mathscr{A})$, we have 
\begin{align*}
\lim_{n \to \infty} \Pr \big( \sqrt{n} \hat{\xi}_{n,\beta}(c_n) \leq z \big) &= 
\begin{cases} 
\Pr \big( \sup_{b \in \RS_{\beta}^\star} \mathscr{A}(b, \underline{\tau}_{\beta}) \leq z \big) & \text{if } z \ge 0 \\
0 & \text{if } z < 0 
\end{cases} \\
&= \Pr \big( \max \big\{ 0, \sup_{b \in \RS_{\beta}^\star} \mathscr{A}(b, \underline{\tau}_{\beta}) \big\} \leq z \big).
\end{align*}
Hence, Lemma \ref{lem:xin_beta_convergence} holds. 
\end{proof}

\begin{proof}[\bf Proof of Theorem \ref{thm:xin_beta_convergence}]
Consider any $x \in \mathbb{R}$ such that the distribution function of $\max \big\{ 0,$ $\sup_{b \in \RS_{\beta}^\star} \mathscr{A}(b, \underline{\tau}_{\beta})\}$ is continuous at $x$. 
By Lemma \ref{lem:xin_beta_convergence}, we can then  find two sequences $\{z_k^+\}$ and $\{z_k^-\}$ converging to $x$ such that  $z_k^+>x>z_k^-$ for each $k$ and 
\begin{align*}
\lim_{n\to\infty}\Pr \big\{ \sqrt{n} \, \hat \xi_{n,\beta}(c_n) \leq z_k^+ \big\}= \Pr\Big(\max \big\{ 0, \sup_{b \in \RS_{\beta}^\star} \mathscr{A}(b, \underline{\tau}_{\beta}) \big\}\leq z_k^+\Big), 
\\
\lim_{n\to\infty}\Pr \big\{ \sqrt{n} \, \hat \xi_{n,\beta}(c_n) \leq z_k^- \big\}= \Pr\Big(\max \big\{ 0, \sup_{b \in \RS_{\beta}^\star} \mathscr{A}(b, \underline{\tau}_{\beta}) \big\}\leq z_k^-\Big). 
\end{align*}  
These then imply that, for any $k$, 
\begin{align*}
    \varlimsup_{n\to\infty} \Pr \big\{ \sqrt{n} \, \hat \xi_{n,\beta}(c_n) \leq x \big\} & \le \lim_{n\to\infty}\Pr \big\{ \sqrt{n} \, \hat \xi_{n,\beta}(c_n) \leq z_k^+ \big\}
    = \Pr\big(\max \big\{ 0, \sup_{b \in \RS_{\beta}^\star} \mathscr{A}(b, \underline{\tau}_{\beta}) \big\}\leq z_k^+\big), 
    \\
    \varliminf_{n\to\infty} \Pr \big\{ \sqrt{n} \, \hat \xi_{n,\beta}(c_n) \leq x \big\} & \ge \lim_{n\to\infty}\Pr \big\{ \sqrt{n} \, \hat \xi_{n,\beta}(c_n) \leq z_k^- \big\}= \Pr\big(\max \big\{ 0, \sup_{b \in \RS_{\beta}^\star} \mathscr{A}(b, \underline{\tau}_{\beta}) \big\}\leq z_k^-\big). 
\end{align*}
Because the distribution of $\max \big\{ 0, \sup_{b \in \RS_{\beta}^\star} \mathscr{A}(b, \underline{\tau}_{\beta})\}$ is continuous at $x$, 
letting $k \to \infty$, we then have 
\begin{align*}
\lim_{n\to\infty}\Pr \big\{ \sqrt{n} \, \hat \xi_{n,\beta}(c_n) \leq x \big\}= \Pr\big(\max \big\{ 0, \sup_{b \in \RS_{\beta}^\star} \mathscr{A}(b, \underline{\tau}_{\beta}) \big\}\leq x\big). 
\end{align*} 
From the above, Theorem \ref{thm:xin_beta_convergence} holds.
\end{proof}

\begin{proposition}\label{prop:sp_A_continuity}
Suppose that Assumption \ref{asmp:iid} holds, and define $\mathscr{A}(b,\underline{\tau}_\beta)$ the same as in \eqref{eq:GP_power}.
For any non-empty compact set $S \subseteq \mathbb{R}$,
the random variable $M = \sup_{b\in S} \mathscr{A}(b,\underline{\tau}_\beta)$ has a continuous distribution function, i.e.\ $\Pr(M = c) = 0$ for every $c \in \mathbb{R}$.
\end{proposition}

\begin{proof}[Proof of Proposition \ref{prop:sp_A_continuity}]
Let $u=\min_{b\in S}F_0(b)$ and $v=\max_{b\in S}F_0(b).$
Because $F_0$ is a continuous, strictly increasing distribution
function and $S$ is nonempty and compact, we must have 
$
0<u\leq v<1.
$
Note that the covariance kernel of a standard Brownian bridge is
\[
K(s,t)=s\wedge t-st,
\qquad s,t\in[0,1].
\]
Define
$
\eta =(1-v)\mathscr B_0(u)+u\mathscr B_0(v).
$
For every $t\in[u,v]$,
$
K(t,u)=u(1-t),
$
$K(t,v)=t(1-v),$
and hence
\begin{align*}
\Cov\big(\mathscr B_0(t), \eta \big)
&=(1-v)K(t,u)+uK(t,v)
=(1-v)u(1-t)+ut(1-v) =u(1-v).
\end{align*}
Consequently,
\begin{align*}
\sigma_\eta^2 \equiv \Var(\eta)
&=(1-v)\Cov(\mathscr B_0(u), \eta )
  +u\Cov(\mathscr B_0(v), \eta)=u(1-v)(1+u-v)>0.
\end{align*}
Let 
\[
\tilde{\eta} = {\eta}/{\sigma_{\eta}},
\quad 
\text{ and } \quad 
\rho={u(1-v)}/{\sigma_{\eta}}
     =\sqrt{{u(1-v)}/{(1+u-v)}}>0.
\]
Then $\tilde{\eta} \sim N(0,1)$. 
We further define, for $t\in[u,v]$,
$
\tilde{\mathscr B}_0(t)
=\mathscr B_0(t)-\rho \tilde{\eta}.
$
From the calculations before, we know 
$
\Cov(\mathscr B_0(t),\tilde{\eta})
=u(1-v)/\sigma_{\eta}
=\rho,
$
which implies that 
\[
\Cov\big(\tilde{\mathscr B}_0(t), \tilde{\eta} \big)=0
\qquad (t\in[u,v]).
\]
Thus, $\tilde{\eta}$ must be independent of the process
$\{\tilde{\mathscr B}_0(t):t\in[u,v]\}$, as they are jointly Gaussian. Since
$\mathscr B_1$ is independent of $\mathscr B_0$, $\tilde{\eta}$ is also
independent of $\mathscr B_1$.

By definition, we then have the following equivalent form for $\mathscr{A}(b,\underline{\tau}_\beta)$: 
\begin{align*}
    \mathscr{A}(b,\underline{\tau}_\beta) 
    & =
    \frac{1}{\sqrt{1-\lambda}}\,\mathscr{B}_0(F_0(b)) 
    - \frac{1}{\sqrt{\lambda}}\,\mathscr{B}_1(F_1(b+\underline{\tau}_\beta)) 
    + h\cdot f_1(b+\underline{\tau}_\beta)\\
    & = 
    \frac{\rho }{\sqrt{1-\lambda}} \tilde{\eta} + 
    \frac{1}{\sqrt{1-\lambda}} \tilde{\mathscr{B}}_0 (F_0(b)) 
    - \frac{1}{\sqrt{\lambda}} \mathscr{B}_1 (F_1(b+\underline{\tau}_\beta) ) 
    + h\cdot f_1(b+\underline{\tau}_\beta).
\end{align*}
Let $\kappa = \rho/\sqrt{1-\lambda}>0$. 
Because $F_0(b)\in[u,v]$ for every $b\in S$, this further implies that 
$M = \sup_{b\in S} \mathscr{A}(b,\underline{\tau}_\beta) = \kappa \tilde{\eta} + W$, where 
\[
W \equiv \sup_{b\in S}
\left\{
\frac{\tilde{\mathscr B}_0(F_0(b))}{\sqrt{1-\lambda}}
-\frac{\mathscr B_1(F_1(b+\underline{\tau}_\beta))}{\sqrt{\lambda}}
+h f_1(b+\underline{\tau}_\beta)
\right\}
\]
is a finite random variable since the stochastic part has
continuous sample paths on the compact set $S$ and $f_1$ is
bounded. 
Moreover, $W$ and $\tilde{\eta}$ are independent. Thus, conditionally on
$W$, the random variable $M$ is Gaussian with mean $W$ and variance
$\kappa^2>0$. 
By the law of iterated expectation, we then have 
$
    \Pr(M = c)
    = \E\{ \Pr(M = c \mid W) \} = 0. 
$

From the above, Proposition \ref{prop:sp_A_continuity} holds. 
\end{proof}

\begin{proof}[\bf Proof of Corollary \ref{cor:asymptotic_power}]
For descriptive convenience, let 
$M =
\sup_{b\in\RS_\beta^\star}
\mathscr{A}(b,\underline{\tau}_\beta)$ and 
$X_n=
\sqrt{{n_1n_0}/{n}} \cdot 
\hat{\xi}_{n,\beta}(c_n).$
By Theorem \ref{thm:xin_beta_convergence} and Slutsky's theorem,
\begin{align*}
X_n
\converged
X
\equiv
\sqrt{\lambda(1-\lambda)}
\max\{0,M\}.
\end{align*}
For $x\geq 0$, define the rescaled randomization tail function
$
\tilde G_n(x)=
G_n(
\sqrt{{n}/{(n_1n_0)}} \cdot x
).
$
By Lemma \ref{lem:null_limit}, for every fixed $x\geq 0$,
\begin{align*}
\tilde G_n(x)
\rightarrow
g(x)
\equiv
\exp(-2x^2).
\end{align*}
Moreover, by definition,  
\begin{align*}
\overline p_{\lceil n_1\beta\rceil,c_n}^{\t}=
G_n\left(\hat{\xi}_{n,\beta}(c_n)\right)=
\tilde G_n(X_n).
\end{align*}

Fix any $\alpha\in(0,1)$ and let
$
q_\alpha
=
\sqrt{-{1}/{2}\cdot \log\alpha}.
$
For any $0<\varepsilon<q_\alpha$, the strict monotonicity of $g$ implies
$
g(q_\alpha-\varepsilon)
>
\alpha
>
g(q_\alpha+\varepsilon).
$
Therefore, for all sufficiently large $n$,
\begin{align*}
\tilde G_n(q_\alpha-\varepsilon) > \alpha > \tilde G_n(q_\alpha+\varepsilon).
\end{align*}
Since $\tilde G_n$ is nonincreasing, it follows that
\begin{align*}
\{X_n\geq q_\alpha+\varepsilon\}
\subseteq
\big\{
\overline p_{\lceil n_1\beta\rceil,c_n}^{\t}
\leq\alpha
\big\}
\subseteq
\{X_n \ge q_\alpha-\varepsilon\}.
\end{align*}
By Proposition \ref{prop:sp_A_continuity}, $M$ has a continuous
distribution function. Thus, the distribution function of $X$ is
continuous at every positive point. In particular, it is continuous at
$q_\alpha-\varepsilon$, $q_\alpha$, and $q_\alpha+\varepsilon$. Hence,
\begin{align*}
\Pr(X\geq q_\alpha+\varepsilon)
&\leq
\liminf_{n\rightarrow\infty}
\Pr\left(
\overline p_{\lceil n_1\beta\rceil,c_n}^{\t}
\leq\alpha
\right)
\leq
\limsup_{n\rightarrow\infty}
\Pr\left(
\overline p_{\lceil n_1\beta\rceil,c_n}^{\t}
\leq\alpha
\right)
\leq
\Pr(X \ge q_\alpha-\varepsilon).
\end{align*}
Letting $\varepsilon\downarrow0$ and using the continuity of the
distribution function of $X$ at $q_\alpha>0$, we have 
\begin{align*}
\lim_{n\rightarrow\infty}
\Pr\left(
\overline p_{\lceil n_1\beta\rceil,c_n}^{\t}
\leq\alpha
\right)
&=\Pr(X>q_\alpha)
=
\Pr\left(
\sqrt{\lambda(1-\lambda)}
\max\{0,M \}
>
\sqrt{-\frac{1}{2}\log\alpha}
\right)\\
&=
\Pr\left(
\sqrt{2\lambda(1-\lambda)}\,M 
>
\sqrt{-\log\alpha}
\right)\\
&=
\Pr\left(
\sqrt{2\lambda(1-\lambda)}
\sup_{b\in\RS_\beta^\star}
\mathscr{A}(b,\underline{\tau}_\beta)
>
\sqrt{-\log\alpha}
\right).
\end{align*}
Therefore, Corollary \ref{cor:asymptotic_power} holds.
\end{proof}

\begin{proof}[\bf Comment on $f_1(b + \underline{\tau}_{\beta})
= f_0(b)$ for $b \in \RS_\beta^\star$]
    Fix $b\in \RS_\beta^\star$. By Lemma \ref{lem:S0*property},
$
    F_0^{-1}(1-\beta)<b<F_1^{-1}(\beta)-\underline{\tau}_\beta,
$
and hence $F_1(b+\underline{\tau}_\beta)<\beta$. 
Thus, 
$\psi_\beta(b,\underline{\tau}_\beta)=0$ implies that 
\begin{align*}
    F_0(b)=1-\beta+F_1(b+\underline{\tau}_\beta).
\end{align*}
Moreover, for $x$ in a sufficiently small neighborhood of $b$, we also have 
\begin{align*}
    \psi_\beta(x,\underline{\tau}_\beta)
    =
    F_0(x)-1+\beta-F_1(x+\underline{\tau}_\beta).
\end{align*}
Because $b$ maximizes $\psi_\beta(\cdot,\underline{\tau}_\beta)$, 
taking derivatives gives 
\begin{align*}
    0
    =
    \frac{\partial}{\partial x}
    \psi_\beta(x,\underline{\tau}_\beta)\bigg|_{x=b}
    =
    f_0(b)-f_1(b+\underline{\tau}_\beta).
\end{align*}
Thus, $f_1(b+\underline{\tau}_\beta)=f_0(b).$
\end{proof}

\begin{proof}[\bf Comment on the mean, variance and covariance for $\mathscr{A}(\cdot, \underline{\tau}_{\beta})$]
For a standard Brownian bridge $\mathscr{B}_0$, we know $\E(\mathscr{B}_0(u)) = 0$ and 
$\Cov\{\mathscr{B}_0(u),\mathscr{B}_0(v)\}
    =
    u(1-v)$ for  $0\le u\leq v \le 1$. 
By definition and the previous comment, we then have, for any $b\in \RS_\beta^\star$, 
\begin{align*}
    \E\{\mathscr{A}(b,\underline{\tau}_\beta)\}
    &=
    h f_1(b+\tau_\beta)
    =
    h f_0(b),
\end{align*}
and for any $b \le b'$, by the mutual independence of $\mathscr{B}_0$ and $\mathscr{B}_1$, 

\begin{align*}
    & \quad \ \Cov\{
        \mathscr{A}(b,\tau_\beta),
        \mathscr{A}(b',\tau_\beta)
    \}\\
    & =
    \frac{1}{1-\lambda}
    \Cov\big\{ \mathscr{B}_0(F_0(b)), \mathscr{B}_0(F_0(b')) \big\}
    + 
    \frac{1}{\lambda} \Cov\big\{ \mathscr{B}_1(F_1(b + \underline{\tau}_{\beta})), \mathscr{B}_1(F_1(b' + \underline{\tau}_{\beta})) \big\}\\
    & = \frac{
        F_0(b)\{1-F_0(b')\}
    }{1-\lambda}
    +
    \frac{
        F_1(b+\tau_\beta)
        \{1-F_1(b'+\tau_\beta)\}
    }{\lambda}.
\end{align*}
In particular, when $b=b' \in \RS_\beta^\star$, we have 
\begin{align*}
    \Var\{\mathscr{A}(b,\tau_\beta)\}
    &=
    \frac{F_0(b)\{1-F_0(b)\}}{1-\lambda}
    +
    \frac{
        F_1(b+\tau_\beta)
        \{1-F_1(b+\tau_\beta)\}
    }{\lambda}\\
    &=
    \frac{F_0(b)\{1-F_0(b)\}}{1-\lambda}
    +
    \frac{
        \{F_0(b)+\beta-1\}
        \{2-\beta-F_0(b)\}
    }{\lambda},
\end{align*}
where the last equality follows from the fact that $F_0(b)=1-\beta+F_1(b+\underline{\tau}_\beta)$ as shown in the previous comment.
\end{proof}

We have studied the asymptotic local power for Gaussian outcomes in Appendix \ref{ssec:Gaussian_local_power}. The corresponding proofs are provided below.

\begin{proof}[\bf Proof of Theorem \ref{thm:asymptotic_power_normal}]
From the comment after the proof of Corollary \ref{cor:asymptotic_power}, for $b_0 \in \RS_{\beta}^\star$, 
\begin{align}
\label{eq:t0_sharp_bound_relation}
F_0(b_0) = 1-\beta+F_1(b_0+\underline{\tau}_{\beta}), \quad f_0(b_0) = f_1(b_0+\underline{\tau}_{\beta}).
\end{align}

Since we assume that $F_0$ and $F_1$ are both Gaussian distribution functions, equality (\ref{eq:t0_sharp_bound_relation}) can be simplified as
\begin{align*}
& \Phi\left( \frac{b_0 - \mu_0}{\sigma_0} \right) - \Phi\left( \frac{b_0 + \underline{\tau}_{\beta} - \mu_1}{\sigma_1} \right) = 1 - \beta , \\
& 
    \left( \frac{b_0 - \mu_0}{\sigma_0} \right)^2 -\left( \frac{b_0 + \underline{\tau}_{\beta} - \mu_1}{\sigma_1} \right)^2 =  \log \frac{\sigma_1^2}{\sigma_0^2}.
\end{align*}
Let 
\begin{align}\label{eq:uv}
    u = \frac{b_0 - \mu_0}{\sigma_0}, \quad v = \frac{b_0 + \underline{\tau}_{\beta} - \mu_1}{\sigma_1}.
\end{align}
Then \eqref{eq:t0_sharp_bound_relation} becomes equivalent to
\begin{align}
\label{eq:u_v_relation}
\Phi(u) - \Phi(v) = 1 - \beta , \quad u^2 -v^2 =  \log \frac{\sigma_1^2}{\sigma_0^2}. 
\end{align}

We first show that $\RS_{\beta}^\star$ only contains a single point. 
From  Lemma \ref{lem:S0*property},  $\RS_{\beta}^\star$ is nonempty. 
It suffices to show there exists at most one pair $(u,v)$ satisfying \eqref{eq:u_v_relation}. Note that, under \eqref{eq:u_v_relation}, $u = \Phi^{-1}(\Phi(v) + 1 - \beta)$. Hence, it suffices to show $\{ \Phi^{-1}(\Phi(v) + 1 - \beta) \}^2-v^2$ is strictly increasing in $v$. Since $\Phi(\cdot)$ is a strictly increasing function, it suffices to show $g(y) = \{ \Phi^{-1}(y + 1 - \beta) \} ^2- \{\Phi^{-1}(y)\}^2$ is strictly increasing in $y$, or equivalently $g^\prime(y) > 0$ for $y\in (0,\beta)$.
Note that, by some algebra, 
$$
g^\prime(y) = 2 \Phi^{-1}(y + 1 - \beta)/\phi(\Phi^{-1}(y + 1 - \beta)) - 2 \Phi^{-1}(y)/\phi(\Phi^{-1}(y )), 
$$ 
where $\phi$ is the density function of the standard Gaussian distribution.
Since $\Phi^{-1}(y + 1 - \beta) > \Phi^{-1}(y)$, 
it suffices to show that $k(x) = x/\phi(x) = \sqrt{2\pi} \ x\exp(x^2/2)$ is strictly increasing in $x\in \mathbb{R}$. 
By some algebra, 
$k^\prime(x) = \sqrt{2\pi}(1+x^2)\exp(x^2/2)>0$ for all $x\in \mathbb{R}$. 
Therefore, $\RS_{\beta}^\star$ only contains one element $b_0$. This then proves (i).   

We next simplify the local power 
\begin{align}\label{eq:asymptotic_power}
    \Gamma_\beta \equiv \Pr\Big(\sqrt{2 \lambda(1-\lambda)}  \mathscr{A}(b_0, \underline{\tau}_{\beta}) > \sqrt{-\log\alpha}\Big) = 
    \Pr\Big(\mathscr{A}(b_0, \underline{\tau}_{\beta}) > c \Big), 
\end{align}
where $\mathscr{A}(b_0, \underline{\tau}_{\beta})$ is defined as in \eqref{eq:GP_power} and $c = \sqrt{-\log\alpha}/\sqrt{2 \lambda(1-\lambda)} > 0$. 
From \eqref{eq:Gaussian_process_mean_variance}, 
$\mathscr{A}(b_0, \underline{\tau}_{\beta})$ is Gaussian with
mean 
\begin{align*}
\mu(b_0) = h \cdot f_1(b_0 + \underline{\tau}_{\beta})
=\frac{h}{\sigma_1} \cdot \phi\Big(\frac{b_0 + \underline{\tau}_{\beta} -\mu_1}{\sigma_1}\Big) 
=\frac{h}{\sigma_1} \cdot \phi(v)
\end{align*}
and variance 
\begin{align*}
\{\sigma(b_0)\}^2 &= \frac{1}{1-\lambda} F_0(b_0)(1-F_0(b_0)) + \frac{1}{\lambda} F_1(b_0 + \underline{\tau}_{\beta})(1-F_1(b_0 + \underline{\tau}_{\beta})) \notag\\
 &= \frac{1}{1-\lambda} \Phi\Big(\frac{b_0-\mu_0}{\sigma_0}\Big)\Big(1-\Phi\Big(\frac{b_0-\mu_0}{\sigma_0}\Big)\Big) + \frac{1}{\lambda} \Phi\Big(\frac{b_0 + \underline{\tau}_{\beta} -\mu_1}{\sigma_1}\Big)\Big(1-\Phi\Big(\frac{b_0 + \underline{\tau}_{\beta} -\mu_1}{\sigma_1}\Big)\Big) \notag\\
&= \frac{1}{1-\lambda} \Phi(u)(1-\Phi(u)) + \frac{1}{\lambda} \Phi(v)(1-\Phi(v)),
\end{align*}
where $u$ and $v$ are defined as in \eqref{eq:uv}. 
Therefore, $\Gamma_\beta$ can be equivalently written as 
\begin{align}
\label{eq:p_beta}
\Gamma_\beta \equiv \Pr\big(\mathscr{A}(b_0, \underline{\tau}_{\beta}) \geq c \big) = \Phi\Big(\frac{\mu(b_0)-c}{\sigma(b_0)}\Big),
\end{align}
which, with fixed $(h, \sigma_1, \lambda, \alpha)$, depends only on $(u,v)$ solved from equation \eqref{eq:u_v_relation}, .

We now prove (ii). Consider the special case where $\sigma_0=\sigma_1=\sigma$. In this case, \eqref{eq:u_v_relation} has the following form: 
\begin{align*}
\Phi(u) - \Phi(v) = 1 - \beta , \quad u^2 -v^2 = 0.
\end{align*}
We can now directly solve $u$ and $v$: 
\begin{align*}
u = \Phi^{-1}\Big(1-\frac{\beta}{2}\Big) = -\Phi^{-1}\Big(\frac{\beta}{2}\Big), \quad v = \Phi^{-1}\Big(\frac{\beta}{2}\Big).
\end{align*}
From \eqref{eq:uv}, this then implies 
\begin{align*}
    \underline{\tau}_\beta=\mu_1-\mu_0+2\sigma \Phi^{-1}(\beta/2), \quad b_0= \mu_0-\sigma\Phi^{-1}(\beta/2).
\end{align*}
It also follows that $\mathscr{A}(b_0, \underline{\tau}_{\beta})$ has mean and variance
\begin{align*}
\mu(b_0) &=  \frac{h}{\sigma} \cdot \phi(v) =
\frac{h}{\sigma} \cdot \phi\Big(\Phi^{-1}\Big(\frac{\beta}{2}\Big)\Big) ,\\
\{\sigma(b_0)\}^2 &= \frac{1}{1-\lambda} \Phi(u)(1-\Phi(u)) + \frac{1}{\lambda} \Phi(v)(1-\Phi(v)) = \frac{1}{(1-\lambda)\lambda} \Big(1-\frac{\beta}{2} \Big)\frac{\beta}{2}.
\end{align*}
Hence, by \eqref{eq:p_beta},
\begin{align}\label{eq:gamma_beta_power}
\Gamma_\beta 
= \Phi \left( \frac{\sqrt{2 \lambda(1-\lambda)} \cdot h/\sigma \cdot \phi(\Phi^{-1}({\beta}/{2}) ) - \sqrt{- \log \alpha}}{\sqrt{\beta (1-{\beta}/{2})}} \right).
\end{align}
Therefore, (ii) holds. 

We finally prove (iii). Fix first $\beta\in(0,1)$. 
Because $\Phi$ is a strictly increasing function and 
\begin{align*}
\frac{\partial}{\partial \sigma}
\left\{
\frac{
\sqrt{2\lambda(1-\lambda)}(h/\sigma)
\phi\{\Phi^{-1}(\beta/2)\}
-
\sqrt{-\log\alpha}
}{
\sqrt{\beta(1-\beta/2)}
}
\right\}
=
-
\frac{
\sqrt{2\lambda(1-\lambda)}h
\phi\{\Phi^{-1}(\beta/2)\}
}{
\sigma^2\sqrt{\beta(1-\beta/2)}
}
<0, 
\end{align*}
$\Gamma_\beta $ in \eqref{eq:gamma_beta_power} must be strictly decreasing in $\sigma$. 
Next, fix $\sigma>0$ and make the substitution
\begin{align*}
x
=
\Phi^{-1}\left(\frac{\beta}{2}\right)
\in(-\infty,0),
\qquad
c_0
=
\frac{
\sqrt{-\log\alpha}
}{
\sqrt{2\lambda(1-\lambda)} \cdot h/\sigma
}
>0.
\end{align*}
Up to a positive multiplicative constant, the argument inside the function $\Phi$ in
\eqref{eq:gamma_beta_power} can be written as
\begin{align*}
W(x)
=
\frac{\phi(x)-c_0}
{\sqrt{\Phi(x)\{1-\Phi(x)\}}}.
\end{align*}
Because $x=\Phi^{-1}(\beta/2)$ is strictly increasing in $\beta$,
to prove that $\Gamma_{\beta}$ is strictly increasing in $\beta$, it
suffices to show that $W'(x)>0$ for every $x<0$.
Letting $p=\Phi(x) = \beta/2 \in (0, 1/2)$ and using the fact that $\phi'(x) = \phi(x)\cdot (-x)$, 
we have 
\begin{align*}
W'(x)
&=
\frac{
\phi(x)
}{
2\{p(1-p)\}^{3/2}
}
\left[
-2x p(1-p)
-
\{\phi(x)-c_0\}(1-2p)
\right]\\
&=
\frac{
\phi(x)
}{
2\{p(1-p)\}^{3/2}
}
\left[
-2x p(1-p)
-
\phi(x)(1-2p)
+
c_0(1-2p)
\right].
\end{align*}
Since $p\in (0, 1/2)$, we have 
$c_0(1-2p)>0$.
Note that 
\begin{align*}
-2x p(1-p)-\phi(x)(1-2p) & = -2x p(1-p)-\phi(x)(1-p) + p\phi(x)
 \\
 & = 
 p(1-p) \Big( -2x - \frac{\phi(x)}{p} + \frac{\phi(x)}{1-p} \Big).
\end{align*}
It remains to show that
\begin{align*}
 -2x - \frac{\phi(x)}{p} + \frac{\phi(x)}{1-p} > 0. 
\end{align*}
Let $t=-x>0$. By symmetry, $p=1-\Phi(t)$ and $\phi(x)=\phi(t)$. It then remains to show, for any $t>0$, 
\begin{align*}
    D(t) \equiv 2t - m(t) + r(t)>0, 
    \quad 
    \text{where } \quad 
    m(t)
=
\frac{\phi(t)}{1-\Phi(t)},
\ \ \text{ and } \ \ 
r(t)
=
\frac{\phi(t)}{\Phi(t)}. 
\end{align*}
Differentiating $m$ and $r$ gives
\begin{align*}
m'(t)=m(t)\{m(t)-t\}, \qquad
r'(t)=-r(t)\{t+r(t)\}.
\end{align*}
Consequently,
\begin{align*}
D'(t)
&=
2-m(t)\{m(t)-t\}-r(t)\{t+r(t)\}\\
&=
\left[1-m(t)\{m(t)-t\}\right]
+
\left[1-r(t)\{t+r(t)\}\right].
\end{align*}
Let $\eta \sim N(0,1)$. 
By the properties of the truncated Gaussian distribution, the two terms on the right-hand side are equivalently
\begin{align*}
1-m(t)\{m(t)-t\}
=
\Var(\eta \mid \eta>t)>0,
\quad 
1-r(t)\{t+r(t)\}=
\Var(\eta \mid \eta \leq t)>0.
\end{align*}
Thus, $D'(t)>0$ for every $t>0$. Moreover,
$
D(0)
=
-m(0)+r(0)
=
0.
$
It follows that $D(t)>0$ for every $t>0$. 
Hence, we prove that 
$\Gamma_\beta$ is strictly increasing in $\beta$.
This completes the proof of part (iii).

From the above, Theorem \ref{thm:asymptotic_power_normal} holds.
\end{proof}

\begin{proof}[\bf Proof of Proposition \ref{prop:local_power_maximum}]
We consider the parameter region $0<\beta<1$ and $\sigma_1>0$.
The boundary values $\beta=1$ and $\sigma_1=0$ below are understood as
the limits $\beta\to1^-$ and $\sigma_1\to0^+$, respectively.

From Theorem \ref{thm:asymptotic_power_normal}, the contact set is a singleton, i.e.,
$\mathcal S_\beta^\star=\{b_0\}$. 
Let $u=F_1(b_0+\underline{\tau}_\beta)$ and $v=F_0(b_0)$.
From \eqref{eq:t0_sharp_bound_relation}, we can verify that 
\begin{align*}
\beta=1+u-v,
\ \ \ 
f_0(b_0)=\frac{\phi\{\Phi^{-1}(v)\}}{\sigma_0}, 
\ \ \ 
f_1(b_0+\underline{\tau}_\beta) = \frac{\phi\{\Phi^{-1}(u)\}}{\sigma_1}, 
\ \ \ 
\sigma_1=\sigma_0
\frac{\phi\{\Phi^{-1}(u)\}}
     {\phi\{\Phi^{-1}(v)\}},
\end{align*}
and 
$\mathscr{A}(b_0, \underline{\tau}_{\beta})$ is Gaussian with
mean 
\begin{align*}
\mu(b_0) = h \cdot f_0(b_0)
= h \frac{\phi\{\Phi^{-1}(v)\}}{\sigma_0}
\end{align*}
and variance 
\begin{align*}
\{\sigma(b_0)\}^2 &= \frac{1}{1-\lambda} F_0(b_0)(1-F_0(b_0)) + \frac{1}{\lambda} F_1(b_0 + \underline{\tau}_{\beta})(1-F_1(b_0 + \underline{\tau}_{\beta})) 
\\
& = \frac{1}{1-\lambda} v(1-v) + \frac{1}{\lambda} u(1-u) 
\\
& = \frac{1}{\lambda(1-\lambda)} \big\{ \lambda v(1-v)+(1-\lambda)u(1-u) \big\}. 
\end{align*}
Let
\begin{align*}
c_\alpha=\sqrt{-1/2 \cdot \log\alpha},
\qquad
a=\frac{\sqrt{\lambda(1-\lambda)}\,h}
        {\sigma_0\sqrt{2\pi}},
\qquad
r(v)=\frac{\phi\{\Phi^{-1}(v)\}}{\phi(0)}.
\end{align*}
By definition, we can verify that $\Gamma_\beta=\Phi\{T(u,v)\}$ with 
\begin{align*}
T(u,v)=\frac{a r(v)-c_\alpha}
{\sqrt{\lambda v(1-v)+(1-\lambda)u(1-u)}}.
\end{align*}

We prove first that 
\begin{align*}
r(v)\le 2\sqrt{v(1-v)}\le1,
\qquad \text{for } 0<v<1,
\end{align*}
with equality at $v=1/2$. 
Let 
$x=\Phi^{-1}(v)$. Then we have 
\begin{align*}
    4 v(1-v) & = 4 \Phi(x) \{ 1 - \Phi(x)\} = 4 \Phi(x) \Phi(-x) = 4 \Phi(|x|) \Phi(-|x|)
    = 
    4 \Phi(|x|) \{1 - \Phi(|x|)\}\\
    & = 1 - \{2\Phi(|x|) - 1\}^2 = 1 - \big\{ \text{erf}\big(|x|/\sqrt{2}\big) \big\}^2
    \ge \exp(-2x^2/\pi) \ge \exp(-x^2)\\
    & = \{ \phi(x)/\phi(0) \}^2 = \{ r(v) \}^2.
\end{align*}
where we use the definition of $\text{erf}(\cdot)$ in (7.2.1) in \citet{NIST:DLMF} and the inequality in (7.8.8) in \citet{NIST:DLMF}. 
This then proves the above inequality.

We consider first the case where 
$a < c_\alpha$.
Then $a r(v)-c_\alpha\le a -c_\alpha < 0$, and
\begin{align*}
\lambda v(1-v)+(1-\lambda)u(1-u)\le\frac14 \{\lambda +(1-\lambda) \} = \frac14.
\end{align*}
Therefore,
\begin{align*}
T(u,v)
\le
2\{a r(v)-c_\alpha\}
\le
2(a-c_\alpha).
\end{align*}
The equality is approached when $u=v=1/2$, which, by Theorem \ref{thm:asymptotic_power_normal}(ii), corresponds to
\begin{align*}
\beta\to1^-
\qquad\text{and}\qquad
\sigma_1\to\sigma_0.
\end{align*}
Thus, $\Gamma_\beta$ is maximized in the limit
$(\sigma_1,\beta)\to(\sigma_0,1^-)$.

We next consider the case where 
$a>c_\alpha$. When 
$a r(v)-c_\alpha\le0$, then $T(u,v)\le0$ and such a point cannot
maximize the local power. When 
$a r(v)-c_\alpha > 0$, we have 
\begin{align*}
\lambda v(1-v)+(1-\lambda)u(1-u)
\ge \lambda v(1-v).
\end{align*}
Letting $s(v)=\sqrt{v(1-v)}$ and using
$r(v)\le2s(v)$ and $2s(v)\le1$, we obtain
\begin{align*}
T(u,v) 
\le \frac{a r(v)-c_\alpha}{\sqrt{\lambda}\,s(v)}  
\le \frac{2a s(v)-c_\alpha}{\sqrt{\lambda}\,s(v)} 
= \frac{2a}{\sqrt{\lambda}} - \frac{c_\alpha}{\sqrt{\lambda}\,s(v)} 
\le \frac{2(a-c_\alpha)}{\sqrt{\lambda}}.
\end{align*}
The equality is approached when $v=1/2$ and $u\to0^+$, which corresponds to 
\begin{align*}
\beta=1+u-v\to\frac12
\qquad\text{and}\qquad
\sigma_1
=
\sigma_0
\frac{\phi\{\Phi^{-1}(u)\}}
     {\phi\{\Phi^{-1}(v)\}}
\to0^+.
\end{align*}
Thus, $\Gamma_\beta$ is maximized in the limit
$(\sigma_1,\beta)\to(0^+,1/2)$.

Finally, we note that 
\begin{align*}
    a - c_\alpha
    & = \frac{\sqrt{\lambda(1-\lambda)}\,h}
        {\sigma_0\sqrt{2\pi}} - \sqrt{-1/2 \cdot \log\alpha}
    = \frac{\sqrt{\lambda(1-\lambda)}}{\sigma_0\sqrt{2\pi}}
    \left\{ 
    h - \frac{\sigma_0\sqrt{ -\pi\log\alpha}}{\sqrt{\lambda(1-\lambda)}}
    \right\}
    \\
    & = \frac{\sqrt{\lambda(1-\lambda)}}{\sigma_0\sqrt{2\pi}}
    (h - \gamma),
\end{align*}
which implies that $a - c_\alpha$ has the same sign as $h - \gamma$.

From the above, Proposition \ref{prop:local_power_maximum} holds. 
\end{proof}

\section{Proof of Remark \ref{rmk:adjustment_infty}, Theorem \ref{thm:bar_F_covariate_adjustment}, Remark \ref{rmk:adjustment_normal}, and Theorems \ref{thm:sharp_bound_consist_covariate}}
\label{ssec:covariate_adjust}

\begin{proof}[\bf Technical details for Remark \ref{rmk:adjustment_infty}]
Let $\mathcal A=\{i:s(X_i)=+\infty\}$.
Given the prespecified adjustment function, define, for $z=0,1$, 
\begin{align*}
\tilde{Y}_i(z)= Y_i(z)-s(X_i) = 
\begin{cases}
Y_i(z)-s(X_i), & i\notin\mathcal A,\\
-\infty, & i\in\mathcal A. 
\end{cases}
\end{align*}
Although
$\tilde{Y}_i(1)-\tilde{Y}_i(0)$ is not well defined for
$i\in\mathcal A$, the original treatment effect
$\tau_i=Y_i(1)-Y_i(0)$ still satisfies, for $1\le i \le n$, 
\begin{align*}
\tilde{Y}_i(1)
=
\tilde{Y}_i(0)+\tau_i.
\end{align*}
For \(i\notin\mathcal A\), the equality holds naturally, whereas
for \(i\in\mathcal A\), both sides equal \(-\infty\). Consequently, the
adjusted observed outcome still satisfies, for $1\le i \le n$, 
\begin{align*}
\tilde{Y}_i
=
\tilde{Y}_i(0)+Z_i\tau_i,
\quad \text{and} \quad 
\tilde{Y}_i(0) = 
\tilde{Y}_i-Z_i\tau_i. 
\end{align*}
As a result, the imputation of the control potential outcomes can still be formulated using the original
treatment-effect vector \(\bs{\tau}\), without defining the
adjusted ITEs.

Moreover, regardless of the hypothesized ITEs $\bs{\delta}\in\mathbb{R}^n$, the units in $\mathcal{A}$ always have the smallest ranks in terms of the imputed adjusted control potential outcomes.
Thus, to minimize the ranks of the treated units, it suffices to minimize the ranks of treated units among units in $\mathcal{A}^\complement$.
Let $\ell=\min\{n_1-k,\sum_{i\in\mathcal{A}^\complement}Z_i\}$.
By the same logic as the proof for \citet[][Theorem 3]{caughey2023randomisation}, the ranks of treated units are minimized over $\bs{\delta}\in\mathcal{H}_{k,c}^{\t}$ when sufficiently large finite ITEs are assigned to the $\ell$ treated units with the largest adjusted observed outcomes in $\mathcal{A}^\complement$ and constant ITEs of size $c$ are assigned to all remaining treated units.
Specifically, the selected units' imputed adjusted control potential outcomes can be taken below all finite adjusted control outcomes.
This also minimizes the RKS statistic, which is nondecreasing in ranks of the treated units.
By the effect-increasing property in Proposition \ref{prop:rks_properties}, this minimum is bounded below by the statistic obtained by further decreasing the selected units' imputed adjusted control potential outcomes to $-\infty$.
Thus, both tests in  Theorem \ref{thm:pval_RKS} and Corollary \ref{cor:pval_KS}  remain valid in the presence of infinite adjustments.

\end{proof}

\begin{lemma}\label{lem:bar_F_CDF}
    $\bar{F}(y)=1+\E\big[\inf_b\{F_1(b\mid X)-F_0(b-y\mid X)\}\big]$
    defined in \eqref{eq:quantile_lower_bound_covariate} is a distribution function. Moreover, it is a continuous function under Assumption \ref{asmp:iid_covariate}.
\end{lemma}

\begin{proof}[\bf Proof of Lemma \ref{lem:bar_F_CDF}]
    For each $x$ and $y\in\mathbb{R}$, define
    \begin{align*}
        \bar{F}(y|x)
        =1+\inf_{b\in\mathbb{R}}
        \big\{F_1(b\mid x)-F_0(b-y\mid x)\big\} 
        =1+\inf_{c\in\mathbb{R}}
        \big\{F_1(c+y\mid x)-F_0(c\mid x)\big\}.
    \end{align*}
    Then $\bar{F}(y)=\E[\bar{F}(y|X)].$
    Because $b\mapsto F_1(b\mid x)-F_0(b-y\mid x)$ is right-continuous,
    its infimum over $\mathbb{R}$ equals its infimum over $\mathbb{Q}$, the set of rational numbers.
    That is,
    \begin{align*}
        \bar{F}(y|x)
        =1+\inf_{b\in\mathbb{Q}}
        \big\{F_1(b\mid x)-F_0(b-y\mid x)\big\},
    \end{align*}
    which is measurable in $x$. 
    Hence, $\bar{F}(y)$ is well defined. 
    Moreover, we can verify that $\bar{F}(y|x)$ is nondecreasing for any $x$. This implies that $\bar{F}(y)=\E[\bar{F}(y|X)]$ is nondecreasing. Note that, by definition, we can verify that $\bar{F}(y|x) \in [0, 1]$ and thus $\bar{F}(y) \in [0, 1]$.

    To prove that $\lim_{y \to + \infty}\bar{F}(y)=1$ and $\lim_{y \to - \infty}\bar{F}(y)=0$, we will first show that for any $x$, $\lim_{y \to + \infty}\bar{F}(y|x)=1$ and $\lim_{y \to - \infty}\bar{F}(y|x)=0$ hold.
    
    On the one hand, we will prove $\lim_{y \to - \infty}\bar{F}(y|x)=0$. Fix any $\varepsilon>0$ and then there exists
$c_\varepsilon\in\mathbb R$ such that
$1-F_0(c_\varepsilon\mid x)<{\varepsilon}/{2}$ since
$\lim_{c\to+\infty}F_0(c\mid x)= 1$. 
Moreover, since
$
   \lim_{y\to-\infty} F_1(c_\varepsilon+y\mid x)= 0,
$
there exists $y_\varepsilon\in\mathbb R$ such that
$
    F_1(c_\varepsilon+y\mid x)<{\varepsilon}/{2}
$
holds for any $y\leq y_\varepsilon$. 
It follows by definition that, for any $y\leq y_\varepsilon$, 
\begin{align*}
    0
    \leq \bar F(y\mid x)
    &\leq
    1+F_1(c_\varepsilon+y\mid x)
    -F_0(c_\varepsilon\mid x)  
    <\varepsilon.
\end{align*}
Thus, we have 
$
    \lim_{y\to-\infty}\bar F(y\mid x)=0.
$

On the other hand, we will prove $\lim_{y\to+\infty} \bar{F}(y|x)=1$. Fix any $\varepsilon>0$. There exists
$c_\varepsilon\in\mathbb R$ such that
$
    F_0(c_\varepsilon\mid x)<\varepsilon
$
since $\lim_{c\to-\infty} F_0(c\mid x)=0$.
Moreover, since
$
   \lim_{y\to+\infty} F_1(c_\varepsilon+y\mid x)= 1,
$
there exists $y_\varepsilon\in\mathbb R$ such that $ F_1(c_\varepsilon+y\mid x)>1-\varepsilon$ holds for any
$y\geq y_\varepsilon$.
Now suppose that $y\geq y_\varepsilon$. 
For any $c\leq c_\varepsilon$, by the monotonicity of $F_0$, 
\[
    F_0(c\mid x)-F_1(c+y\mid x)
    \leq F_0(c\mid x)
    \leq F_0(c_\varepsilon\mid x)
    <\varepsilon.
\]
For any $c>c_\varepsilon$, by the monotonicity of $F_1$, 
\begin{align*}
    F_0(c\mid x)-F_1(c+y\mid x)
    &\leq 1-F_1(c+y\mid x)  \leq 1-F_1(c_\varepsilon+y\mid x) 
    <\varepsilon.
\end{align*}
Consequently, for any $y\geq y_\varepsilon$, we have 
\begin{align}\label{eq:1minusF_barx}
     0
    \leq 1-\bar F(y\mid x)
    =
    \sup_{c\in\mathbb R}
    \left\{
        F_0(c\mid x)-F_1(c+y\mid x)
    \right\}
    \leq\varepsilon.   
\end{align}
Thus, we have $\lim_{y\to+\infty}\bar F(y\mid x)=1.$

By the bounded convergence theorem, 
$\lim_{y \to + \infty}\bar{F}(y)=1$ and $\lim_{y \to - \infty}\bar{F}(y)=0$ hold.

    Next, we prove the right-continuity of $\bar{F}$. Fix $y\in\mathbb{R}$ and let
    $y_n\downarrow y$. Because $\bar{F}(y|x)$ is nondecreasing,
    \begin{align*}
        \liminf_{n\to\infty}\bar{F}(y_n|x)\ge \bar{F}(y|x).
    \end{align*}
    Fix any $\varepsilon>0$. By the definition of
    the infimum, we can find $c_\varepsilon\in\mathbb{R}$ such that
    \begin{align*}
        1+F_1(c_\varepsilon+y\mid x)-F_0(c_\varepsilon\mid x)
        \le \bar{F}(y|x)+\varepsilon.
    \end{align*}
    Since $c_\varepsilon+y_n\downarrow c_\varepsilon+y$ and
    $F_1(\cdot\mid x)$ is right-continuous, we have 
    \begin{align*}
        \limsup_{n\to\infty}\bar{F}(y_n|x)
        &\le
        \lim_{n\to\infty}
        \{
            1+F_1(c_\varepsilon+y_n\mid x)
            -F_0(c_\varepsilon\mid x)
        \} \\
        &=1+F_1(c_\varepsilon+y\mid x)
        -F_0(c_\varepsilon\mid x) \\
        &\le \bar{F}(y|x)+\varepsilon.
    \end{align*}
    Letting $\varepsilon\downarrow0$ and combining the previous inequalities, for any $x$, 
    \begin{align*}
        \bar{F}(y_n|x)\to \bar{F}(y|x) \text{ as } n\to \infty.
    \end{align*}
    Note that $0\le \bar{F}(y_n|x)\le1$.  By the bounded convergence theorem, we then have 
    \begin{align*}
        \bar{F}(y_n)
        =\E[\bar{F}(y_n|X)]
        \to\E[\bar{F}(y|X)]
        =\bar{F}(y).
    \end{align*}
    Thus, $\bar{F}$ is right-continuous. 

    Finally, we prove the continuity of $\bar{F}$ under Assumption \ref{asmp:iid_covariate}. 
    For any $y,y'\in\mathbb R$,
    by Lemma \ref{lem:absolute_sup},
    \begin{align*}
        |\bar{F}(y'|x)-\bar{F}(y|x)|
        &= \Big|\inf_{c\in\mathbb{R}}
        \big\{F_1(c+y'\mid x)-F_0(c\mid x)\big\} - \inf_{c\in\mathbb{R}}
        \big\{F_1(c+y\mid x)-F_0(c\mid x)\big\}\Big|\\
        &= \Big|\sup_{c\in\mathbb{R}}
        \big\{F_0(c\mid x)-F_1(c+y'\mid x)\big\} - \sup_{c\in\mathbb{R}}
        \big\{F_0(c\mid x)-F_1(c+y\mid x)\big\}\Big|\\
        &\leq
        \sup_{c\in\mathbb R}
        |F_1(c+y'\mid x)-F_1(c+y\mid x)|.
    \end{align*}
    Because every continuous distribution function is uniformly continuous
    on $\mathbb R$, the right-hand side converges to zero as $y'\to y$.
    Thus, $\bar{F}(\cdot |x)$ is continuous for almost every $x$. Note that $0\leq \bar{F}(y|x)\leq1$. By the bounded
    convergence theorem, $\bar{F}(y)=\mathbb{E}[\bar{F}(y|X)]$ is continuous.

    From the above, Lemma \ref{lem:bar_F_CDF} holds. 
\end{proof}

\begin{lemma} \label{lem:bar_F_covariate_adjustment1}
    Under Assumption \ref{asmp:iid_covariate}, the sharp upper bound $\bar{F}(\cdot)$ in \eqref{eq:quantile_lower_bound_covariate} on the ITE distribution function can be equivalently written as 
	\begin{align*}
		\bar{F}(y)=1+\inf_b\big\{F_{y,1}(b)-F_{y,0}(b-y )\big\},
	\end{align*}
	where $F_{y,1}$ and $F_{y,0}$ are defined in \eqref{eq:CDF_adj_PO}. 
\end{lemma}

\begin{proof}[\bf Proof of Lemma \ref{lem:bar_F_covariate_adjustment1}]

By Lemma \ref{lem:bar_F_CDF}, \(\bar F\) is a continuous distribution
function. We first show that, for each fixed \(y\in\mathbb R\), the
minimizer in \eqref{eq:cov_adj} can be chosen measurably as a function
of \(x\).
Let \(\mathcal H=[-\pi/2,\pi/2]\), and define
\[
\tilde g_y(h,x)
=
\begin{cases}
F_1(\tan h\mid x)-F_0(\tan h-y\mid x),
    & h\in(-\pi/2,\pi/2),\\
0,  & h\in\{-\pi/2,\pi/2\}.
\end{cases}
\]
For almost every fixed \(x\), the map \(h\mapsto\tilde g_y(h,x)\) is
continuous on \(\mathcal H\): continuity in the interior follows from
Assumption \ref{asmp:iid_covariate}, while continuity at the endpoints
follows from the usual tail limits of distribution functions. For every
fixed \(h\), the map \(x\mapsto\tilde g_y(h,x)\) is measurable.
Since \(\mathcal H\) is compact, the measurable maximum theorem \citep[][Theorem 18.19]{aliprantis2006infinite} yields a measurable
function \(h_y(x)\) satisfying
$
h_y(x)\in
\operatorname*{arg\,min}_{h\in\mathcal H}\tilde g_y(h,x)
$
almost surely.
If \(h_y(x)=-\pi/2\), replace it by \(\pi/2\). This preserves
measurability and the minimizing property because
\(\tilde g_y(-\pi/2,x)=\tilde g_y(\pi/2,x)=0\).
Denote the resulting measurable minimizer by \(\overline h_y(x)\), and
define \(\overline s(x;y)=\tan\{\overline h_y(x)\}\) when
\(\overline h_y(x)<\pi/2\), and \(\overline s(x;y)=+\infty\) when
\(\overline h_y(x)=\pi/2\). Then \(\overline s(\cdot;y)\) is measurable
and
\[
\overline s(x;y)
\in
\operatorname*{arg\,min}_{b\in\mathbb R\cup\{+\infty\}}
\left\{F_1(b\mid x)-F_0(b-y\mid x)\right\}.
\]

Fix $y\in\mathbb R$ and define
\begin{align*}
m(y|x)
=
\inf_{b\in\mathbb R} \{F_1(b\mid x)-F_0(b-y\mid x)\}
=
F_1(\overline s(x;y)\mid x)-F_0(\overline s(x;y)-y\mid x),
\end{align*}
with the convention that $F_z(+\infty\mid x)=1$ for $z=0,1$. 
We can verify that
the conditional
distribution functions of the adjusted potential outcomes given covariates satisfy
$$
F_{y,z}(c\mid x)
\equiv 
\Pr\{ Y_i(z) - \overline s(X_i;y) \le c \mid X=x \}
=
F_z (c+\overline s(x;y)\mid x ) \ \ \text{ for $z=0,1$.}
$$
Thus, for every $c\in\mathbb R$,
\begin{align*}
F_{y,1}(c)-F_{y,0}(c-y)
&=
\mathbb E\Big[
F_1\big(c+\overline s(X;y)\mid X\big)
-
F_0\big(c+\overline s(X;y)-y\mid X\big)
\Big] 
\geq
\E[m(y|X)].
\end{align*}
Note that the above inequality holds with equality at $c=0$. 
Consequently,
\begin{align*}
\inf_{c\in\mathbb R}
\{F_{y,1}(c)-F_{y,0}(c-y)\}
&=\E[m(y|X)]
= 
\E\left[
\inf_{b\in\mathbb R}
\{F_1(b\mid X)-F_0(b-y\mid X)\}
\right].
\end{align*}
This then implies that 
\begin{align*}
\bar F(y)
= 1 + \mathbb E\left[
\inf_{b\in\mathbb R}
\{F_1(b\mid X)-F_0(b-y\mid X)\}
\right]
=
1+
\inf_{c\in\mathbb R}
\{F_{y,1}(c)-F_{y,0}(c-y)\}.
\end{align*}

From the above, Lemma \ref{lem:bar_F_covariate_adjustment1} holds.
\end{proof}

\begin{lemma}\label{lem:bar_F_covariate_adjustment2}
 Under Assumption \ref{asmp:iid_covariate}, the sharp lower bound $\underline{\tau}_{\beta}^{\textup{co}}$ in \eqref{eq:quantile_lower_bound_covariate} on ITE quantiles can be equivalently written as
    \begin{align*}
		\underline{\tau}_{\beta}^{\textup{co}} =  \sup_{u\in[0,\beta]} \{ F_{\underline{\tau}_{\beta}^{\textup{co}}, 1}^{-1}(u)-F_{\underline{\tau}_{\beta}^{\textup{co}}, 0}^{-1}(1+u-\beta) \},   
	\end{align*}
    where we define $(-\infty)-(-\infty)=-\infty$ in the above expression, and $F_{\underline{\tau}_{\beta}^{\textup{co}}, z}^{-1} (\cdot)$ is  defined as $F_{y, z}^{-1}(\cdot)$ with $y=\underline{\tau}_{\beta}^{\textup{co}}$  for $z=0,1$. 
\end{lemma}

\begin{proof}[\bf Proof of Lemma \ref{lem:bar_F_covariate_adjustment2}]
Fix any $\beta\in(0,1)$, 
by Lemma \ref{lem:bar_F_covariate_adjustment1}, 
let $s_\beta(x)=\bar s(x;\underline{\tau}^{\mathrm{co}}_\beta)$ be a measurable function satisfying \eqref{eq:cov_adj}. For $z=0,1$, let
\begin{align}\label{eq:Gz}
G_z(b)
\equiv
F_{\underline{\tau}^{\mathrm{co}}_\beta,z}(b)
=
\Pr\{Y_i(z)-s_\beta(X_i)\le b\}
=\E[ F_z(b+s_\beta(X)\mid X)],
\qquad b\in\mathbb{R}.
\end{align}
For any $x$ and any $y\in\mathbb{R}$, define
\begin{align*}
m(y|x)
=
\inf_{b\in\mathbb{R}}
\{F_1(b\mid x)-F_0(b-y\mid x)\} \in [-1, 0].
\end{align*}
By the definition of $s_\beta(x)$,
\begin{align*}
m(\underline{\tau}^{\mathrm{co}}_\beta|x)
= 
F_1\big(s_\beta(x)\mid x\big)
-
F_0\big(s_\beta(x)-\underline{\tau}^{\mathrm{co}}_\beta\mid x\big).
\end{align*}

First, we prove that $\E[m(\underline{\tau}^{\mathrm{co}}_\beta|X)] = \beta - 1$. 
From Lemma \ref{lem:bar_F_CDF}, $\bar F$ is a continuous distribution function under Assumption \ref{asmp:iid_covariate}.
From \citet[][p. 304, Lemma 21.1]{van2000asymptotic} and by the definition that $\underline{\tau}^{\mathrm{co}}_\beta=\bar{F}^{-1}(\beta)$, 
we have $\bar F(\underline{\tau}^{\mathrm{co}}_\beta)=\beta$. 
By definition, this further implies that 
\begin{align*}
\E\big[m(\underline{\tau}^{\mathrm{co}}_\beta|X)\big]
=
\E\big[\inf_{b \in \mathbb{R}} \big\{ F_1(b\mid X)-
F_0(b-\underline{\tau}^{\mathrm{co}}_\beta\mid X)\big\}\big]
=
\bar F(\underline{\tau}^{\mathrm{co}}_\beta)-1
=
\beta-1.
\end{align*}

Second, we prove that 
$p_\beta
\equiv 
\Pr\{s_\beta(X)=+\infty\} < \beta.$
Because $m(\underline{\tau}^{\mathrm{co}}_\beta|x) = 0$ when $\{s_\beta(x)=+\infty\}$ holds,
we have 
\begin{align*}
    \beta - 1 & = \E\big[ m(\underline{\tau}^{\mathrm{co}}_\beta|X)\big]
    = \E\big[\I\{s_\beta(X)=+\infty\}m(\underline{\tau}^{\mathrm{co}}_\beta|X)\big]
    + 
    \E\big[\I\{s_\beta(X)<+\infty\}m(\underline{\tau}^{\mathrm{co}}_\beta|X)\big]\\
    & = 
    \E\big[\I\{s_\beta(X)<+\infty\}m(\underline{\tau}^{\mathrm{co}}_\beta|X)\big]
\end{align*}
Because $\beta-1<0$, $\Pr\{s_\beta(X)<+\infty\}=1-p_\beta$ must be positive. 
Moreover, note that, for almost every $x$, whenever
$s_\beta(x)<+\infty$, 
both 
$F_1\big(s_\beta(x)\mid x\big)$ and $F_0\big(s_\beta(x)-\underline{\tau}^{\mathrm{co}}_\beta\mid x\big)$ are strictly between 0 and 1, implying that $m(\underline{\tau}^{\mathrm{co}}_\beta|x)>-1$. 
Therefore, we must have 
\begin{align*}
    \beta - 1 & =
    \E\big[\I\{s_\beta(X)<+\infty\}m(\underline{\tau}^{\mathrm{co}}_\beta|X)\big]
    > - \E\big[\I\{s_\beta(X)<+\infty\} \big] = - (1-p_{\beta}),
\end{align*}
which implies that $p_{\beta}< \beta$.

Third, we prove that, for $z=0,1$, $G_z$ in \eqref{eq:Gz} is  continuous and strictly increasing on $\mathbb{R}$.  
By definition, for $z=0,1$ and $b\in \mathbb{R}$, 
\begin{align*}
G_z(b) & = \Pr\{Y_i(z)-s_\beta(X_i)\le b, s_\beta(X_i) = + \infty\} + \Pr\{Y_i(z)-s_\beta(X_i)\le b, s_\beta(X_i) < + \infty\}
\\
& =
p_\beta+
\mathbb{E}\left[
\mathbbm{1}\{s_\beta(X)<+\infty\}
F_z\big( b+s_\beta(X)\mid X \big)
\right]. 
\end{align*}
By the dominated convergence theorem and the continuity of $F_z(\cdot\mid x)$ for almost every $x$, we can show that
$G_z$ is continuous on $\mathbb{R}$. In addition, for any
$b_1<b_2$,
\begin{align*}
G_z(b_2)-G_z(b_1)=
\mathbb{E}\left[
1\{s_\beta(X)<+\infty\}
\left\{
F_z(b_2+s_\beta(X)\mid X)
-
F_z(b_1+s_\beta(X)\mid X)
\right\}
\right]>0.
\end{align*}
Thus, $G_z$ is strictly increasing on $\mathbb{R}$.

Fourth, from the discussion before and by the dominated convergence theorem, we have
\begin{align*}
\lim_{b\to-\infty}G_z(b)=p_\beta,
\qquad
\lim_{b\to+\infty}G_z(b)=1.
\end{align*}
In addition, $G_z^{-1}(u)=-\infty$ for $u\in [0,p_\beta]$ and $G_z^{-1}(u) \in\mathbb{R}$ for $u \in (p_\beta,1)$.

Fifth, we show that $\underline{\tau}^{\mathrm{co}}_\beta
=
\inf\{d\in\mathbb{R}:H_\beta(d)\ge\beta\}
\equiv
H_\beta^{-1}(\beta)$, where 
\begin{align}
H_\beta(d)
=
1+\inf_{b\in\mathbb{R}}
\{G_1(b)-G_0(b-d)\},
\qquad d\in\mathbb{R}.
\label{eq:H_beta_adjusted}
\end{align}
For every $b\in\mathbb{R}$, by definition and the law of iterated expectation, 
\begin{align*}
G_1(b)-G_0(b-\underline{\tau}^{\mathrm{co}}_\beta)
&=
\mathbb{E}\left[
F_1\big(b+s_\beta(X)\mid X\big)-
F_0\big(b+s_\beta(X)-\underline{\tau}^{\mathrm{co}}_\beta\mid X\big)\right]\\
&\ge
\mathbb{E}\big[m(\underline{\tau}^{\mathrm{co}}_\beta|X)\big]=
\beta-1.
\end{align*}
Moreover, the above equality holds at $b=0$. 
Thus, 
\begin{align}
H_\beta(\underline{\tau}^{\mathrm{co}}_\beta)
=
1+\inf_{b\in\mathbb{R}}\{G_1(b)-G_0(b-\underline{\tau}^{\mathrm{co}}_\beta)\}
=
1+G_1(0)-G_0(-\underline{\tau}^{\mathrm{co}}_\beta)
=
\beta.
\label{eq:H_at_q_beta}
\end{align}
For any $d<\underline{\tau}^{\mathrm{co}}_\beta$, 
since $G_0$ is strictly
increasing as shown earlier, 
we have 
\begin{align*}
H_\beta(d)
\le 1+G_1(0)-G_0(-d)
< 1+G_1(0)-G_0(-\underline{\tau}^{\mathrm{co}}_\beta)
= \beta.
\end{align*}
Combining this inequality with \eqref{eq:H_at_q_beta} gives
\begin{align*}
H_\beta^{-1}(\beta) \equiv \inf\{d\in\mathbb{R}:H_\beta(d)\ge\beta\}
= 
\underline{\tau}^{\mathrm{co}}_\beta. 
\end{align*}

Sixth, we define 
\begin{align*}
L_\beta
=
\sup_{u\in[0,\beta]}
\left\{
G_1^{-1}(u)
-
G_0^{-1}(1+u-\beta)
\right\},
\end{align*}
and show that it can be equivalently written as 
\begin{align}
L_\beta
=
\sup_{u\in(p_\beta,\beta)}
\left\{
G_1^{-1}(u)
-
G_0^{-1}(1+u-\beta)
\right\}.
\label{eq:L_beta_restricted}
\end{align}
As shown earlier, $G_z^{-1}(u)=-\infty$ for $u\in [0,p_\beta]$. 
Thus, by the convention that $(-\infty)-(-\infty)=-\infty$, 
$G_1^{-1}(u)
-
G_0^{-1}(1+u-\beta)$ is always $-\infty$ for $u\in [0,p_\beta]$. 
In addition, when $u=\beta$, 
$G_1^{-1}(u)
-
G_0^{-1}(1+u-\beta) = G_1^{-1}(\beta)
-
G_0^{-1}(1)$ is also $-\infty$. Thus, \eqref{eq:L_beta_restricted} must hold.

Finally, we prove that, for every $d\in\mathbb{R}$,
\begin{align*}
H_\beta(d)\ge\beta
\quad\Longleftrightarrow\quad
L_\beta\le d.
\end{align*}
By definition,
\begin{align*}
H_\beta(d)\ge\beta
&\Longleftrightarrow
G_1(b)-G_0(b-d)\ge\beta-1
\quad\text{for every }b\in\mathbb{R}\\
&\Longleftrightarrow
G_0(a)\le1-\beta+G_1(a+d)
\quad\text{for every }a\in\mathbb{R}\\
&\Longleftrightarrow
G_0(a)\le1-\beta+G_1(a+d)
\quad\text{for every }a\in\mathbb{R} \text{ such that } G_0(a) > 1-\beta+p_{\beta},
\end{align*}
where the second equivalence follows by setting $a=b-d$, and the last equivalence follows from the fact that $G_1(a+d) \ge p_{\beta}$ as shown earlier. 
Let $u = G_0(a) -1+\beta$. 
By the fact shown earlier that $G_1$ and $G_0$ are both continuous and strictly increasing, we have 
\begin{align*}
H_\beta(d)\ge\beta
&\Longleftrightarrow
G_0(a)\le1-\beta+G_1(a+d)
\quad\text{for every }a\in\mathbb{R} \text{ such that } G_0(a) > 1-\beta+p_{\beta}\\
&\Longleftrightarrow 
u + 1 - \beta \le 1 - \beta + G_1(G_0^{-1}(u+1-\beta) + d) \quad\text{for every } u \in (p_{\beta}, \beta)\\
&\Longleftrightarrow 
G_1^{-1} (u)  \le  G_0^{-1}(1+u-\beta) + d \quad\text{for every } u \in (p_{\beta}, \beta)\\
&\Longleftrightarrow  L_\beta \le d, 
\end{align*}
where the last equivalence follows from \eqref{eq:L_beta_restricted}.

From the above, we then have 
\begin{align*}
    \underline{\tau}^{\mathrm{co}}_\beta
& =
\inf\{d\in\mathbb{R}:H_\beta(d)\ge\beta\} = \inf\{d\in\mathbb{R}:L_\beta \le d\} = L_{\beta}
\\
& = \sup_{u\in[0,\beta]}
\left\{
G_1^{-1}(u)
-
G_0^{-1}(1+u-\beta)
\right\}
= \sup_{u\in[0,\beta]}
\left\{
F_{\underline{\tau}^{\mathrm{co}}_\beta,1}^{-1}(u)
-
F_{\underline{\tau}^{\mathrm{co}}_\beta,0}^{-1}
(1+u-\beta)
\right\},
\end{align*}
where the last equality follows from the definition in \eqref{eq:Gz}. 
Therefore, Lemma \ref{lem:bar_F_covariate_adjustment2} holds. 
\end{proof}

\begin{proof}[\bf Proof of Theorem \ref{thm:bar_F_covariate_adjustment}]
    Theorem \ref{thm:bar_F_covariate_adjustment}(i) follows from Lemmas \ref{lem:bar_F_CDF} and \ref{lem:bar_F_covariate_adjustment1}, 
    and Theorem \ref{thm:bar_F_covariate_adjustment}(ii) follows from Lemma \ref{lem:bar_F_covariate_adjustment2}.
\end{proof}

\begin{proof}[\bf Technical details for Remark \ref{rmk:adjustment_normal}]
Let $\Phi$ and $\phi$ be the distribution and density functions of the standard Gaussian distribution.
Because $Y(z)\mid X\sim \mathcal{N}(\mu_z(X),\sigma^2(X))$ for $z=0,1$, the optimal adjustment \eqref{eq:cov_adj} has the following form:
\begin{align*}
    \bar{s}(x;y) \in \argmin_{b \in \mathbb{R}\cup\{+\infty\}} g(b), \quad \text{where} \ \  g(b)=\Phi\Big(\frac{b-\mu_1(x)}{\sigma(x)}\Big)-\Phi\Big(\frac{b-y-\mu_0(x)}{\sigma(x)}\Big).
\end{align*}
We can verify that
\begin{align*}
    g'(b)=\frac{1}{\sigma(x)} \bigg\{\phi\Big(\frac{b-\mu_1(x)}{\sigma(x)}\Big)-\phi\Big(\frac{b-y-\mu_0(x)}{\sigma(x)}\Big)\bigg\}.
\end{align*}

Consider first the case where $-\mu_1(x) > -y-\mu_0(x)$ or equivalently $\mu_1(x)-\mu_0(x)< y$. 
In this case, $g(b) > 0$ for any $b \in \mathbb{R}$ since $\Phi$ is  strictly increasing. Thus, $g(b)$ is minimized at $b=+\infty$, and $\bar{s}(x;y)=+\infty$. 

Consider then the case where $\mu_1(x)-\mu_0(x) > y$. 
By the property of $\phi$, we can verify that $g'(b)<0$ if $b<(\mu_1(x)+\mu_0(x)+y)/2$, $g'(b)>0$ if $b>(\mu_1(x)+\mu_0(x)+y)/2$, and $g'(b)=0$ if $b=(\mu_1(x)+\mu_0(x)+y)/2$. Thus, $g(b)$ achieves its minimum at $b=(\mu_1(x)+\mu_0(x)+y)/2$, and $\bar{s}(x;y)=(\mu_1(x)+\mu_0(x)+y)/2$.  

Consider finally the case where $\mu_1(x)-\mu_0(x) = y$. In this case, $g(b)=0$ for all $b$, and we can choose $\bar{s}(x;y)=(\mu_1(x)+\mu_0(x)+y)/2$.  

From the above, up to the additive constant $y/2$, the optimal adjustment function equals $(\mu_1(x)+\mu_0(x))/2$ if $\mu_1(x)-\mu_0(x)\ge y$ and $+\infty$ otherwise. 
\end{proof}

\begin{proof}[\bf Proof of Theorem \ref{thm:sharp_bound_consist_covariate}]

Consider either of the two adjustment functions
\begin{align*}
s_a(x)=\bar s(x;a),
\qquad
a\in\{c,\underline{\tau}_{\beta}^{\co}\}.
\end{align*}
Let $F_{a,1}$ and $F_{a,0}$ denote the marginal distribution functions
of the corresponding adjusted potential outcomes 
$Y_i(1)-s_a(X_i)$ and $Y_i(0)-s_a(X_i),$ 
respectively. For any $d\in\mathbb{R}$, define 
\begin{align}
\Delta_{a}(d)
=
\sup_{b\in\mathbb{R}}
\big\{
F_{a,0}(b)
-
\min\left\{
1,1-\beta+F_{a,1}(b+d)
\right\}
\big\}.
\label{eq:population_gap_two_adjustments}
\end{align}
We will first show that
\begin{align*}
\Delta_a(c)>0
\qquad
\text{for both }a=c\text{ and }a=\underline{\tau}_{\beta}^{\co}.
\end{align*}

Consider first adjustment $a=c$. Because
$c<\underline{\tau}_{\beta}^{\co}=\bar F^{-1}(\beta)$ and $\bar F$ is a distribution
function, by Lemma \ref{lem:quntile_property}(ii), we have $\bar F(c)<\beta$.
By Theorem \ref{thm:bar_F_covariate_adjustment}(i),
\begin{align*}
\beta > \bar F(c)
&=
1+\inf_{b\in\mathbb{R}}
\left\{
F_{c,1}(b+c)-F_{c,0}(b)
\right\}.
\end{align*}
Therefore, there exists $b_c\in\mathbb{R}$ such that
\begin{align}
F_{c,0}(b_c)-F_{c,1}(b_c+c)>1-\beta.
\label{eq:strict_gap_adjustment_c}
\end{align}
Because $F_{c,0}(b_c)\leq 1$, \eqref{eq:strict_gap_adjustment_c}
implies $F_{c,1}(b_c+c)<\beta$.
Thus, 
\begin{align*}
\Delta_c(c)
&\geq 
F_{c,0}(b_c)
-
\min\left\{
1,1-\beta+F_{c,1}(b_c+c)
\right\} \\
&= F_{c,0}(b_c)-\left\{1-\beta+F_{c,1}(b_c+c)\right\}
>0.
\end{align*}

Next, consider adjustment $a=\underline{\tau}_{\beta}^{\co}$.
By Theorem \ref{thm:bar_F_covariate_adjustment}(ii),
\begin{align*}
\underline{\tau}_{\beta}^{\co}
=
\sup_{u\in[0,\beta]}
\big\{
F_{\underline{\tau}_{\beta}^{\co},1}^{-1}(u)
-
F_{\underline{\tau}_{\beta}^{\co},0}^{-1}(1+u-\beta)
\big\},
\end{align*}
with the convention $(-\infty)-(-\infty)=-\infty$.
Following the proof of Theorem \ref{thm:tau_nbeta}(i), we have
\begin{align}
\underline{\tau}_{\beta}^{\co}
=
\inf\{
d\in\mathbb{R}:
\Delta_{\underline{\tau}_{\beta}^{\co}}(d)=0
\}.
\label{eq:oracle_adjusted_cdf_gap}
\end{align}
We can verify that the function $\Delta_{\underline{\tau}_{\beta}^{\co}}(\cdot)$ is nonnegative and monotone
nonincreasing. 
Because $c<\underline{\tau}_{\beta}^{\co}$, \eqref{eq:oracle_adjusted_cdf_gap} implies
$\Delta_{\underline{\tau}_{\beta}^{\co}}(c)>0$.

We now prove the convergence of the p-value based on adjustment $s_a(x)$ with either $a=c$ or $a=\underline{\tau}_{\beta}^{\co}$.  
Let $\hat F_{n,a}^{\t}$ and
$\hat F_{n,a}^{\c}$ denote the empirical distribution functions of
the adjusted observed outcomes in the treated and control groups,
respectively, and define
\begin{align*}
\hat\Delta_{n,a}(c)
=
\sup_{b\in\mathbb{R}}
\bigg\{
\hat F_{n,a}^{\c}(b)
-
\min\Big\{
1,
1-\frac{\lceil n_1\beta\rceil}{n_1}
+\hat F_{n,a}^{\t}(b+c)
\Big\}
\bigg\}.
\end{align*}
Note that the adjusted outcomes may take the value $-\infty$, but
their empirical distribution functions remain well-defined on
$\overline{\mathbb{R}}$. The Glivenko--Cantelli theorem continues to
hold because $\overline{\mathbb{R}}$ can be mapped order-preservingly
into a bounded real interval. Therefore, under Assumption
\ref{asmp:iid_covariate} and the CRE with $n_1/n \to \lambda \in (0,1)$,
\begin{align} \label{eq:gc_adjusted}
\sup_{b\in\mathbb{R}}
\big|
\hat F_{n,a}^{\t}(b)-F_{a,1}(b)
\big|\convergeas 0,
 \qquad
\sup_{b\in\mathbb{R}}
\big|
\hat F_{n,a}^{\c}(b)-F_{a,0}(b)
\big|\convergeas 0.
\end{align}
Using Lemmas \ref{lem:absolute_min} and \ref{lem:absolute_sup} and by the same logic as the proof of Lemma \ref{lem:xin_converge}, 
\begin{align*}
\big|
\hat\Delta_{n,a}(c)-\Delta_a(c)
\big|
\leq
\sup_{b\in\mathbb{R}}
\left|
\hat F_{n,a}^{\c}(b)-F_{a,0}(b)
\right|
+
\sup_{b\in\mathbb{R}}
\left|
\hat F_{n,a}^{\t}(b)-F_{a,1}(b)
\right|
+
\frac{1}{n_1}.
\end{align*}
It follows from \eqref{eq:gc_adjusted} that
\begin{align}
\hat\Delta_{n,a}(c)
\convergeas
\Delta_a(c)>0. 
\label{eq:empirical_gap_positive_limit}
\end{align}
Because $n_1/n\to\lambda\in(0,1)$, we also have
${n_1n_0}/{n} \to \infty$.
Combining this with \eqref{eq:empirical_gap_positive_limit} gives
\begin{align}
\sqrt{\frac{n_1n_0}{n}} \,
\hat\Delta_{n,a}(c)
\convergeas
+\infty. 
\label{eq:scaled_gap_diverges}
\end{align}
By the same logic as the proof of Theorem \ref{thm:sharp_bound_consist_fin}, we can then derive that 
$\overline p_{\lceil n_1\beta\rceil,c}^{\t,\co}=G_n\big(\hat\Delta_{n,a}(c)\big)
\convergeas 0$.

From the above, Theorem \ref{thm:sharp_bound_consist_covariate} holds. 
\end{proof}

\section{Proofs of Theorems \ref{thm:ci_adj_ITE} and \ref{thm:ci_adj_quanitle}}\label{ssec:proof_covariate_interval}

\begin{proof}[\bf Proof of Theorem \ref{thm:ci_adj_ITE}]

Fix $\beta\in(0,1)$ and let $k_n=\lceil n_1\beta\rceil$.
The monotonicity of the p-value $p_{k_n,c}^{\t,\co}$ follows immediately from the effect-increasing property of the RKS statistic in \eqref{prop:rks_properties} and the fact that $G_n(\cdot)$ in \eqref{eq:rand_p_val} is nonincreasing. 

We then prove the validity of the prediction interval $\mathcal I_{k_n,\alpha}^{\t,\co}$. 
By definition, 
\begin{align*}
\tau_{\t(k_n)} \notin\mathcal I_{k_n,\alpha}^{\t,\co}
\quad\Longleftrightarrow\quad
p_{k_n,\tau_{\t(k_n)}}^{\t,\co}\leq\alpha.
\end{align*}
From Remark \ref{rmk:adjustment_infty}, 
we know that  
$
p_{k_n,\tau_{\t(k_n)}}^{\t,\co}
\geq
G_n( t_{\mathrm{RKS}}
(
\boldsymbol Z,
\tilde{\boldsymbol Y}(0)
) ),
$
which, by the validity of the usual randomization test, is stochastically larger than or equal to the uniform distribution on $(0,1)$. 
This then implies the validity of the prediction interval $\mathcal I_{k_n,\alpha}^{\t,\co}$.

We finally prove asymptotic sharpness under Assumption
\ref{asmp:iid_covariate} and the oracle adjustment
$
s(x)
=
\bar s(x;\underline{\tau}_\beta^{\co}).
$
From Remark \ref{rmk:pre_int_cn}, 
$$
p_{\lceil n_1 \beta \rceil,c}^{\t,\co} \le \alpha
\Longleftrightarrow
\mathcal{I}^{\t,\co}_{\lceil n_1 \beta \rceil, \alpha} \subseteq (c, \infty)
\Longrightarrow 
c \le  \inf  \mathcal{I}_{\lceil n_1 \beta \rceil,\alpha}^{\t, \co}. 
$$
The asymptotic sharpness then follows directly from 
Theorem \ref{thm:sharp_bound_consist_covariate}, since  $\overline{p}_{\lceil n_1 \beta \rceil,c}^{\t, \co} \to 0$ almost surely for $c<\underline{\tau}_{\beta}^{\textup{co}}$.

From the above, Theorem \ref{thm:ci_adj_ITE} holds. 
\end{proof}

\begin{proof}[\bf Proof of Theorem \ref{thm:ci_adj_quanitle}]

Fix $c\in\mathbb{R}$. 
The monotonicity of the p-value $p_{\lceil n_1\beta\rceil,c}^{\t,\co}$ follows immediately from the effect-increasing property of the RKS statistic in \eqref{prop:rks_properties} and the fact that $G_n(\cdot)$ in \eqref{eq:rand_p_val} is nonincreasing.

We then prove the validity of the prediction interval $\mathcal J_{c,\alpha}^{\t,\co}$. 
By definition, 
\begin{align*}
F_{n\tau}^{\t}(c) \notin  \mathcal J_{c,\alpha}^{\t,\co}
\quad\Longleftrightarrow\quad
p_{\lceil n_1 F_{n\tau}^{\t}(c)\rceil,c}^{\t,\co}\leq\alpha.
\end{align*}
Let
$
S_{n\tau}^{\t}(c)
=
\sum_{i=1}^n
Z_i\mathbf 1(\tau_i\leq c).
$
We can verify that 
$
\left\lceil
n_1F_{n\tau}^{\t}(c)
\right\rceil
=
S_{n\tau}^{\t}(c).
$
By definition of $S_{n\tau}^{\t}(c)$, the true treatment-effect vector satisfies
$
\boldsymbol\tau
\in
\mathcal H_{S_{n\tau}^{\t}(c),c}^{\t}.
$
From Remark \ref{rmk:adjustment_infty}, 
we know that  
$
p_{S_{n\tau}^{\t}(c),c}^{\t,\co}
\geq
G_n\big( t_{\mathrm{RKS}}
\bigl(
\boldsymbol Z,
\tilde{\boldsymbol Y}(0)
\bigr) \big).
$
By the validity of the usual randomization test, we can then derive the validity of the prediction interval $\mathcal J_{c,\alpha}^{\t,\co}$.

We finally prove the asymptotic sharpness under Assumption
\ref{asmp:iid_covariate} 
and the adjustment
$
s(x)=\bar s(x;c).
$
By the monotonicity of the p-value, 
$$
p_{\lceil n_1 \beta \rceil,c}^{\t, \co} \le \alpha
\Longleftrightarrow
\mathcal J_{c,\alpha}^{\t,\co} \subseteq [0, \beta)
\Longrightarrow 
\beta \ge \sup  \mathcal{J}_{c,\alpha}^{\t,\co}. 
$$
The asymptotic sharpness then follows directly from 
Theorem \ref{thm:sharp_bound_consist_covariate}.

From the above, Theorem \ref{thm:ci_adj_quanitle} holds. 
\end{proof}

\end{document}